\documentclass{elsarticle}

\pdfoutput=1
\usepackage{amsmath, array, amsthm, amssymb, multirow, boldline, hyperref, setspace, titlesec, algorithm2e, color}
\usepackage[margin = 1in]{geometry}
\usepackage{graphics, graphicx}
\graphicspath{{Images/}}

\makeatletter
\def\ps@pprintTitle{%
 \let\@oddhead\@empty
 \let\@evenhead\@empty
 \def\@oddfoot{}%
 \let\@evenfoot\@oddfoot}
\makeatother

\DeclareMathOperator*{\argmin}{arg\,min}

\titleformat{\subsection}
{\bfseries\slshape}
{\thesubsection}
{0.5em}
{}

\begin{document}

	\begin{frontmatter}

\title{Geometric Sensitivity Measures for Bayesian Nonparametric Density Estimation Models}

\author[add1]{Abhijoy Saha}
\ead{saha.58@osu.edu}
\author[add1]{Sebastian Kurtek}
\ead{kurtek.1@stat.osu.edu}
\address[add1]{Department of Statistics, The Ohio State University}

	\begin{abstract}
	
We propose a geometric framework to assess global sensitivity in Bayesian nonparametric models for density estimation. We study sensitivity of nonparametric Bayesian models for density estimation, based on Dirichlet-type priors, to perturbations of either the precision parameter or the base probability measure. To quantify the different effects of the perturbations of the parameters and hyperparameters in these models on the posterior, we define three geometrically-motivated global sensitivity measures based on geodesic paths and distances computed under the nonparametric Fisher-Rao Riemannian metric on the space of densities, applied to posterior samples of densities: (1) the Fisher-Rao distance between density averages of posterior samples, (2) the log-ratio of Karcher variances of posterior samples, and (3) the norm of the difference of scaled cumulative eigenvalues of empirical covariance operators obtained from posterior samples. We validate our approach using multiple simulation studies, and consider the problem of sensitivity analysis for Bayesian density estimation models in the context of three real datasets that have previously been studied.

	\end{abstract}


	\end{frontmatter}

\section{Introduction}
\label{sec:Intro}

There has been an increased interest in applying rich classes of Bayesian nonparametric and semiparametric methods to complex applied problems due to recent advances in efficient computational techniques \cite{dey1998, hjort2010}. Parametric models, based on finite-dimensional parameter sets, in certain cases restrict the scope of inference. In contrast, Bayesian nonparametric models allow us to do inference on a bigger class of infinite-dimensional parameter spaces, e.g., function spaces \cite{muller2013, muller2015}, and are more robust to model misspecifications than their parametric counterparts. In this paper, we consider the problem of assessing global sensitivity to the choices of prior parameter specifications in the context of nonparametric density estimation from the Bayesian viewpoint. In the next two sections, we review the literature on (1) Bayesian nonparametric density estimation methods, and (2) sensitivity analysis of Bayesian models (parametric and nonparametric).

\subsection{Bayesian Nonparametric Density Estimation}

The standard density estimation problem typically starts with a random sample that is presumed to have been generated from some unknown distribution. The general setting for Bayesian inference in this context then requires a probability model for this unknown distribution. Under a nonparametric framework, a prior probability model is specified in an appropriate infinite-dimensional function space. An important nonparametric prior considered in many works is the Dirichlet process introduced by Ferguson \cite{ferguson1973}. However, samples drawn from the Dirichlet process are discrete in nature, and are hence unsuitable for density estimation. To address this, the popular Dirichlet process mixture model \cite{escobar1995, maceachern1994} was introduced, which is the preferred method for continuous density estimation in Bayesian nonparametric literature. Theoretical results for the Dirichlet process mixture model in the context of density estimation were developed by Ghosal and van der Vaart \cite{ghosal2001}, Ghosh and Ramamoorthi \cite{ghosh2003}, and Walker \cite{walker2004}. The limiting behavior of this model was studied by Bush et al. \cite{bush2010}. Additionally, Lee et al. \cite{lee2014} extended this work and studied the interpretability of the parameters of this model.

Considerable improvements have been made to the conventional Dirichlet process mixture model in the context of density estimation. In particular, Griffin \cite{griffin2010} considered a univariate setup using the infinite mixture of normals model, and suggested an alternative hierarchical structure. He showed that the proposed modification, based on partitioning of the total variance, improves density estimation. Further, based on the work by Yang and Marron \cite{yang1999}, Bean et al. \cite{bean2016} proposed a transformation for estimating heavy-tailed and skewed densities. An outline of other methods for Bayesian density estimation can be found in Dey et al. \cite{dey1998}, Walker et al. \cite{walker1999}, and M{\"u}ller and Quintana \cite{muller2004}. Recently, MacEachern \cite{maceachern2016} provided a brief overview of nonparametric Bayesian methods in general.

To perform inference in the Bayesian density estimation problem, we often approximate the posterior distribution, or some functional thereof, via Markov chain Monte Carlo (MCMC) sampling techniques. The rapid development of such Bayesian models is a direct result of significant advances in computational methods, which make extensive Monte Carlo simulation studies possible. Many authors focused on developing efficient MCMC algorithms for posterior sampling from the Dirichlet process mixture model \cite{escobar1994,maceachern1998_2,dey1998,neal2000}. The underlying goal of these algorithms is to define a Markov chain with the target posterior distribution of interest as the stationary distribution, and then to generate samples from this chain. Most of these works are based on the Gibbs sampling technique or some modified version of it, e.g., collapsed Gibbs sampler. Ishwaran and James \cite{ishwaran2001} provide two Gibbs sampling methods for fitting Bayesian nonparametric models in a more general setting. Other methods for fitting models with nonparametric priors include sequential importance sampling \cite{maceachern1999}, predictive recursions \cite{newton1999}, and particle filtering \cite{fearnhead2004}.

\subsection{Sensitivity Analysis for Bayesian Models}

Assessment of sensitivity to model assumptions is an important part of statistical inference. Since our interest primarily lies in the posterior induced by the prior on the space of probability measures, it is important to study the extent to which inferences based on the posterior distribution are sensitive to the key components of Bayesian analysis: the prior and sampling distributions. To date, multiple studies have been considered to evaluate the influence of these components; see Kass et al. \cite{kass1989}, Gustafson and Wasserman \cite{gustafson1995}, Ruggeri and Sivaganesan \cite{ruggeri2000}, Oakley and O'Hagan \cite{oakley2004}, and Millar and Stewart \cite{millar2007}. As shown by Berger \cite{berger1990, berger1994}, assessment of sensitivity to the specification of prior distributions over the parameter space is a crucial part of the inferential process in the Bayesian setting. Specifically, a Bayesian model is non-robust and sensitive with respect to the prior distribution if its posterior distribution varies significantly when the parameters of the prior are perturbed slightly \cite{roos2015}. Gustafson \cite{gustafson1996} investigated the sensitivity of posterior expectations to perturbations of prior marginals. Ruggeri \cite{ruggeri2008} suggested using a robustness measure to check the influence of perturbations of the prior distribution, chosen from a class of suitable priors. For a detailed account of sensitivity analysis in Bayesian models, and the impact of prior distributions on posterior inference, see Insua and Ruggeri \cite{insua2000}.

Primarily, there exist two main ways to assess Bayesian robustness and influence of prior specification on the model output: global and local sensitivity analysis. In global sensitivity analysis \cite{berger1982}, one is interested in deriving variational measures emerging from a large class of prior perturbations. Alternatively, local sensitivity analysis \cite{gustafson2000} uses differential calculus to determine the rate of influence on the posterior with respect to small perturbations of the prior. Sivaganesan \cite{sivaganesan2000} provides a comparison of the global and local approaches to sensitivity analysis. Berger \cite{berger1982} also proposed the idea of using global methods to assess posterior uncertainty. As a measure of robustness, the intuitive nature of the global approach also makes it interpretable and very appealing.

Infinitesimal perturbations of a statistical model are closely linked to the geometry of the space of probability density functions (PDFs). Zhu et al. \cite{zhu2011} defined a geometric perturbation model in the Bayesian setting, which was used to analyze various perturbation schemes. Recently, Kurtek and Bharath \cite{kurtek2015} proposed a general geometric framework to assess local and global sensitivity of Bayesian procedures under a parametric setup. In the nonparametric setting, Nieto--Barajas and Pr{\"u}nster \cite{nieto2009} perform a sensitivity analysis for density estimators by perturbing priors based on a new prior process, referred to as normalized random measures with independent increments. However, their approach to measure sensitivity of a perturbed model did not have any geometrical considerations.

\subsection{Our Contributions and Paper Organization}

Bayesian nonparametric density estimation models are based on probability measures over the space of absolutely continuous cumulative distribution functions (CDFs) or the space of PDFs. Under the duality of CDFs and PDFs, the set of nonparametric densities provides a natural setting for the study of sensitivity to prior specification. Hence, a geometric structure on this space of densities is an appropriate setting to study the problem in its natural habitat. The main contribution of this paper is to unify model perturbation and global sensitivity assessment for Bayesian nonparametric models under a computationally tractable Riemannian geometric framework. As in \cite{kurtek2015}, our framework for assessing sensitivity is based on the nonparametric Fisher--Rao Riemannian geometry of the space of PDFs. Noting that the output of MCMC techniques in this problem is a random sample of PDFs, we define three distinct global sensitivity measures, based on summaries of the posterior distribution, which can be used to evaluate deviations of posterior density samples arising from various perturbations of the baseline model. Based on these geometric sensitivity measures, we quantify the impact of the prior distribution on density estimation in various Bayesian nonparametric models based on Dirichlet-type priors. Specifically, we consider the following nonparametric priors: (1) Dirichlet process, (2) Dirichlet process mixture of Gaussians, (3) common component variance model, and (4) different component variance model. The last two models are different parameterizations of the hierarchical Dirichlet process mixture model as proposed by Griffin \cite{griffin2010}. All three of the measures have natural geometric calibration, and we provide easy to implement tools for their application in real PDF estimation problems. We additionally believe that the three measures have intuitive meaning in terms of the generated posterior sample densities.

Section \ref{sec:Back} provides a brief review of the nonparametric Bayesian density estimation models used in this work, and a description of the Riemannian geometric tools on which our measures are based. In Section \ref{sec:GeomSensAn}, we define the three geometric global sensitivity measures. Section \ref{sec:Ex} presents the results of simulation studies along with an analysis of three classical datasets using the proposed method. Finally, Section \ref{sec:Discuss} contains a brief discussion and directions for future work.

	\section{Modeling and Geometry Background}
	\label{sec:Back}
	
We begin by providing relevant background details on which the proposed framework for global sensitivity analysis is built. In Section \ref{subsec:NPBDensity}, we formalize the problem of density estimation and provide a concise description of the Bayesian nonparametric models under consideration. We then introduce a useful representation space of PDFs in Section \ref{subsec:Geom} and define the necessary geometric tools on this space that are used to define the geometric sensitivity measures.
	
	\subsection{Nonparametric Bayesian Density Estimation}
	\label{subsec:NPBDensity}
Density estimation is a fundamental statistical inference problem. From a nonparametric viewpoint, we are primarily concerned with estimating an unknown distribution on the basis of independent and identically distributed (i.i.d.) observations, with as few assumptions as possible. Let $x_1, \dots, x_n$ denote a sample of size $n$ from an unknown distribution $G \in \mathcal{G}$, where $\mathcal{G}$ denotes the set of all possible distributions. To completely specify the Bayesian model, we need to place a prior $\pi$ on the distribution $G$. This is in contrast to the parametric density estimation problem where we specify a parametric model for the i.i.d. observations, and then place a prior on the parameters of this model. In particular, we consider a random probability measure on $\mathcal{G}$, i.e., a Bayesian nonparametric prior on an infinite-dimensional parameter space \cite{muller2015}. The model can be written as:
\begin{align*}
x_i \mid G &\stackrel{iid}{\sim} G, \quad i = 1, \dots, n,\\
G &\sim \pi.
\end{align*}
The marginal distribution $m$ induced by the model and the prior is given by:
\begin{equation*}
m(A) = \int \mathbb{P}_G(A) d\pi(G),
\end{equation*}
where
\begin{equation*}
\mathbb{P}_G(A) = \int I_A(x_1, \dots, x_n) dG(x_1) \dots dG(x_n).
\end{equation*}
Our final goal is to calculate the posterior distribution $p$:
\begin{equation*}
p(A) \equiv p(G \in A \mid x_1, \dots, x_n).
\end{equation*}
If there exists a dominating measure for $\mathcal{G}$, then the posterior can be computed using Bayes' theorem. However, if there is no dominating measure, the posterior is still well-defined, but cannot be attained using Bayes' theorem directly. In contrast to parametric models, the formula for the nonparametric prior $\pi$ is not available explicitly. To this end, there exist various sampling algorithms for generating samples from the prior and the posterior as discussed in Section \ref{sec:Intro}.

We consider four different density estimation models based on popular Dirichlet-type priors: (1) Dirichlet process, (2) Dirichlet process mixture of Gaussians, (3) common (or constant) component variance model, and (4) different component variance model. The last two models, introduced by Griffin \cite{griffin2010}, are an alternative hierarchical model to the Dirichlet process mixture of Gaussians, and contain hyperparameters that can be interpreted as the location, scale and smoothness of the unknown density. Since all of these models are well-studied in the literature, we do not discuss their properties in great detail. Instead, we specify each model completely in the next section, and present a summary of their important aspects. We also briefly comment on the sampling technique that we use in each of these models to draw density samples from the posterior.
	
	\subsubsection{Dirichlet Process}
	\label{subsubsec:DP}
	
A Dirichlet process (DP) prior for $G$ is a prior on the space of all probability measures and was introduced by Ferguson \cite{ferguson1973} in 1973. Since each draw from a DP is itself a distribution, it is often viewed as a distribution over distributions. This was the first Bayesian nonparametric model to be studied extensively, and has several representation schemes, which make it useful for different model building approaches \cite{blackwell1973}. The DP is determined by two parameters: (1) a positive scalar concentration/precision parameter $\alpha$, and (2) a probability measure $G_0$ that defines the location of the DP around which distributions are drawn. The DP has an appealing property, which makes it a convenient choice as the prior in Bayesian nonparametric density estimation problems: with a DP as the prior $\pi$, the posterior distribution for $G$ is a DP as well. In other words, the DP is conjugate with respect to i.i.d. sampling and it results in a straightforward closed-form for the posterior in the same family:
\begin{align}
\label{eq:DPModel}
	x_i \mid G &\stackrel{iid}{\sim} G, \quad i = 1, \dots, n \nonumber \\
	G &\sim DP(\alpha, G_0), \nonumber \\
	G \mid x_1, \dots, x_n &\sim DP\Big(\alpha + n, \dfrac{\alpha}{\alpha + n} G_0 + \dfrac{n}{\alpha + n} F_n \Big),
\end{align}
where $F_n$ is the empirical distribution function.

To draw density samples from the posterior, we use the stick-breaking process representation introduced by Sethuraman \cite{sethuraman1994}. Using this approach, the distribution is composed of two independent sequences of draws: (1) locations of atoms of mass, and (2) the amount of mass associated with each of these locations. By construction, each draw from the posterior produces discrete distributions with probability one. Since our global sensitivity measures are based on the assumption that the posterior samples are PDFs (as we will see in Section \ref{subsec:Geom}), we convolve the weights obtained from the stick-breaking process with a Gaussian kernel; this generates continuous posterior density samples, which are in general more useful for density estimation problems.
	
	\subsubsection{Dirichlet Process Gaussian Mixture Model}
	\label{subsubsec:DPGMM}

The limitation of DP is that it generates discrete distributions, which are not preferable for continuous density estimation problems. Independent draws from a DP with a continuous probability measure $G_0$, result in distributions that are singular with respect to each other. This problem can be fixed by a slight modification: instead of using the DP directly as a prior, one can use it as a mixing measure in a simple parametric mixture model \cite{ferguson1983, escobar1995}. The technique of modeling a complex distribution as a mixture of simpler parametric distributions is a convenient way of identifying latent classes that account for dependencies between variables \cite{neal2000}. For the purpose of our study, we concentrate on the parametric Gaussian mixture model formulation, wherein each observation is normally distributed with its own set of parameters. This is also referred to as the Dirichlet process Gaussian mixture model (DPGMM). In this section, index $i$ is used to denote observations $i = 1, \dots, n$, and index $j$ is used to denote components of the mixture model $j = 1, \dots, J$.

The PDF of an observation $x$, corresponding to a finite Gaussian mixture model with $J$ components can be written as:
\begin{equation}
\label{eq:GaussMixture}
	f(x \mid \theta_1, \dots, \theta_J) = \sum_{j = 1}^{J} w_j \mathcal{N}(x \mid \mu_j, R_j^{-1}),
\end{equation}
where $\theta_j = \{ \mu_j, R_j, w_j\}$ is the set of parameters for component $j \in \{1, \dots J\}$. Here, $w_j$ denotes the mixing proportion, and $\mu_j$ and $R_j$ denote the mean vector and precision matrix for component $j$, respectively. Note that $w_j>0$ for all $j$ and $\sum_{j = 1}^{J} w_j=1$.

Let $x_1, \dots, x_n$ denote a set of $n$ independent observations. To obtain the DPGMM as an infinite limit of the finite mixture model in (\ref{eq:GaussMixture}), we need to introduce arbitrary indicator variables that encode the mixture component to which a particular observation $x_i$ belongs \cite{gorur2010}. Specifically, each $x_i$ is assumed to belong to a cluster indexed by $j$. Thus, instead of estimating parameters for each observation separately, we estimate parameters for each of the $J$ clusters parameterized by $\theta_j$. We define a joint prior distribution $G$ on the model parameters $\mu_j, R_j$, which is drawn from a DP, and place a symmetric Dirichlet distribution with parameter $\frac{\alpha}{J}$ on the mixing proportions $w_1, \dots, w_J$. Then, taking the infinite limit of the number of mixture components, i.e., $J \rightarrow \infty$, the DPGMM model can be expressed as:
\begin{align}
\label{eq:DPGMModel}
	x_i \mid \mu_{i}, R_{i} & \stackrel{iid}{\sim} \mathcal{N}(x_i \mid \mu_{i}, R_{i}^{-1}), \quad i = 1, \dots, n \nonumber \\
	(\mu_i, R_i) \mid G & \sim G, \nonumber \\
	G & \sim DP(\alpha, G_0).
\end{align}
To completely define the model, we need to specify the probability measure $G_0(\mu, R)$. This relates to the distribution of the prior on the parameters of the components in the infinite mixture model, i.e., the mean vector and the precision matrix. The prior distribution of the mean vector $\mu$ is Gaussian:
\begin{equation}
\label{eq:DPGMMPriorMean}
	\mu \mid \boldsymbol{m}, r, R \sim \mathcal{N}(\boldsymbol{m}, (rR)^{-1}),
\end{equation}
and the prior distribution of the precision matrix $R$ is Wishart:
\begin{equation}
\label{eq:DPGMMPriorPrecision}
	R \mid \nu, S \sim Wishart( \nu, (\nu S)^{-1}).
\end{equation}
Note that the dependence between the precision of the data and the prior distribution of the mean is needed to establish conjugacy. This model specification is known as conjugate DPGMM, where the mixture components share a common set of hyperparameters $\boldsymbol{m}, r, \nu, S$.

In contrast to DP, posterior samples generated from the DPGMM are continuous, smooth PDFs. To draw posterior density samples from the DPGMM, MCMC sampling techniques are normally used. Neal \cite{neal2000} provides a detailed discussion of such sampling methods. In our implementation, we specifically work with the Chinese restaurant process (CRP) representation of the model \cite{pitman2002}, and build a collapsed Gibbs sampler. Using a CRP to approximate the posterior over all of the parameters has two advantages \cite{gershman2012}: (1) we can examine how the data is grouped and the number of components it is grouped into, and (2) we can use the resulting posterior predictive distribution from the CRP representation to form additional components based on new data.

\subsubsection{Common Component Variances Model}
\label{subsubsec:CCV}

Griffin \cite{griffin2010} proposed the common component variances (CCV) model as an alternative parameterization of the univariate Gaussian mixture model, wherein a different prior distribution of the mean and variance components is considered. As shown in \cite{griffin2010}, this model admits a hierarchical structure, which has good predictive performance for Bayesian density estimation problems. In the CCV model, we assume that all individual component variances are equal:
\begin{align}
\label{eq:CCVModel}
x_i \mid \mu_i & \stackrel{iid}{\sim} \mathcal{N}(x_i \mid \mu_{i}, a \sigma^2), \quad i = 1, \dots, n \nonumber \\
\mu_i \mid G & \sim G, \nonumber \\
G & \sim DP(\alpha, G_0),
\end{align}
where $ 0 < a < 1$. The prior distribution $G_0(\mu)$ of the component means is Gaussian:
\begin{equation}
	\mu \mid \mu_0, a, \sigma^2 \sim \mathcal{N}(\mu \mid \mu_0, (1-a) \sigma^2),
\end{equation}
where the hyperparameters are distributed as:
\begin{equation*}
	\mu_0 \sim \mathcal{N}(\mu_{00}, \lambda_0^{-1}),\quad \sigma^{-2} \sim  Ga(s_0, s_1), \quad a \sim Be(a_0, a_1).
\end{equation*}
Parameters $\mu_0$ and $\sigma^2$ can be interpreted as the location and scale of the marginal distribution, respectively. The parameter $a$ can be interpreted as a smoothness parameter that reflects confidence in the smoothness of the unknown density. When $a$ is close to one, all of the individual component means $\mu_i$ will be close to $\mu_0$. On the other hand, if $a$ is close to zero, the component means will be normally distributed with an approximate variance of $\sigma^2$, and the component variances will be very small. In most applications, the precision parameter $\alpha$ of the Dirichlet process is unknown. There have been several suggestions for choosing an appropriate prior for $\alpha$ \cite{escobar1995, walker1997}. To implement this model, we chose the prior suggested by Griffin and Steel \cite{griffin2004}:
\begin{equation}
	\pi(\alpha) = \gamma^\eta \dfrac{\Gamma(2\eta)}{(\Gamma(\eta))^2} \dfrac{\alpha^{\eta - 1}}{(\alpha + \gamma)^{2\eta}}.
\end{equation}
They suggest interpreting the hyperparameters $\gamma$ as prior sample size, and $\eta$ as a variance parameter (larger values of $\eta$ lead to higher concentration of the prior).

Posterior samples from this model are generated using a standard Gibbs sampler for the usual conjugate Dirichlet process mixture model \cite{maceachern1998}. As discussed in Section \ref{subsubsec:DPGMM}, here we also introduce indicator variables to implement the MCMC sampler. For a complete description of the full conditional distributions for each parameter, see \cite{griffin2010}.

	\subsubsection{Different Component Variances Model}
	\label{subsubsec:DCV}

The CCV model fails to capture distributions which have multiple modes with different variance structure for each component. To overcome this limitation, Griffin \cite{griffin2010} introduced the different component variances (DCV) model. The DCV model is defined as:
\begin{align}
\label{eq:DCVModel}
x_i \mid \mu_i, \zeta_i & \stackrel{iid}{\sim} \mathcal{N}(x_i \mid \mu_{i}, a (\phi - 1) \zeta_i \sigma^2), \quad i = 1, \dots, n \nonumber \\
(\mu_i, \zeta_i) \mid G & \sim G, \nonumber \\
G & \sim DP(\alpha, G_0),
\end{align}
where $ 0 < a < 1$. The prior distribution $G_0(\mu, \zeta)$ is chosen to be Gaussian inverse gamma, which is the conditionally conjugate form. Specifically,
\begin{equation}
\mu \mid \mu_0, a, \sigma^2 \sim \mathcal{N}(\mu \mid \mu_0, (1-a) \sigma^2),
\end{equation}
and
\begin{equation}
\zeta^{-1} \mid \phi \sim Ga(\zeta^{-1} \mid \phi, 1).
\end{equation}
The hyperparameters $\mu$, $\sigma^2$ and $a$ have the same distribution as in the CCV model. Note that the shape parameter $\phi$ should be greater than one. If $\phi$ is large, the distribution will be close to normal. The prior specification for $\alpha$ and the interpretation of other parameters remain the same as in the previously described CCV model.

Since this is a nonconjugate Dirichlet process mixture model, we use a well-known algorithm proposed by Neal \cite{neal2000} to generate posterior samples. A full description of the algorithm and the implemented Gibbs sampler is provided in \cite{griffin2010}.

	\subsection{Geometry of PDF Space}
	\label{subsec:Geom}

The proposed framework for sensitivity analysis utilizes the geometry of the space of PDFs. In the following, we provide explicit analytical expressions for different geometric quantities of interest; most of these tools have been previously defined in \cite{kurtek2015}.

\subsubsection{Fisher--Rao Riemannian Geometry of the PDF Space}
	\label{subsubsec:FRMetric}
	
Without loss of generality, we restrict our attention to univariate densities on $[0,1]$; the framework is equally valid for all finite dimensional distributions. We define $\mathcal{P} = \{p:[0, 1] \rightarrow \mathbb{R}_{\geq0} \mid \int_{0}^{1} p(x) dx = 1\}$ as the space of all PDFs. For a PDF $p$ on the interior of $\mathcal{P}$, we also define the tangent space $T_{p}(\mathcal{P}) = \{\delta p: [0, 1] \rightarrow \mathbb{R} \mid \int_{0}^{1} \delta p(x) p(x) dx = 0 \}$. Intuitively, the tangent space $T_{p}(\mathcal{P})$ at any point $p$ contains all possible perturbations of the PDF $p$. Under this setup, for any $p \in \mathcal{P}$ and any two tangent vectors $\delta p_1, \delta p_2 \in  T_{p}(\mathcal{P})$, the nonparametric Fisher--Rao Riemannian metric (simply referred to as FR metric hereafter) is given by the inner product \cite{rao1945, kass1997}:
\begin{equation}
\label{eq:fr}
	\left\langle\langle \delta p_1, \delta p_2 \right\rangle\rangle_{p} = \int_{0}^{1} \delta p_1(x) \delta p_2(x) \dfrac{1}{p(x)} dx.
\end{equation}
In 1982, \v Cencov \cite{cencov1982} showed an important property of this metric: it is invariant to re-parameterizations, making it attractive for use in various statistical tasks.

However, we observe a drawback of the FR metric: it depends on $p \in \mathcal{P}$, and hence changes from point to point on the space of PDFs. This leads to cumbersome computations, requiring numerical methods to approximate various geometric quantities of interest in practice (e.g., geodesic distances). Bhattacharya \cite{bhattacharyya1943} proposed a convenient square-root transformation, which provides a drastic simplification of the Riemannian geometry of the representation space of PDFs. Define a continuous mapping $\phi: \mathcal{P} \rightarrow \Psi$, where $\phi(p) = \psi = +\sqrt{p}$ is the square-root density (SRD) of a PDF $p$. Then, the inverse mapping is given by $\phi^{-1}(\psi) = p = \psi^2$. The corresponding space of SRDs is $\Psi = \{ \psi: [0, 1] \rightarrow \mathbb{R}_{\geq0} \mid \int_{0}^{1} \psi^2(x) dx = 1 \}$, i.e., the positive orthant of the unit Hilbert sphere \cite{lang2012}.

\begin{figure}[!t]
	\begin{center}
			\begin{tabular}{|c|c|}
				\hline
				(a) & (b)\\
                \hline
				\includegraphics[width = 3.75in]{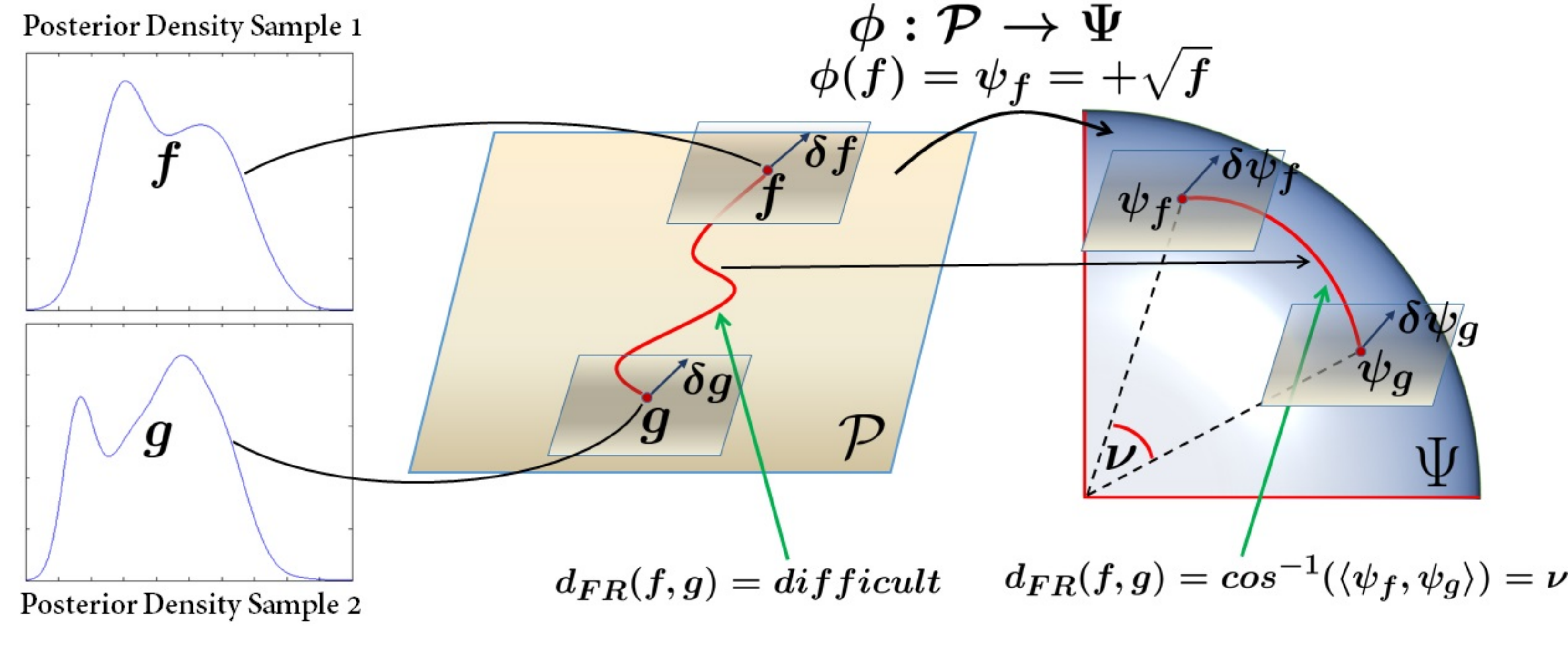}&\includegraphics[width = 2.25in]{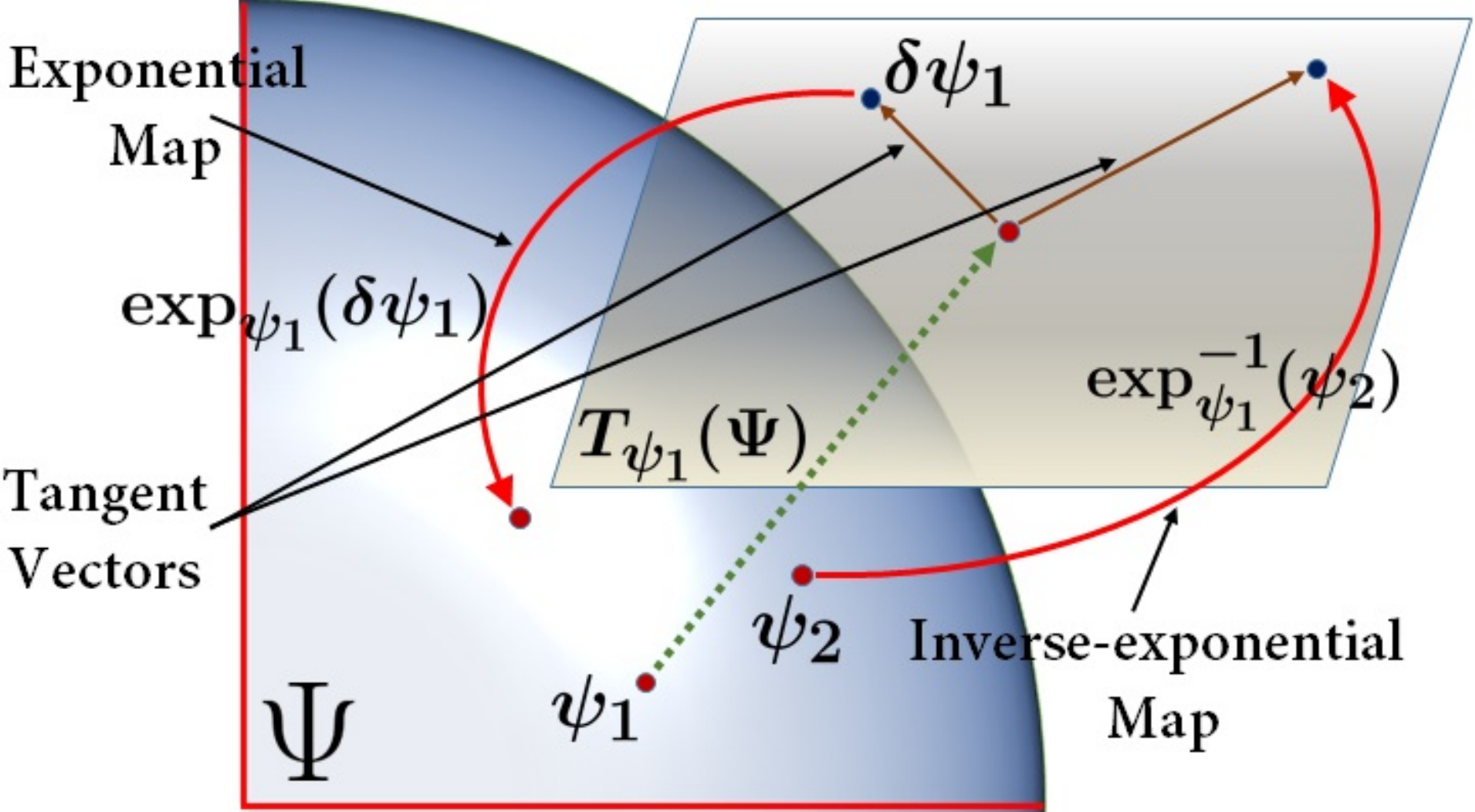}\\
                \hline
			\end{tabular}
	\end{center}
	\caption{(a) Square-root transform from $\mathcal{P}$ to the positive orthant of the unit Hilbert sphere $\Psi$, where $f, g$ represent two posterior density samples and $\upsilon$ is the FR geodesic distance between them. (b) Description of the exponential and inverse-exponential maps, where $\psi_1$ is a point on $\Psi$ and $\delta \psi_1$ is a tangent vector in $T_{\psi_1} (\Psi)$.}
	\label{fig:Transformation}
\end{figure}

\begin{figure}[!t]
	\begin{center}
		\resizebox{\columnwidth}{!}{
			\begin{tabular}{|c|c|c|}
				\hline
				\includegraphics[scale = 0.11]{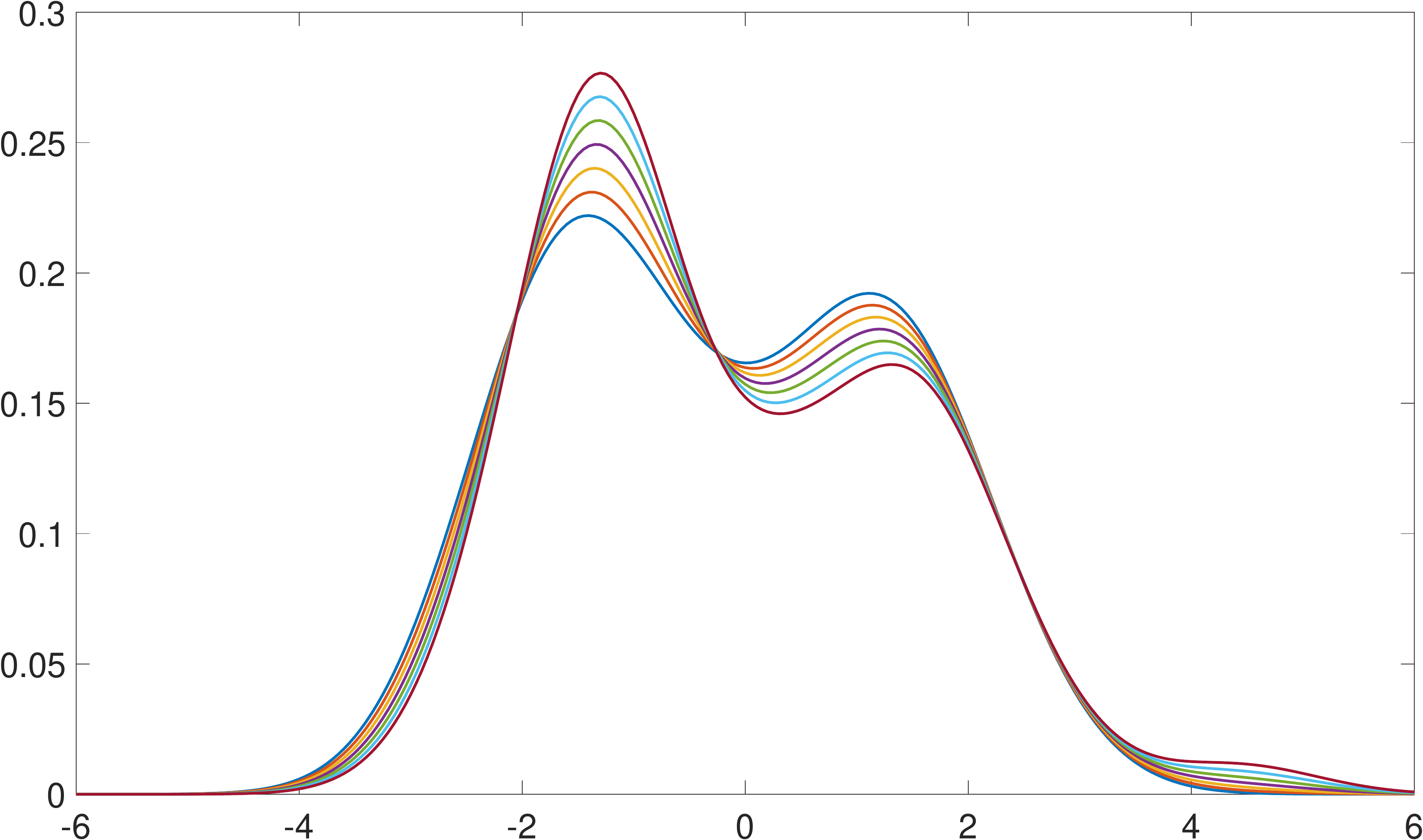}  & \includegraphics[scale = 0.11]{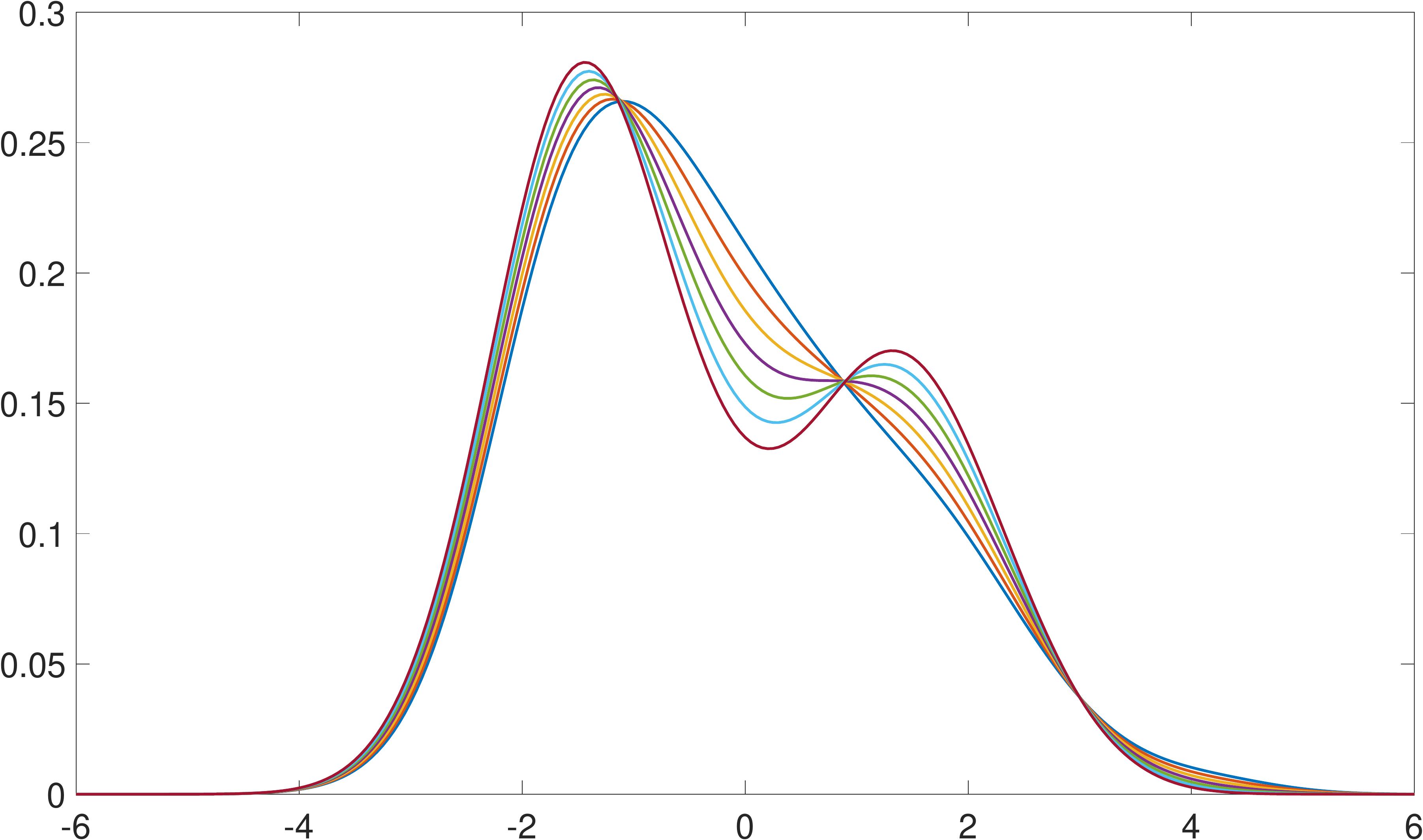}  & \includegraphics[scale = 0.11]{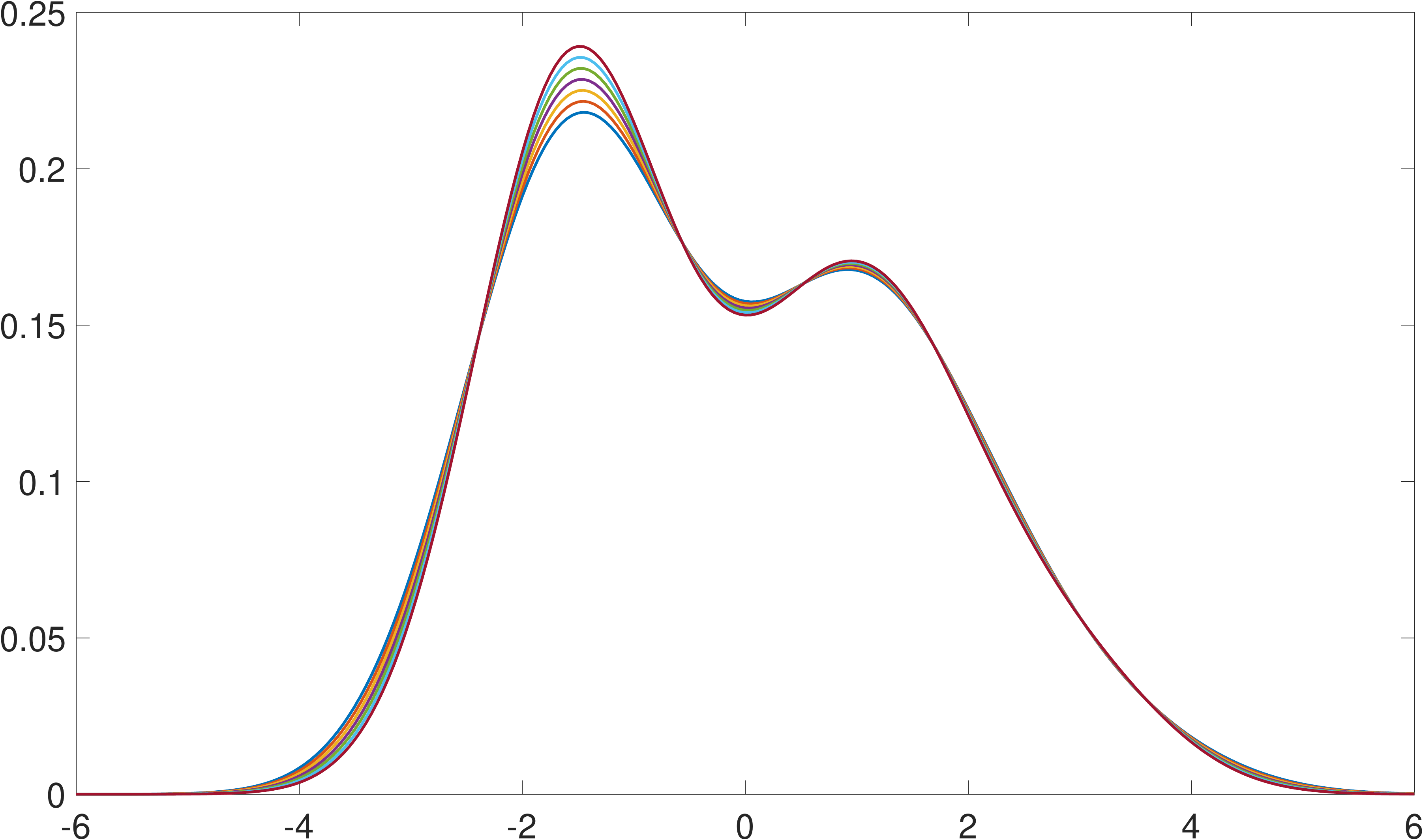}\\ \hline
                $d_{FR}=0.1368$&$d_{FR}=0.1346$&$d_{FR}=0.0510$\\
                \hline
			\end{tabular}
		}
	\end{center}
	\caption{Three examples of geodesic paths between PDFs with the corresponding geodesic distances. The path is sampled using seven equally spaced points.}
	\label{fig:ExampleFRgeods}
\end{figure}

Furthermore, for an element $\psi \in \Psi$ not lying on the boundary, let $T_\psi (\Psi) = \{ \delta \psi \mid \left\langle \delta \psi, \psi \right\rangle = 0 \}$ denote the tangent space at $\psi$, where $\langle \cdot , \cdot \rangle$ denotes the standard ${\mathbb{L}^2}$ inner product. Under the SRD representation, one can then show that, for any two vectors $\delta\psi_1, \delta\psi_2 \in T_\psi (\Psi)$, the FR metric defined in (\ref{eq:fr}) becomes the standard ${\mathbb{L}^2}$ Riemannian metric:
\begin{equation}
\label{eq:innprod}
	\left\langle \delta \psi_1, \delta \psi_2 \right\rangle = \int_{0}^{1} \delta \psi_1(x) \delta \psi_2(x) dx.
\end{equation}
Since the differential geometry of the sphere under the standard $\mathbb{L}^2$ metric is well-known, one can define various geometric tools for analyzing PDFs analytically. For example, the geodesic distance between two PDFs $p_1, p_2 \in \mathcal{P}$ under the FR metric, now represented using their SRDs $\psi_1, \psi_2 \in \Psi$, is simply given by the angle between them on $\Psi$:
\begin{equation}
\label{eq:upsilon}
	d_{FR}(p_1,p_2) = d_{\mathbb{L}^2}(\psi_1,\psi_2) = \cos^{-1}(\left\langle \psi_1, \psi_2 \right\rangle) = \upsilon.
\end{equation}
Note that since we are restricted to the positive orthant of the unit sphere, the geodesic distance $\upsilon$ between any two transformed PDFs on $\Psi$ is bounded above by $\pi/2$. Figure \ref{fig:Transformation}(a) provides a graphical description of the SRD-based simplification. Furthermore, in Figure \ref{fig:ExampleFRgeods}, we display three toy examples of geodesic paths between different PDFs in each panel, along with the associated geodesic distances. Note that the distances measure the length of these geodesic paths; thus, when we observe large deformations, the geodesic distances are also larger.

	\subsubsection{Statistical Analysis of Transformed PDFs using Geometric Tools}
	\label{subsubsec:Analysis}

We use several standard geometric tools in the proposed framework for sensitivity analysis. First, we define two maps which can transfer points from the SRD space $\Psi$ to the respective tangent space associated with any point $\psi \in \Psi$, $T_{\psi}(\Psi)$, and vice versa. The exponential map, $\exp_\psi: T_\psi (\Psi) \rightarrow \Psi$, is used to map points from the tangent space $T_\psi (\Psi)$ to the representation space $\Psi$, and is defined as (for $\psi \in \Psi$ and $\delta\psi \in T_\psi (\Psi)$):
\begin{equation}
\label{eq:exp}
	\exp_\psi(\delta\psi) = \cos(\left\| \delta\psi \right\|)\psi + \sin(\left\| \delta\psi \right\|)\dfrac{\delta\psi}{\left\| \delta\psi \right\|},
\end{equation}
where $\left\| \cdot \right\|$ denotes the usual ${\mathbb{L}^2}$ norm. Similarly, the inverse-exponential map, $\exp^{-1}_{\psi} : \Psi \rightarrow T_\psi (\Psi)$, takes a point from the representation space $\Psi$ and maps it to an element of the tangent space $T_{\psi}(\Psi)$, and is given by (for $\psi_1, \psi_2 \in \Psi$):
\begin{equation}
\label{eq:invexp}
	\exp^{-1}_{\psi_1}(\psi_2) = \frac{\upsilon}{\sin(\upsilon)}\left( \psi_2 - \cos(\upsilon) \psi_1 \right), \quad \upsilon = d_{FR}(p_1,p_2).
\end{equation}
Figure \ref{fig:Transformation}(b) provides an illustration of these two maps.

Using the simple definitions of the exponential and inverse-exponential maps, we can exploit the geometry of $\Psi$ to compute various sample statistics of PDFs. For example, given a posterior sample of PDF estimates, we can compute an average PDF as a representative of this set; we can additionally compute a measure of overall variance in this sample. We use the notion of the Karcher mean, which is a generalized version of an average on a metric space. Let $\psi_1, \psi_2, \dots, \psi_n$ denote a collection of SRDs. Then $\bar{\psi}$, the sample Karcher mean, is the minimizer of the Karcher variance $\rho$ defined as:
\begin{equation}
\label{eq:karcherVar}
	\rho(\bar{\psi}) = \frac{1}{n}\sum_{i=1}^{n} d_{\mathbb{L}^2}(\psi_i, \bar{\psi})^2.
\end{equation}
In particular,
\begin{equation}
\label{eq:karcherMean}
\bar{\psi} = \argmin_{\psi \in \Psi} \sum_{i=1}^{n} d_{\mathbb{L}^2}(\psi_i, \psi)^2 = \argmin_{\psi \in \Psi} \sum_{i=1}^{n} \{ \cos^{-1}(\left\langle \psi_i, \psi \right\rangle) \}^2.
\end{equation}
A gradient-based algorithm for computing the Karcher mean on $\Psi$ is presented as Algorithm \ref{algo:KarcherMean} \cite{dryden1998, kurtek2017}. This algorithm can be initialized using either any element from the given sample or the extrinsic average.

\RestyleAlgo{boxruled}
\begin{algorithm}[!t]
	\label{algo:KarcherMean}
	\SetAlgoLined
	Initialize estimate of Karcher mean: $\bar{\psi}_j$. Set $j = 0$ and $\epsilon_1, \epsilon_2$ to be small positive values.\\
	\vspace{0.5em}
	\begin{enumerate}
		\itemsep0.5em
		\item For each $i = 1,\dots, n$, compute $\delta u_i = \exp^{-1}_{\bar{\psi}_j}(\psi_i)$.
		\item Compute the average direction in the tangent space $\delta \bar{u} = \frac{1}{n}\sum_{i=1}^{n}\delta u_i$.
		\item If $\|\delta \bar{u}\| < \epsilon_1$, stop and return $\bar{\psi}_j$ as the Karcher mean.\\
		Otherwise, update using $\bar{\psi}_{j+1} = \exp_{\bar{\psi}_j}(\epsilon_2\delta\bar{u})$.
		\item Set $j = j+1$.
		\item Return to step $1$.
	\end{enumerate}
	\caption{Karcher mean on $\Psi$.}
\end{algorithm}

Furthermore, we can explore the variability in a collection of PDFs using tangent principal component analysis (tPCA). Given a sample of $n$ SRDs and their Karcher mean $\bar{\psi}$, we can use the inverse-exponential map to obtain tangent vectors $v_i = \exp^{-1}_{\bar{\psi}}(\psi_i)$, $i = 1, \dots, n$, $v_i \in T_{\bar{\psi}} (\Psi)$. To determine the dominant modes of variation, we can then perform eigendecomposition of the covariance operator for the tangent vectors and obtain the corresponding eigenfunctions.

At the implementation stage, PDFs are typically sampled using $N$ equally-spaced points. This results in an $N\times N$ finite-dimensional sample covariance matrix $C$. Since the sample size $n$ is usually less than $N$, consequently, $n$ controls the degree of variation in the data. The steps to compute tPCA of $n$ SRDs are presented in Algorithm \ref{algo:TangentPCA}. $\Sigma$ is a diagonal matrix (also of size $N\times N$) whose elements are the principal component variances (eigenvalues of $C$) ordered from largest to smallest. The columns of $U$ contain the eigenvectors of $C$ and represent the corresponding principal modes of variation in the given sample.

\RestyleAlgo{boxruled}
\begin{algorithm}[!t]
	\label{algo:TangentPCA}
	\SetAlgoLined
	\begin{enumerate}
		\itemsep0.5em
		\item Compute $\bar{\psi}$, the Karcher mean of $\psi_1, \dots, \psi_n$ using Algorithm \ref{algo:KarcherMean}.
		\item For each $i = 1, \dots, n$, compute $v_i = \exp^{-1}_{\bar{\psi}}(\psi_i)$ using the inverse-exponential map (\ref{eq:invexp}).
		\item Compute the sample covariance matrix $C=\dfrac{1}{n-1}\sum_{i=1}^{n}v_i v_i^T$.
		\item Perform singular value decomposition (SVD) of $C$,  $C = U \Sigma U^T$.
	\end{enumerate}
	\caption{Tangent PCA}
\end{algorithm}

\begin{figure}[!t]
	\begin{center}
		\resizebox{\columnwidth}{!}{
			\begin{tabular}{|c|c|c|c|}
                \hline
                (a)&(b)&(c)&(d)\\
				\hline
				\includegraphics[scale = 0.11]{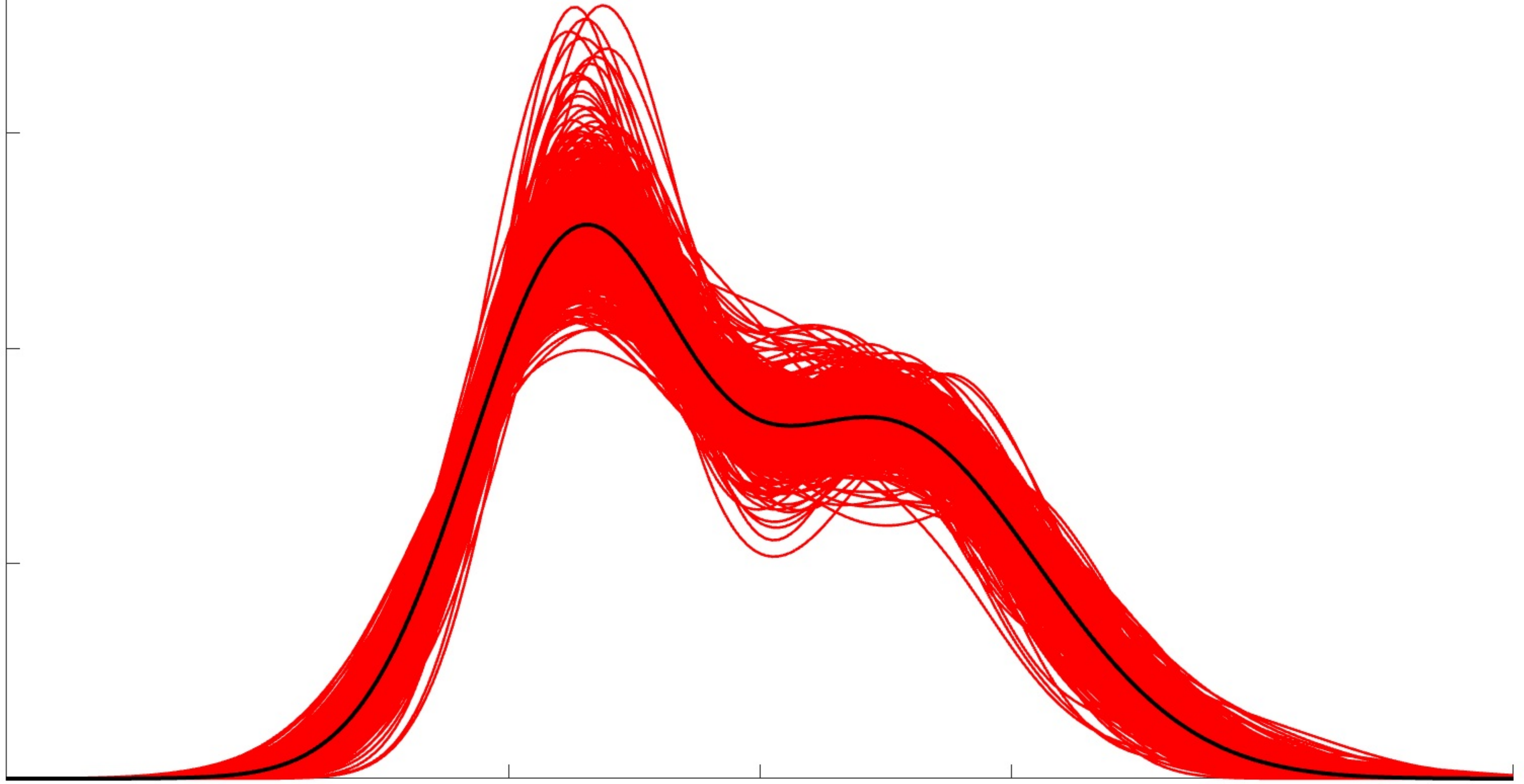}  & \includegraphics[scale = 0.11]{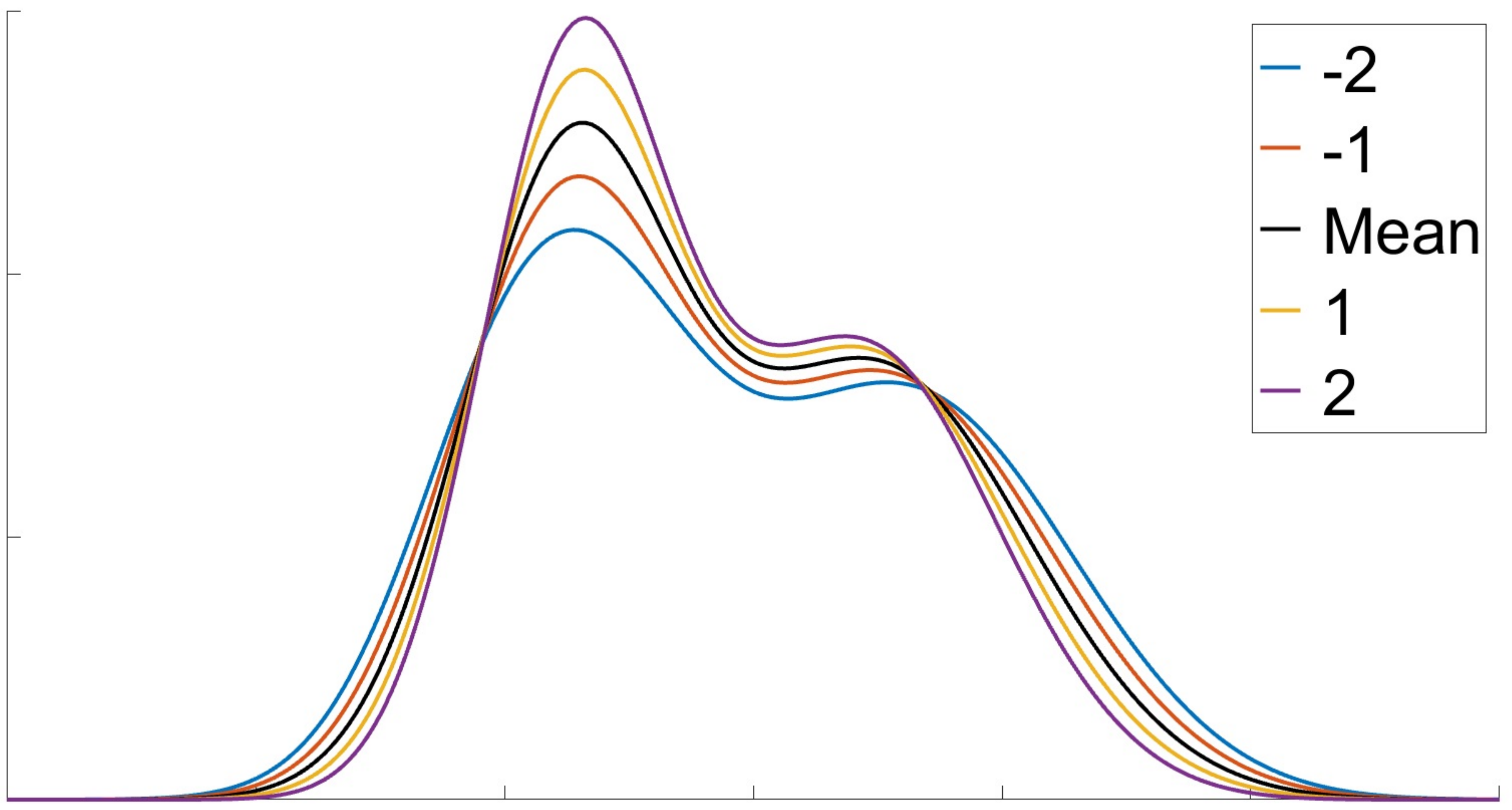}  & \includegraphics[scale = 0.11]{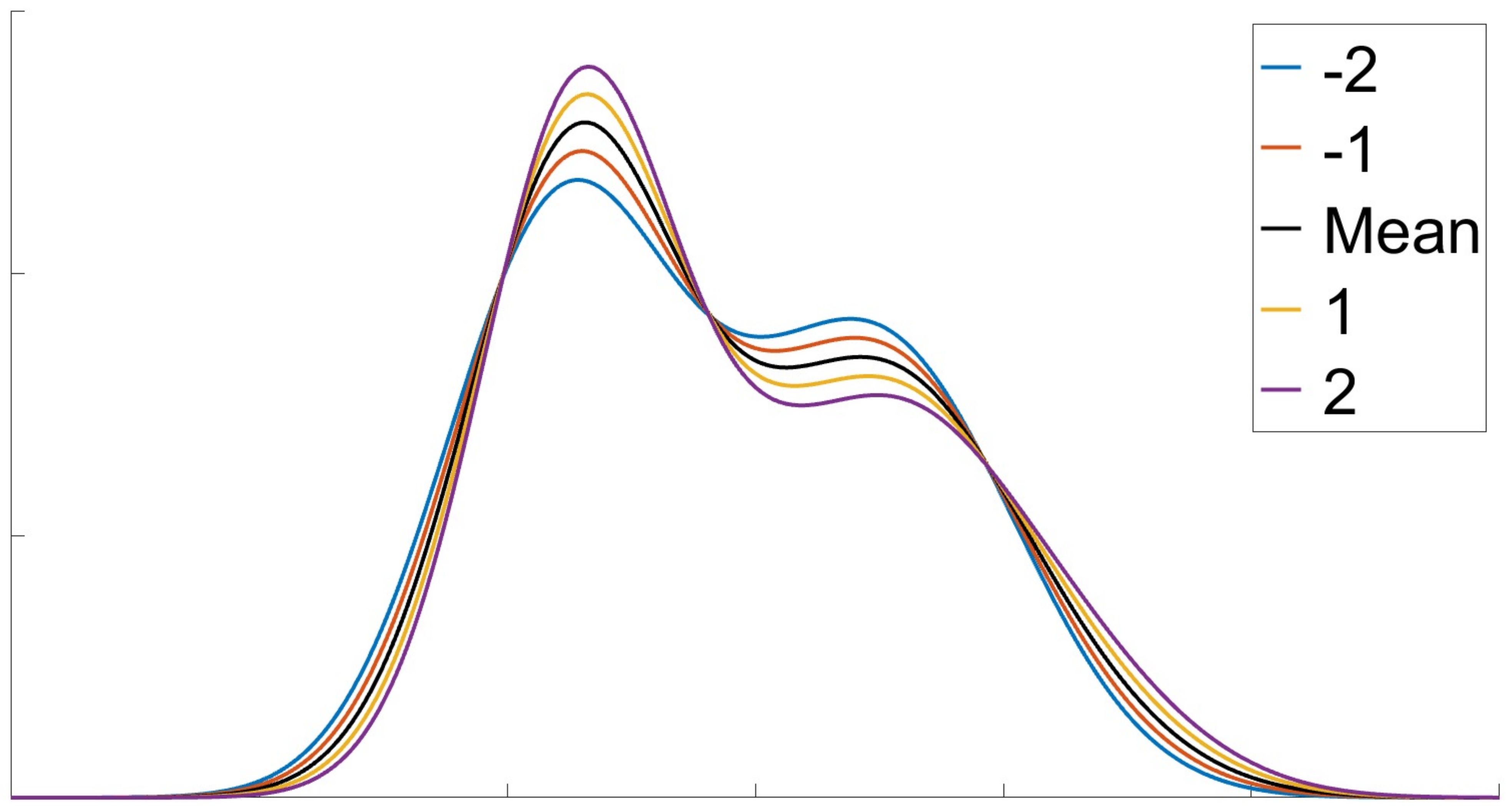} & \includegraphics[scale = 0.11]{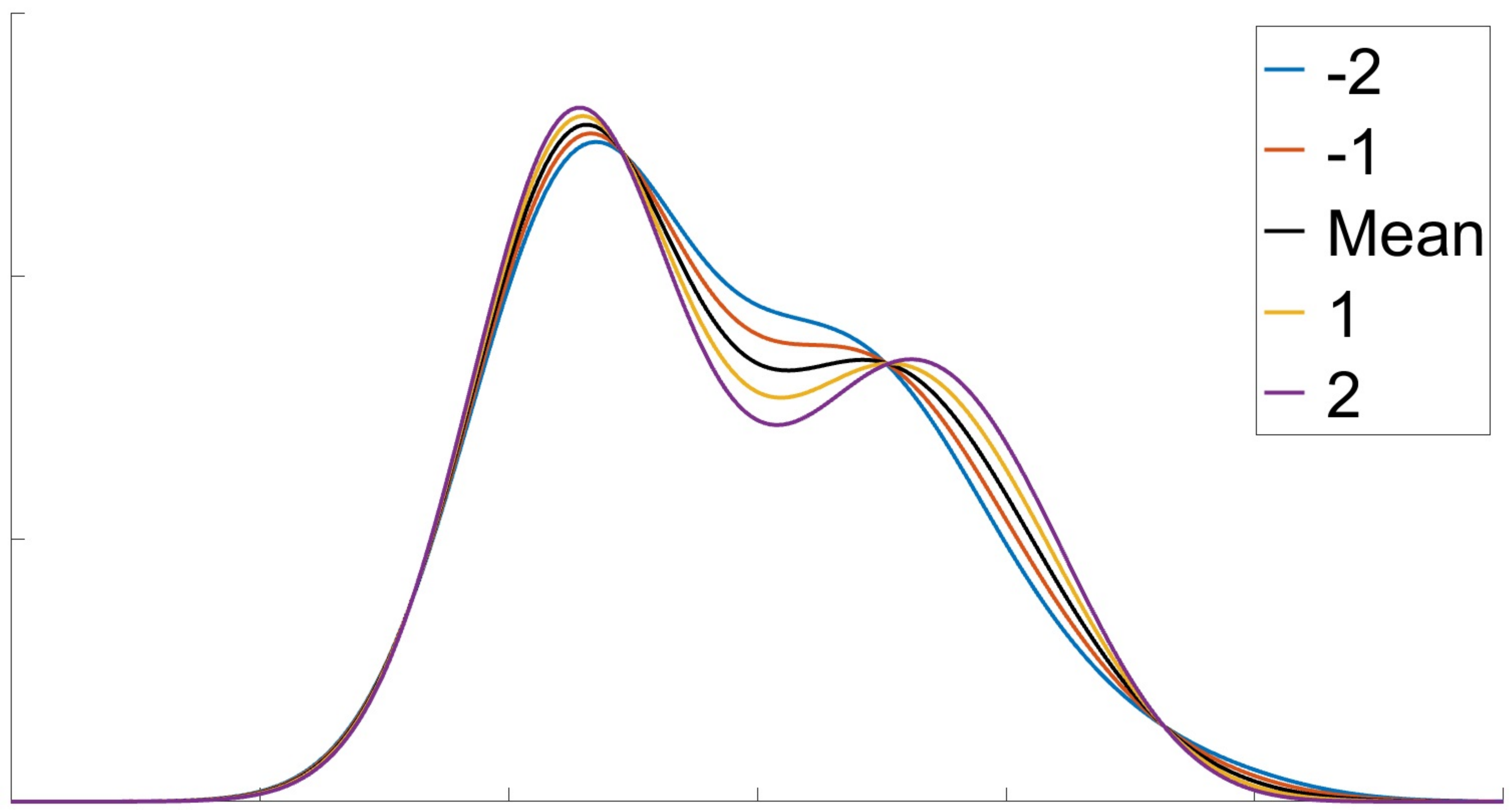}\\ \hline
                \hline
			\end{tabular}
		}
	\end{center}
	\caption{Tangent PCA for a sample of PDFs. (a) PDF sample in red and Karcher mean in black. (b) First principal direction (PD) of variability. (c) Second PD of variability. (d) Third PD of variability. All PDs are displayed as a path from $-2$ standard deviations to $+2$ standard deviations around the Karcher mean.}
	\label{fig:ExampleFRPCA}
\end{figure}

Figure \ref{fig:ExampleFRPCA} presents an example of applying Algorithms \ref{algo:KarcherMean} and \ref{algo:TangentPCA} to a collection of PDFs. To obtain a single PDF, 100 random numbers are generated from a Gaussian mixture distribution with two components and then smoothed using a Gaussian kernel. This procedure is repeated multiple times to get a sample of PDFs as shown in panel (a) of Figure \ref{fig:ExampleFRPCA}. The sample is displayed in red with the Karcher mean overlayed in black. The Karcher mean appears to be a good representative of this sample. Then, using tPCA, we explore the principal directions of variability in the given sample. We plot the three principal directions in panels (b)-(d), respectively, as a path sampled at $-2, -1, 0, +1$ and $+2$ standard deviations from the mean. The displayed paths reflect natural variability in the given sample. In particular, they capture the relative sizes of the two modes as well as a transformation from a bimodal density to a unimodal one.

	\section{Geometric Measures of Global Sensitivity}
	\label{sec:GeomSensAn}
	
In this section, we define novel, geometrically-motivated global measures that can be used to assess sensitivity of nonparametric Bayesian models for density estimation. We specifically consider perturbations of the prior structures discussed in Section \ref{subsec:NPBDensity}. However, one could use the same measures under other setups, e.g., case influence analysis. The prior perturbations broadly take two forms: (1) changes in the precision parameter $\alpha$, or (2) changes in the parameters and hyperparameters for the probability measure $G_0$. Note that from here on, we do not make a distinction between model parameters and hyperparameters for simplicity.

	\subsection{Notation and Problem Definition}
	\label{subsec:NotaProbSet}
	
Let $\pi_0$ denote the baseline prior, and let $\ \pi_\mathcal{S} = \{ \pi_{s_1}, \dots, \pi_{s_K} \}$ denote a finite class of prior perturbations or contaminants. Rather than working with the resulting posteriors directly, we will use MCMC-generated samples to assess global sensitivity. Thus, let $p_0^1, \dots, p_0^n$ denote a baseline posterior sample of nonparametric density estimates (of size $n$), and let $\psi_0^1, \dots, \psi_0^n$ denote their SRDs. Similarly, $p_{s_k}^1, \dots, p_{s_k}^m$ denote a posterior sample of size $m$ generated using the contaminated prior $\pi_{s_k}$, and $\psi_{s_k}^1, \dots, \psi_{s_k}^m$ denote their corresponding SRDs. Now, to assess sensitivity we will use the various geometric tools defined on the SRD representation space. In particular, we can use Algorithm \ref{algo:KarcherMean} to calculate the baseline and perturbed posterior averages based on the given samples. Note that the posterior average for both models is an intrinsic average that directly uses the FR Riemannian geometry of the PDF sapce; this geometry is used to define other sensitivity measures as well. Additionally, as will becomes clear in subsequent sections, once posterior samples are generated, they are easy to compute making the proposed framework interpretable and computationally efficient.

	\subsection{Global Sensitivity Measures}
	\label{subsec:GlobSensMeas}

\begin{figure}[!t]
	\begin{center}
		\includegraphics[width = 5in]{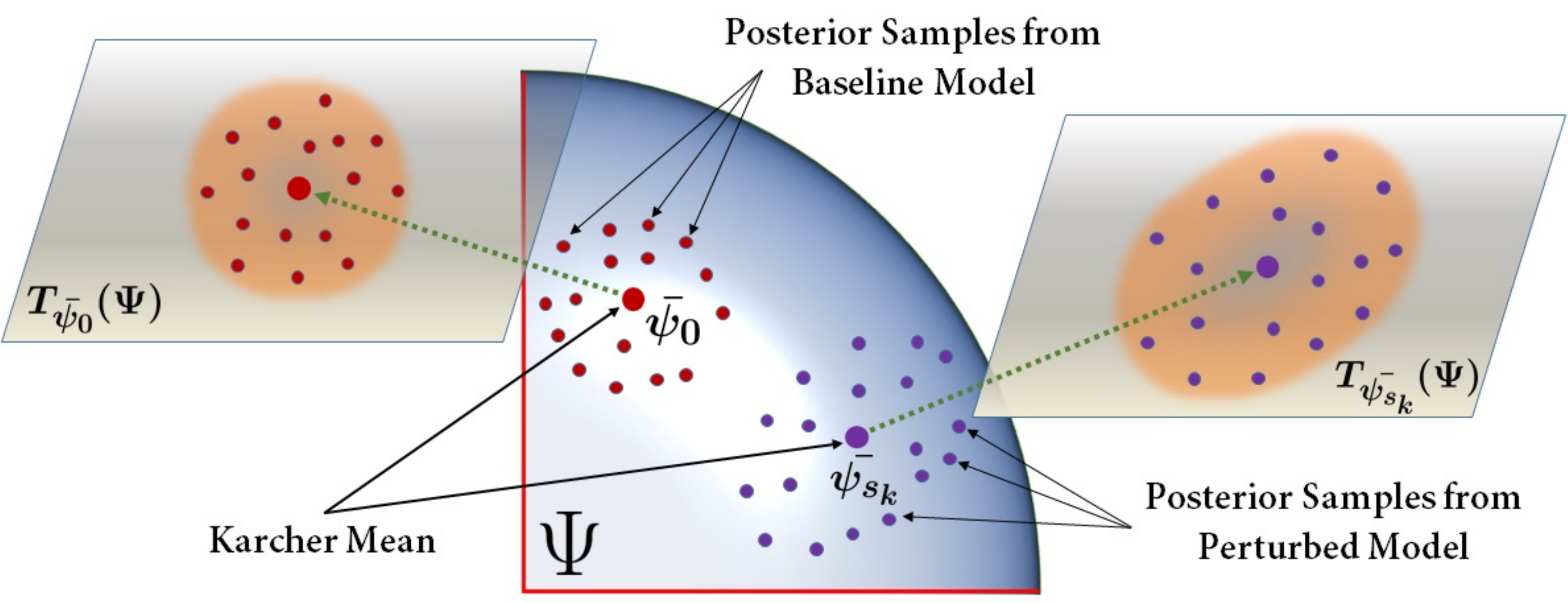}
	\end{center}
	\caption{Our three sensitivity measures capture the difference in shift and covariance structure of the baseline model and a perturbed model based on posterior samples.}
	\label{fig:SensMeas}
\end{figure}

To study the diverse effects of the prior perturbations on the resulting posterior, we define three complementary global measures of sensitivity: (1) the FR distance between probability density averages of posterior samples, (2) difference in logarithm of overall Karcher variances of posterior samples, and (3) norm of the difference of scaled cumulative eigenvalues of the empirical covariance operators obtained from posterior samples. All of these measures take into account the geometry of the space on which the posterior density samples lie.  We provide an overall pictorial summary of the three proposed global sensitivity measures in Figure \ref{fig:SensMeas}. Next, we define them precisely and provide illustrative toy examples to show the utility of each of these measures. Note that the PDFs corresponding to all of the models in Figure \ref{fig:ExampleFR} are smoothed density estimates of random numbers generated from various Gaussian distributions. Similarly, the collection of PDFs in Figures \ref{fig:ExampleKV} and \ref{fig:ExampleCV} are smoothed density estimates of random numbers generated from different simple parametric models. For each of these toy examples and illustrations, the PDFs are not actual posterior samples generated from a Bayesian nonparametric model.

	\subsubsection{Shift Sensitivity Measure}
	\label{subsubsec:MeasureShift}
	
Let $\bar{p}_0$ and $\bar{p}_{s_k}$ denote the averages of posterior samples from the baseline model and a perturbed model, respectively, with corresponding SRDs $\bar{\psi}_0$ and $\bar{\psi}_{s_k}$. Again, these can be easily computed using Algorithm \ref{algo:KarcherMean}. The two averages can be used as a valid (first moment) characterization of the density samples from their respective posteriors. Then, the geodesic distance between the posterior sample averages provides a valid measure of ``shift" between the two posterior samples. Thus, to define our first measure of global sensitivity, we compute the FR distance between $\bar{p}_0$ and $\bar{p}_{s_k}$:
\begin{equation}
	\mathbb{D}(\pi_0, \pi_{s_k}) = d_{FR}(\bar{p}_0, \bar{p}_{s_k}) = d_{\mathbb{L}^2}(\bar{\psi}_0,\bar{\psi}_{s_k}) = \cos^{-1}(\left\langle \bar{\psi}_0, \bar{\psi}_{s_k} \right\rangle).
\end{equation}
An advantage of the measure $\mathbb{D}$ is that it is bounded below by zero and above by $\pi/2$, providing a natural scale for sensitivity analysis.

\begin{figure}[!t]
	\begin{center}
		\resizebox{\columnwidth}{!}{
			\begin{tabular}{|c|c|c|c|}
				\hline
				\multicolumn{2}{|c|}{(a)} & \multicolumn{2}{c|}{(b)} \\ \hline
				\includegraphics[scale = 0.11]{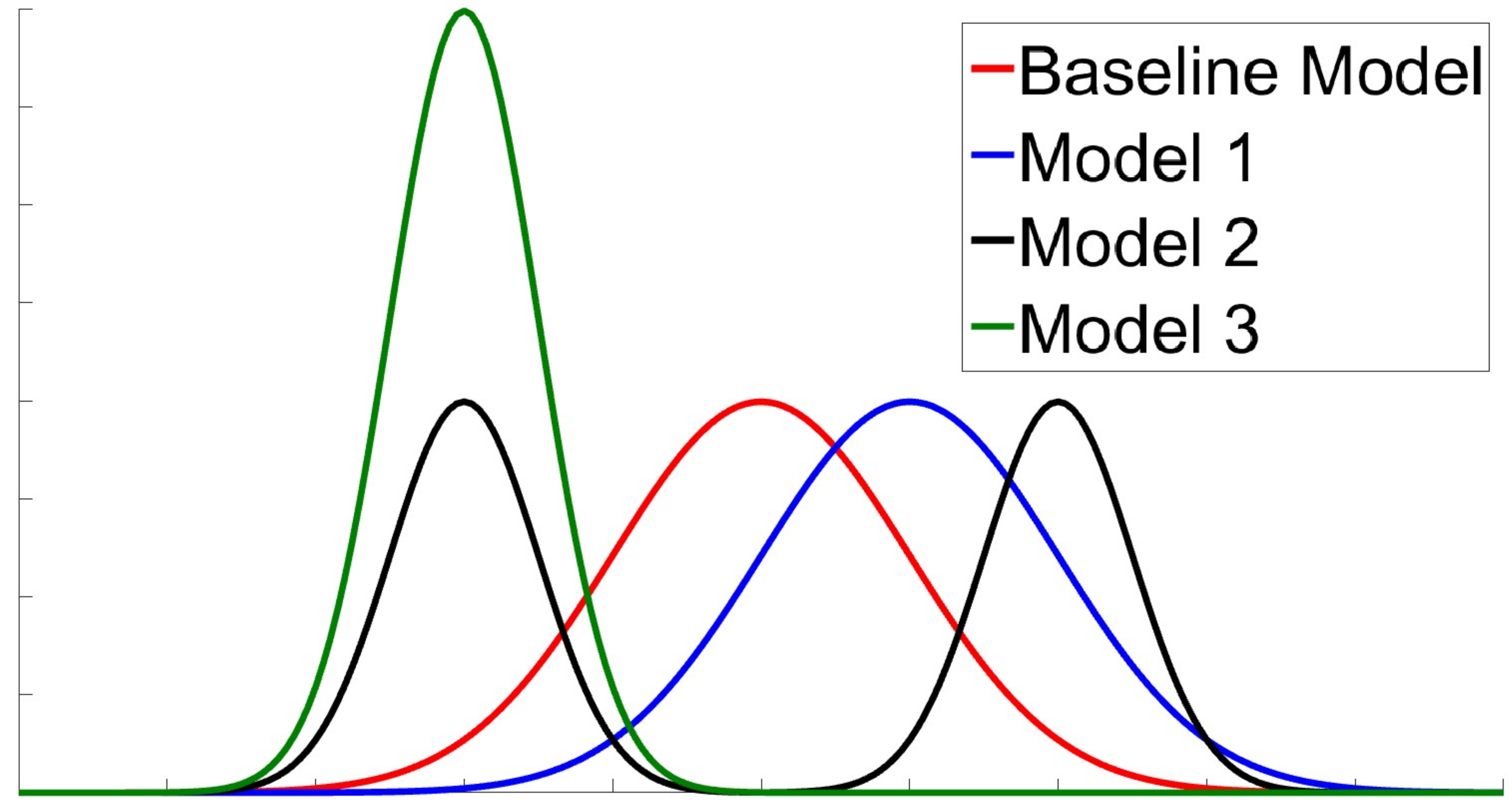}         & \includegraphics[width = 1.45in]{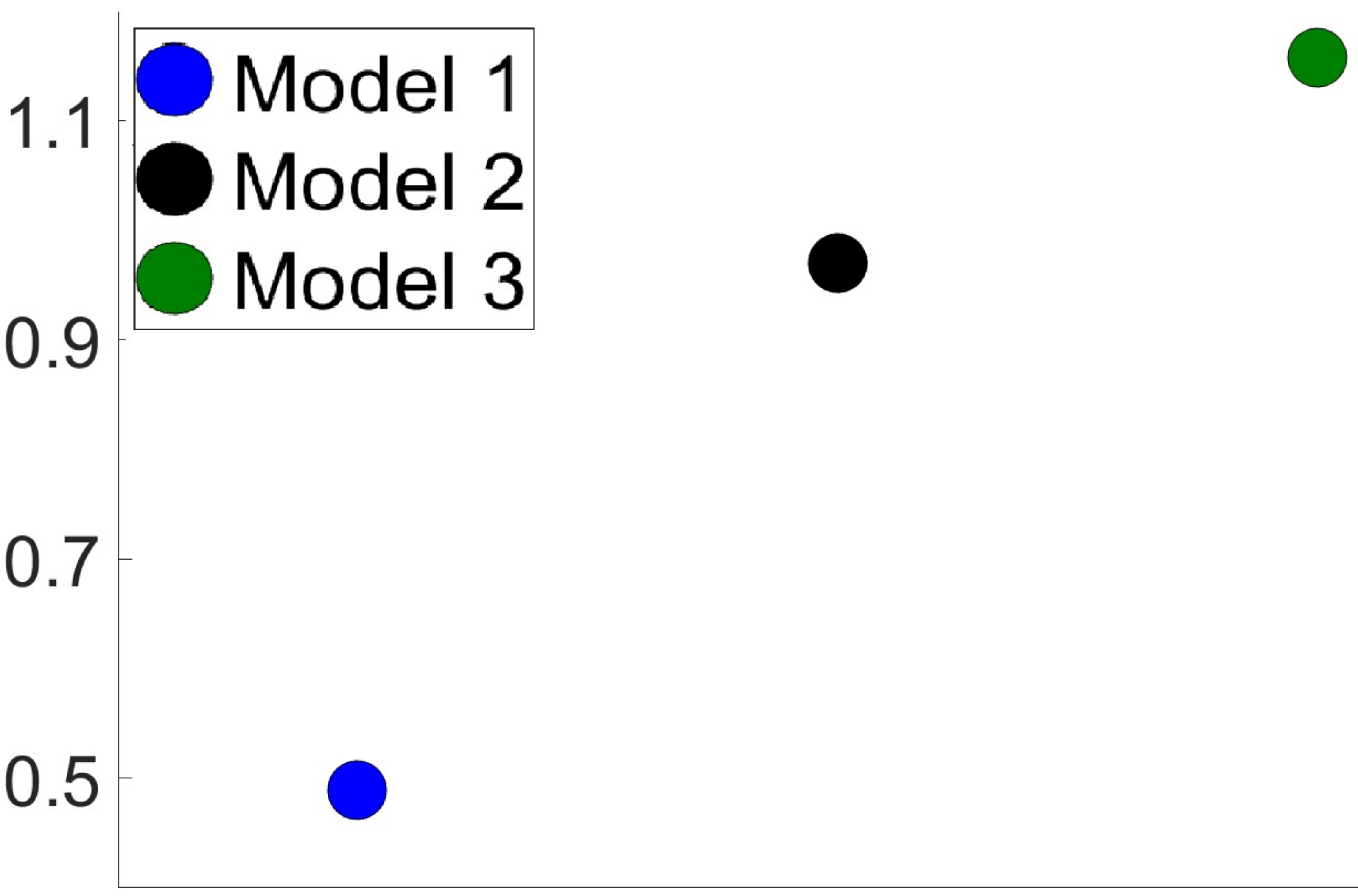}        & \includegraphics[scale = 0.11]{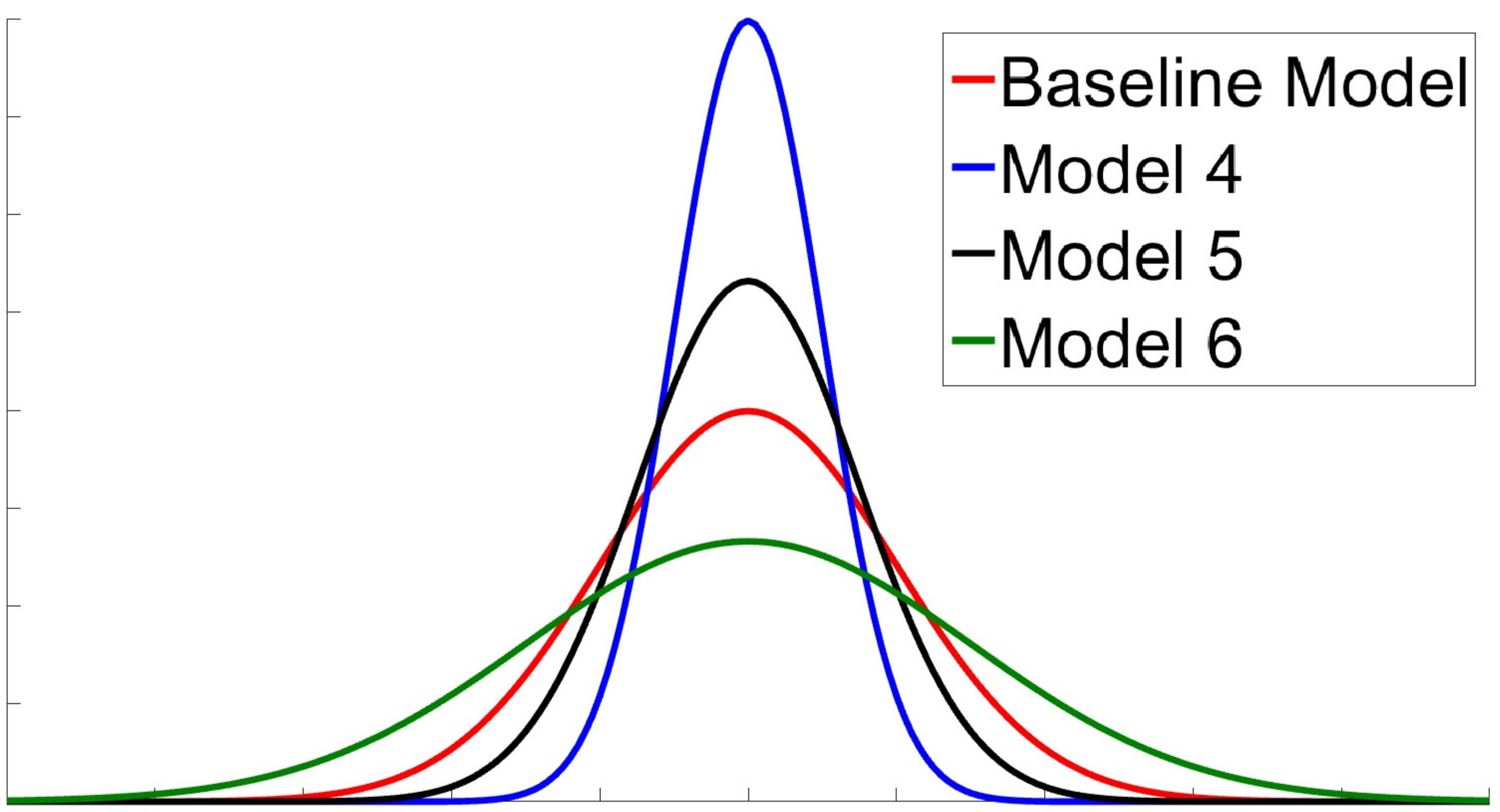}         & \includegraphics[width = 1.45in]{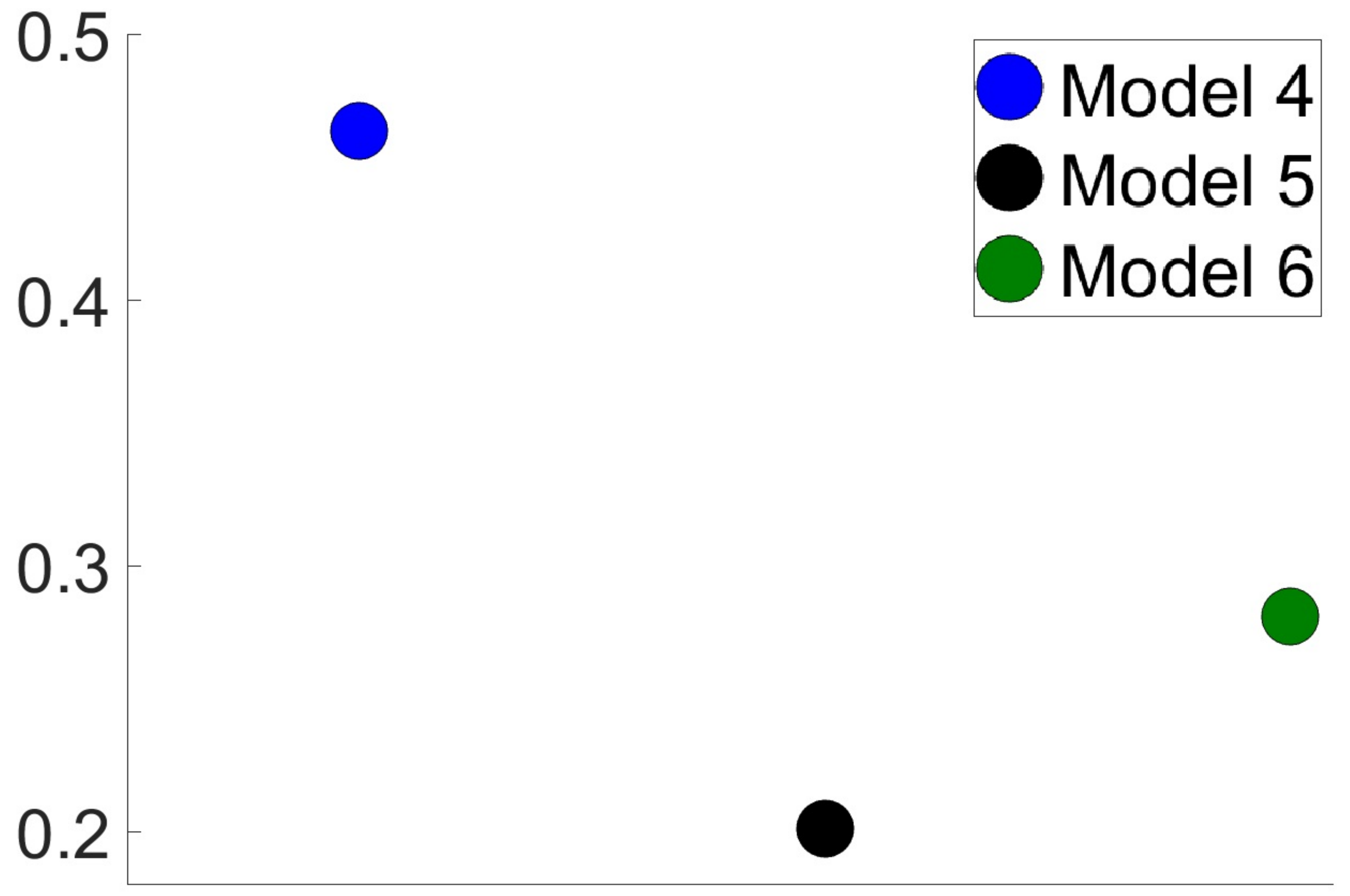}        \\ \hline
			\end{tabular}
		}
	\end{center}
	\caption{Example to show the utility of the sensitivity measure $\mathbb{D}$ to capture the shift between posterior density samples. We consider two examples: (a) averages that differ in their mean and variance, and (b) averages that differ in variance only.}
	\label{fig:ExampleFR}
\end{figure}

To show the effectiveness of the measure $\mathbb{D}$, we consider two different setups as seen in Figure \ref{fig:ExampleFR}. In panel (a), we consider a scenario where the posterior sample averages of the baseline and perturbed models have different means and variances. We compute the measure $\mathbb{D}$ for each of the three models, and plot the resulting values. We can clearly see that the measure serves as a good indicator of the extent of deviation of Models 1, 2 and 3 from the baseline model. In panel (b), we consider an example where the posterior sample averages based on the baseline and perturbed models have the same means, but different variances. We use this example to emphasize that \emph{this measure captures the shift between posterior samples rather than the shift between posterior sample averages.} Thus, despite the fact that the posterior sample averages differ only in their variances, we observe an intuitive result. Visually, the baseline model posterior sample average is closest to the perturbed posterior sample averages plotted in green and black. The blue perturbed posterior sample average has much smaller variance than the others. The sensitivity measure $\mathbb{D}$ clearly reflects this, and is nearly double for the blue posterior sample than for the black one (it is much larger than the green one as well).

	\subsubsection{Spread Sensitivity Measures}
	\label{subsubsec:MeasureVar}

We would now like to measure the difference in the overall spread or variability of the posterior samples. This can be achieved via the Karcher variance defined in (\ref{eq:karcherVar}). In particular, we consider the log-ratio of the variances obtained from the perturbed posterior samples and the baseline posterior samples:
\begin{align}
	\mathbb{V} (\pi_0, \pi_{s_k}) &= \log\bigg(\frac{\dfrac{1}{m}\sum_{j=1}^{m}  \Big\{ d_{\mathbb{L}^2}(\psi_{s_k}^j, \bar{\psi}_{s_k}) \Big\}^2}{\dfrac{1}{n}\sum_{i=1}^{n} \Big\{ d_{\mathbb{L}^2}(\psi_0^i, \bar{\psi}_0) \Big\}^2}\bigg), \nonumber \\
	&= \log\Big( \dfrac{1}{m}\sum_{j=1}^{m}  \Big\{ \cos^{-1}(\left\langle \psi_{s_k}^j, \bar{\psi}_{s_k}\right\rangle) \Big\}^2\Big) - \log\Big(\dfrac{1}{n}\sum_{i=1}^{n}\Big\{\cos^{-1}(\left\langle\psi_0^i, \bar{\psi}_0\right\rangle) \Big\}^2\Big).
\end{align}
The measure $\mathbb{V}$ naturally captures the direction of sensitivity (i.e., less vs. more variance in the perturbed posterior samples), which we felt provides useful information in this context. The measure is unbounded due to the log transformation.

\begin{figure}[!t]
	\begin{center}
		\resizebox{\columnwidth}{!}{
			\begin{tabular}{|c|c|c|c|c|}
				\hline
				\multicolumn{5}{|c|}{(a)}  \\ \hline
				\includegraphics[scale = 0.09]{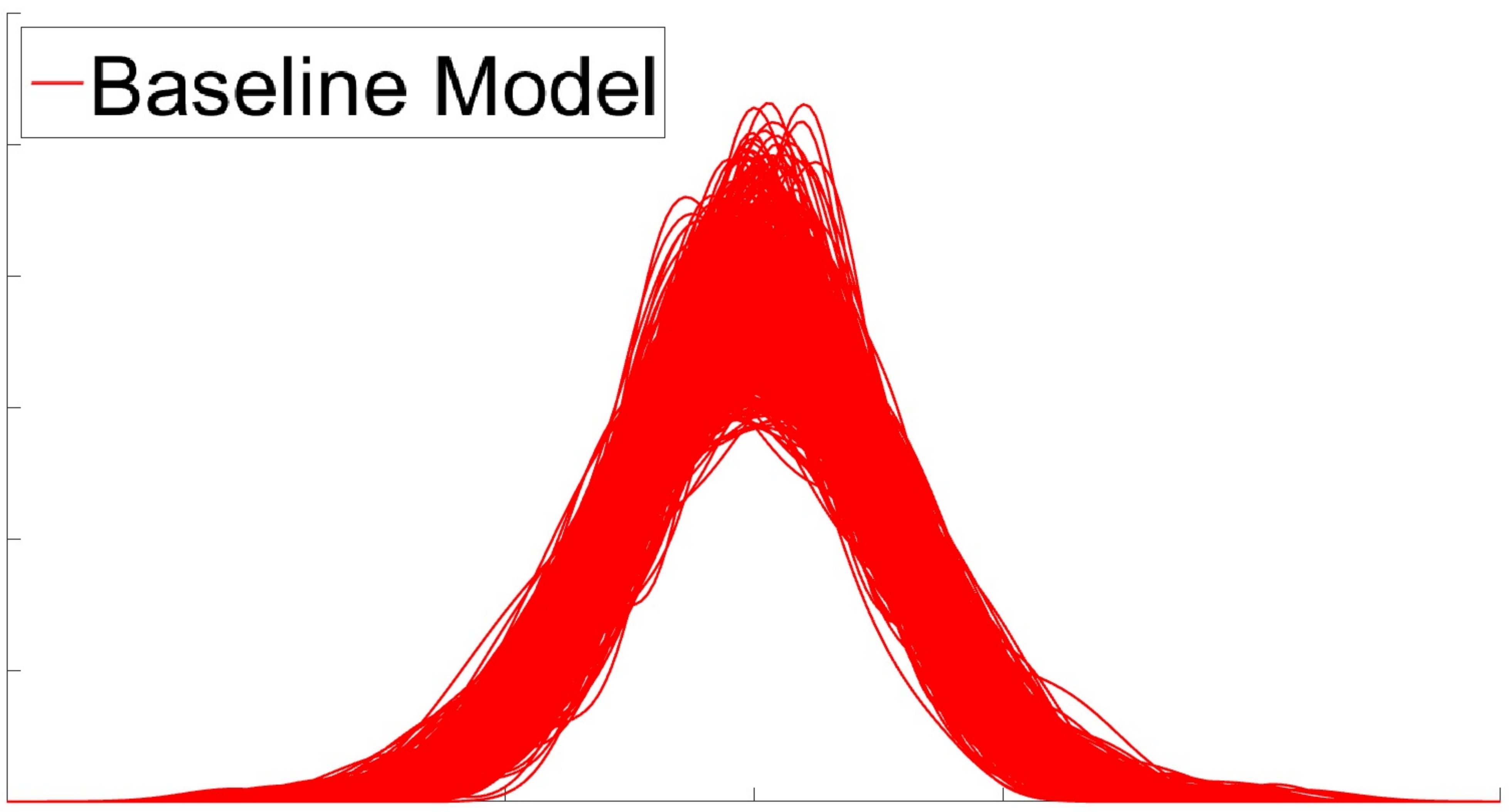}&\includegraphics[scale = 0.09]{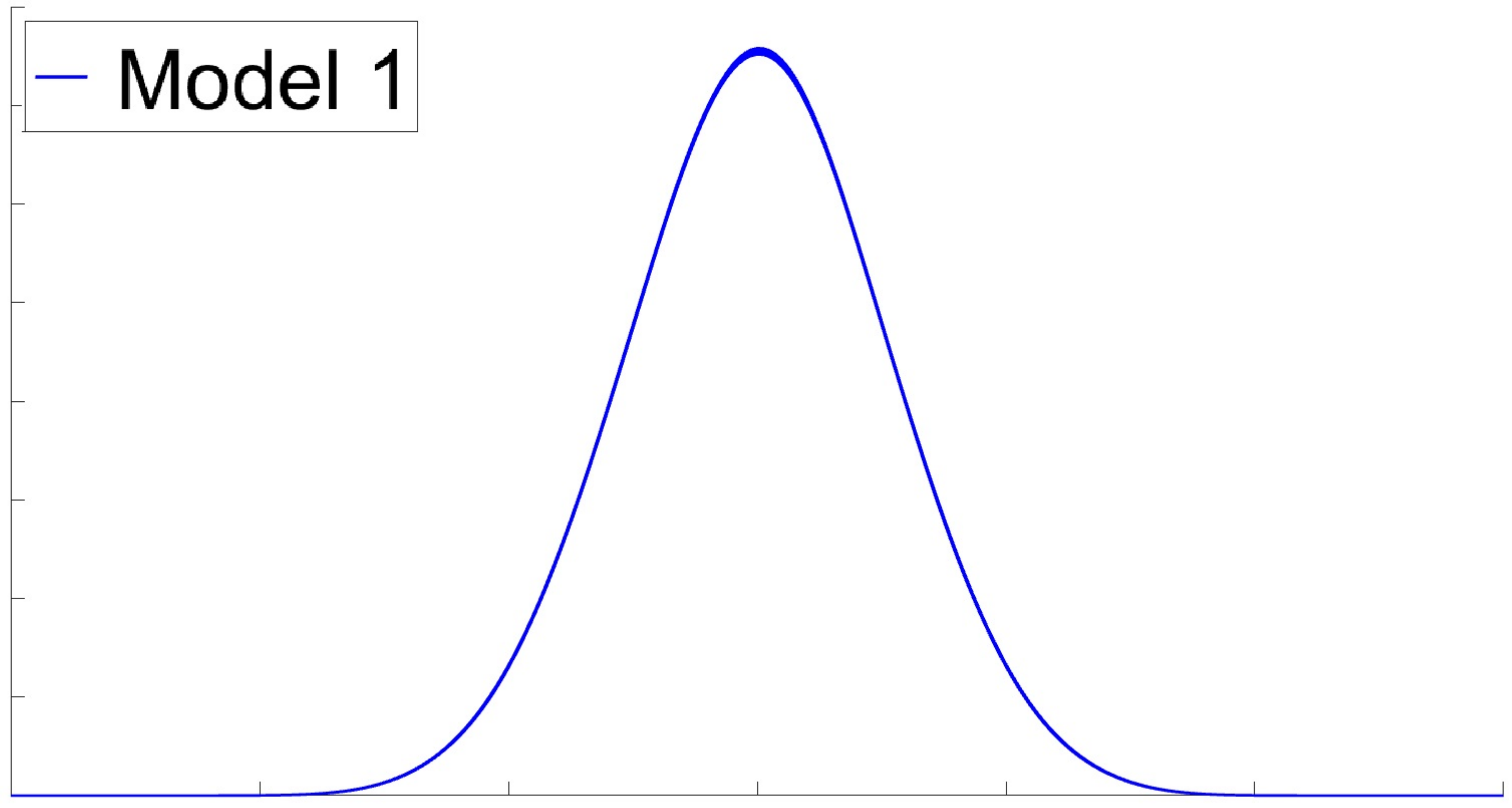}     & \includegraphics[scale = 0.09]{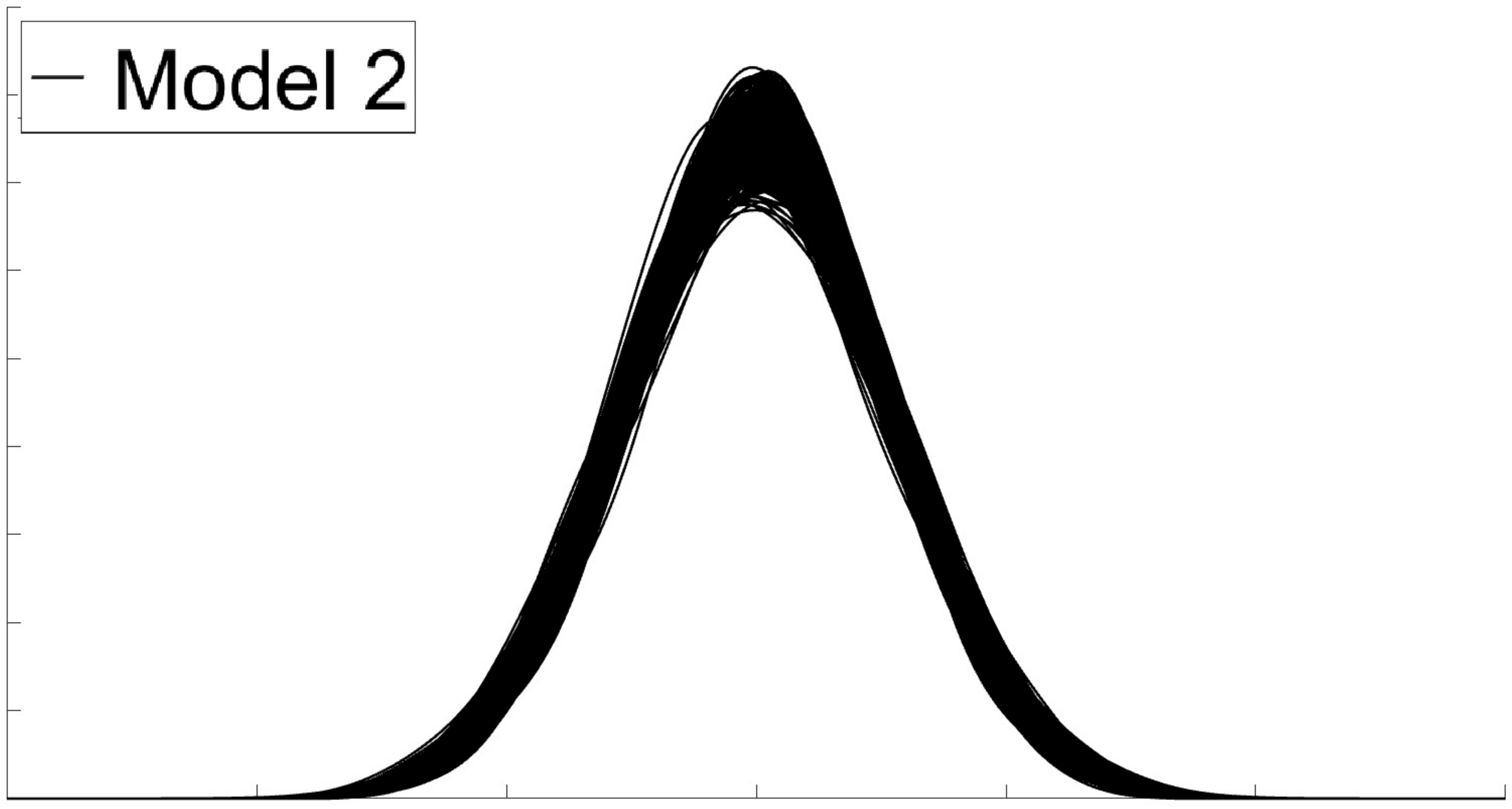}    &
				\includegraphics[scale = 0.09]{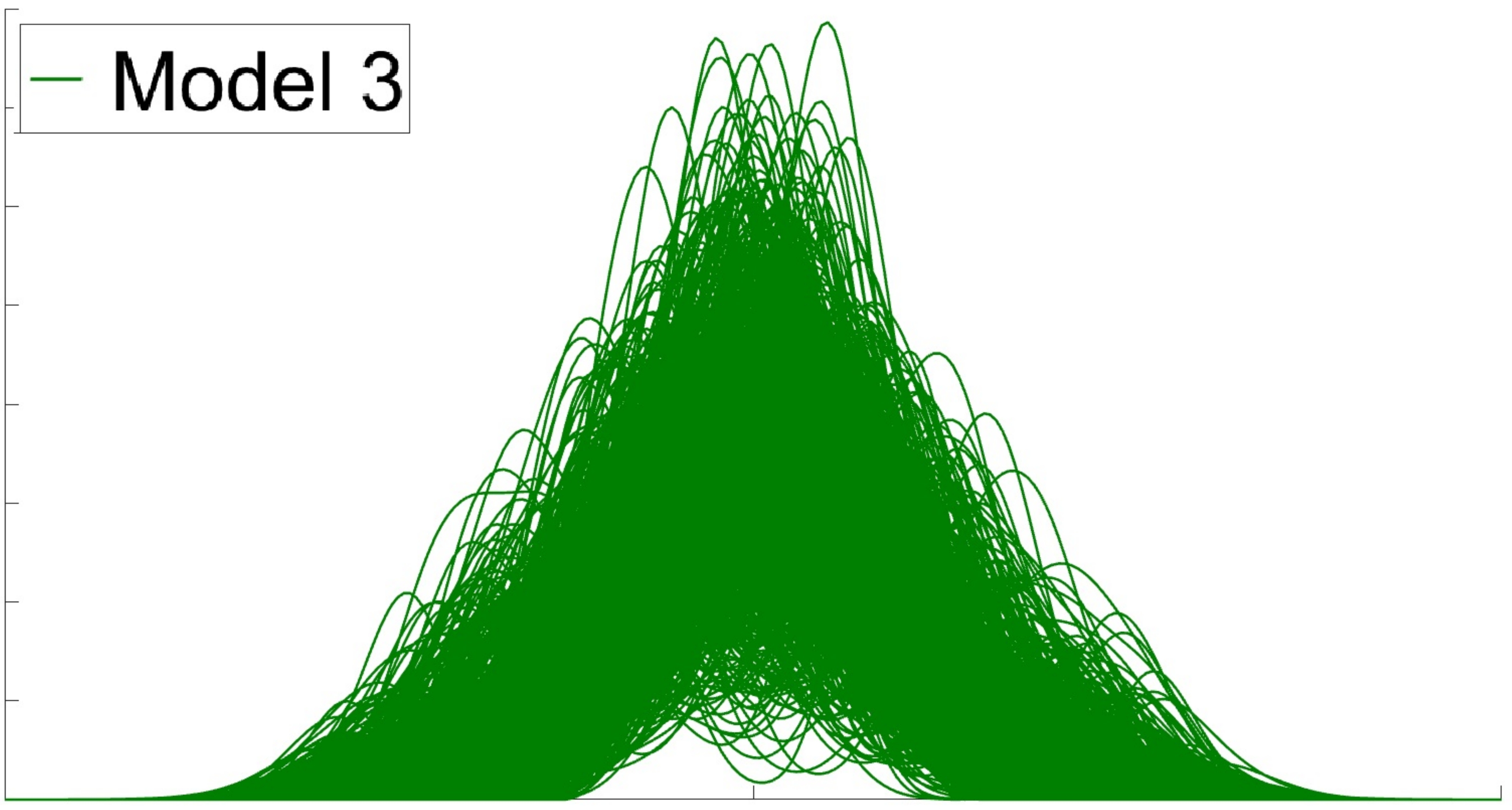}    & \includegraphics[scale = 0.09]{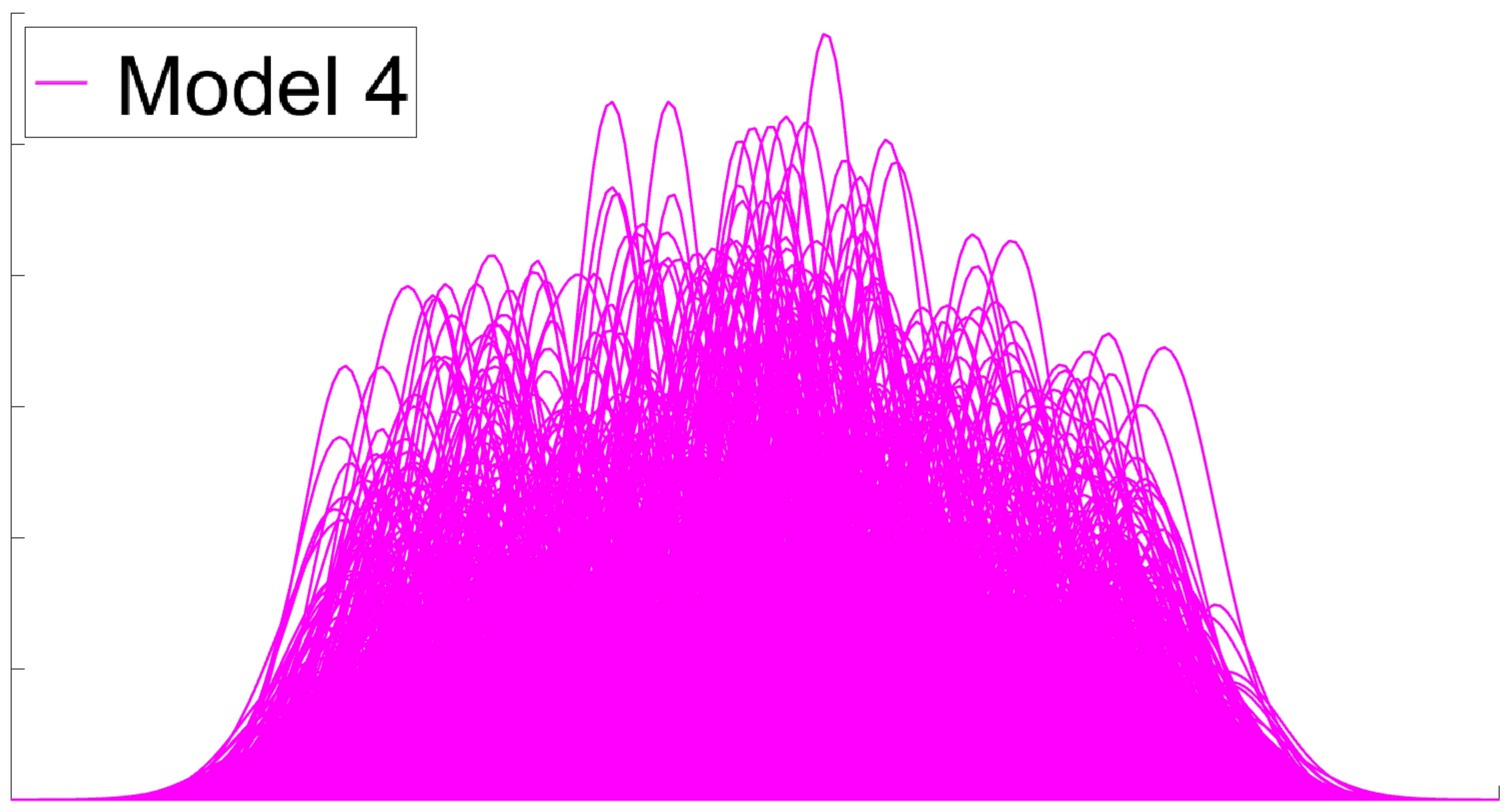} \\ \hline
			\end{tabular}
		}
			\begin{tabular}{|c|c|}
				\hline
				(b)&(c)  \\ \hline
				\includegraphics[scale = 0.15]{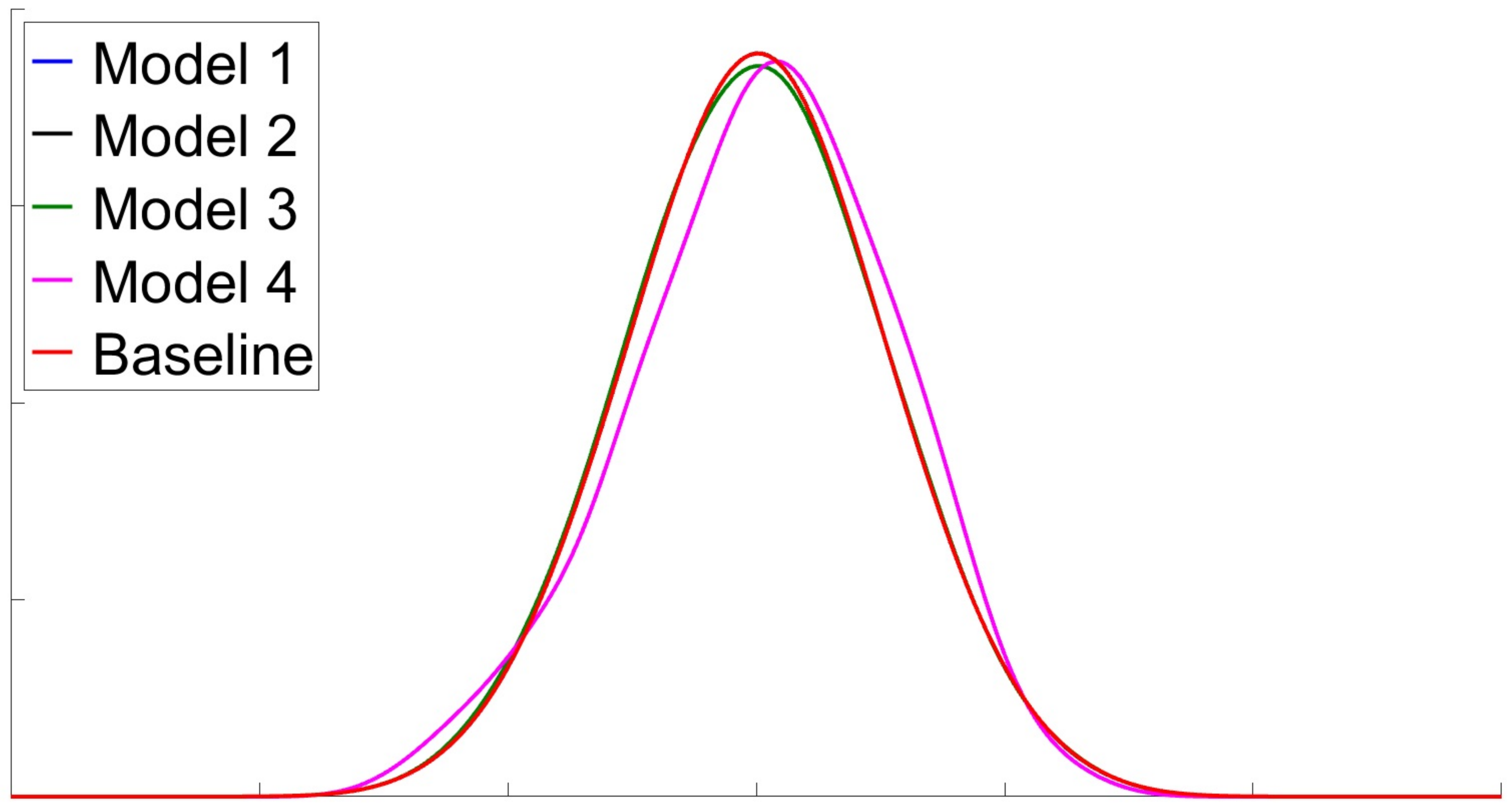}&\includegraphics[scale = 0.15]{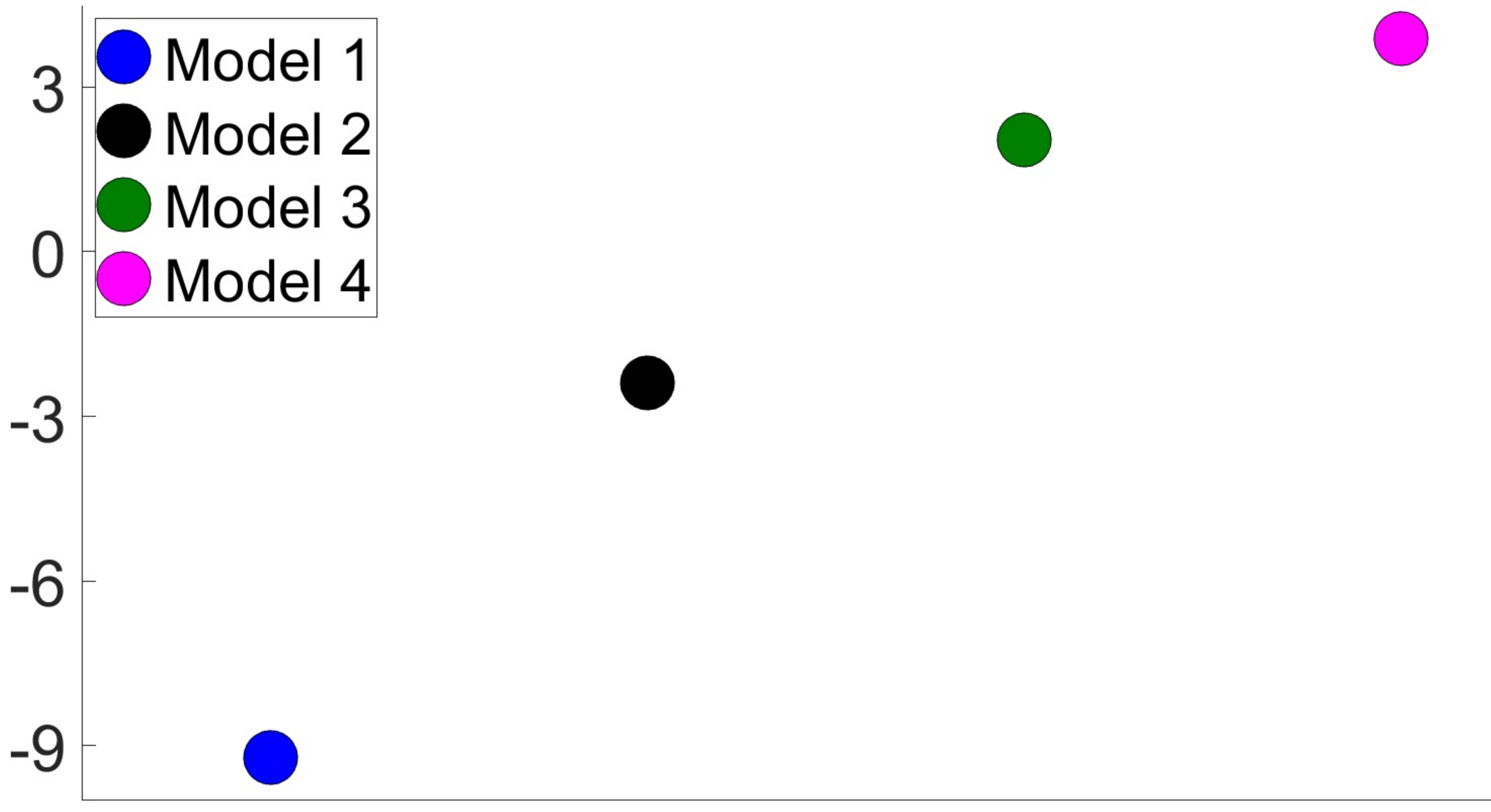}\\ \hline
			\end{tabular}
	\end{center}
	\caption{Example to show the utility of the sensitivity measure $\mathbb{V}$ in capturing the overall variance differences across posterior samples. (a) Posterior samples from baseline model (red) and four perturbed models (blue, black, green and magenta). (b) Sample averages for all five models. (c) Values of sensitivity measure $\mathbb{V}$.}
	\label{fig:ExampleKV}
\end{figure}

To showcase the effectiveness of the measure $\mathbb{V}$, we consider four different perturbed models, labeled as Models 1, 2, 3 and 4. For each of the perturbed models, we have ensured that the posterior sample average is very similar to the posterior sample average of the baseline model; this is confirmed visually in Figure \ref{fig:ExampleKV}(b). Thus, the measure $\mathbb{D}$ introduced in the previous section is always very small. For Models 1 and 2, we decreased the overall posterior sample variance (by factors 0.01 and 0.3, respectively) while for Models 3 and 4 we inflated the overall posterior sample variance (by factors 3 and 10, respectively). Panel (a) in Figure \ref{fig:ExampleKV} shows the posterior samples from the baseline model and each of the four perturbed models. Panel (c) shows the plot of the measure $\mathbb{V}$ for the different perturbed models that we consider. Our expectations are confirmed by this plot: Models 1 and 2 result in negative values of $\mathbb{V}$ (corresponding to less variability in the perturbed models than in the baseline model) while Models 3 and 4 result in positive values of $\mathbb{V}$ (corresponding to more variability in the perturbed model). Furthermore, the magnitude of  for Models 2 and 3 is approximately the same since the variance was scaled by approximately the same factor in either direction. The magnitude of $\mathbb{V}$ for Model 1 is much greater than the others as expected.

While the overall variance in the posterior samples provides an intuitive measure of global spread sensitivity, we may also be interested in capturing finer changes in the covariance structure. To study another aspect of covariance structure changes of the posterior samples generated from the baseline and contaminated models, we can use Algorithm \ref{algo:TangentPCA} to perform tPCA. The eigenvalues obtained from the diagonal matrix $\Sigma$ provide the amount of variation in the different principal directions estimated from the given posterior samples. Let $\lambda_0^{(1)}, \dots, \lambda_0^{(d)}$ and $\lambda_{s_k}^{(1)}, \dots, \lambda_{s_k}^{(d)}$ denote the first $d$ eigenvalues corresponding to the $d$ principal directions of the baseline and perturbed models, respectively. The total variance captured by the $d$ directions is given by the sum of the $d$ eigenvalues in each model, i.e., $TV_0 = \sum_{j=1}^{d} \lambda_0^{(j)}$ and $TV_{s_k} = \sum_{j=1}^{d} \lambda_{s_k}^{(j)}$. The percentage of variance explained by the first $l$ principal components can be found by dividing the sum of the first $l$ eigenvalues by the total sum of all of the eigenvalues, e.g., $\dfrac{\sum_{j=1}^{l} \lambda_0^{(j)}}{TV_0}$ for the baseline model. Let $\boldsymbol{\omega_0} = \Bigg( \dfrac{\lambda_0^{(1)}}{TV_0}, \dfrac{\lambda_0^{(1)} + \lambda_0^{(2)}}{TV_0}, \dots, \dfrac{\sum_{j=1}^{d-1} \lambda_0^{(j)}}{TV_0}, 1 \Bigg)$  and $\boldsymbol{\omega_{s_k}} = \Bigg( \dfrac{\lambda_{s_k}^{(1)}}{TV_{s_k}}, \dfrac{\lambda_{s_k}^{(1)} + \lambda_{s_k}^{(2)}}{TV_{s_k}}, \dots, \dfrac{\sum_{j=1}^{d-1} \lambda_{s_k}^{(j)}}{TV_{s_k}}, 1 \Bigg)$ denote the vectors of such scaled cumulative eigenvalues obtained from the baseline and perturbed posterior models, respectively. Then, the norm of the difference of $\boldsymbol{\omega_0}$ and $\boldsymbol{\omega_{s_k}}$ can be used as a measure of sensitivity to assess the difference in the covariance shape of the two posterior samples:
\begin{equation}
	\mathbb{E} (\pi_0, \pi_{s_k}) = \|\boldsymbol{\omega_0}-\boldsymbol{\omega_{s_k}}\|.
\end{equation}
To show that this measure is bounded above, consider two $d$-dimensional vectors $\boldsymbol{\omega_A}$ and $\boldsymbol{\omega_B}$ of scaled cumulative eigenvalues obtained from models $A$ and $B$ respectively. In model $A$, assume that the first principal direction explains all of the variability, i.e. $\boldsymbol{\omega_A} = (1, 1, \dots, 1)$. For model $B$, consider the other extreme case where each of the $d$ principal directions contributes uniformly to the variance. In other words, each eigenvalue $\lambda_B^{(j)} = \dfrac{1}{d}$ and $\boldsymbol{\omega_B} = \Bigg( \dfrac{1}{d}, \dfrac{2}{d}, \dots, \dfrac{d-1}{d}, 1 \Bigg)$. Then,
\begin{equation}
 \|\boldsymbol{\omega_A}-\boldsymbol{\omega_B}\| = \sqrt{\sum_{j = 1}^{d} \Big(1 - \dfrac{j}{d} \Big)^2} = \sqrt{\sum_{j = 1}^{d-1} \Big(1 - \dfrac{j}{d} \Big)^2}.
\end{equation}
Thus, if we consider $d$ principal directions to compute $\boldsymbol{\omega_0}$ and $\boldsymbol{\omega_{s_k}}$, the measure $\mathbb{E}$ is bounded below by zero and bounded above by $\sqrt{\sum_{j = 1}^{d-1} \Big(1 - \dfrac{j}{d} \Big)^2}$.

\begin{figure}[!t]
	\begin{center}
		\begin{tabular}{|c|c|}
			\hline
			(a) & (b) \\ \hline
			\includegraphics[scale = 0.18]{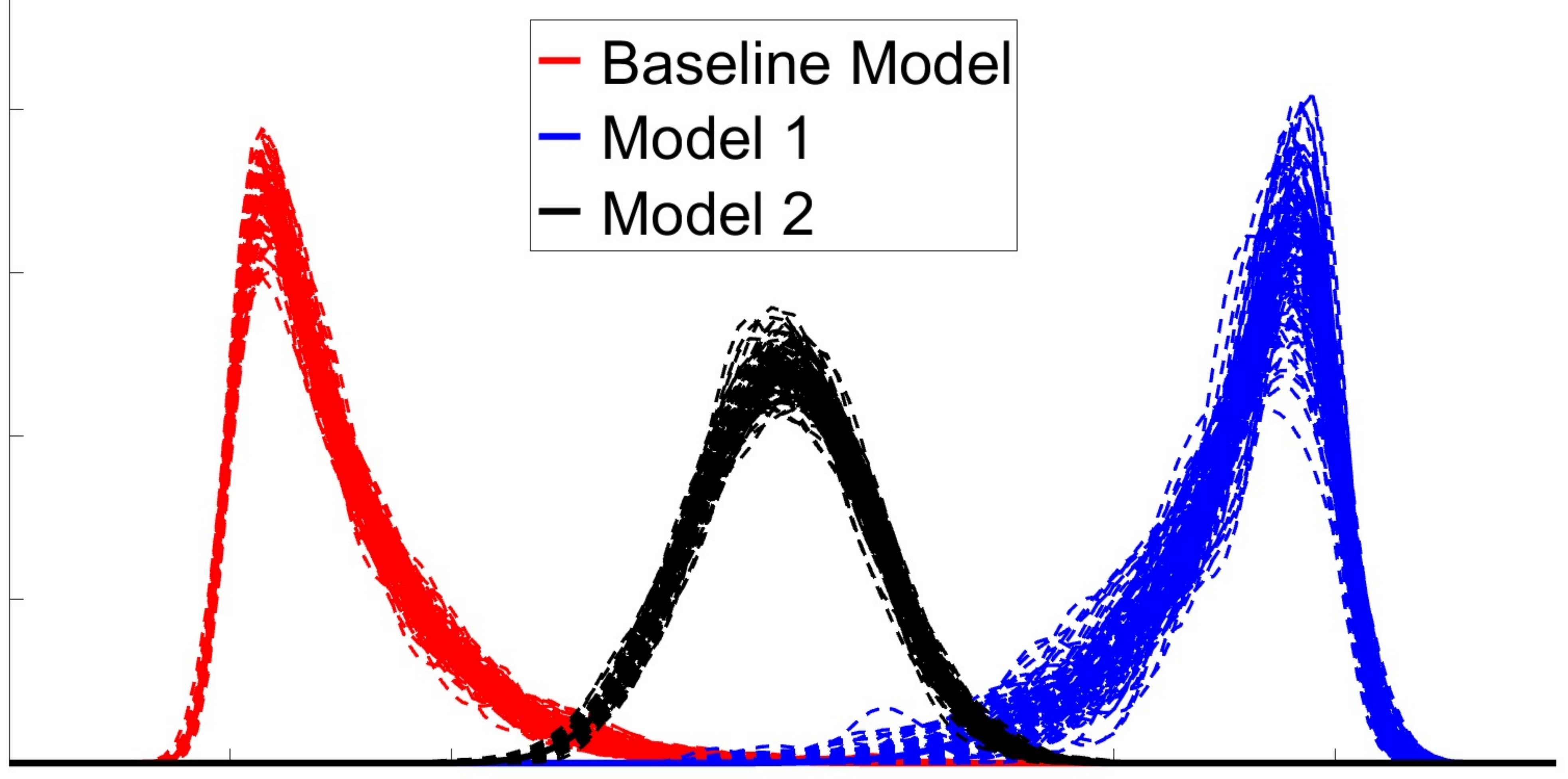} & \includegraphics[scale = 0.18]{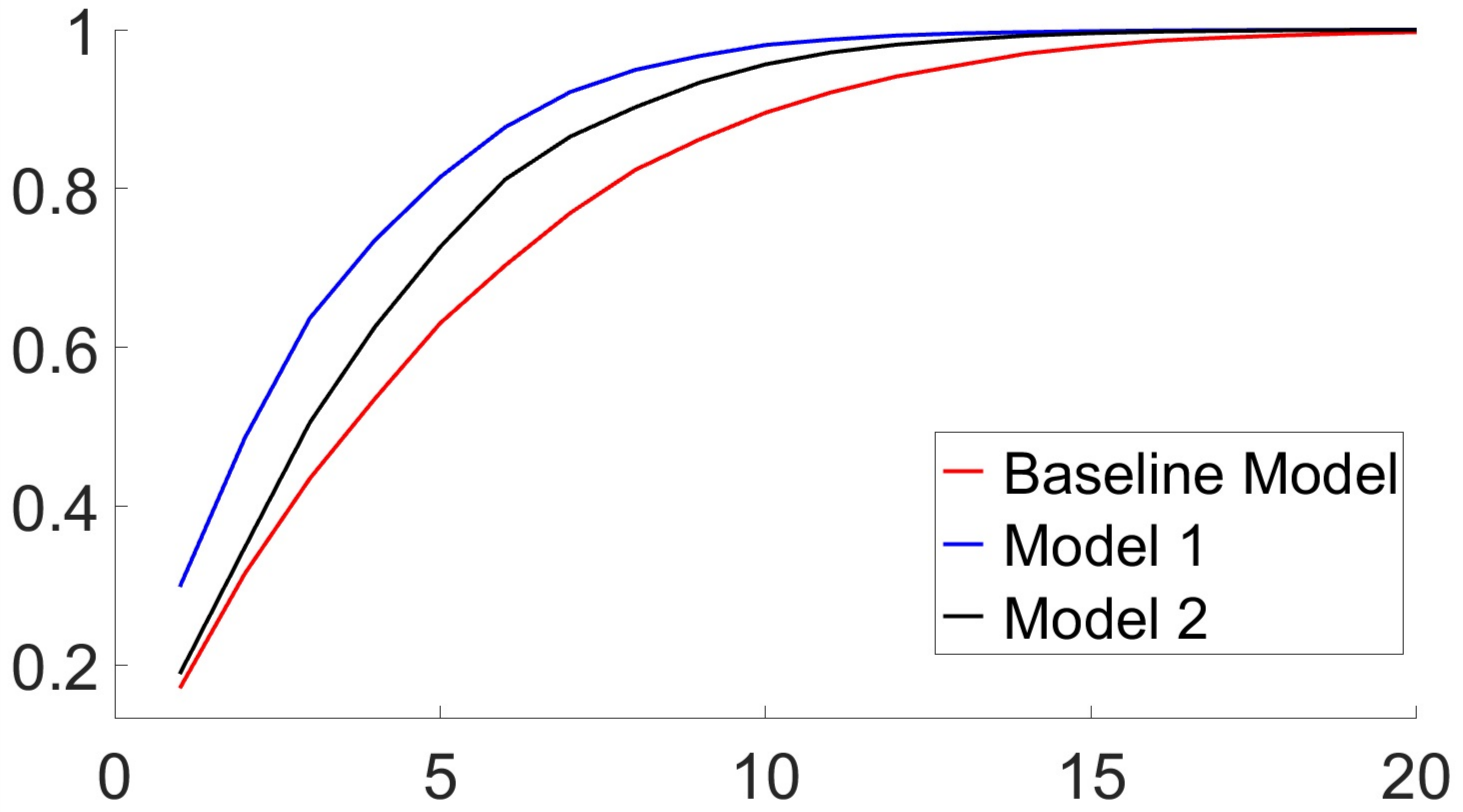} \\ \hline
		\end{tabular}
	\end{center}
	\caption{Example to show the utility of the sensitivity measure $\mathbb{E}$ in capturing the difference in covariance structure between posterior density samples.}
	\label{fig:ExampleCV}
\end{figure}

In Figure \ref{fig:ExampleCV}(a), we consider a baseline model and two perturbed models with different covariance structures. Note that Models 1 and 2 in this example are unrelated to any of the models considered previously. We take into account the first twenty eigenvalues, obtained via tPCA, for each model, and plot the proportion of cumulative variance explained by the corresponding principal components in panel (b). For Models 1 and 2, the measure $\mathbb{E}$ is equal to $0.5077$ and $0.2570$, respectively, i.e., Model 2 is more similar to the baseline model than Model 1, with respect to the covariance structure of the posterior samples.

	\section{Simulated and Real Data Examples}
	\label{sec:Ex}
	
In this section, we demonstrate the use of the proposed global sensitivity measures. In all examples, we consider a baseline model with fixed parameter settings. Then, to assess sensitivity to different perturbations of the model parameters, we perturb a single parameter holding all of the others fixed. Based on the generated posterior samples from the baseline and perturbed models, we use the three proposed geometric measures of global sensitivity to study the effect of prior perturbations on density estimation. We also plot 95\% bands based on the empirical distribution of the measures for the baseline model, and specific values of the perturbed models; these were computed based on 25 different replicates. The values for which the intervals are drawn have been appropriately marked in all of the plots. The size of the intervals provides us with an idea of the sampling variability of the proposed sensitivity measures for specific values of parameters. First, in Section \ref{subsec:Simu}, we perform simulation studies for all of the nonparametric density estimation models discussed in Section \ref{subsec:NPBDensity}. Next, in Section \ref{subsec:RealData}, we fit the DCV model to three real datasets and assess the sensitivity of the posterior to various perturbations of the model parameters. For each of these real datasets, we fit the DPGMM as well, where we only perturb the parameter $\alpha$.
	
	\subsection{Simulation Studies}
	\label{subsec:Simu}
	
	\subsubsection{DP Model}
	\label{subsubec:DPsimu}
	
In this section, we consider density estimation on a bounded support $[0, 1]$ using the DP model. In particular, the probability measure $G_0$ is chosen to be either a highly skewed $Beta$ distribution or the $Uniform$ distribution on $[0,1]$. We perturb parameters $\alpha$ and $G_0$ for this model. For all of the examples in this section, the simulated data was generated from a $Uniform(0, 1)$ distribution.
	
\begin{figure}[!t]
	\begin{center}
		\begin{tabular}{c|c|c|c|}
			\cline{2-4}
			& $Unif (0, 1)$ & $Beta (1, 5)$ & $Beta (5, 1)$ \\ \hline
			\multicolumn{1}{|c|}{n = 10}  & \includegraphics[scale = 0.18]{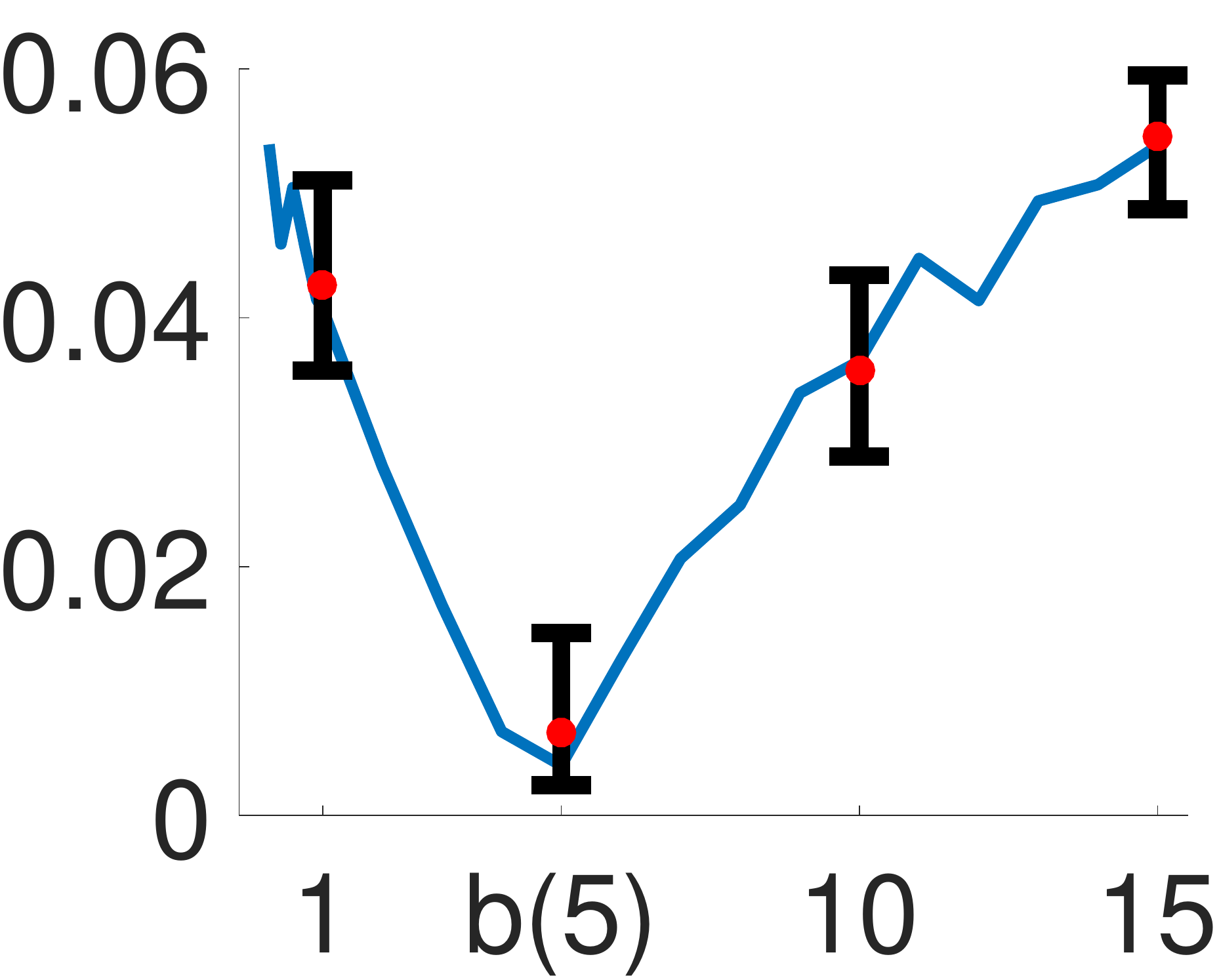}     & \includegraphics[scale = 0.18]{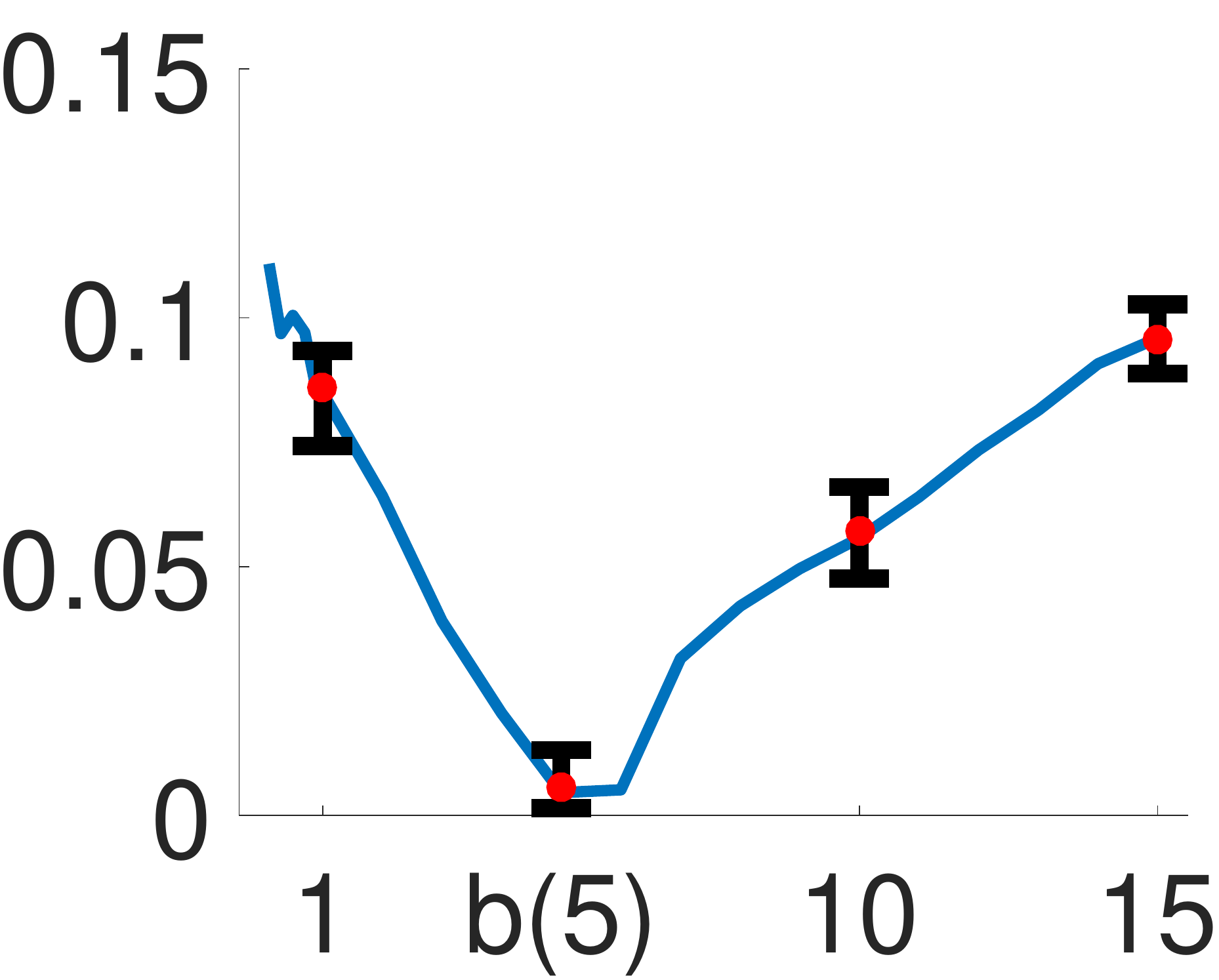}     & \includegraphics[scale = 0.18]{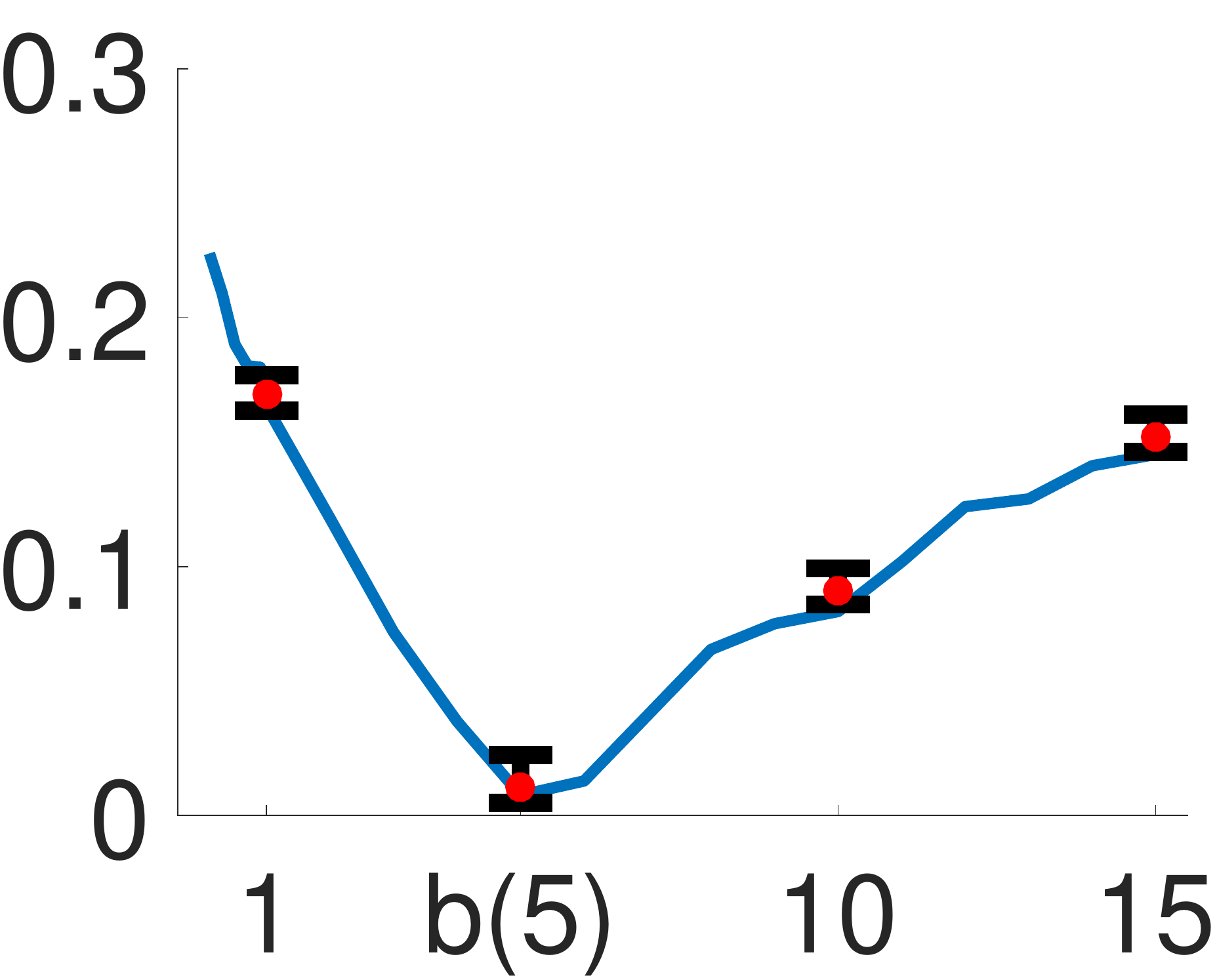}     \\ \hline
			\multicolumn{1}{|c|}{n = 50}  & \includegraphics[scale = 0.18]{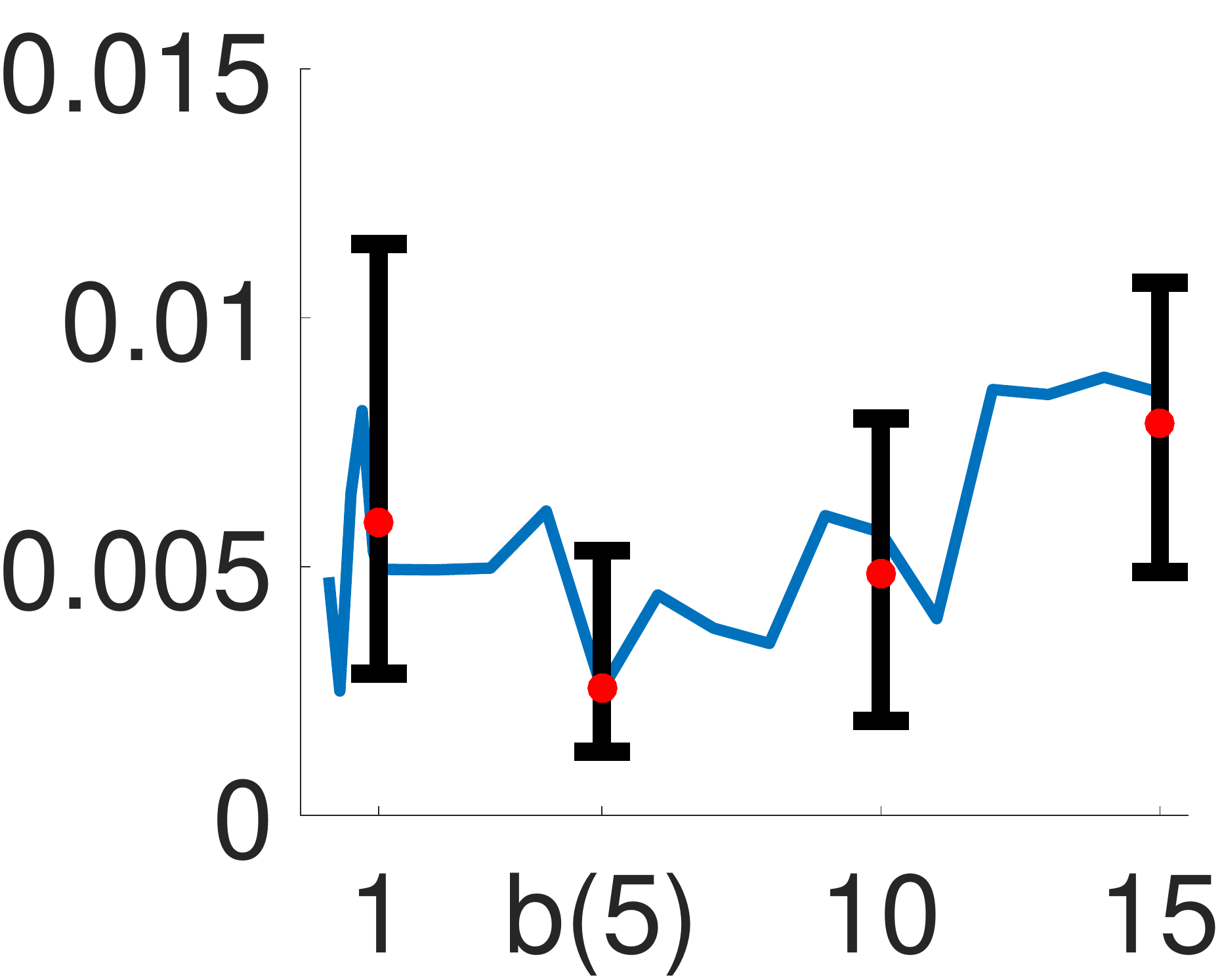}     & \includegraphics[scale = 0.18]{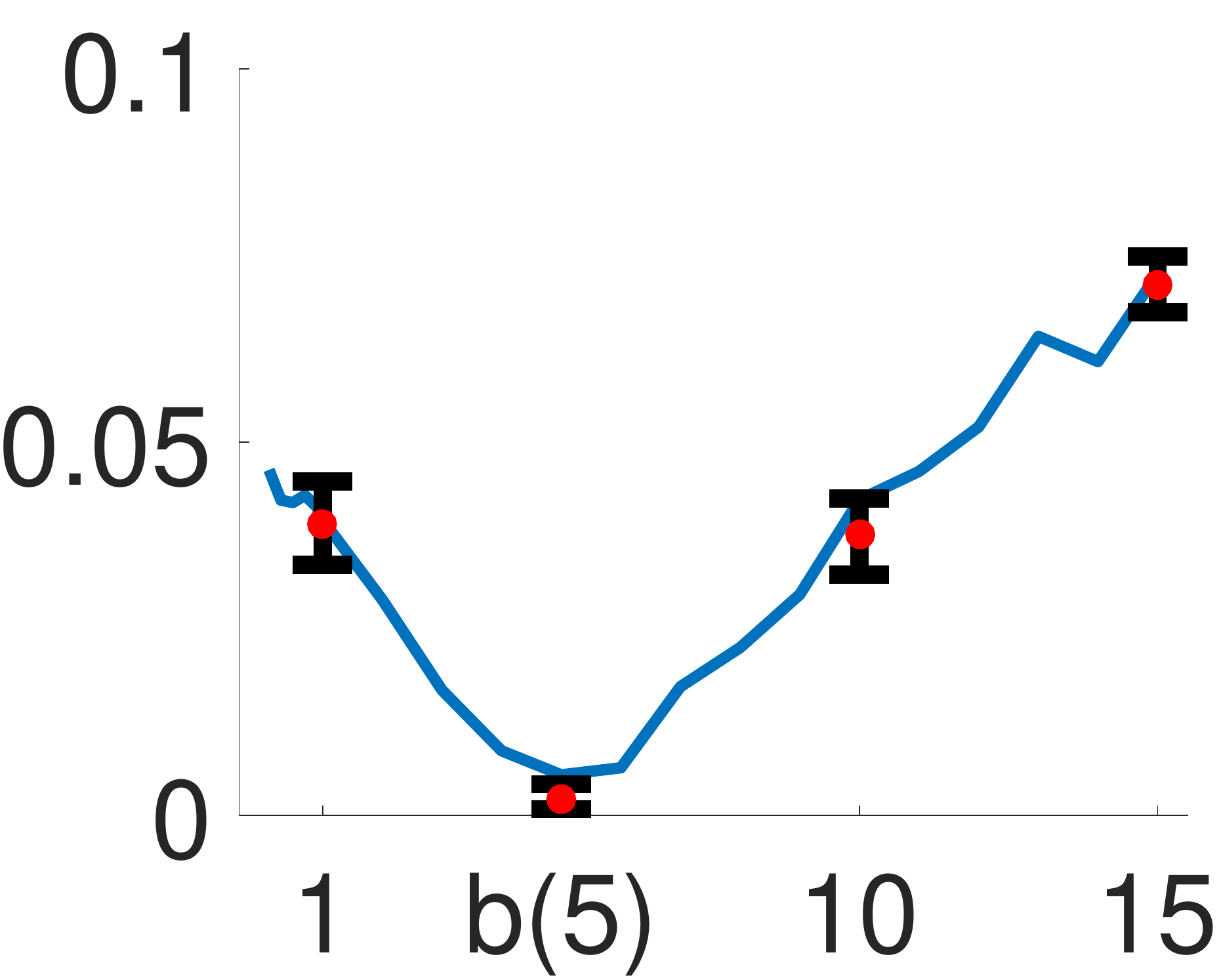}     & \includegraphics[scale = 0.18]{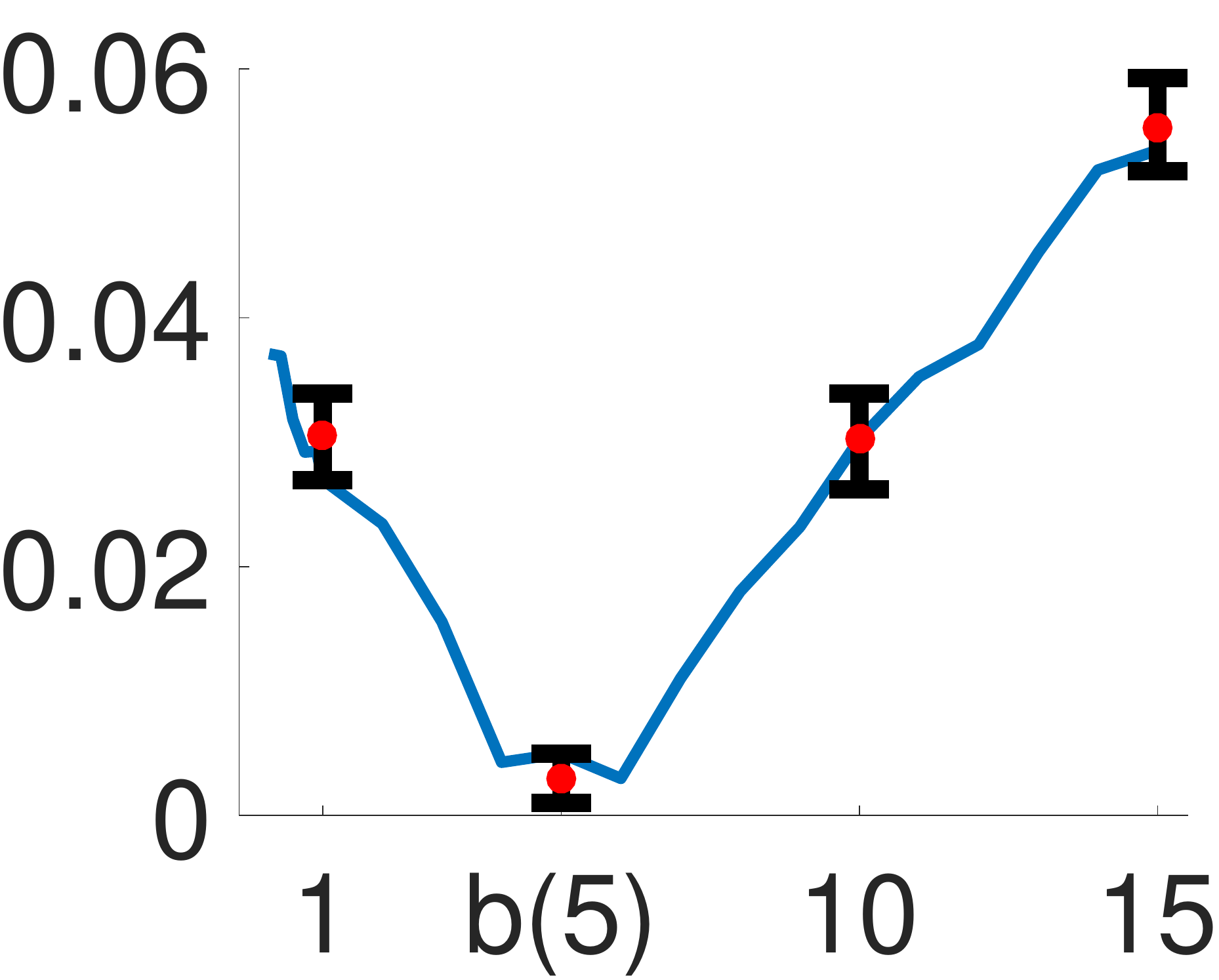}     \\ \hline
			\multicolumn{1}{|c|}{n = 100}  & \includegraphics[scale = 0.18]{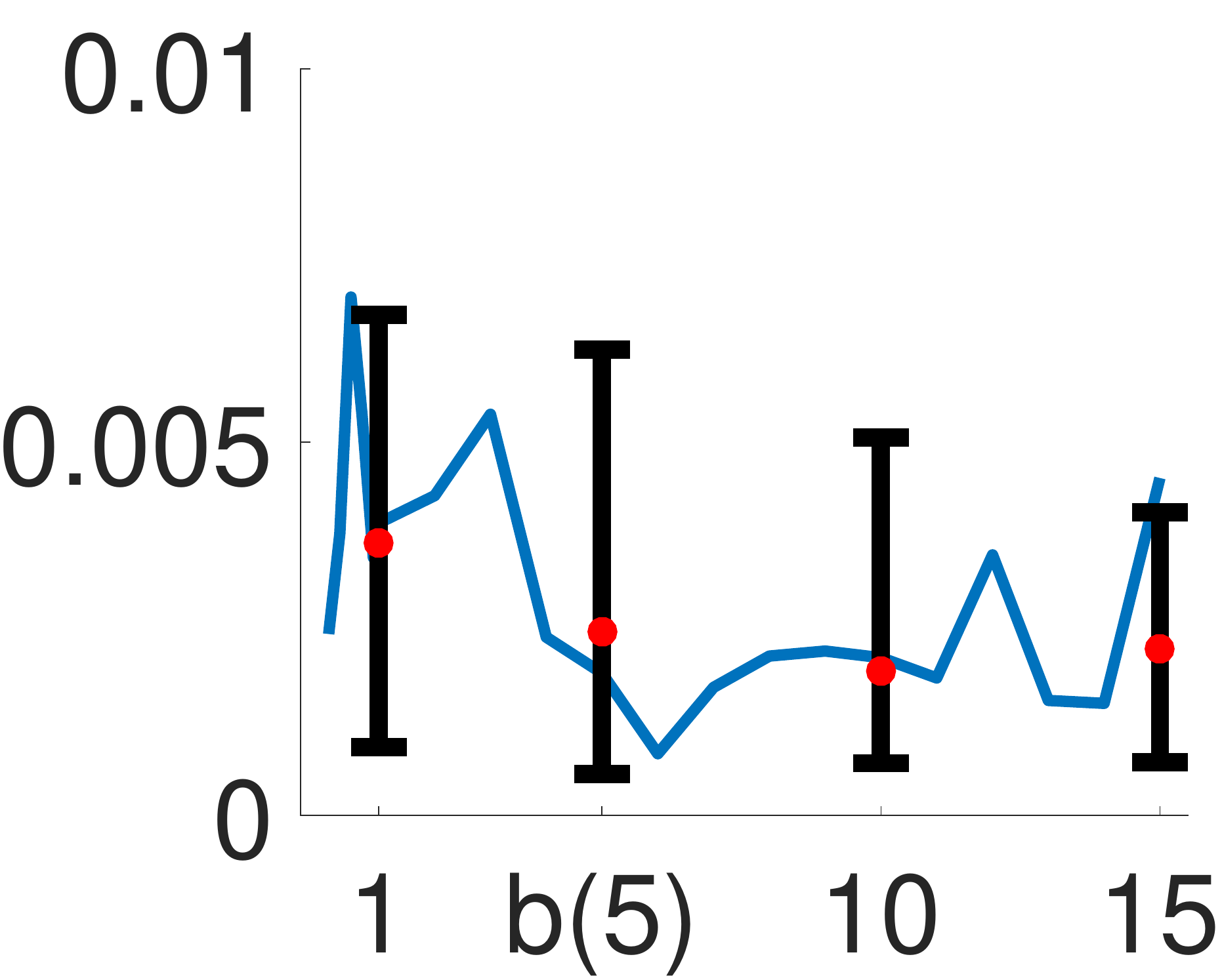}     & \includegraphics[scale = 0.18]{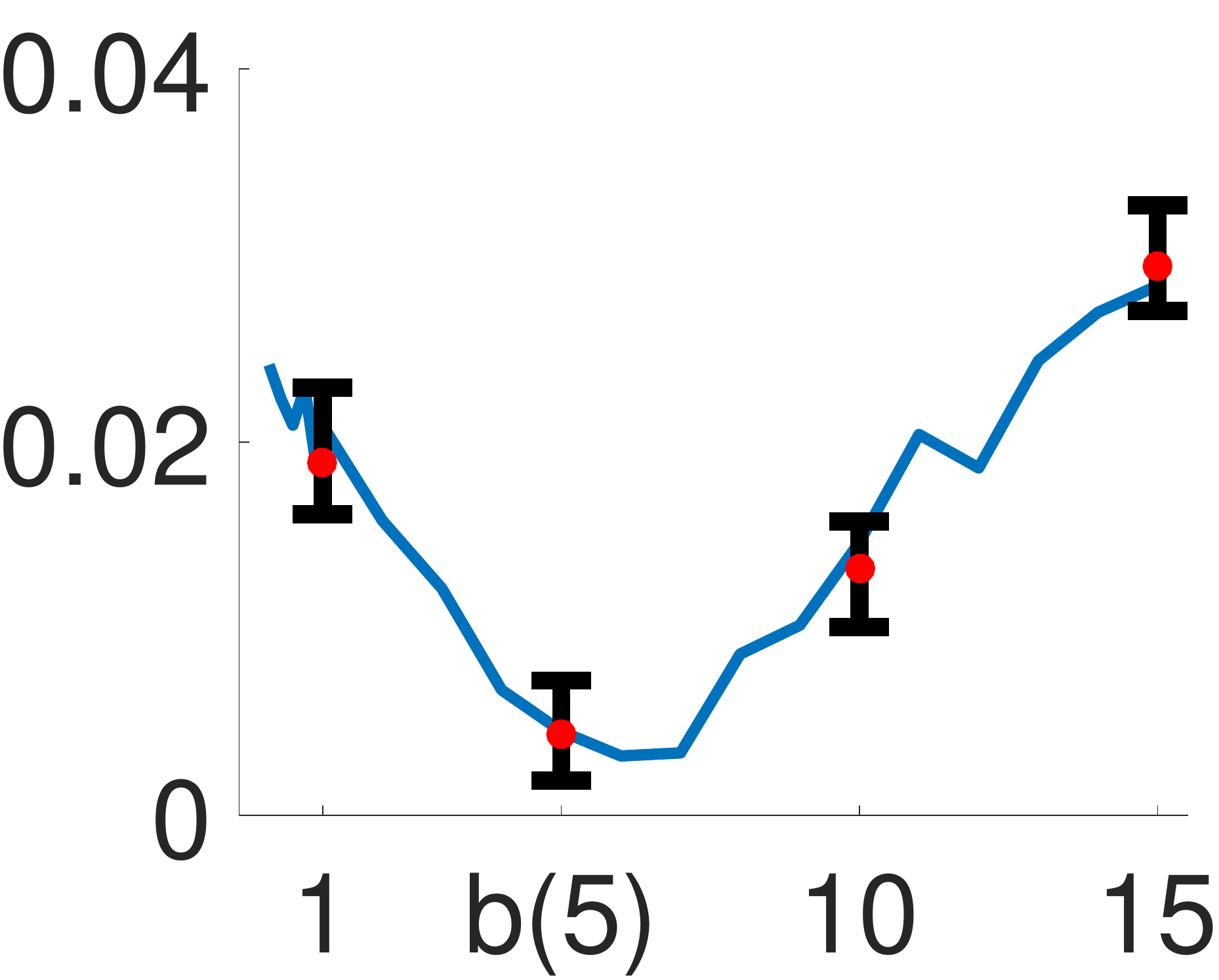}     & \includegraphics[scale = 0.18]{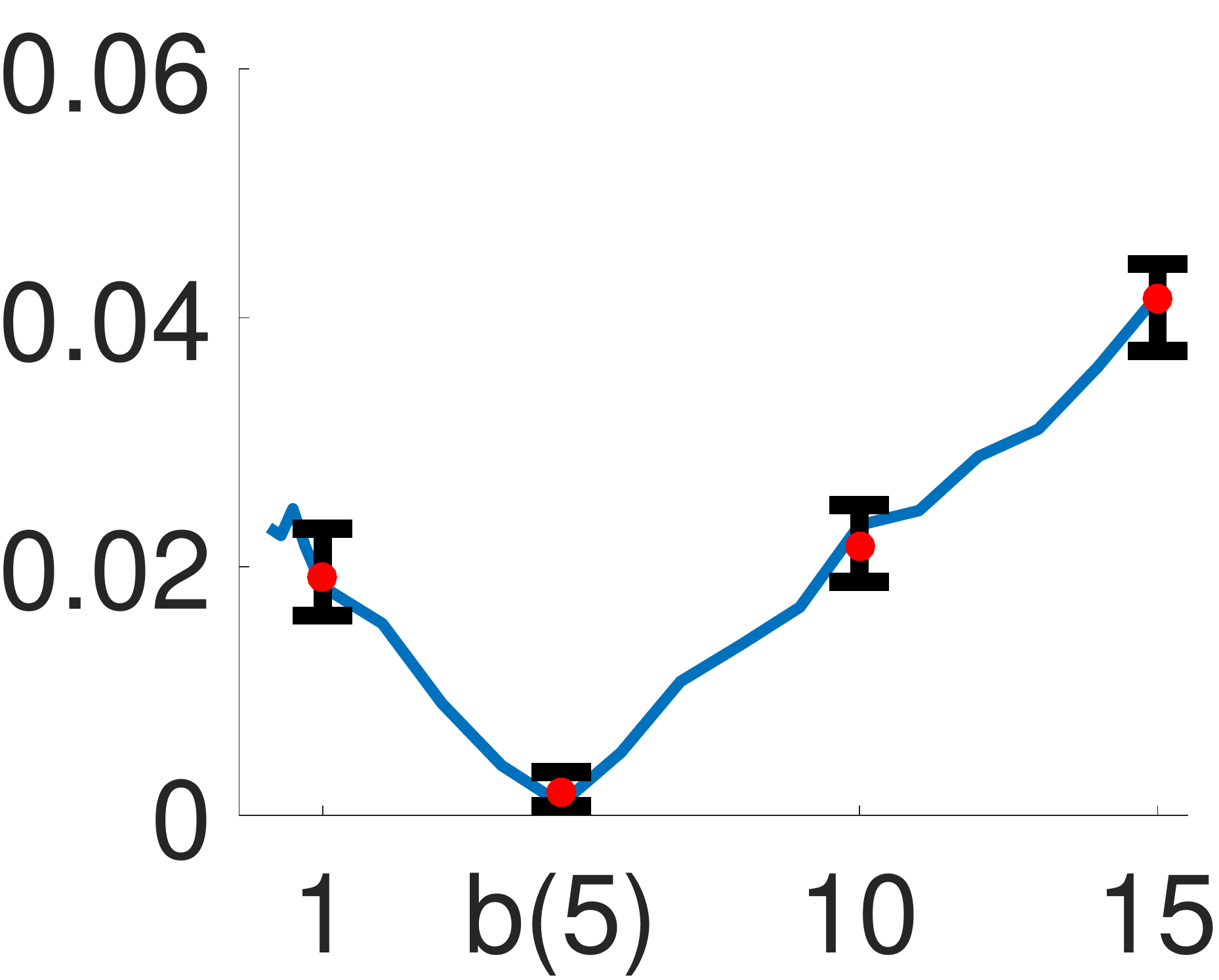}     \\ \hline
		\end{tabular}
	\end{center}
	\caption{Assessment of sensitivity using the shift measure $\mathbb{D}$. We consider perturbations of the parameter $\alpha$ in the DP model for different choices of the probability measure $G_0$ and sample size $n$. The values of $\alpha$ are plotted on the $x$-axis with baseline $\alpha = 5$.}
	\label{fig:DP_alpha}
\end{figure}

\noindent\textbf{Perturbing $\alpha$, $G_0$ fixed}: We choose three different sample sizes $n$, and make three different choices for $G_0$: $Uniform (0, 1)$, $Beta (1, 5)$ and $Beta(5, 1)$. For each choice of $G_0$ and $n$, we perturb the value of $\alpha$ from the baseline model with $\alpha=5$. Figure \ref{fig:DP_alpha} displays the changes in the measure $\mathbb{D}$ for different sample sizes $n$ and probability measures $G_0$. The perturbed values of $\alpha$, shown on the $x$-axis, range from 0.1 to 15. Recall (see (\ref{eq:DPModel})) that the posterior in this case is a DP centered at the probability measure $\dfrac{\alpha}{\alpha + n} G_0 + \dfrac{n}{\alpha + n} F_n$, which is a convex combination of the prior probability measure $G_0$ and the empirical distribution $F_n$. Since the weight associated with the empirical distribution is proportional to the sample size $n$, we expect $F_n$ to dominate the centering probability measure of the posterior as we increase the sample size $n$ relative to $\alpha$ (the converse is true when $\alpha$ is large relative to $n$). In general, for large sample sizes, we expect the posterior DP to be fairly robust to perturbations of $\alpha$. This behavior can be clearly noticed in all examples presented in Figure \ref{fig:DP_alpha}, where we see that the scale of the sensitivity measure $\mathbb{D}$ capturing the difference in shift of the baseline and perturbed posteriors decreases as we increase the sample size. From Figure \ref{fig:DP_alpha}, we also notice that when $G_0 = Uniform (0, 1)$, the DP model is robust to changes in $\alpha$ for all sample sizes (the scale of the $y$-axis is very small). This again makes sense since in this case $G_0$ and $F_n$ (which is the empirical CDF based on observations from $Uniform (0, 1)$) should be similar, and thus the tradeoff between $\alpha$ and $n$ is not as important. When $G_0 = Beta(1, 5)$ or $G_0 = Beta(5, 1)$, the sensitivity measure $\mathbb{D}$ effectively captures the shift differences between the baseline and perturbed model posterior samples when $n=10$. When the sample size is increased, we notice that the DP model becomes more robust to changes in $\alpha$. These trends again can be explained by considering the two weights associated with $G_0$ and $F_n$ in the posterior.

\begin{figure}[!t]
	\begin{center}
		\begin{tabular}{c|c|c|c|}
			\cline{2-4}
			& $Unif (0, 1)$ & $Beta (1, 5)$ & $Beta (5, 1)$ \\ \hline
			\multicolumn{1}{|c|}{n = 10}  & \includegraphics[scale = 0.18]{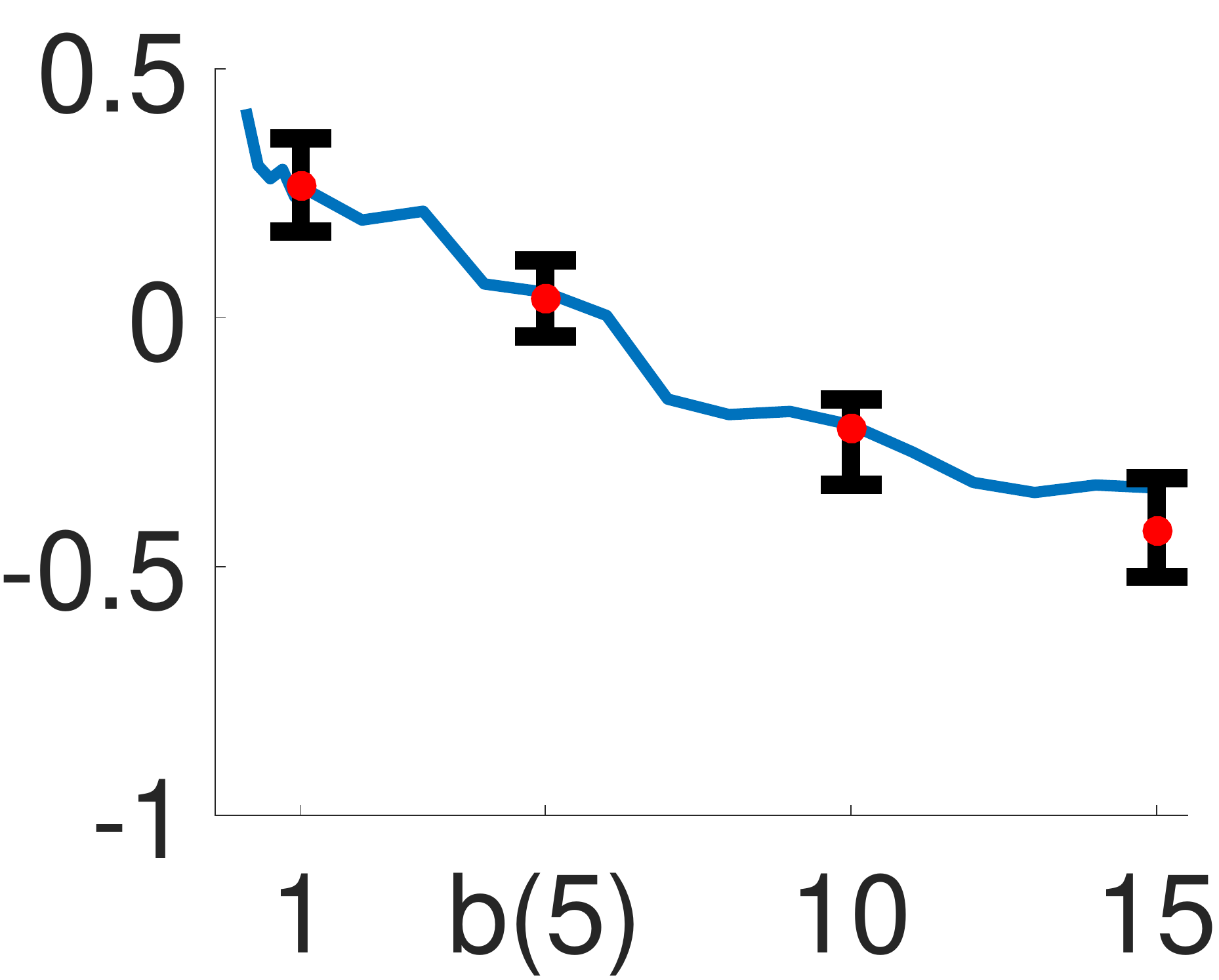}     & \includegraphics[scale = 0.18]{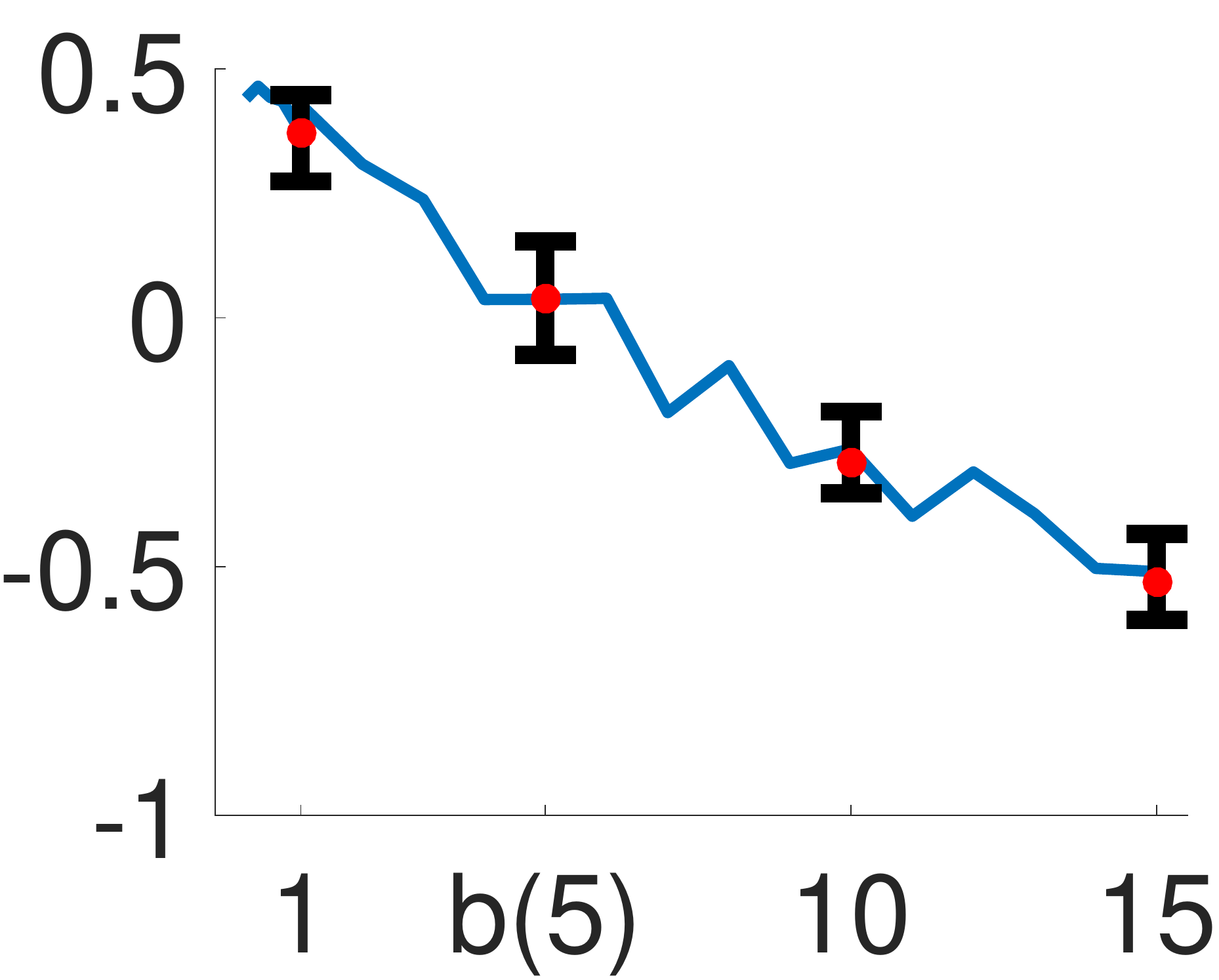}     & \includegraphics[scale = 0.18]{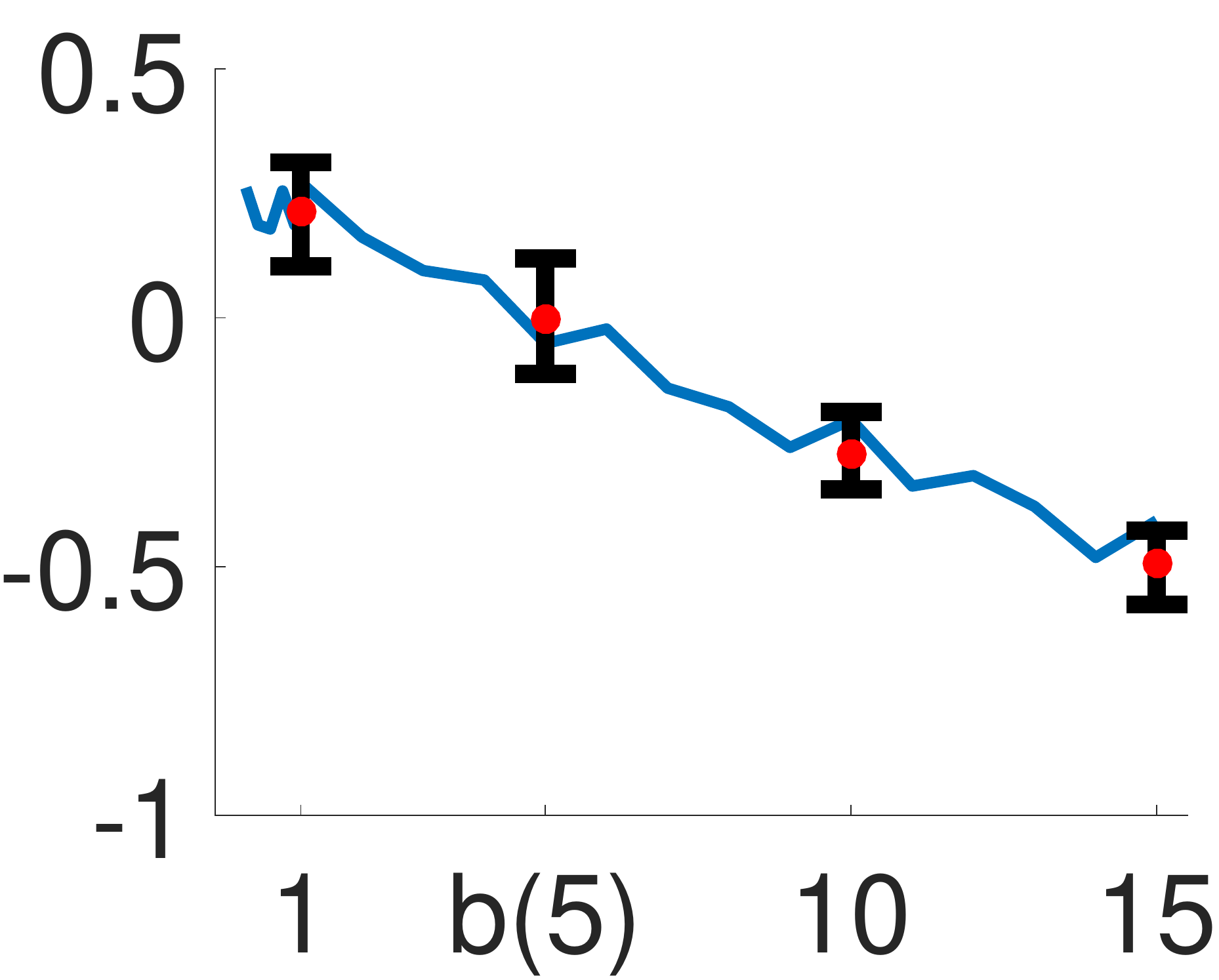}     \\ \hline
			\multicolumn{1}{|c|}{n = 50}  & \includegraphics[scale = 0.18]{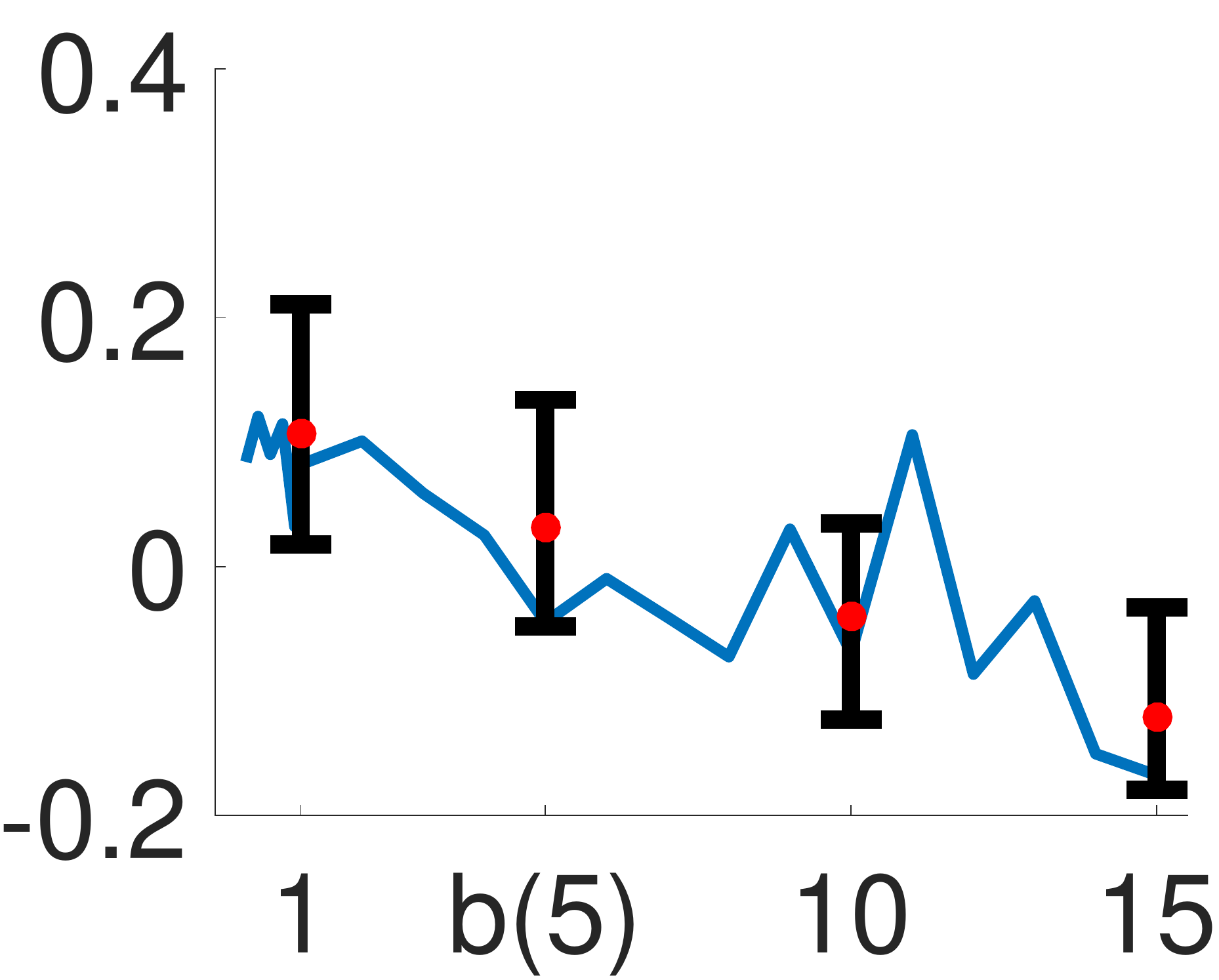}     & \includegraphics[scale = 0.18]{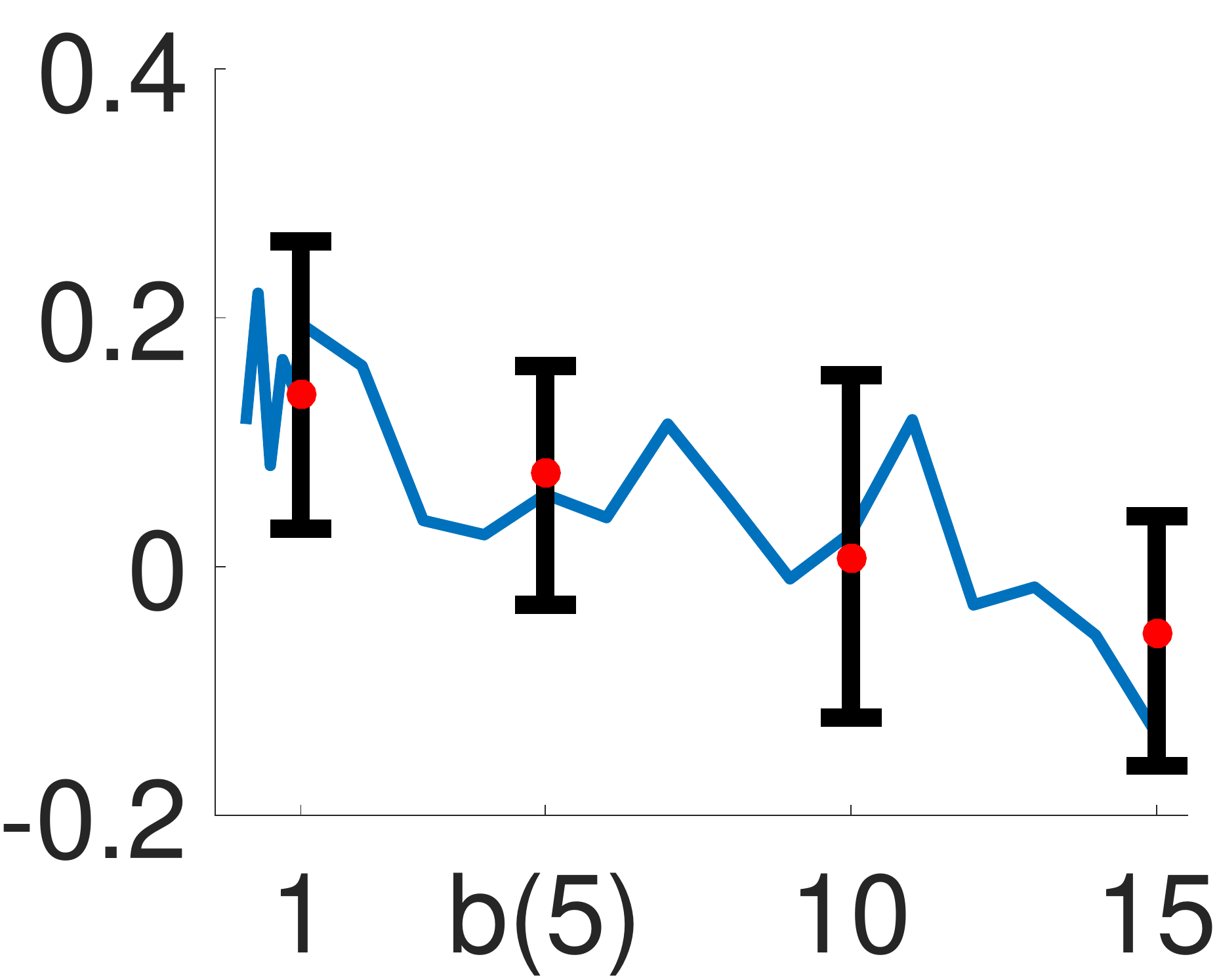}     & \includegraphics[scale = 0.18]{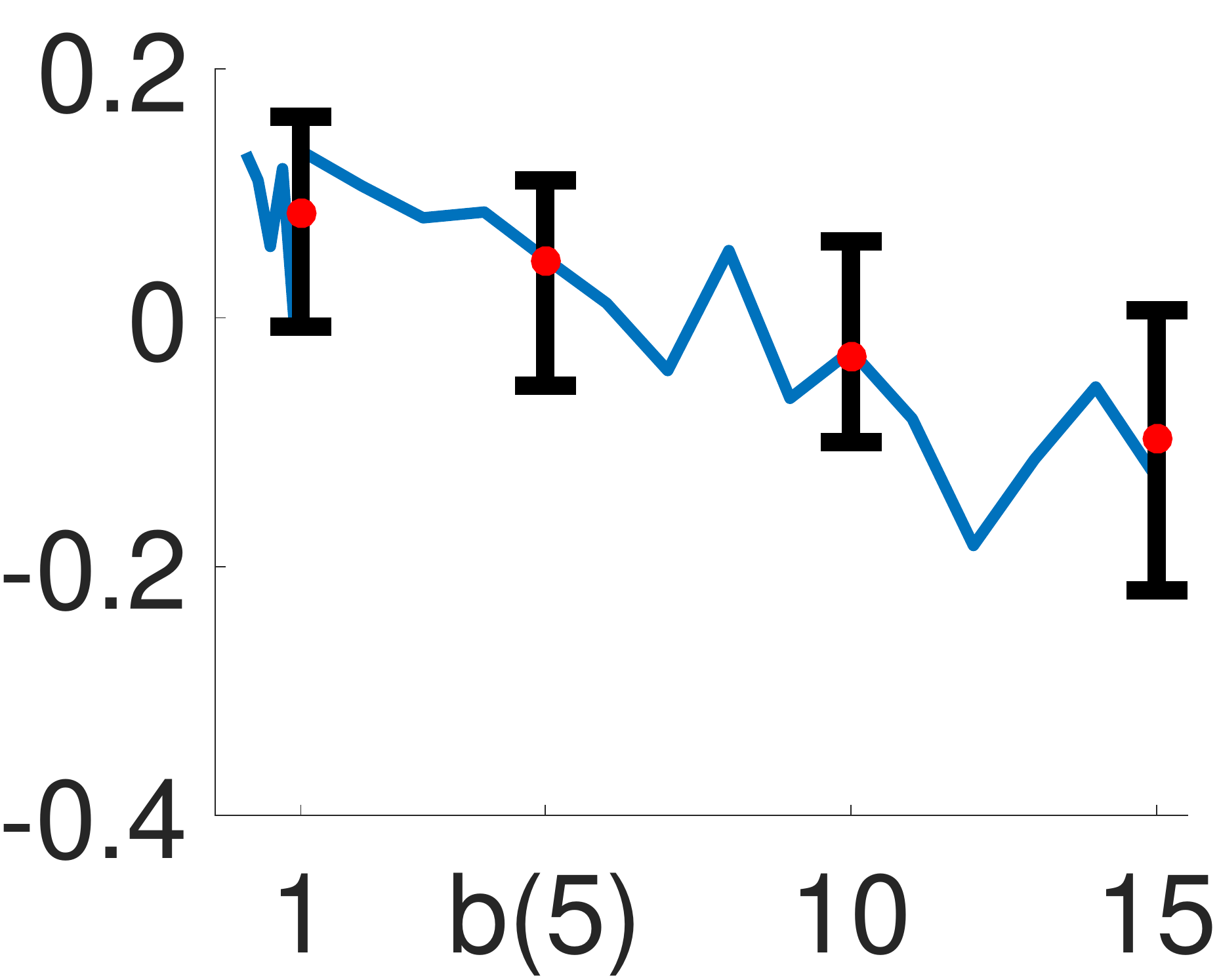}     \\ \hline
			\multicolumn{1}{|c|}{n = 100}  & \includegraphics[scale = 0.18]{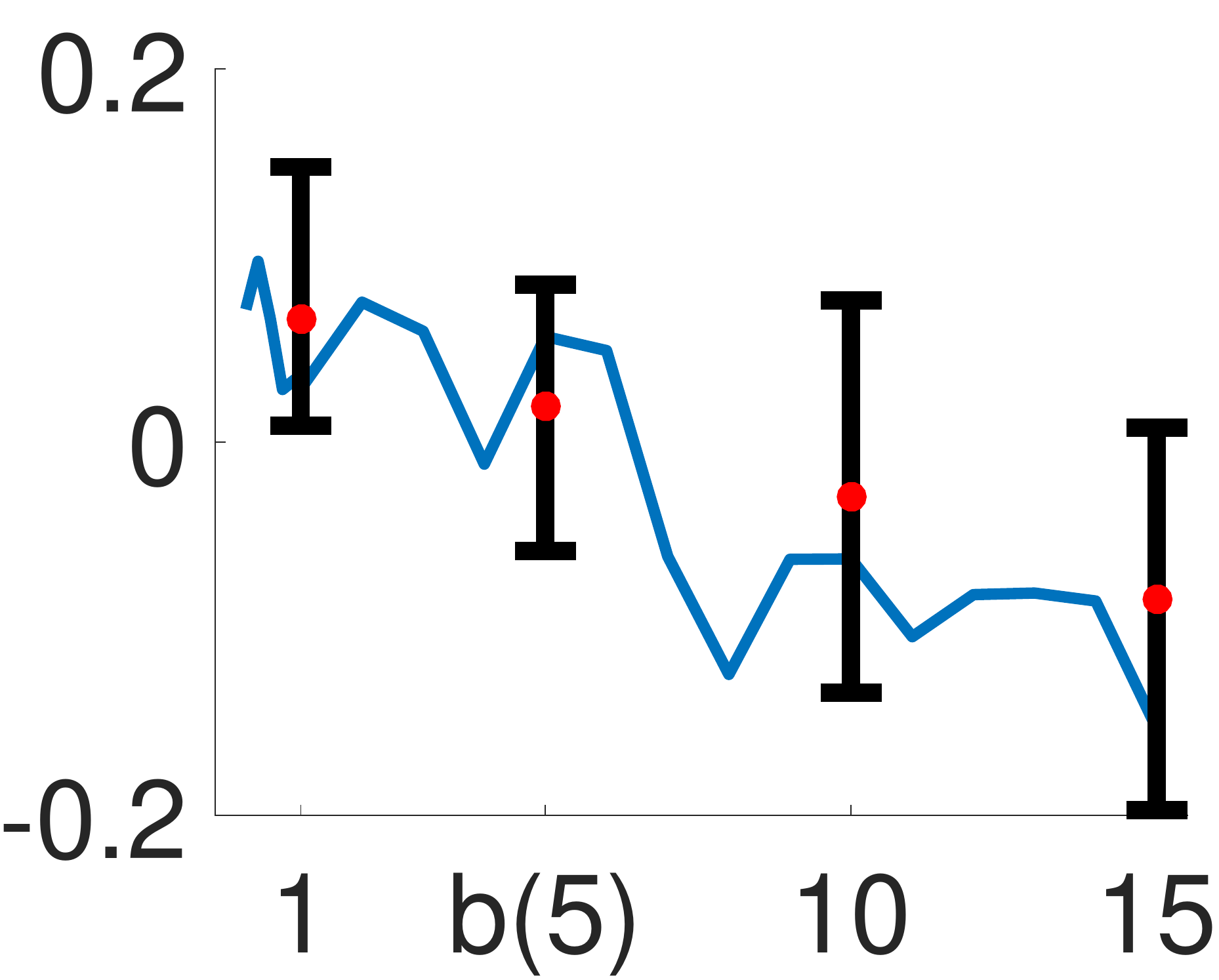}     & \includegraphics[scale = 0.18]{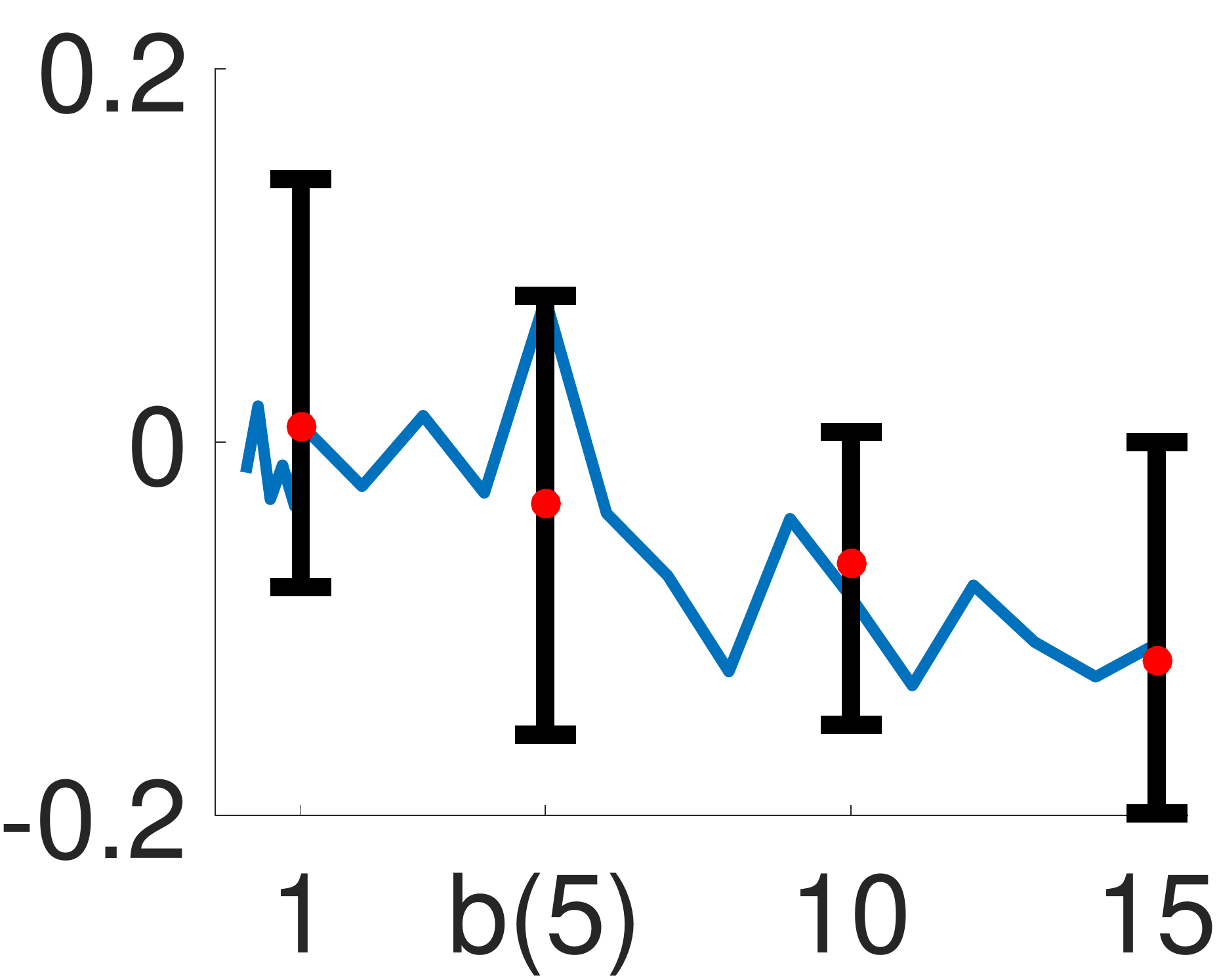}     & \includegraphics[scale = 0.18]{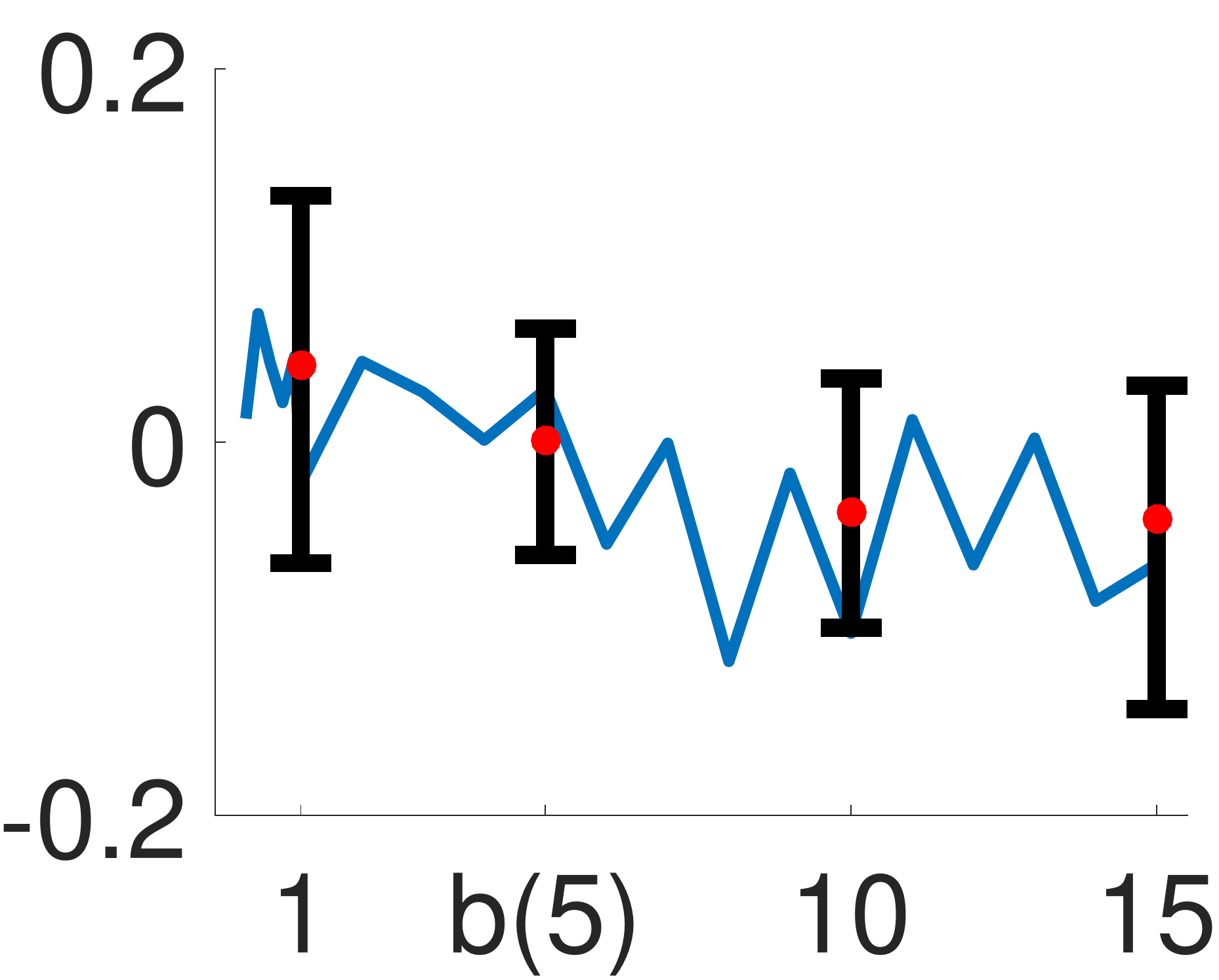}     \\ \hline
		\end{tabular}
	\end{center}
	\caption{Assessment of sensitivity using the spread measure $\mathbb{V}$. We consider perturbations of the parameter $\alpha$ in the DP model for different choices of the probability measure $G_0$ and sample size $n$. The values of $\alpha$ are plotted on the $x$-axis with baseline $\alpha = 5$.}
	\label{fig:DP_alpha_KV}
\end{figure}

\begin{figure}[!t]
	\begin{center}
		\begin{tabular}{c|c|c|c|}
			\cline{2-4}
			& $\alpha = 1$ & $\alpha = 10$ & $\alpha = 25$ \\ \hline
			\multicolumn{1}{|c|}{n = 10}  & \includegraphics[scale = 0.18]{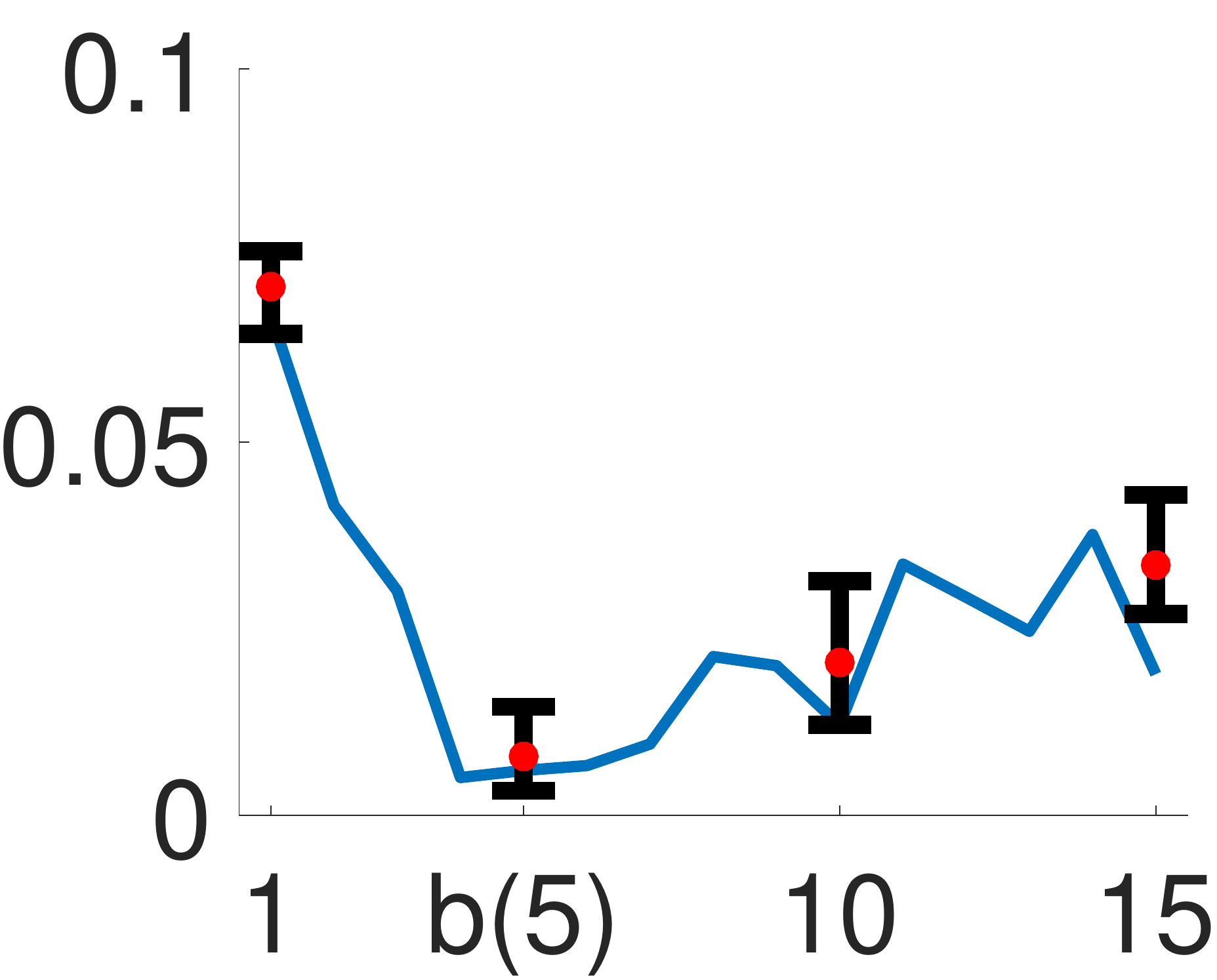}     & \includegraphics[scale = 0.18]{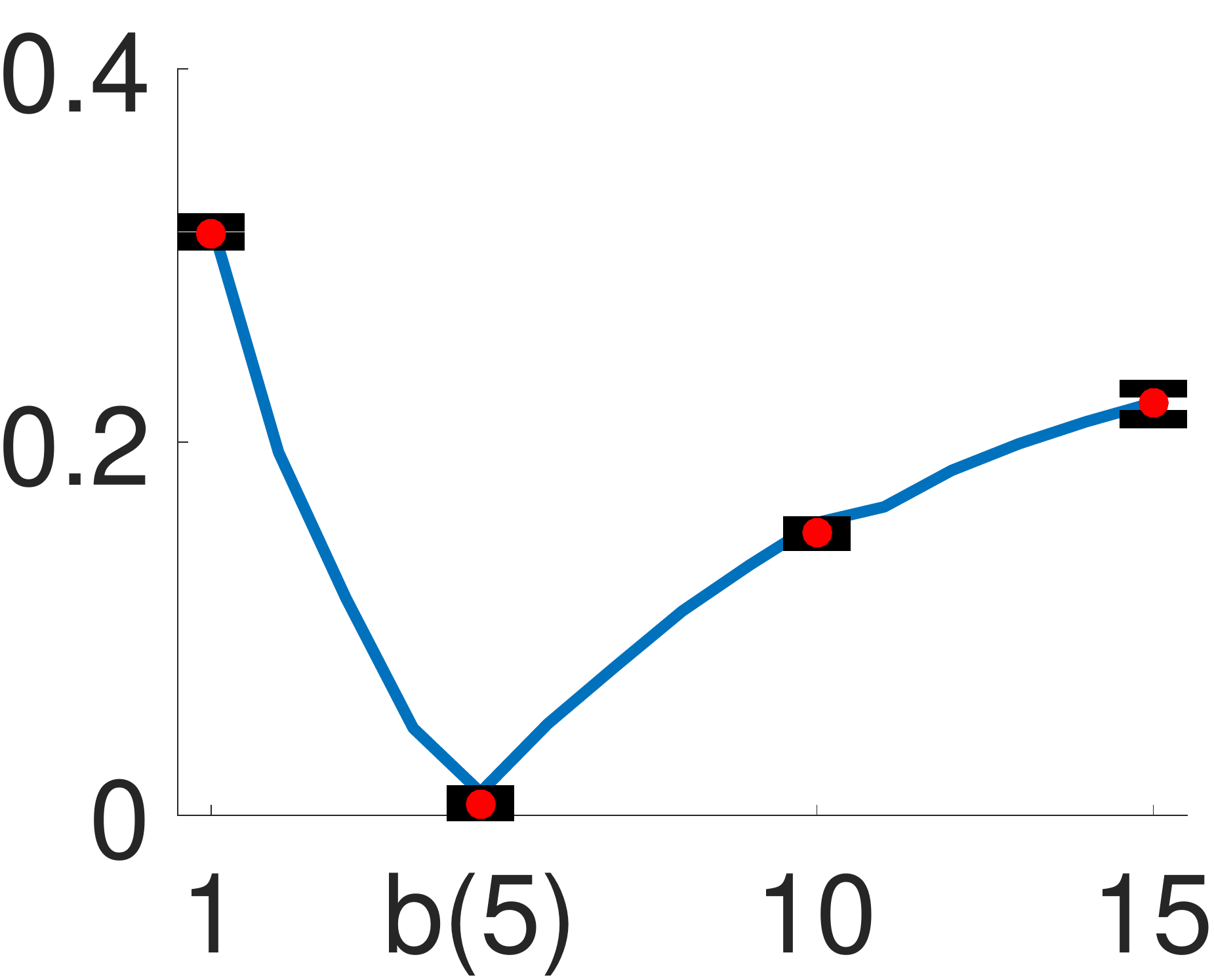}     & \includegraphics[scale = 0.18]{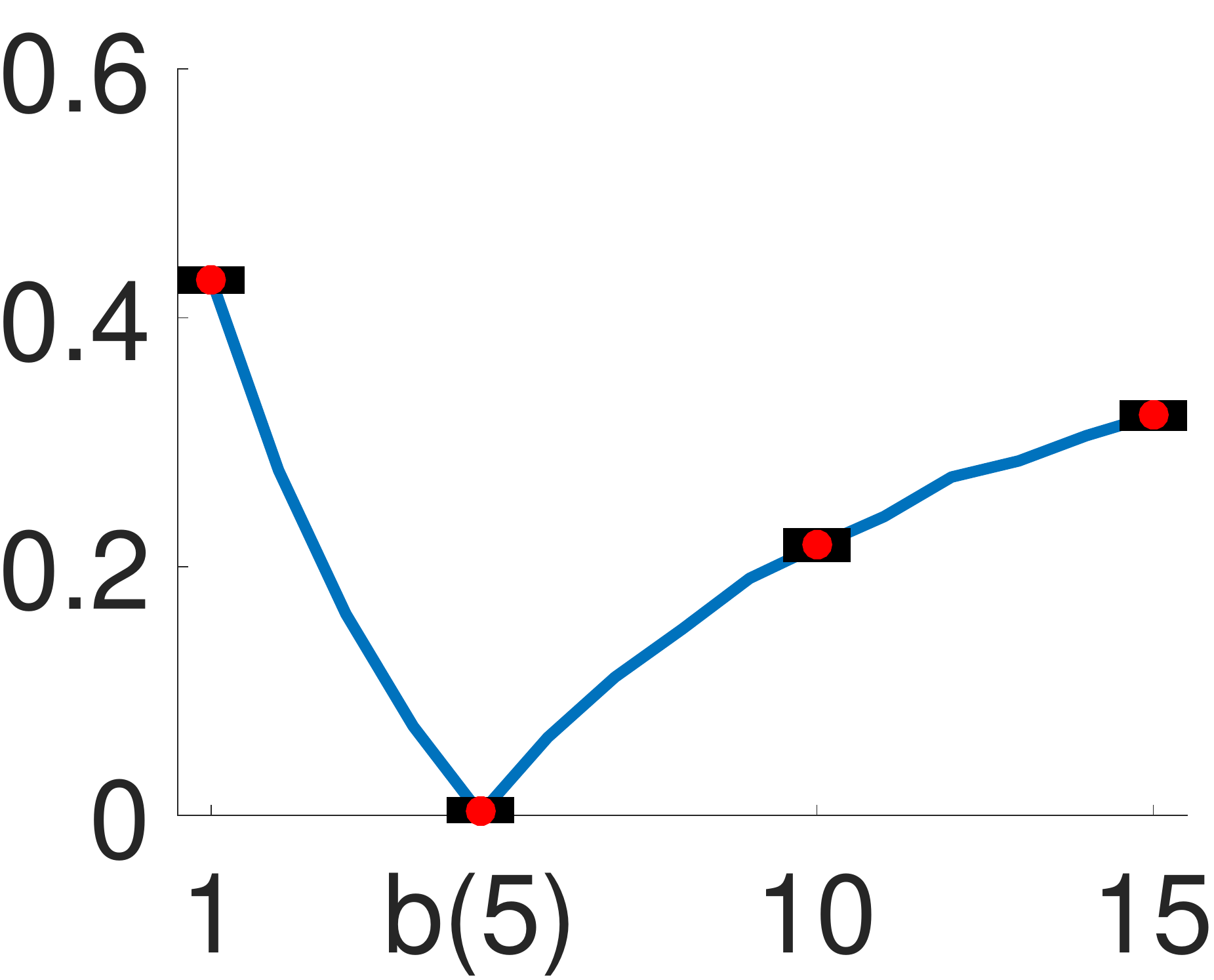}     \\ \hline
			\multicolumn{1}{|c|}{n = 50}  & \includegraphics[scale = 0.18]{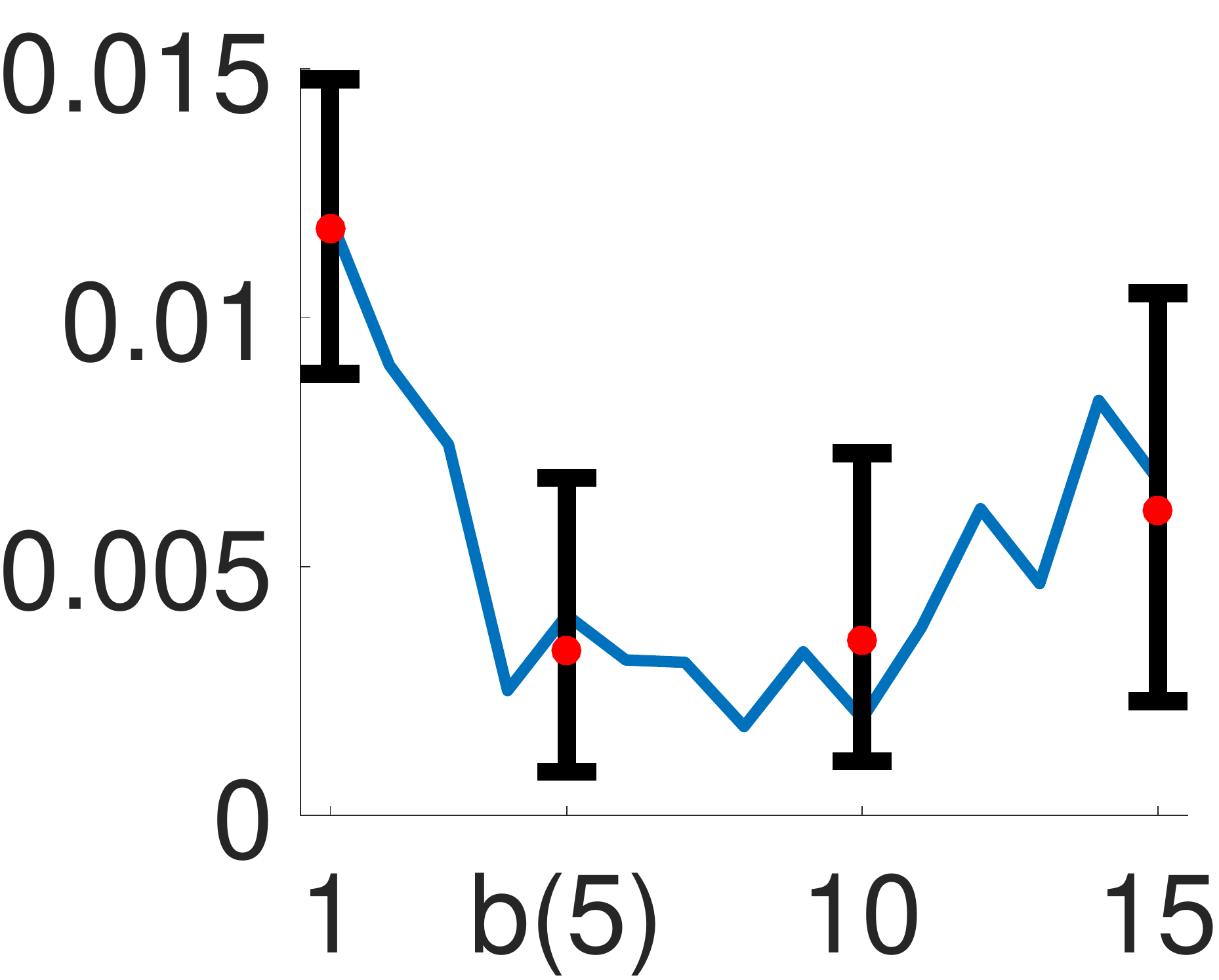}     & \includegraphics[scale = 0.18]{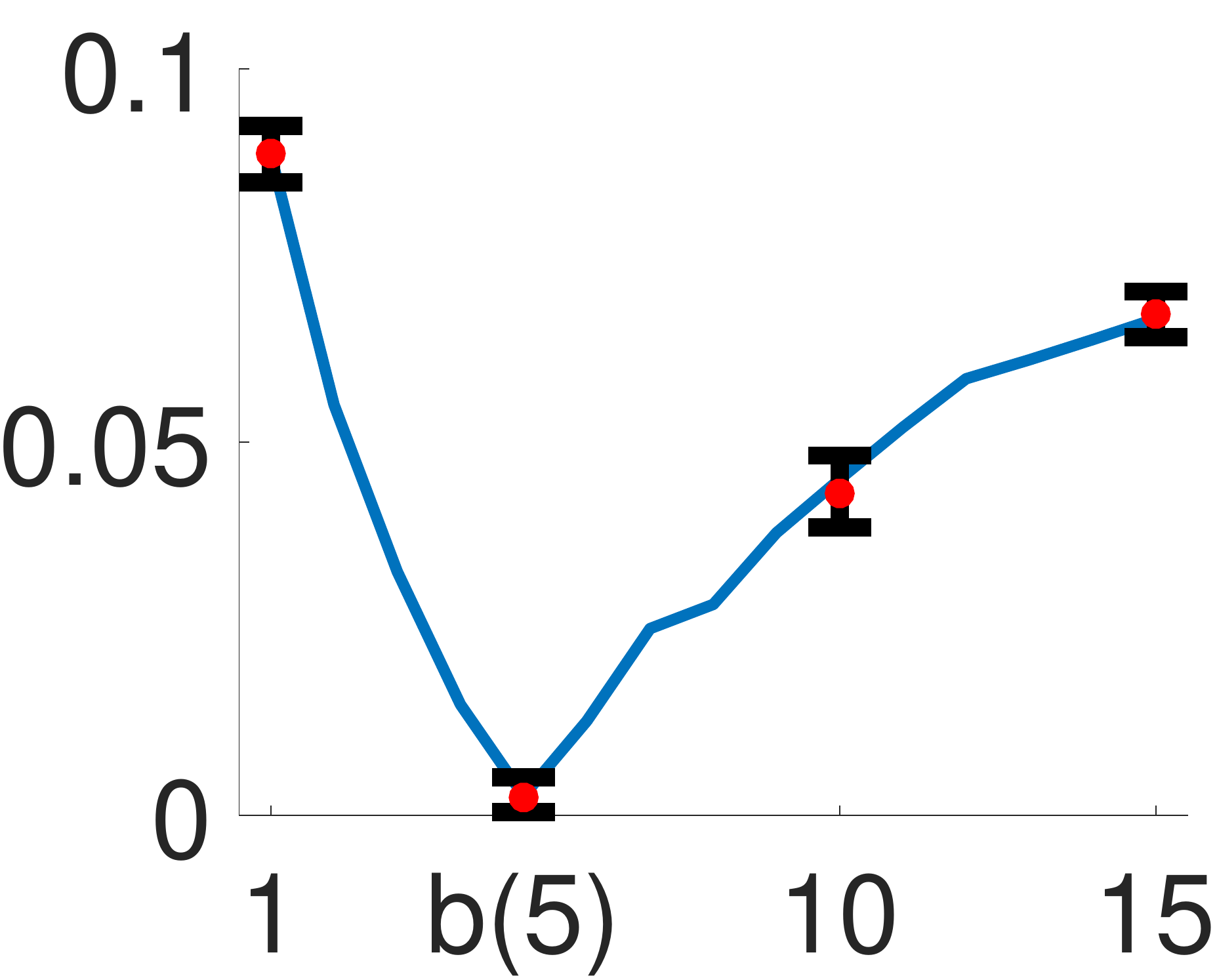}     & \includegraphics[scale = 0.18]{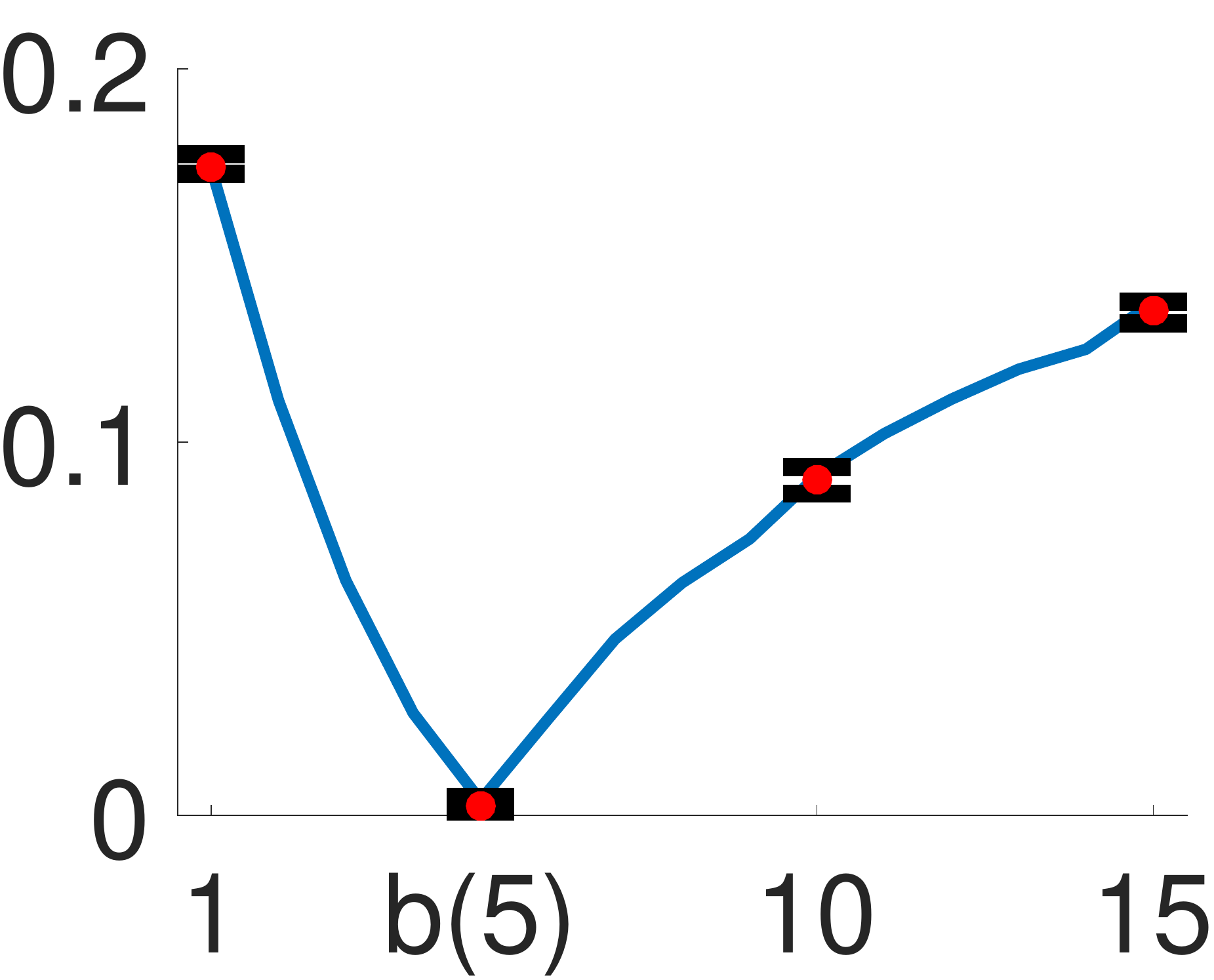}     \\ \hline
			\multicolumn{1}{|c|}{n = 100}  & \includegraphics[scale = 0.18]{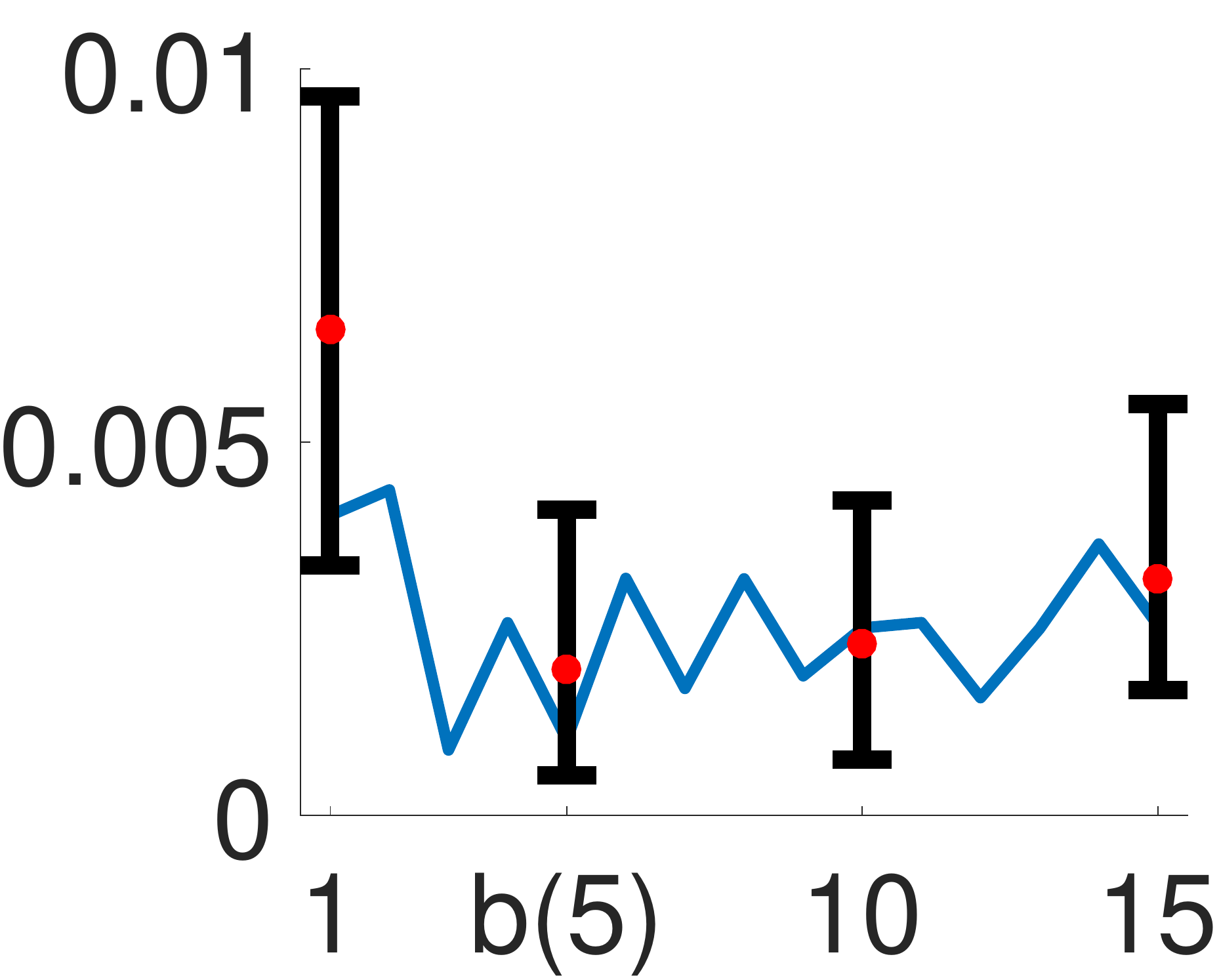}     & \includegraphics[scale = 0.18]{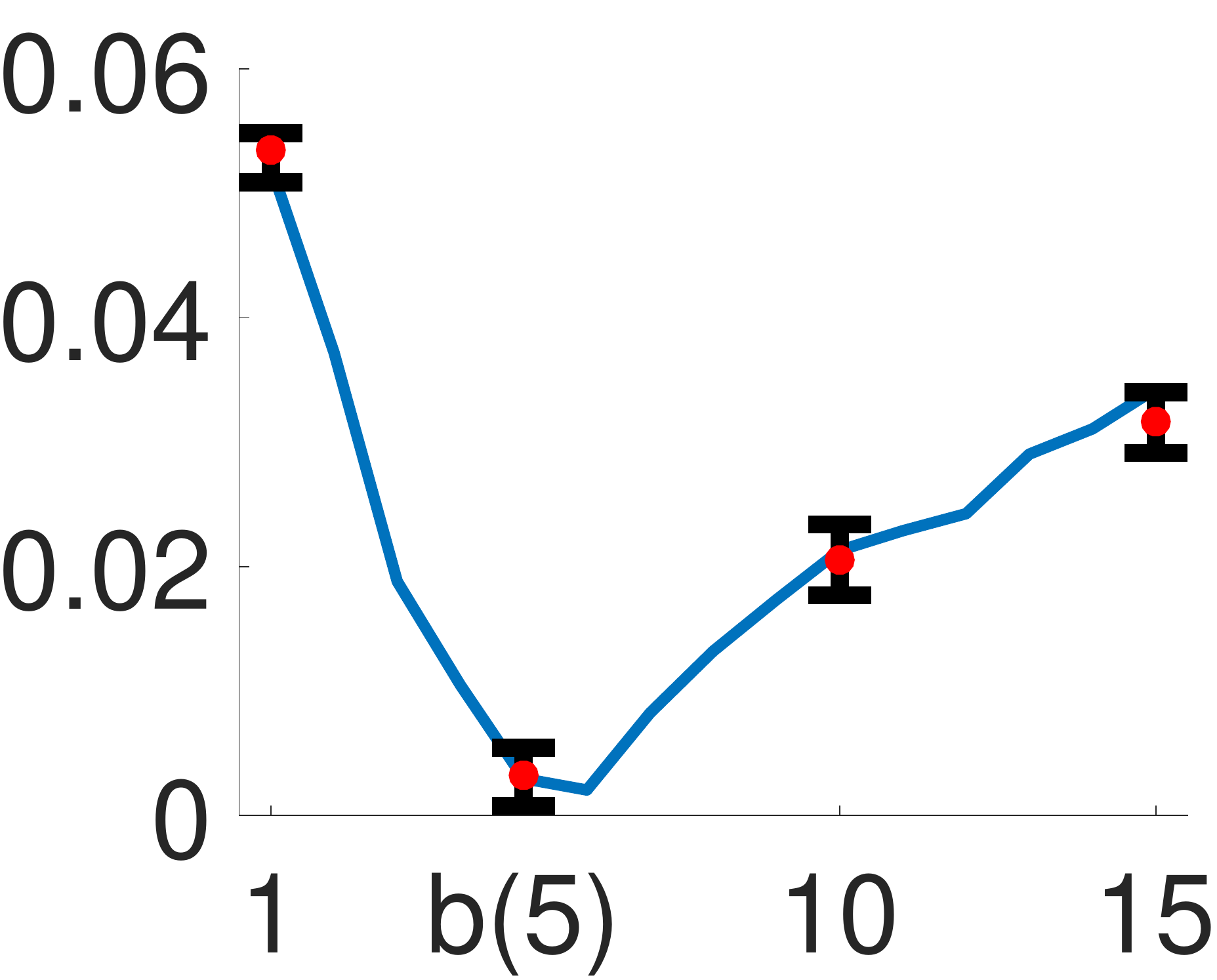}     & \includegraphics[scale = 0.18]{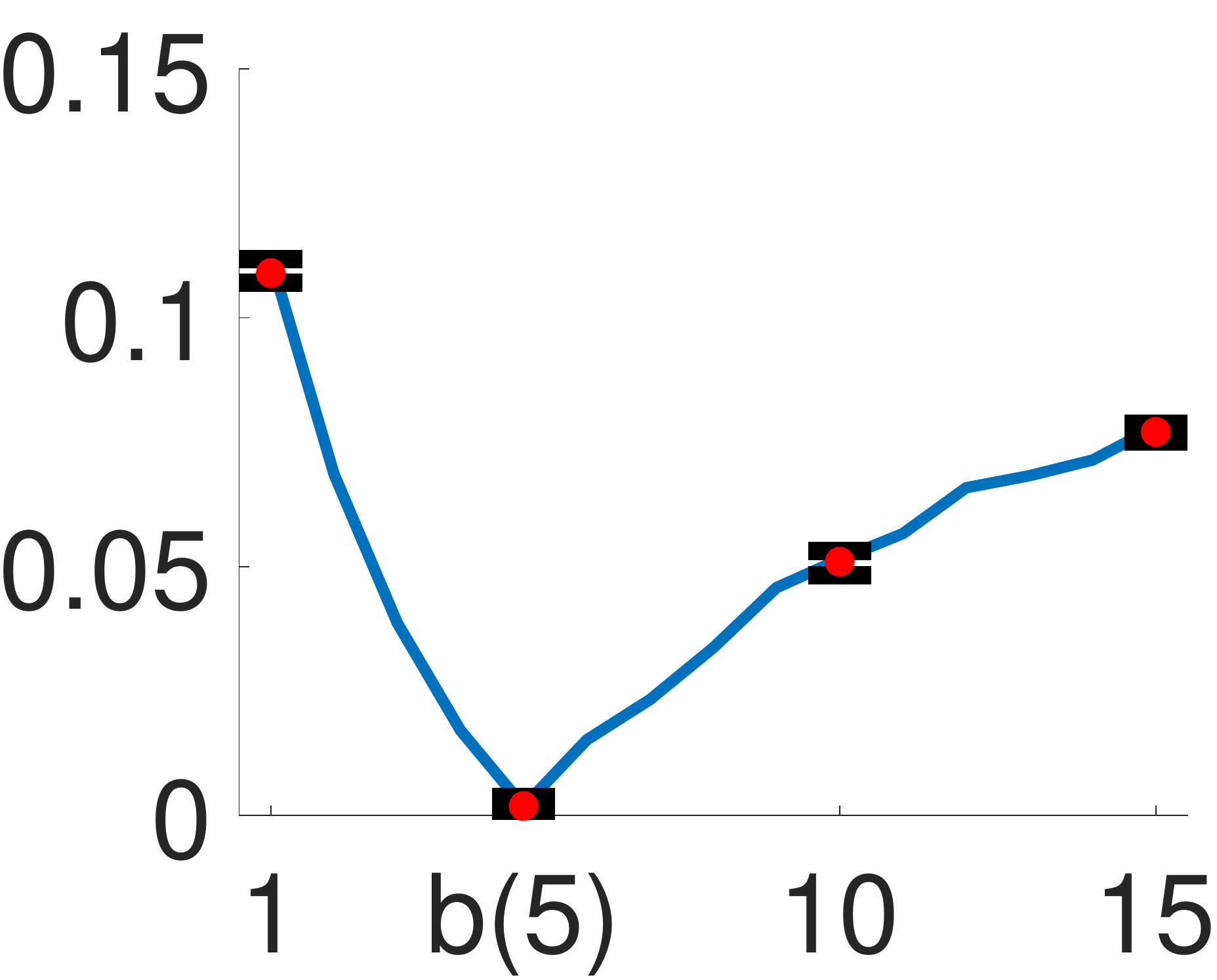}     \\ \hline
		\end{tabular}
	\end{center}
	\caption{Assessment of sensitivity using the shift measure $\mathbb{D}$. We consider perturbations of the centering measure $G_0=Beta(5, b)$ in the DP model for different choices of the the concentration parameter $\alpha$ and sample size $n$. The values of $b$ are plotted on the $x$-axis with baseline $b = 5$.}
	\label{fig:DP_G0}
\end{figure}

\begin{figure}[!t]
	\begin{center}
		\begin{tabular}{c|c|c|c|}
			\cline{2-4}
			& $\alpha = 1$ & $\alpha = 10$ & $\alpha = 25$ \\ \hline
			\multicolumn{1}{|c|}{n = 10}  & \includegraphics[scale = 0.18]{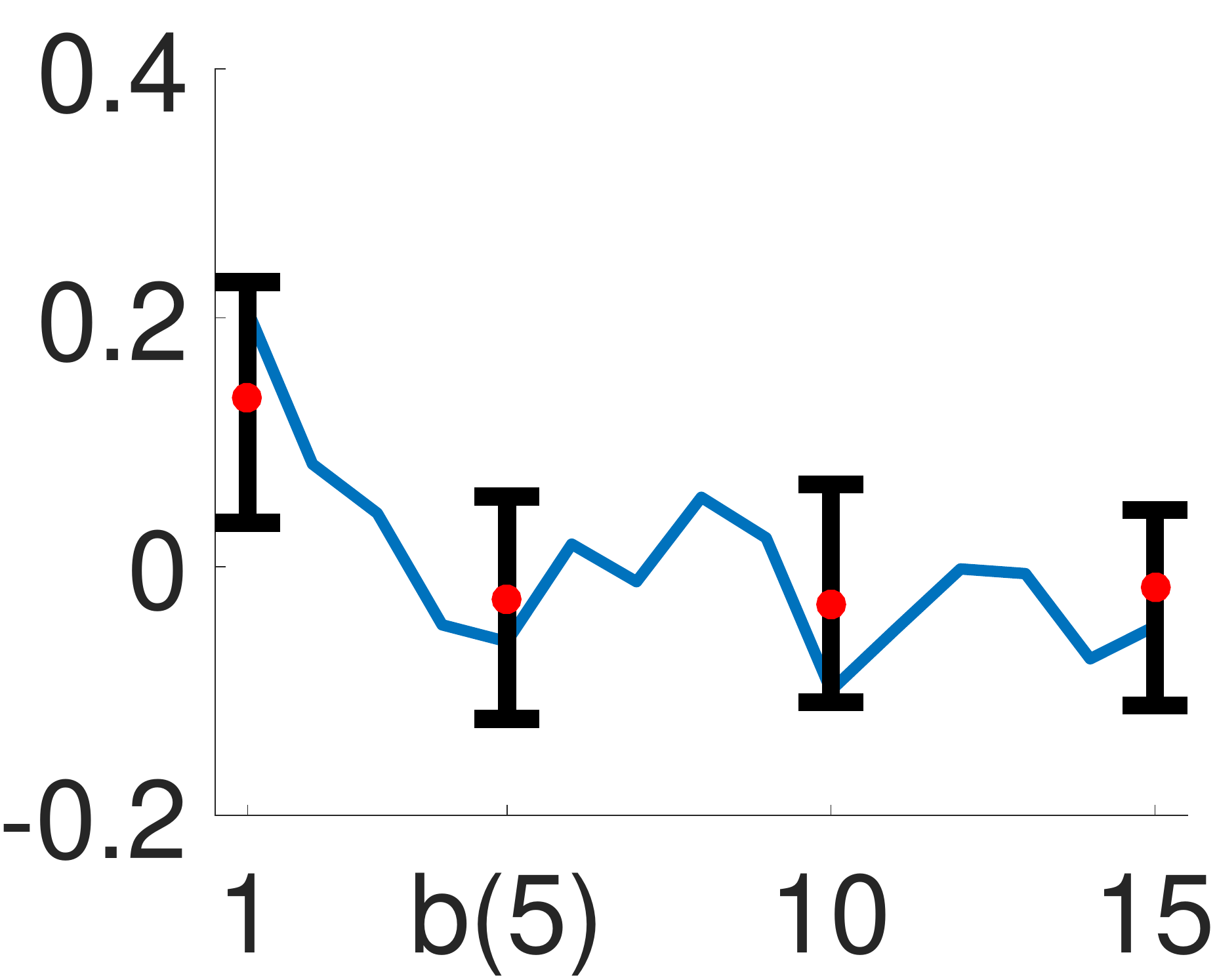}     & \includegraphics[scale = 0.18]{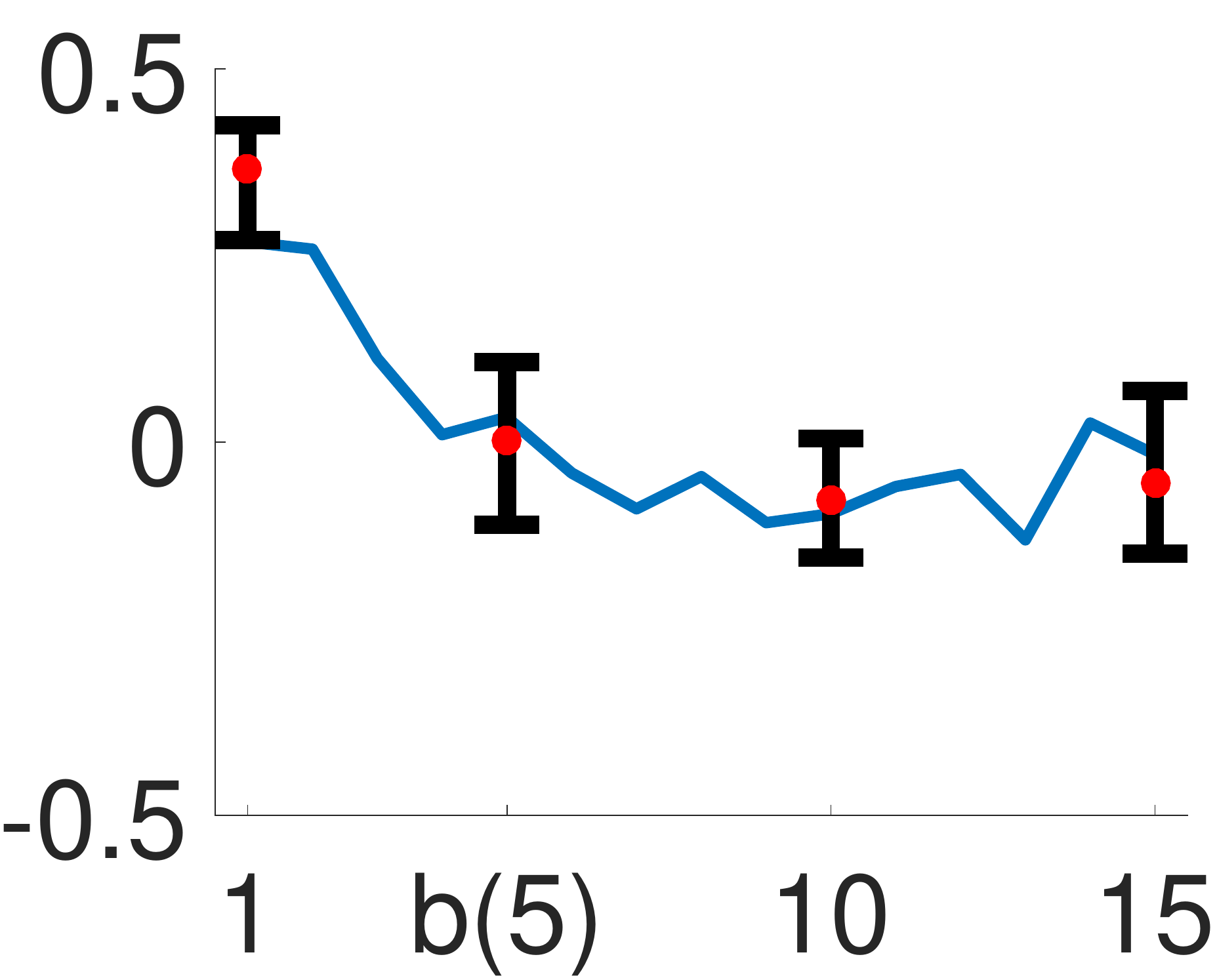}     & \includegraphics[scale = 0.18]{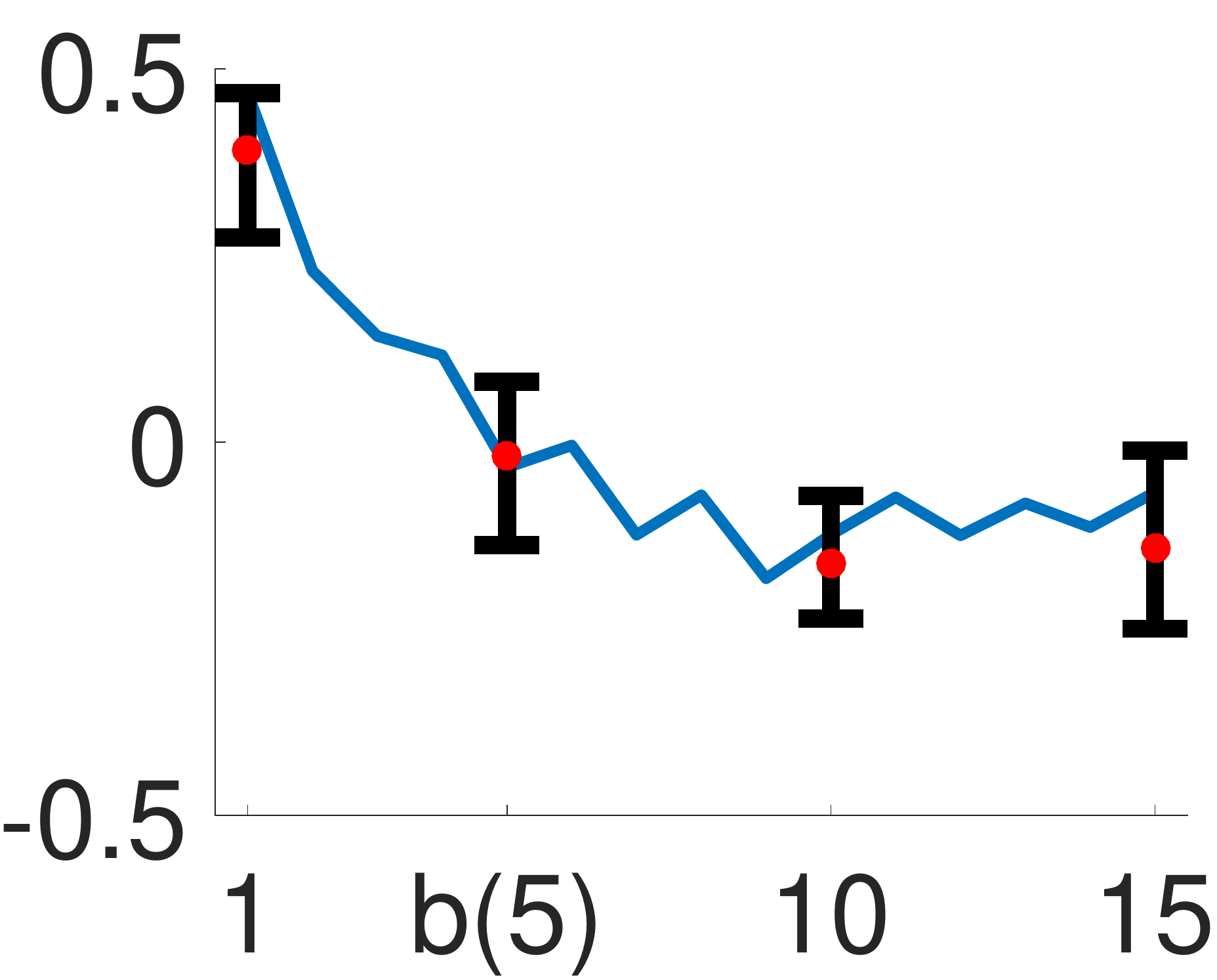}     \\ \hline
			\multicolumn{1}{|c|}{n = 50}  & \includegraphics[scale = 0.18]{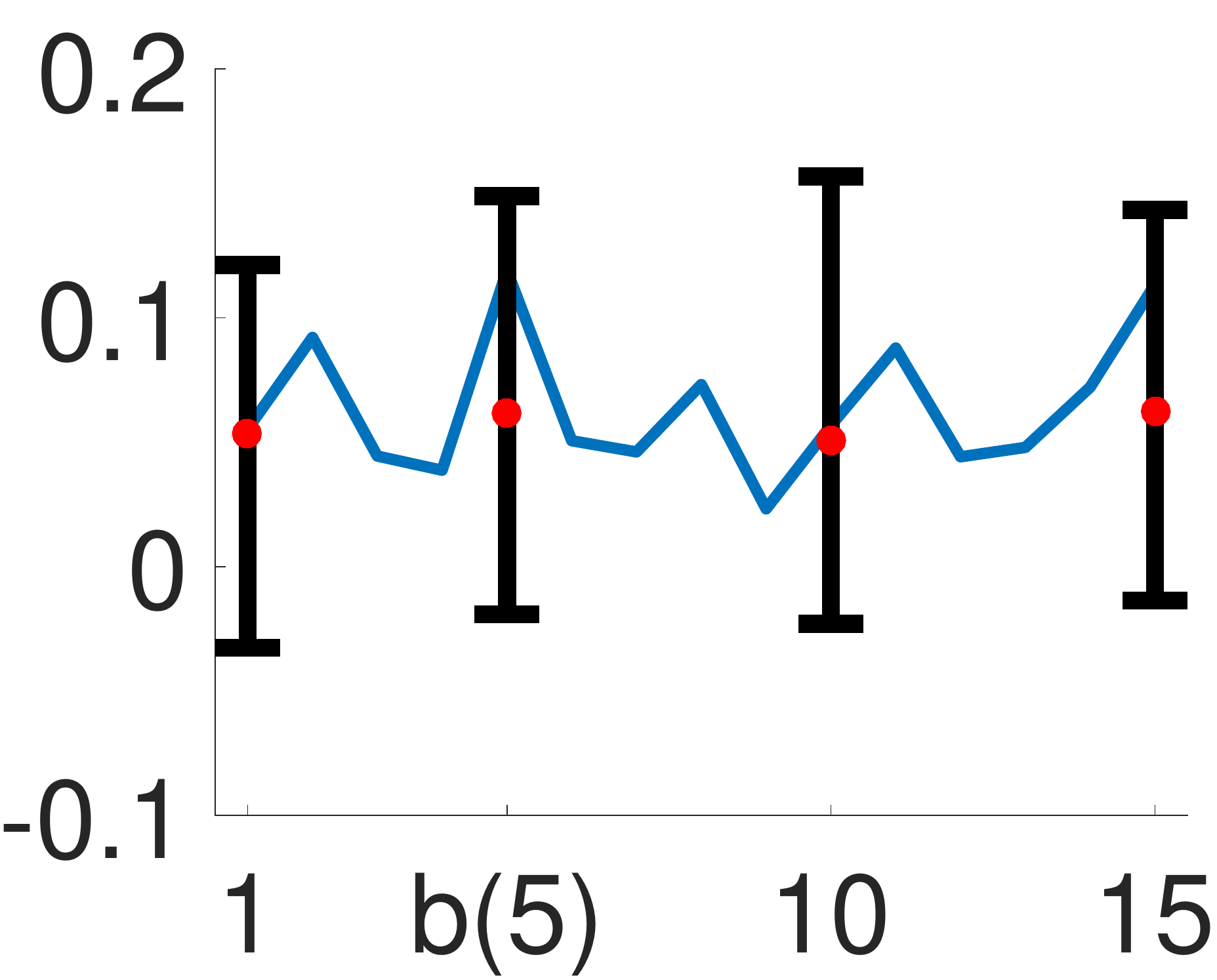}     & \includegraphics[scale = 0.18]{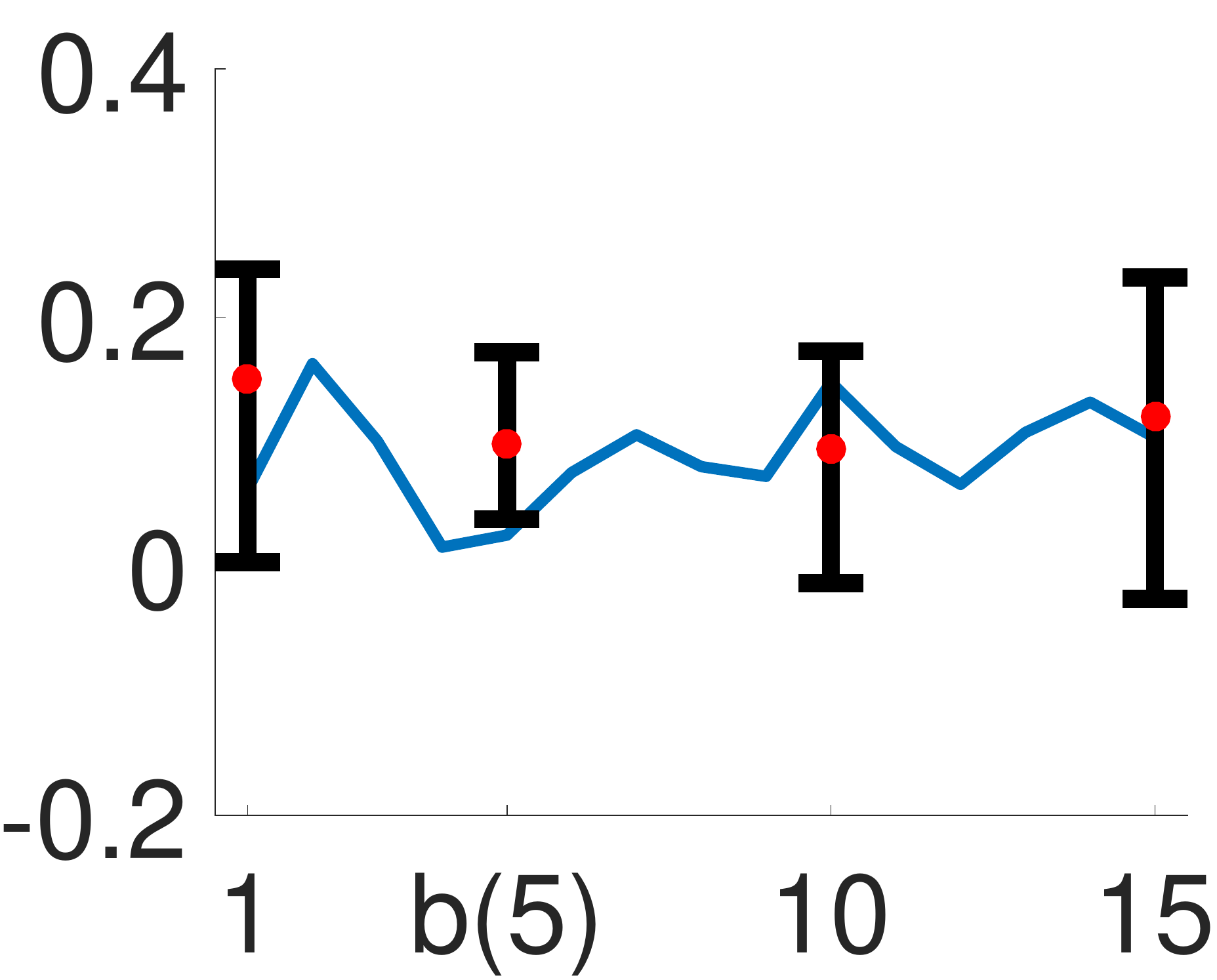}     & \includegraphics[scale = 0.18]{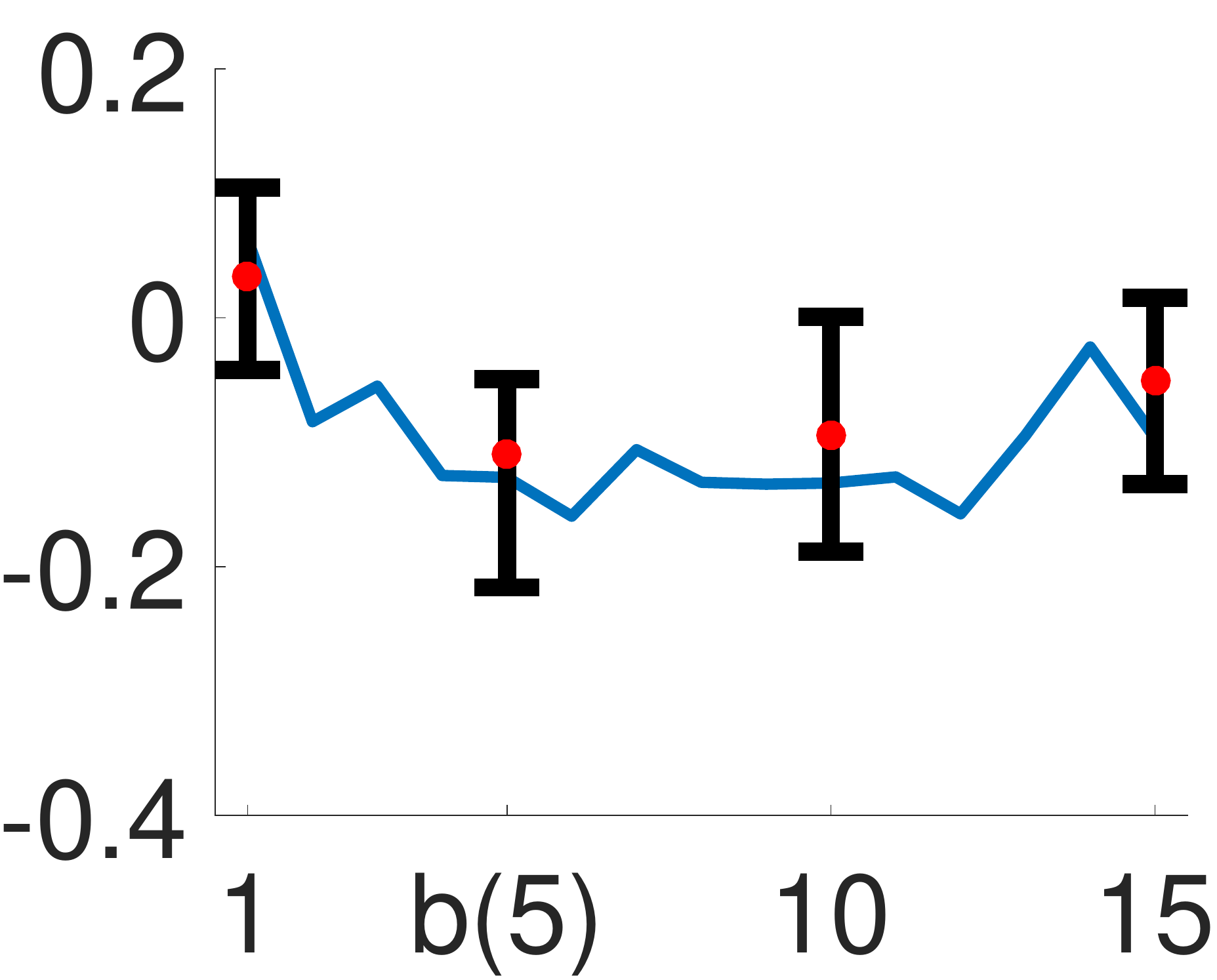}     \\ \hline
			\multicolumn{1}{|c|}{n = 100}  & \includegraphics[scale = 0.18]{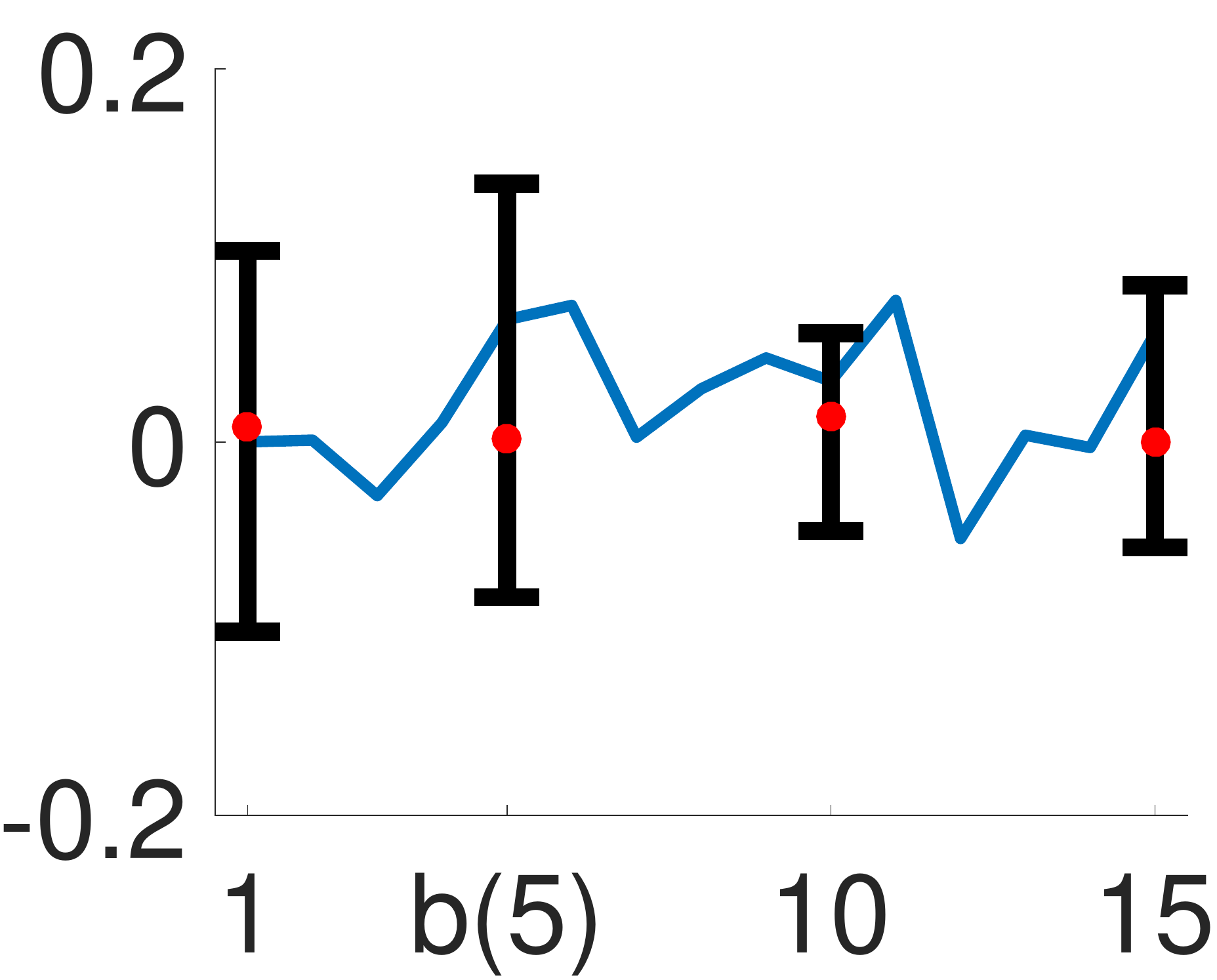}     & \includegraphics[scale = 0.18]{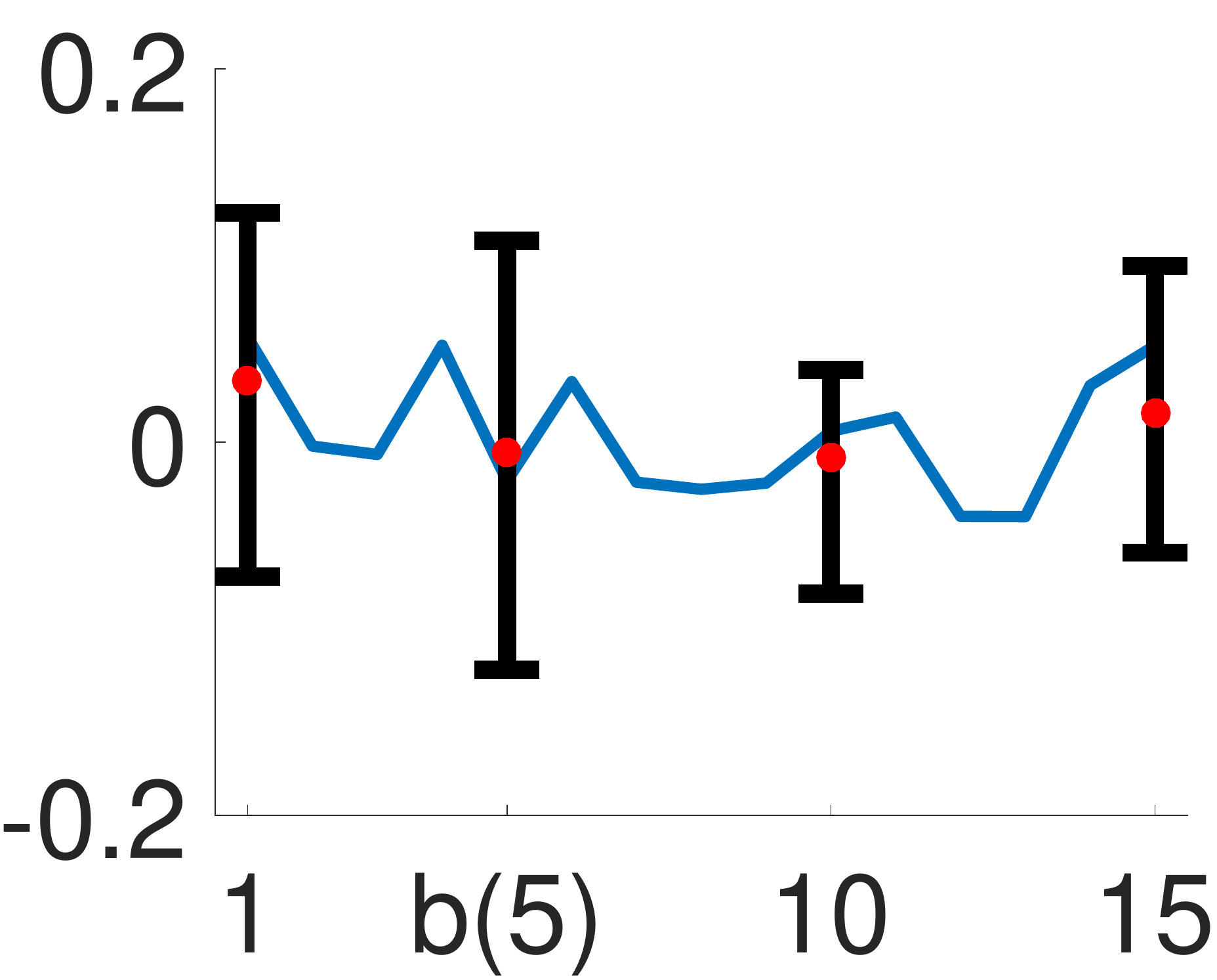}     & \includegraphics[scale = 0.18]{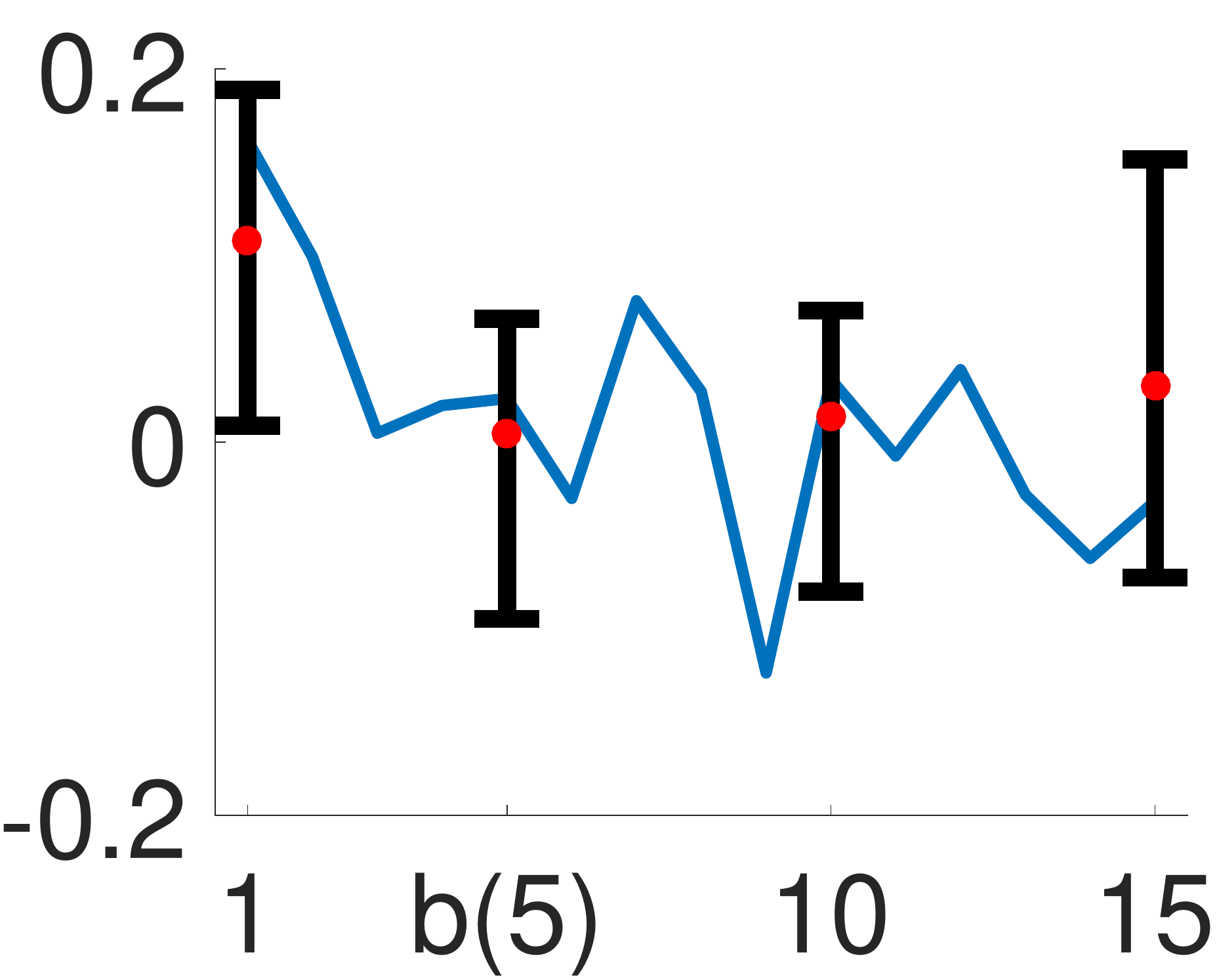}     \\ \hline
		\end{tabular}
	\end{center}
	\caption{Assessment of sensitivity using the spread measure $\mathbb{V}$. We consider perturbations of the centering measure $G_0=Beta(5, b)$ in the DP model for different choices of the the concentration parameter $\alpha$ and sample size $n$. The values of $b$ are plotted on the $x$-axis with baseline $b = 5$.}
	\label{fig:DP_G0_KV}
\end{figure}

The sensitivity measure $\mathbb{V}$ shown in Figure \ref{fig:DP_alpha_KV}, which captures differences in overall posterior variability, shows a downward trend as the value of $\alpha$ is increased. This is not surprising since $\alpha$ acts as the precision parameter for the DP model. Thus, as we increase the precision in the prior (and consequently the posterior), the variance of posterior samples should decrease. Finally, we note that the measure $\mathbb{E}$ corresponding to the covariance structure of the posterior samples does not show any noticeable pattern, indicating robustness to perturbations of $\alpha$ (we omitted this figure for brevity).

\noindent\textbf{Perturbing $G_0$, $\alpha$ fixed}: We again choose three different sample sizes $n$ and values of $\alpha$ for this simulation study. For each choice of $n$ and $\alpha$, we perturb the probability measure $G_0$ in the baseline model, which is chosen as a $Beta (5, 5)$. To do this, we modify the second parameter of the $Beta$ distribution: the perturbed models are centered at a $Beta (5, b)$, with the value of $b$ ranging from 1 to 15. Perturbing the parameter $b$ of the beta distribution leads to a gradual change in skewness (from negative to positive). Note that the FR distance between the PDFs of the baseline and perturbed centering measures behaves monotonically, i.e., as one moves away from the baseline in either direction the distance increases.

Figure \ref{fig:DP_G0} displays the sensitivity measure $\mathbb{D}$. A clear pattern is noticeable for large values of $\alpha$ as we perturb $G_0$; the posterior is sensitive to changes of the centering measure. Additionally, as observed earlier, the DP model becomes more robust with increasing sample size $n$, i.e., the scale of the $y$-axis measuring the shift in the posterior becomes smaller as we move from the top row to the bottom row. For $\alpha = 1$ and $n = 100$, we expect the posterior DP to be dictated by the sample size, and the effect of perturbations of the measure $G_0$ to be overshadowed. This is apparent when we look at the corresponding plot in Figure \ref{fig:DP_G0} (last row, first column) where the measure $\mathbb{D}$ is of order $10^{-3}$. In general, the sensitivity measures $\mathbb{V}$ (see Figure \ref{fig:DP_G0_KV}) and $\mathbb{E}$ (not shown here for brevity) do not exhibit any trends as we vary the baseline probability measure $G_0$ indicating robustness.
	
	\subsubsection{DPGMM}
	\label{subsubsec:DPMsimu}
	
	\begin{figure}[!t]
		\begin{center}
			\begin{tabular}{c|c|c|c|}
				\cline{2-4}
				& $\mathbb{D}$ & $\mathbb{V}$ & $\mathbb{E}$ \\ \hline
				\multicolumn{1}{|c|}{$\alpha$} & \includegraphics[scale = 0.18]{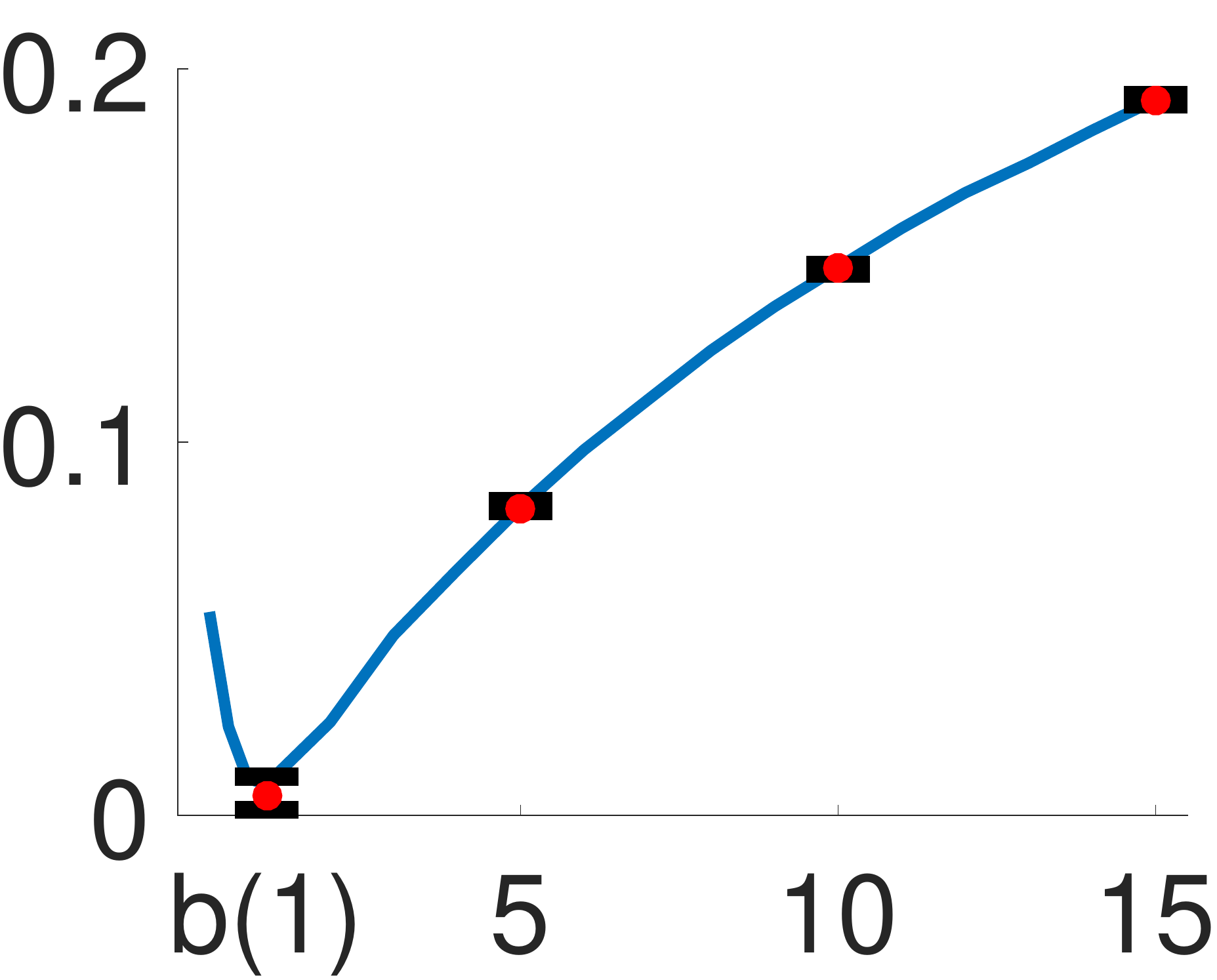}    & \includegraphics[scale = 0.18]{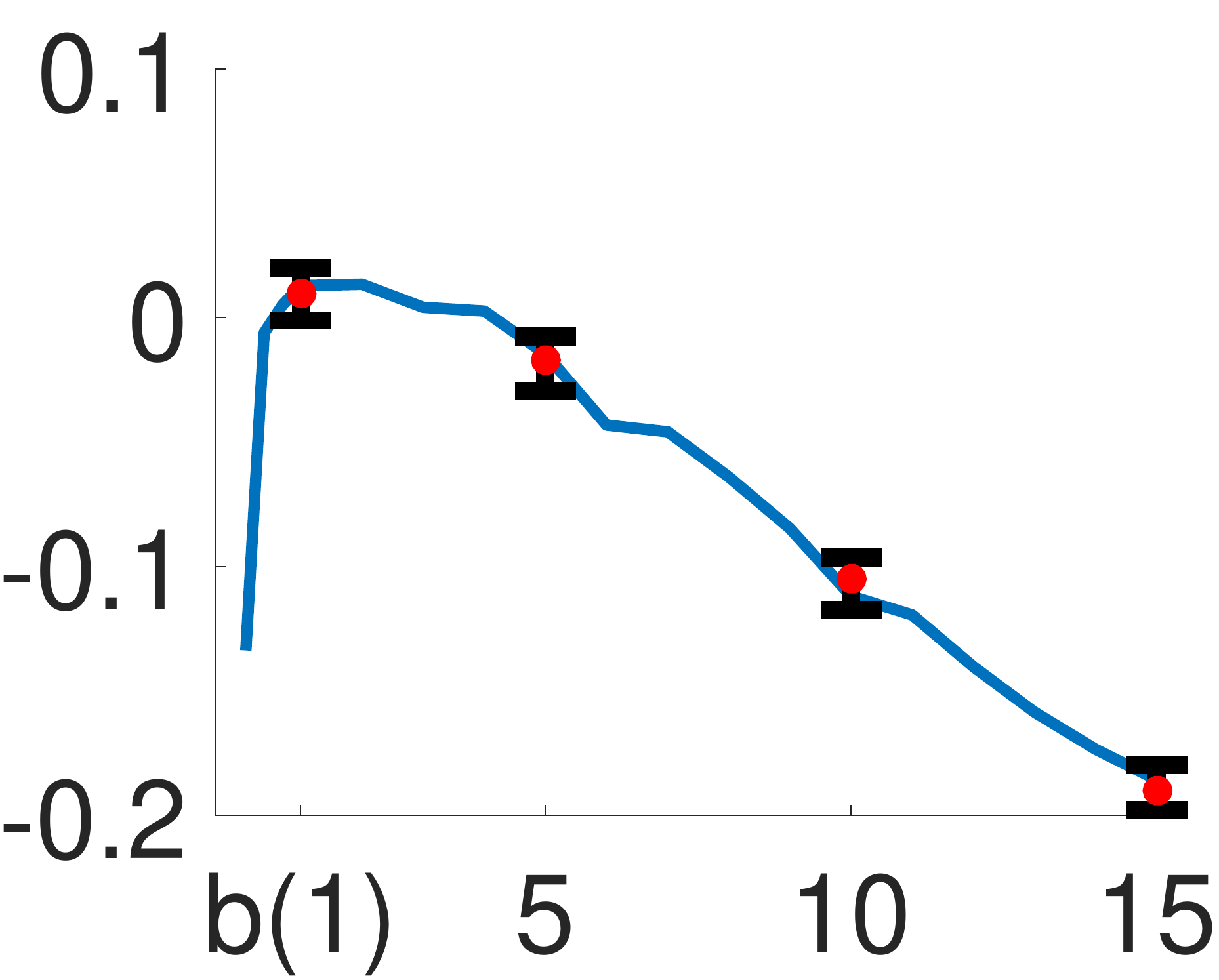}     & 		\includegraphics[scale = 0.18]{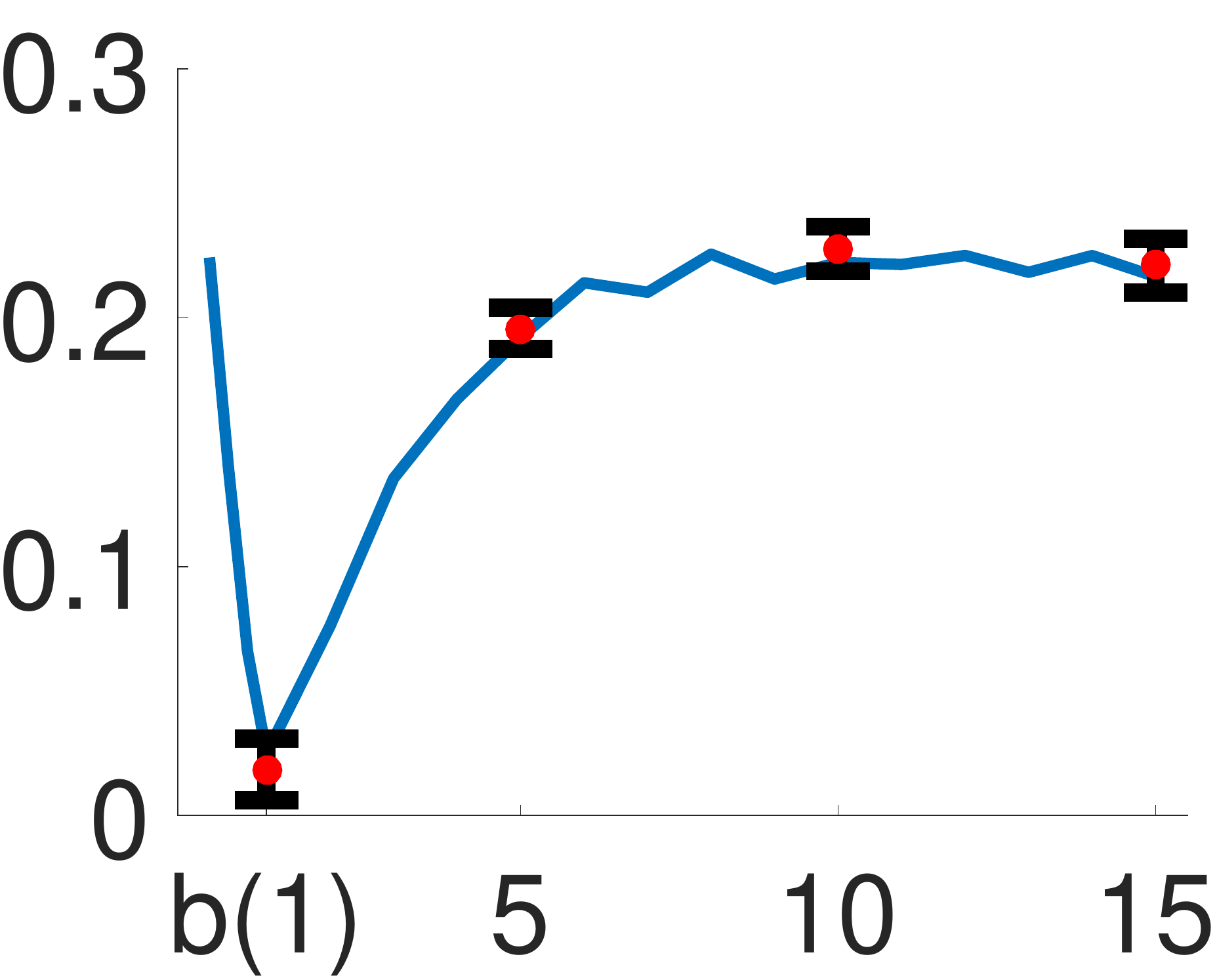}            \\ \hline
				\multicolumn{1}{|c|}{$\boldsymbol{m}$} & \includegraphics[scale = 0.18]{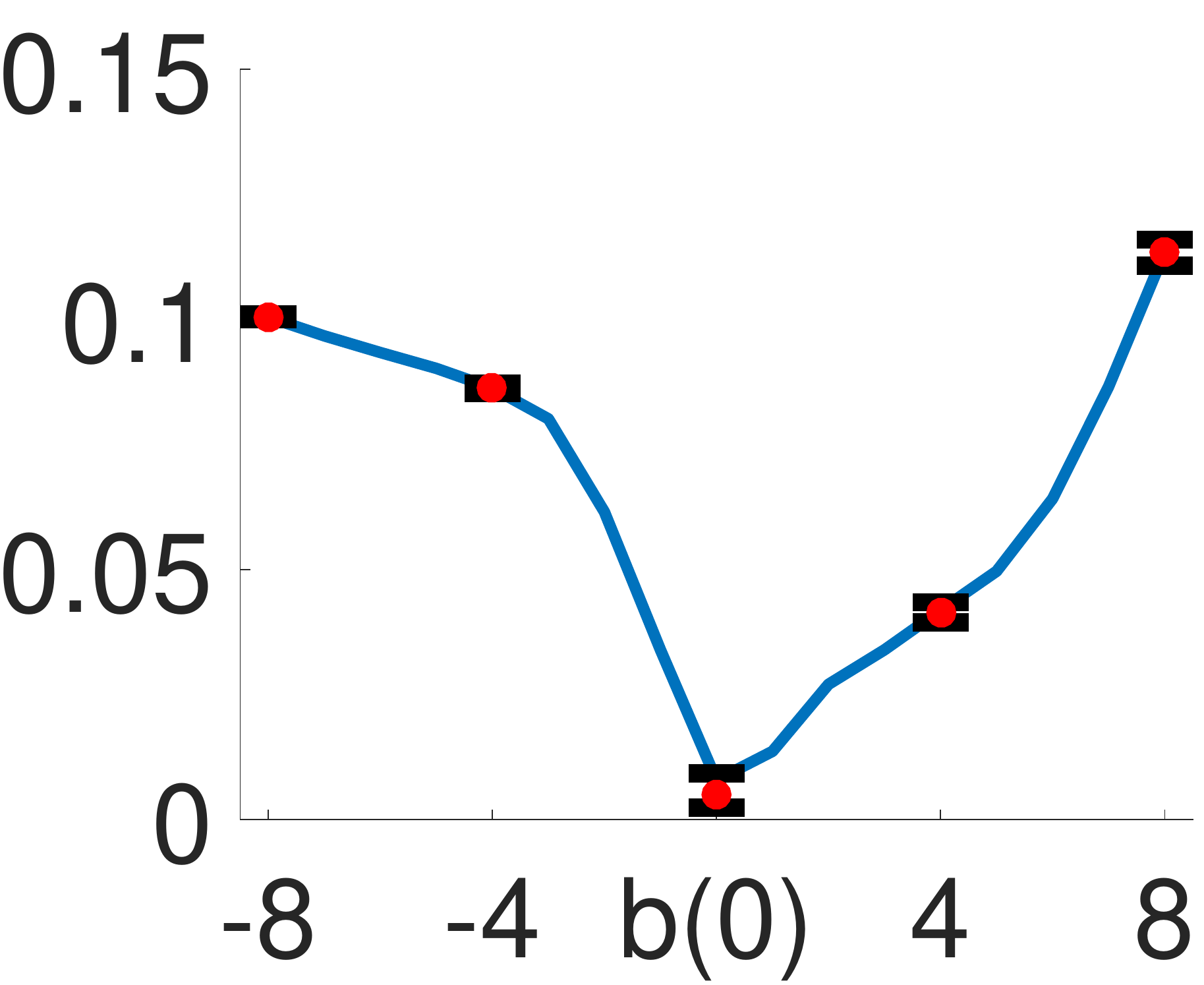}  & \includegraphics[scale = 0.18]{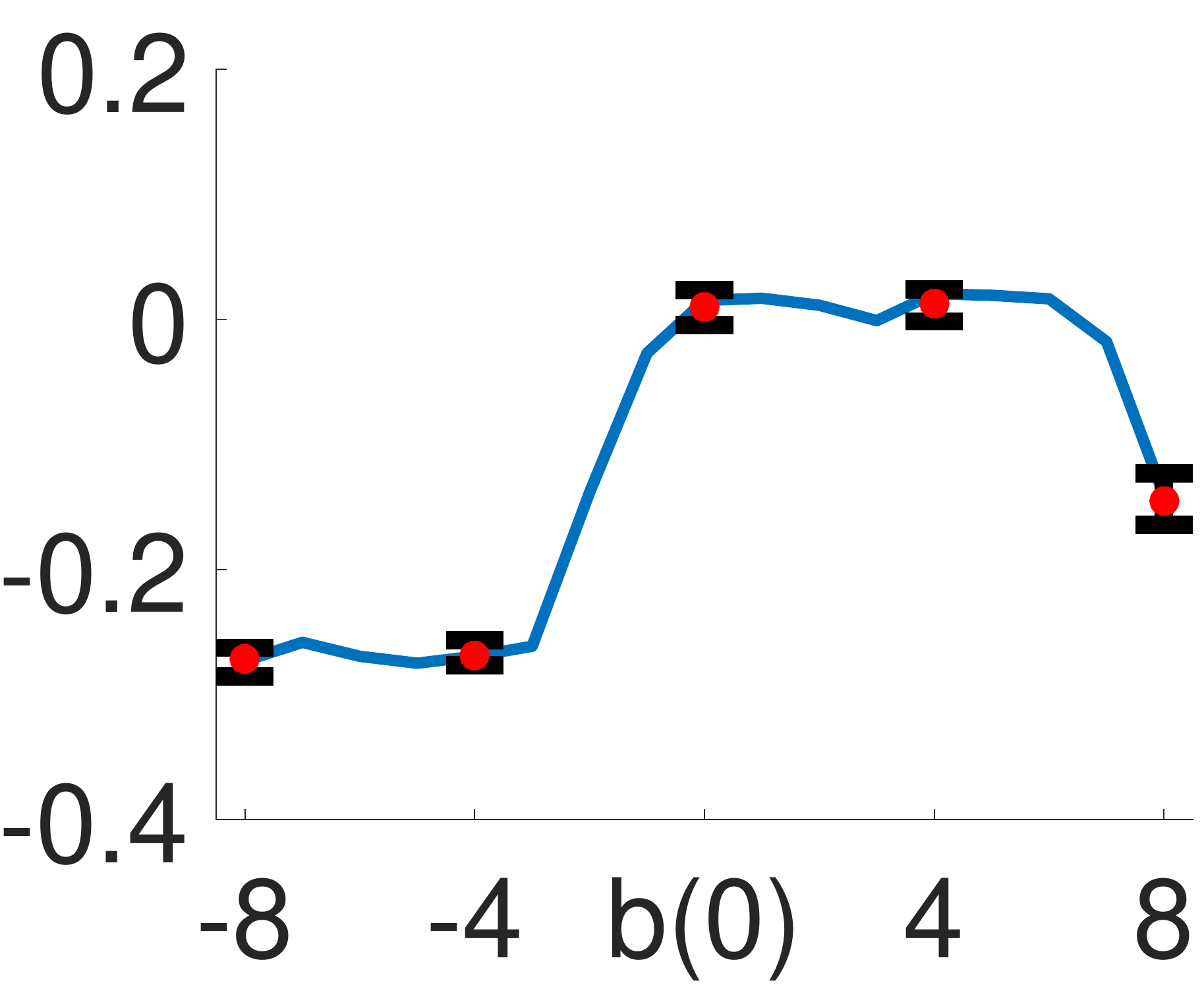}     & 		\includegraphics[scale = 0.18]{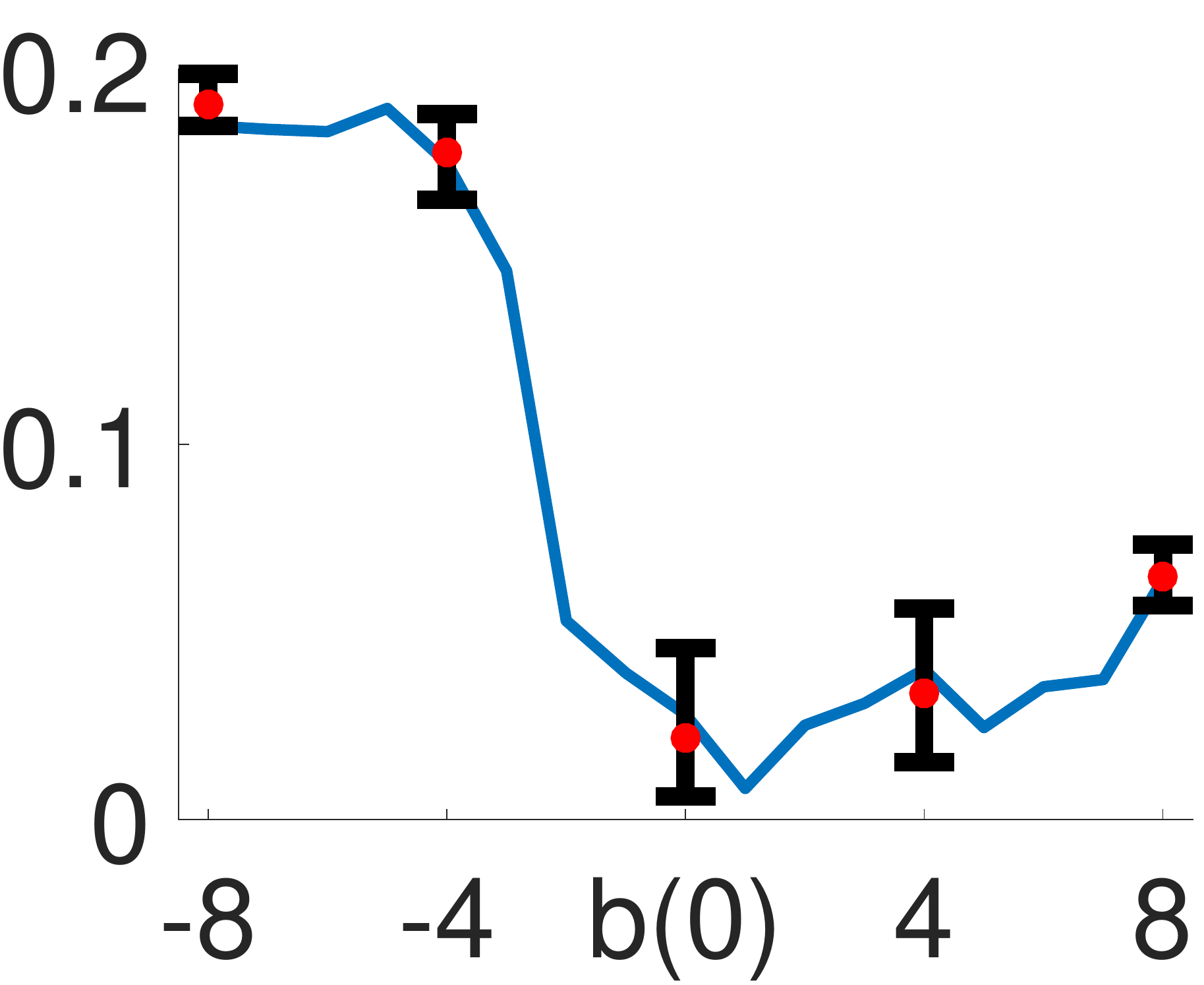}             \\ \hline
				\multicolumn{1}{|c|}{$r$}   &   \includegraphics[scale = 0.18]{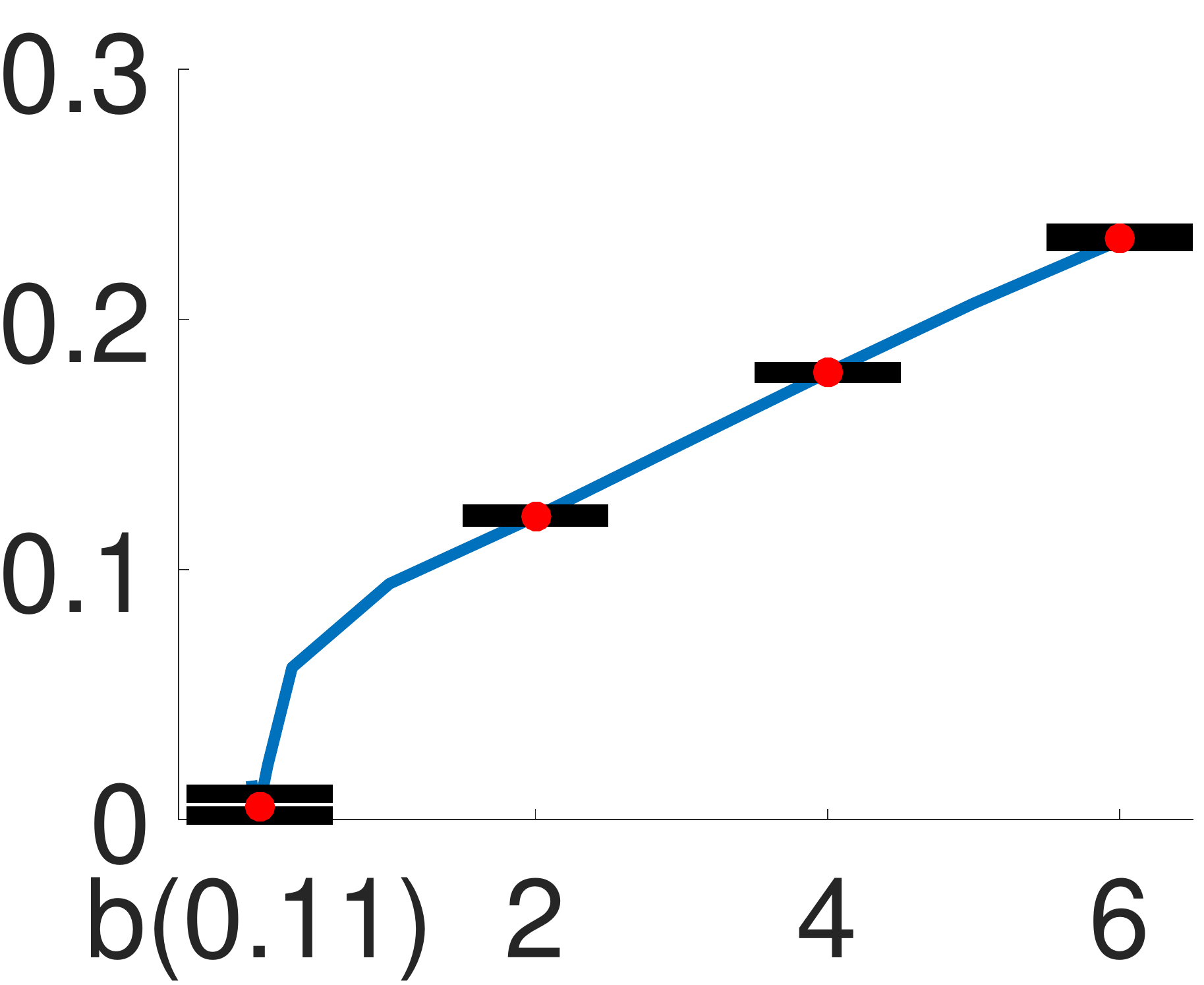}          & \includegraphics[scale = 0.18]{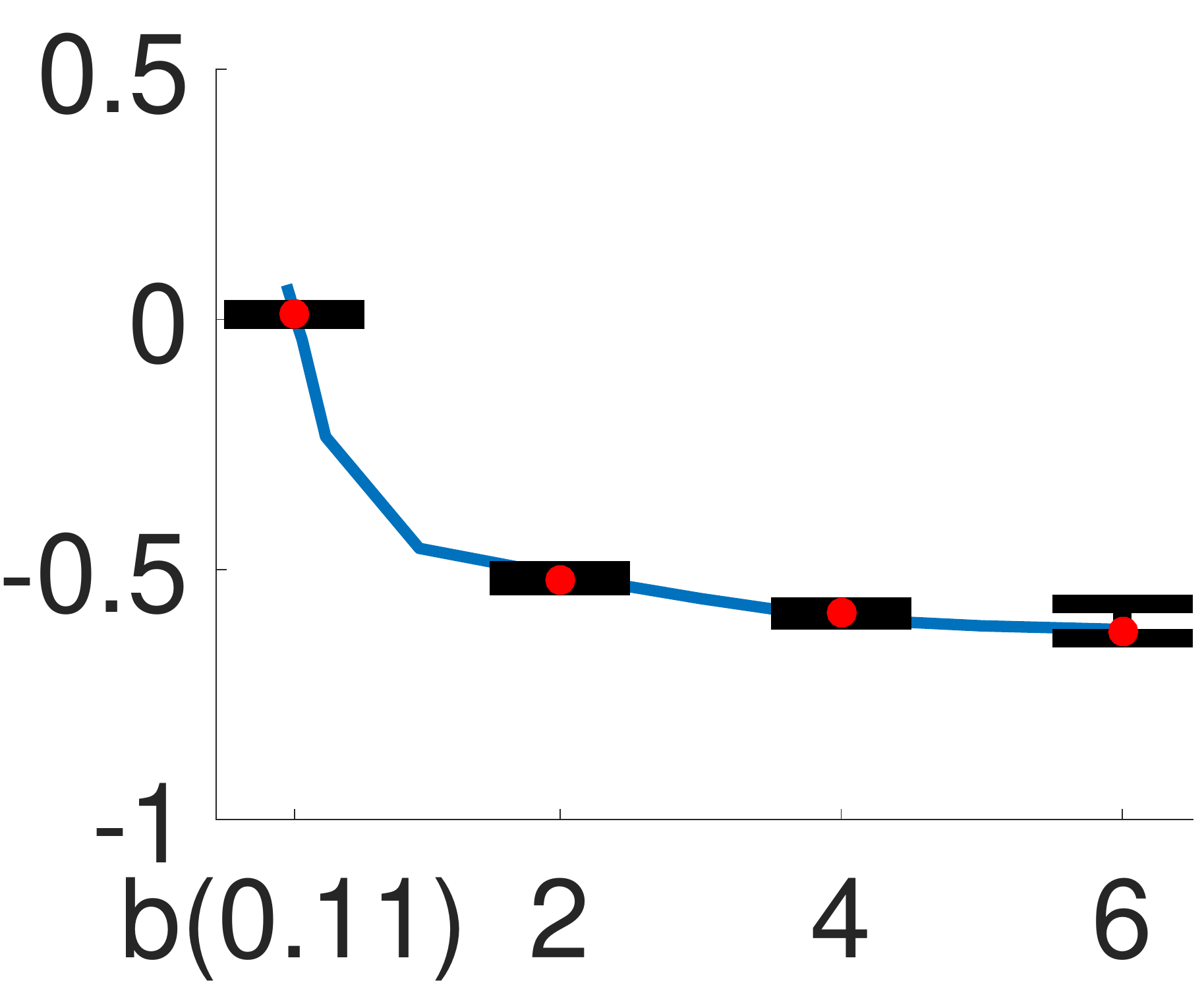}     & 		\includegraphics[scale = 0.18]{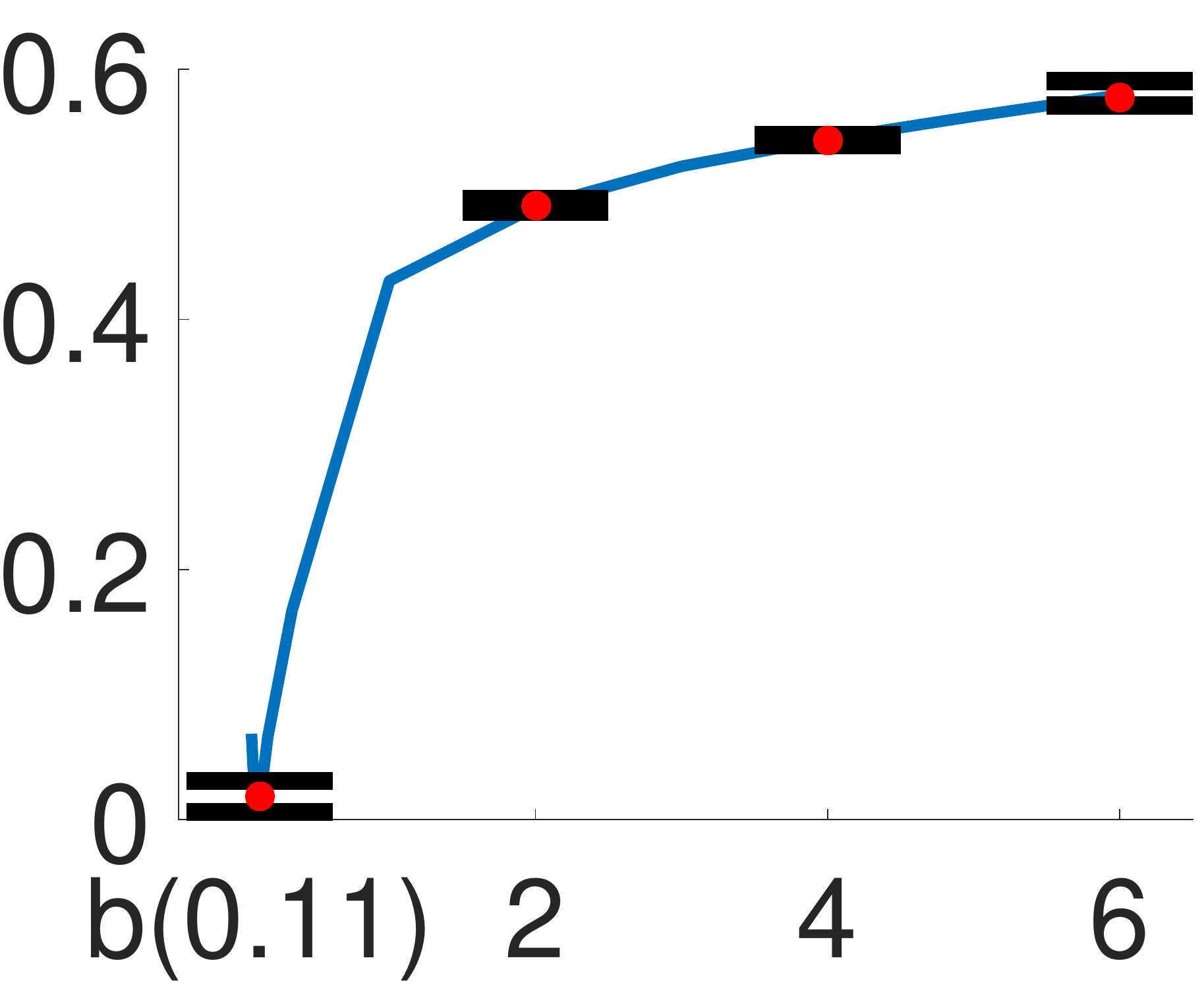}                \\ \hline
				\multicolumn{1}{|c|}{$\nu$} &  \includegraphics[scale = 0.18]{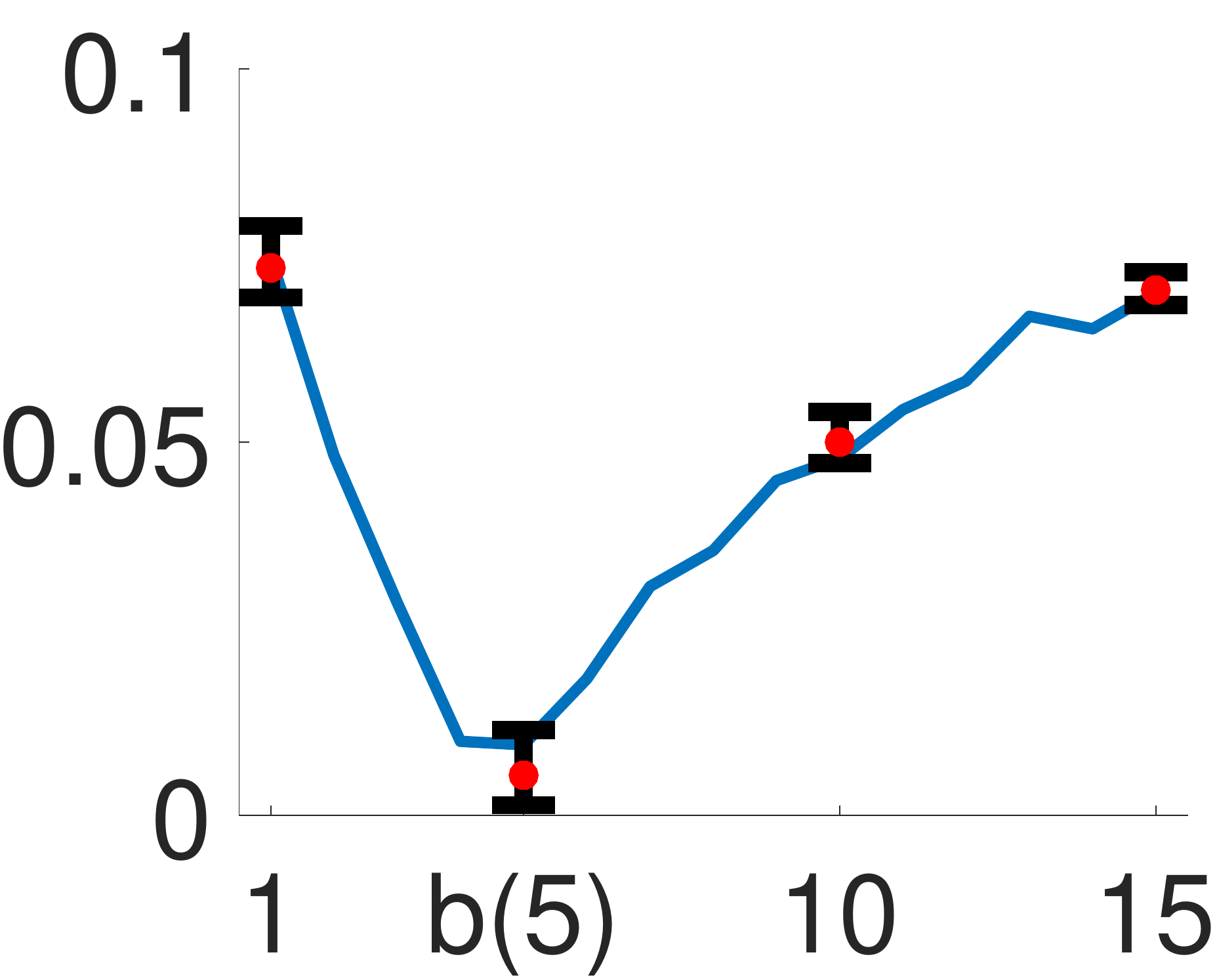}          & \includegraphics[scale = 0.18]{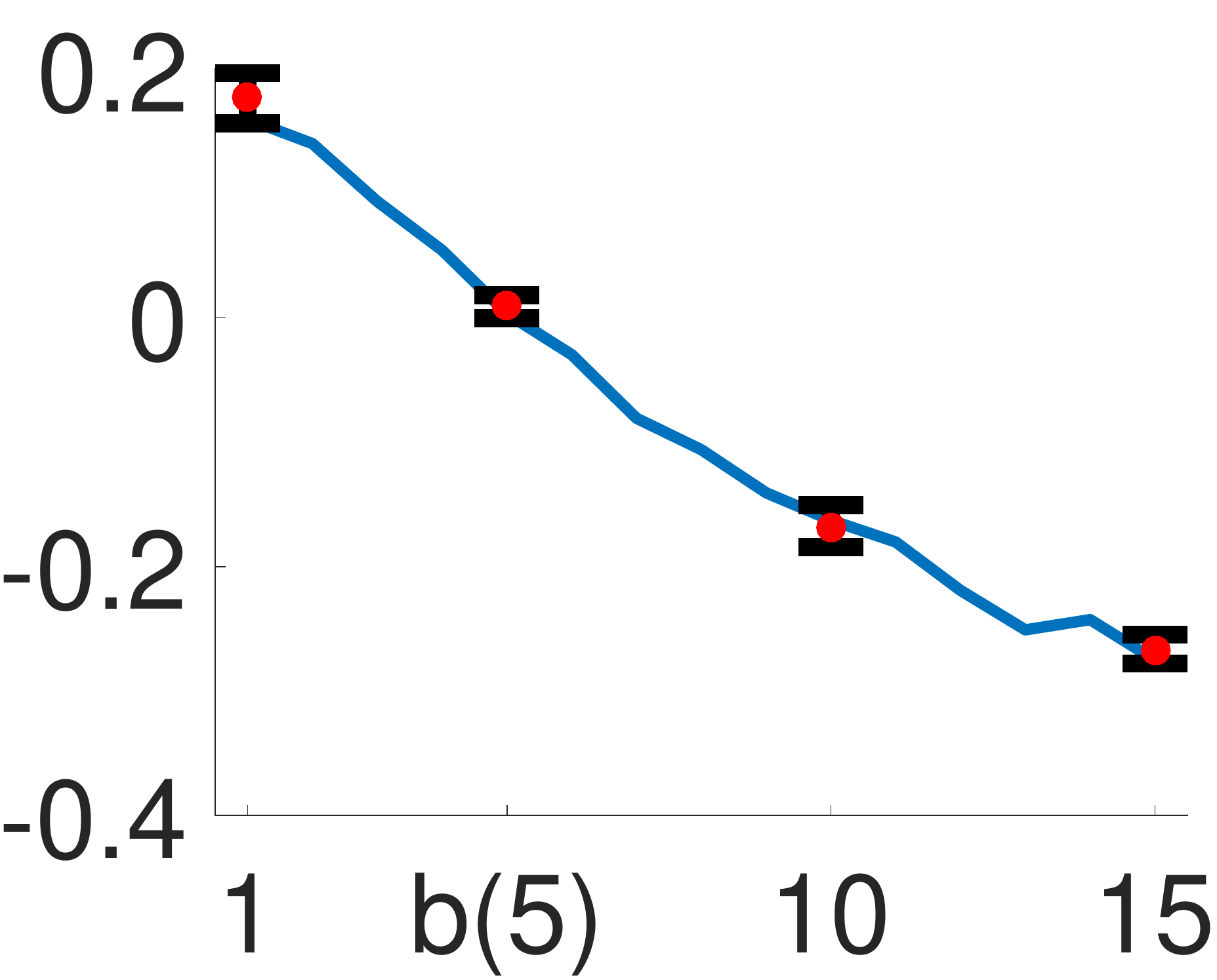}     & 		\includegraphics[scale = 0.18]{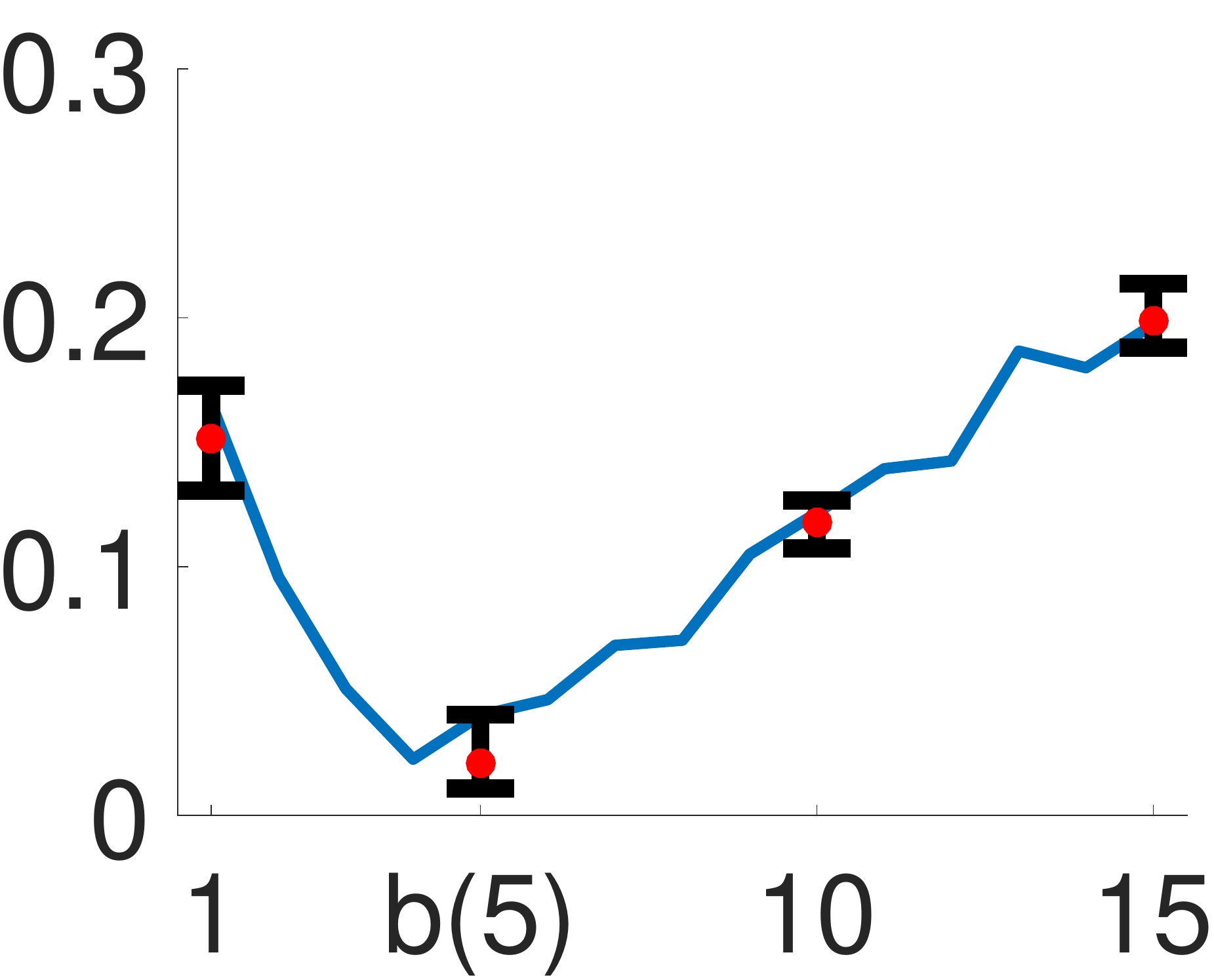}               \\ \hline
				\multicolumn{1}{|c|}{$S$}   &   \includegraphics[scale = 0.18]{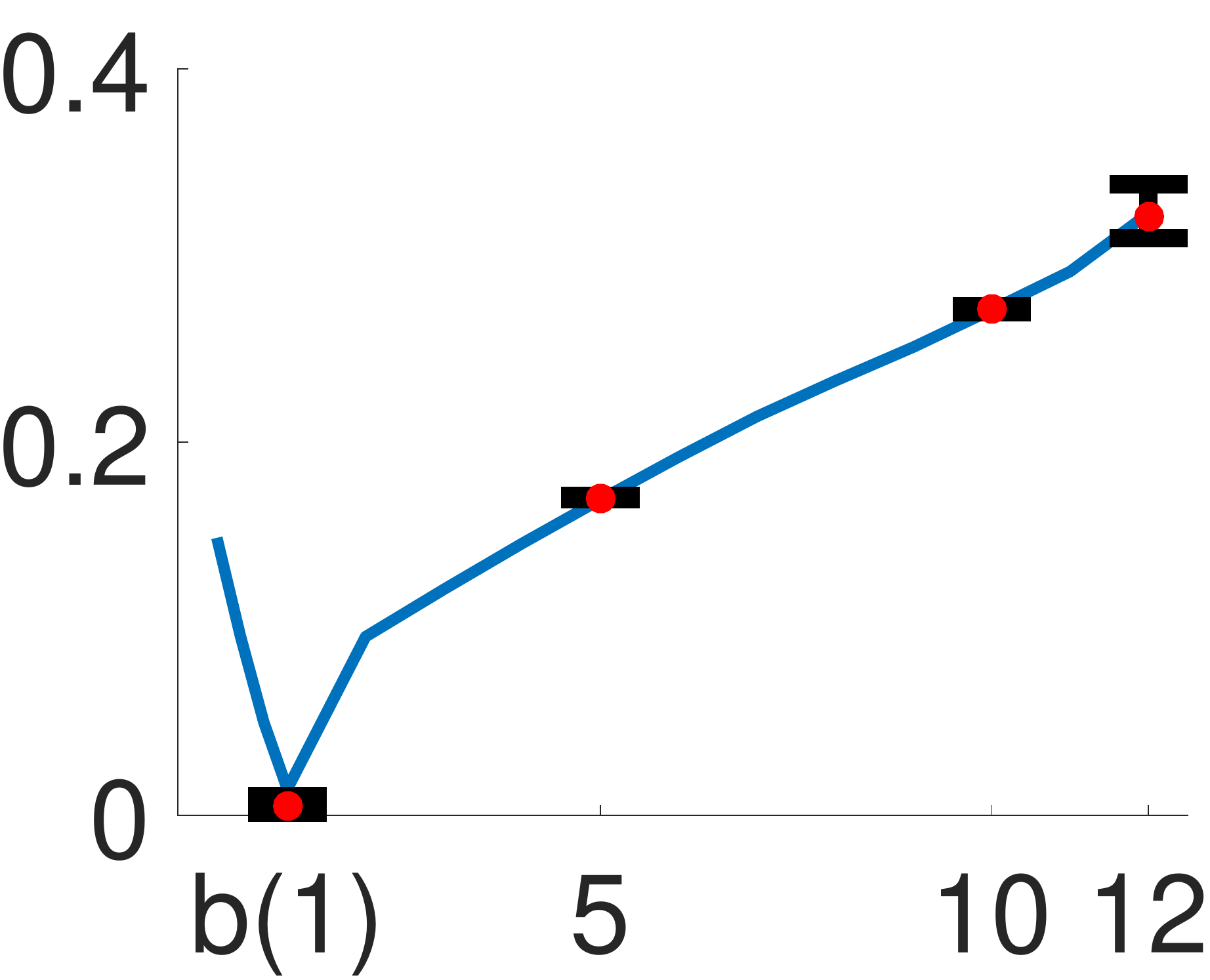}          & \includegraphics[scale = 0.18]{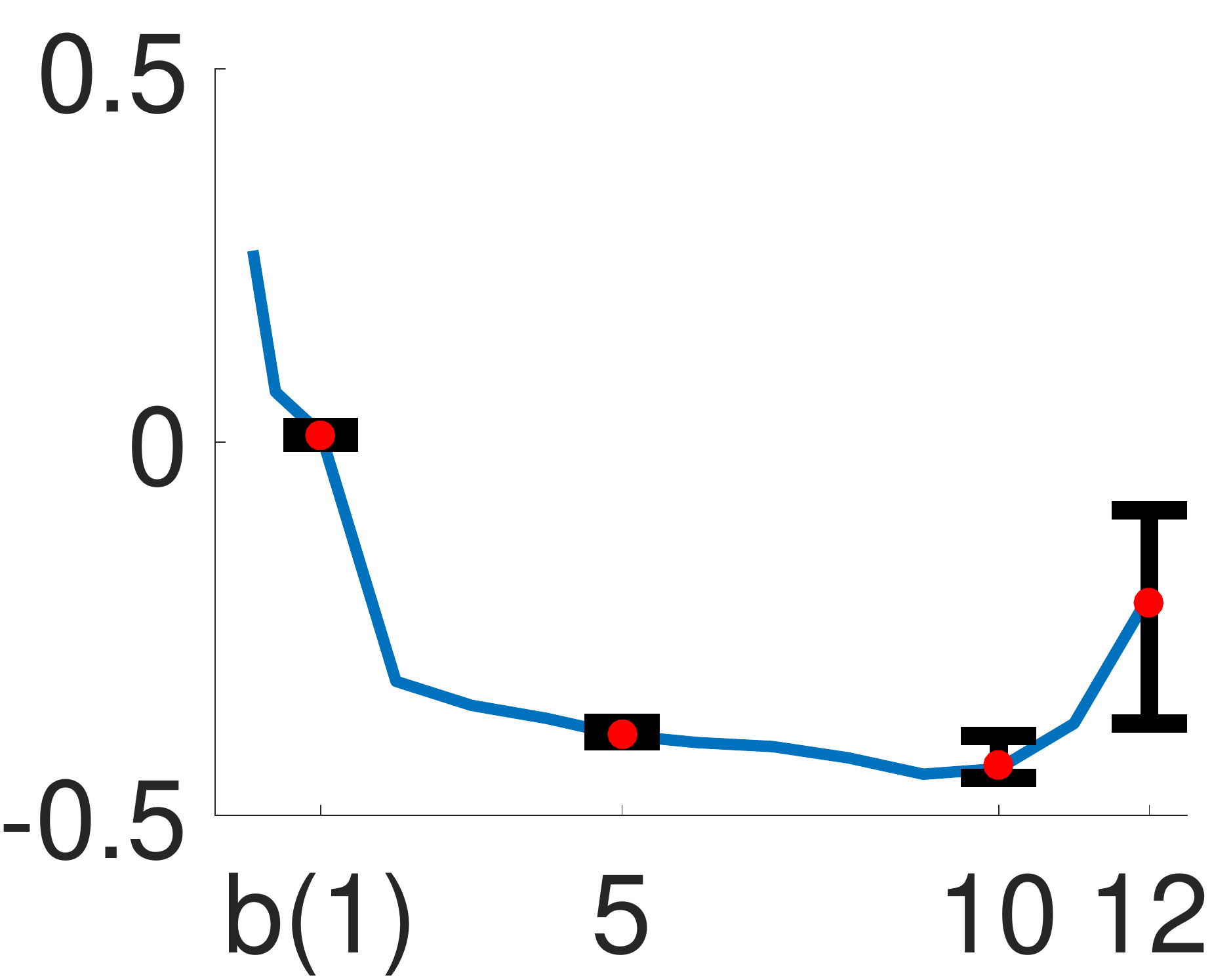}     & 		\includegraphics[scale = 0.18]{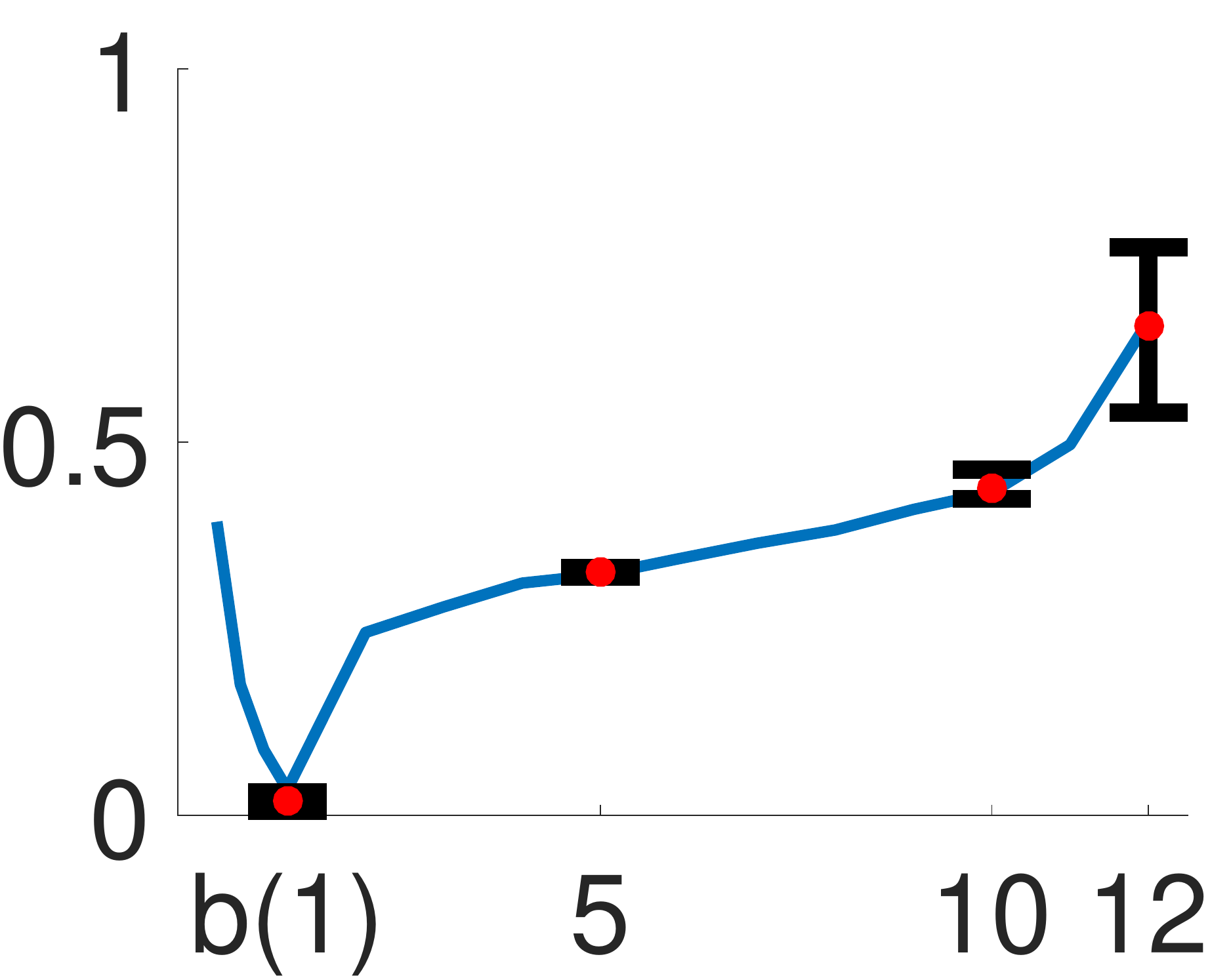}                \\ \hline
			\end{tabular}
		\end{center}
		\caption{Assessment of sensitivity using all of the proposed measures for various perturbations of prior parameters in the DPGMM. The baseline setting is marked with a $b$.}
		\label{fig:DPGMM_combined}
	\end{figure}
	
In this section, we perform simulation studies for the DPGMM and the proposed global measures to assess model sensitivity to prior perturbations of the different parameters in the model. We simulate a dataset from a mixture of three Gaussian components, and then perform sensitivity analysis under various perturbation schemes. Specifically, our simulations are based on perturbing the following set of parameters: (a) concentration parameter $\alpha$, (b) parameter $\boldsymbol{m}$ in the prior distribution of the mean $\mu$, (c) relative precision $r$ of the mean versus data, (d) degrees of freedom $\nu$ for the precision component, and (e) prior variance $S$ for the precision component. The last four perturbations are aimed at altering the probability measure $G_0$, and checking their effect on the posterior samples, as was done for the DP model. Although the parameter notation that we established in Section \ref{subsubsec:DPGMM} is given for the multivariate case, we will only consider univariate density estimation here for simplicity.
		
We plot all of the proposed measures for different perturbations of the DPGMM parameters in Figure \ref{fig:DPGMM_combined}. The first column of the figure identifies the parameter being perturbed. The baseline model for this simulation study has the following parameter settings: $\alpha = 1$, $\boldsymbol{m} = 0$, $r = \frac{1}{9}$, $\nu = 5$ and $S = 1$. For each row in Figure \ref{fig:DPGMM_combined}, only one of the parameters is perturbed while all of the other parameter settings remain fixed at the baseline model. The ranges of the various perturbations are as follows: $\alpha =$ 0.1 to 15, $\boldsymbol{m} =$ $-8$ to 8, $r = \frac{1}{18}$ to 6, $\nu =$ 1 to 15 and $S = $ 0.1 to 12. For each plot, the baseline value of the model parameter being perturbed is indicated separately on the $x$-axis. We see that the DPGMM is sensitive to prior parameter perturbations, and all of the proposed geometric measures provide important information with respect to different aspects of the posterior density samples. For all of the perturbations under consideration, sensitivity measure $\mathbb{D}$ is smallest at the baseline model, and increases as we perturb the parameters away in either direction. Thus, this measure is very useful in capturing global shift deviations in the posterior samples of the DPGMM. The spread sensitivity measure $\mathbb{V}$ shows a clear decreasing trend for perturbations of the parameter $\nu$. This indicates that the variance of the posterior samples steadily decreases as we increase the degrees of freedom for the precision component. Further, we note that $\mathbb{E}$ stabilizes for large values of $\alpha$. This implies that the covariance structure of the posterior samples is sensitive for values of $\alpha$ close to the baseline, and eventually becomes insensitive to further increases in $\alpha$. However, the corresponding plot of $\mathbb{V}$ suggests that the overall variance of the posterior samples obtained from the perturbed models continues to change, even for large values of $\alpha$. Also, perturbations of $\alpha$ cause a fairly significant shift in the posterior samples, which can be clearly seen from the plot of the measure $\mathbb{D}$.
	
	\subsubsection{CCV Model}
	\label{subsubsec:CCVsimu}
	
Next, we assess sensitivity of the CCV model by perturbing model parameters. We generate the same type of dataset that we had used for studying the DPGMM in the previous section. We perturb the following set of parameters for this simulation: (a) shape parameters $a_0$ and $a_1$ in the $Beta$ prior for $a$, and (b) parameters $\eta$ and $\gamma$ in the prior for $\alpha$. The setup of this study is similar to the ones considered before, where we perturb a single parameter and fix all of the other model settings.

	\begin{figure}[!t]
		\begin{center}
			\begin{tabular}{c|c|c|c|}
				\cline{2-4}
				& $\mathbb{D}$ & $\mathbb{V}$ & $\mathbb{E}$ \\ \hline
				\multicolumn{1}{|c|}{$a_0$} & \includegraphics[scale = 0.18]{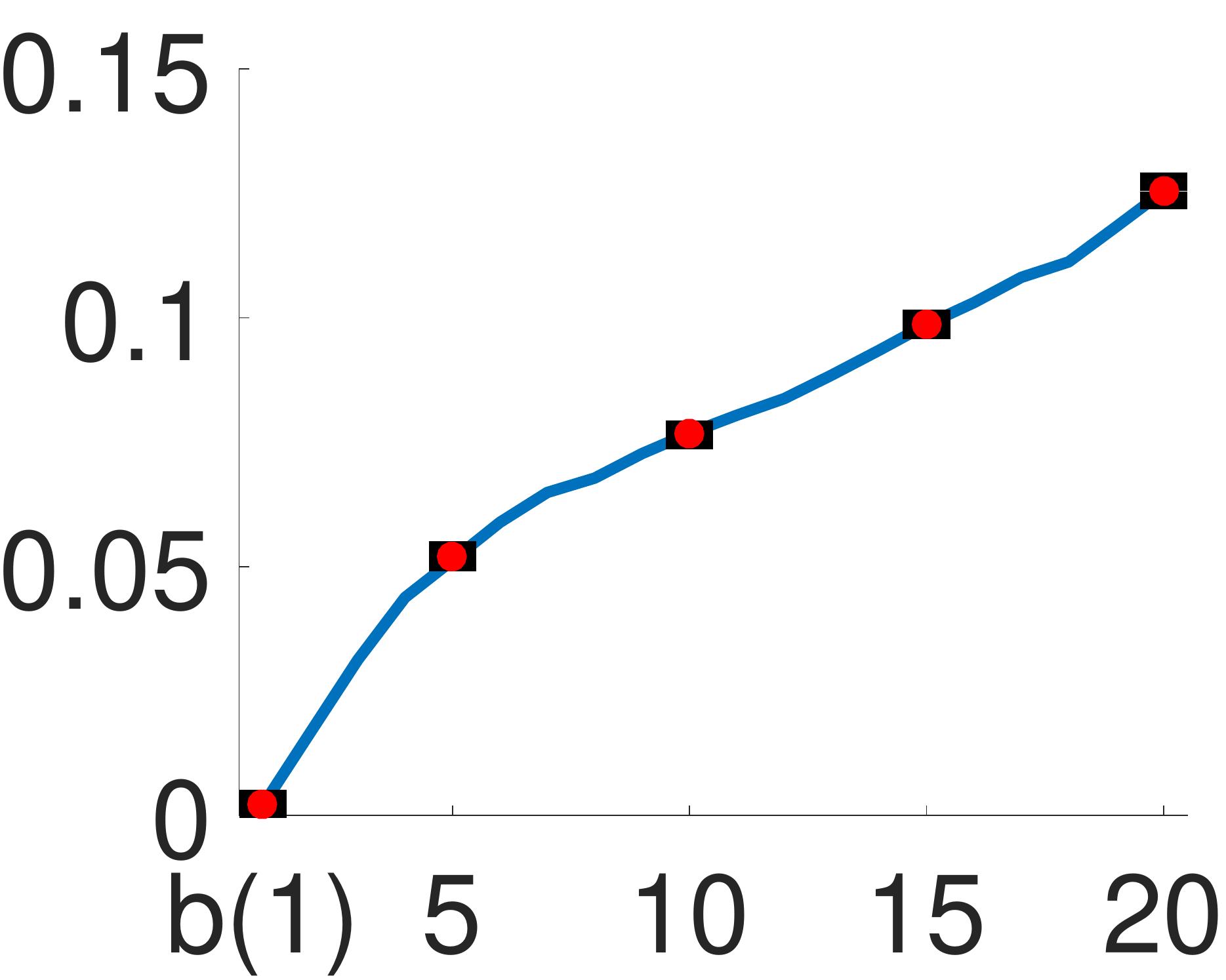}     & \includegraphics[scale = 0.18]{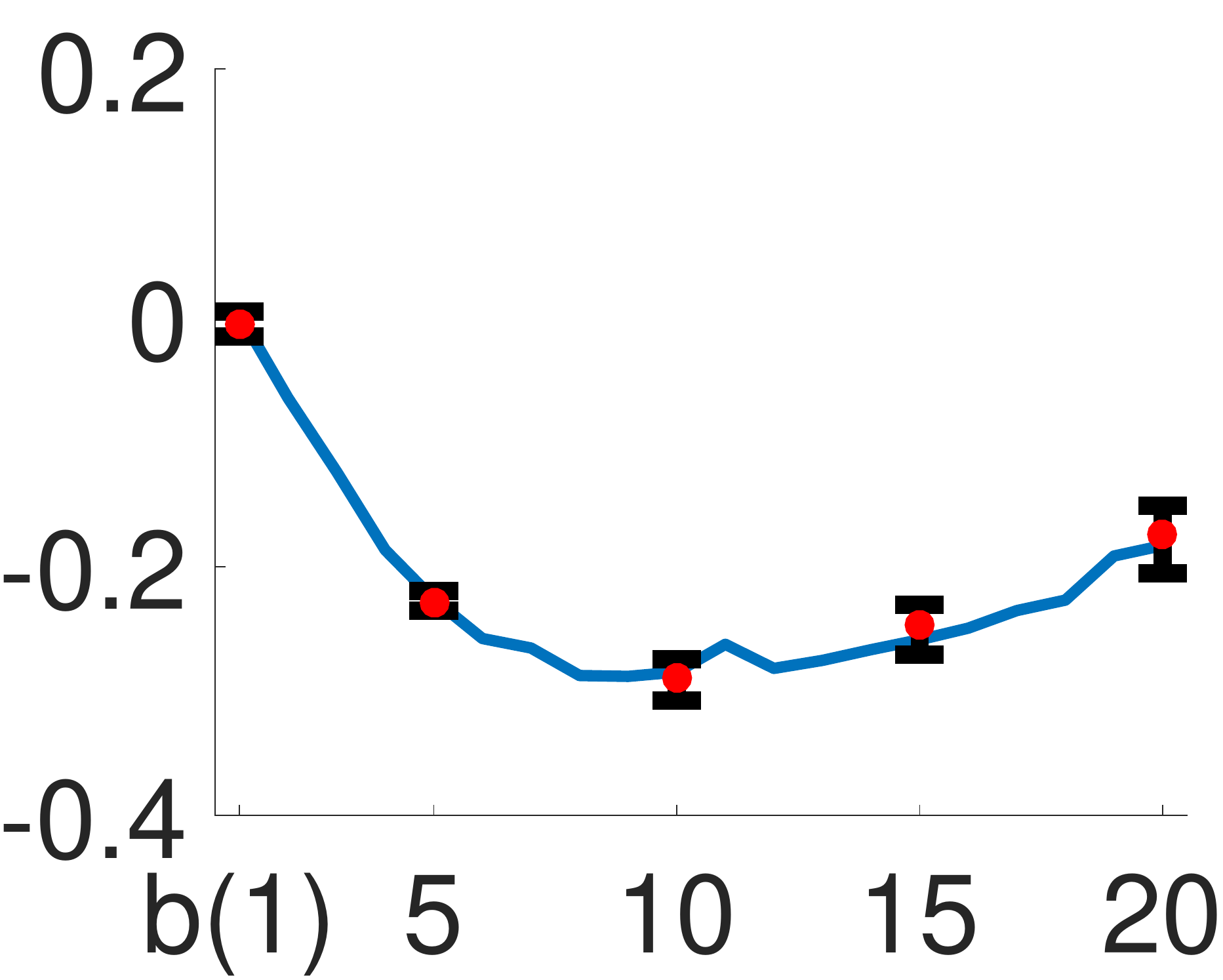}   &	\includegraphics[scale = 0.18]{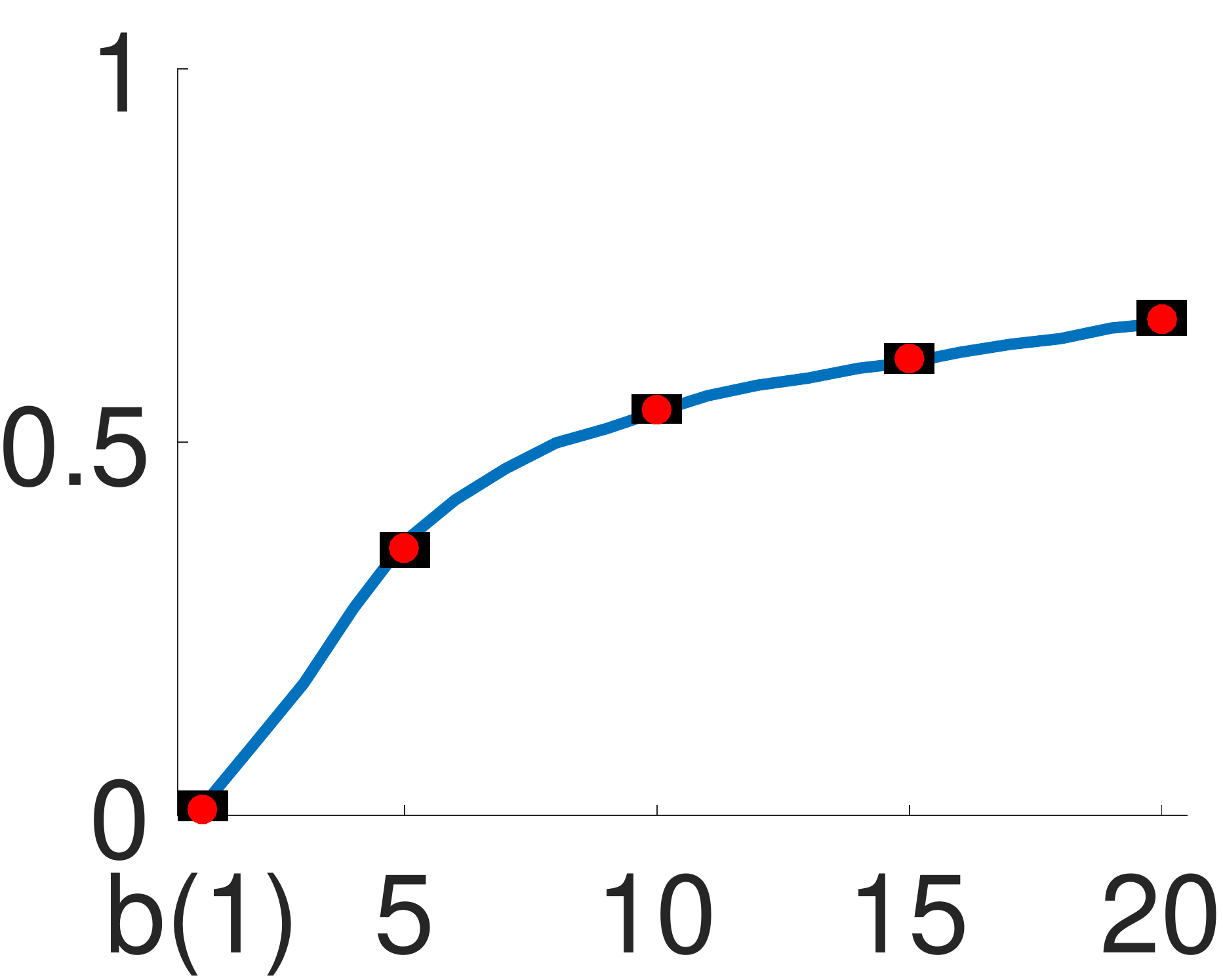}           \\ \hline
				\multicolumn{1}{|c|}{$a_1$} & \includegraphics[scale = 0.18]{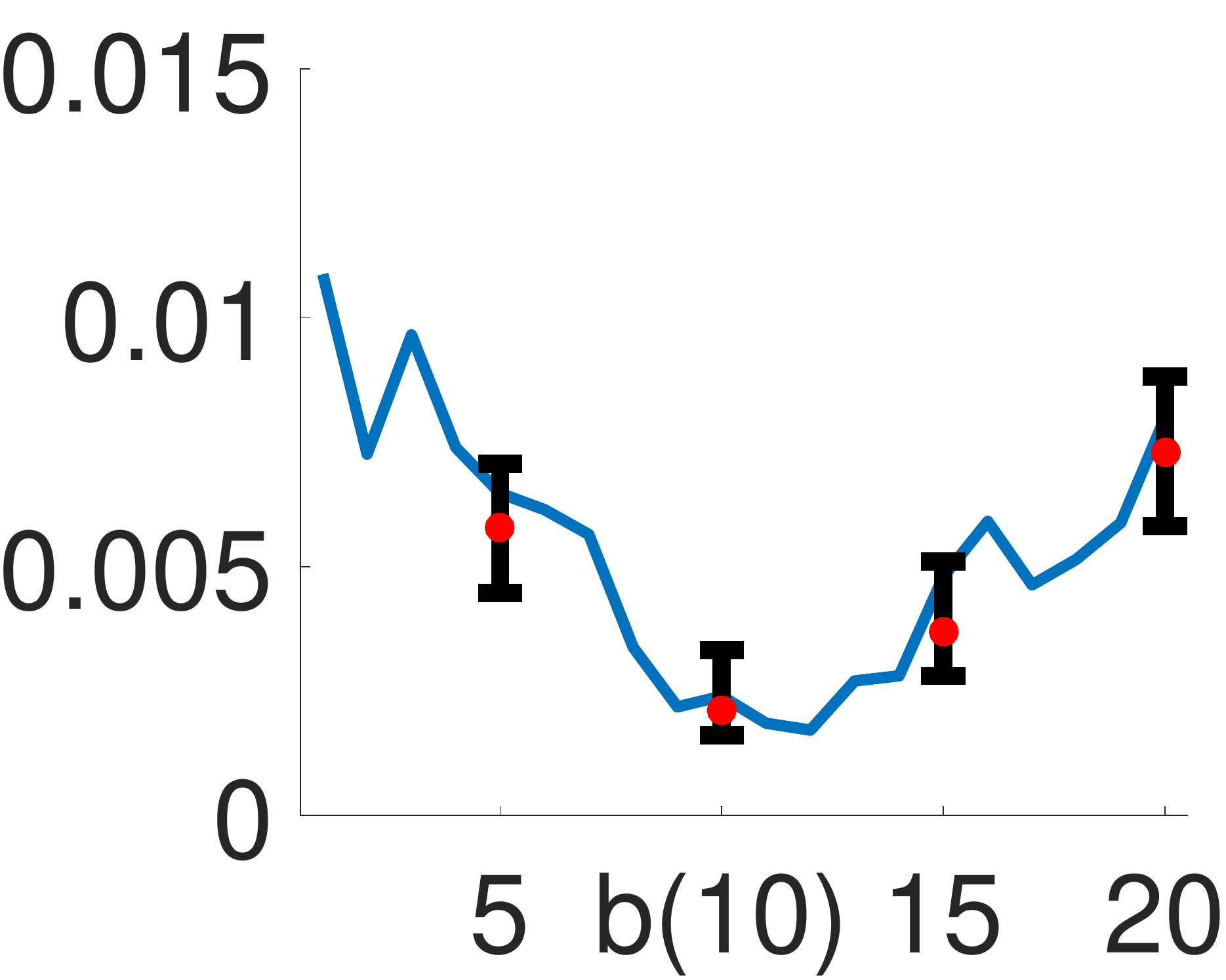}     & \includegraphics[scale = 0.18]{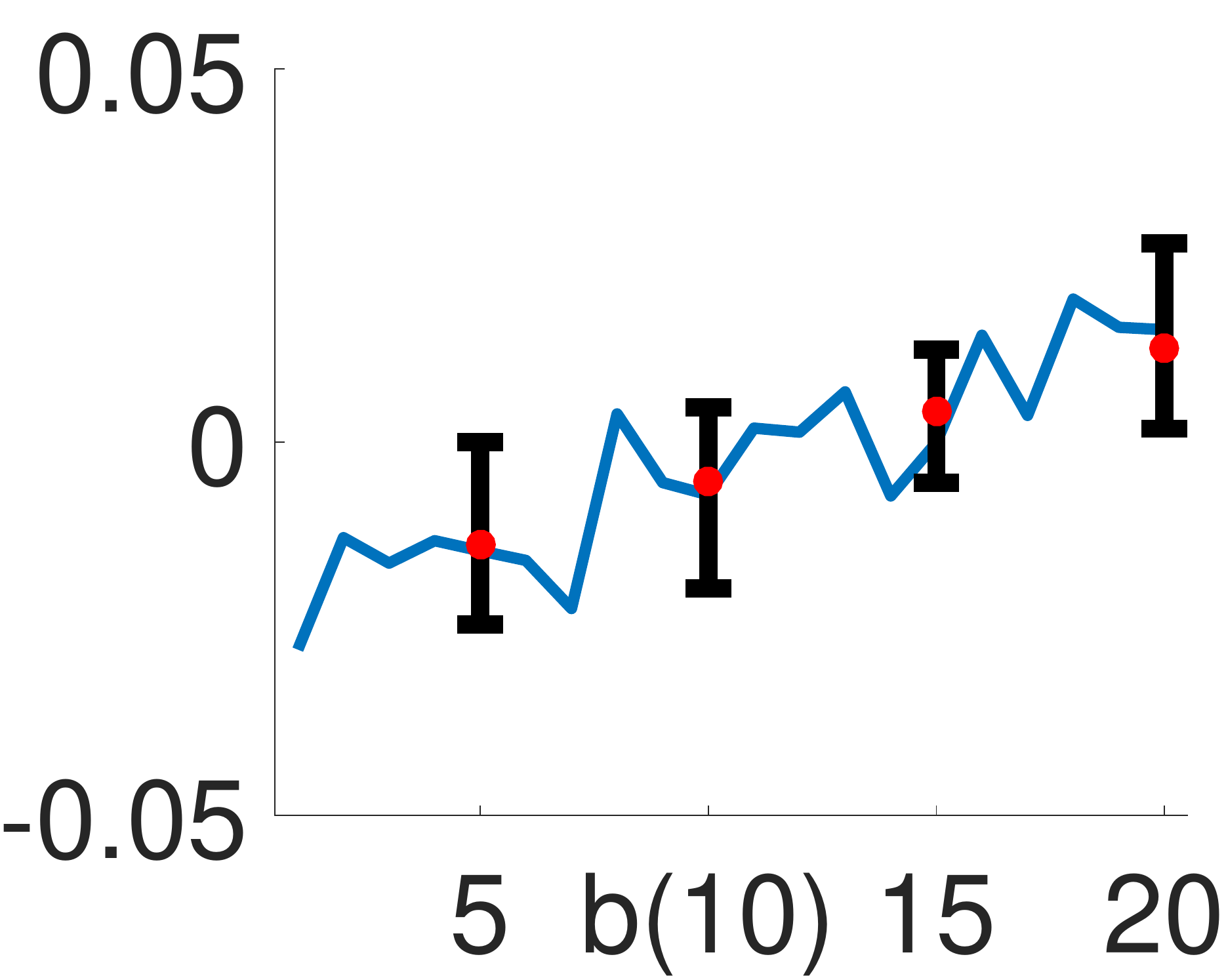}   &	\includegraphics[scale = 0.18]{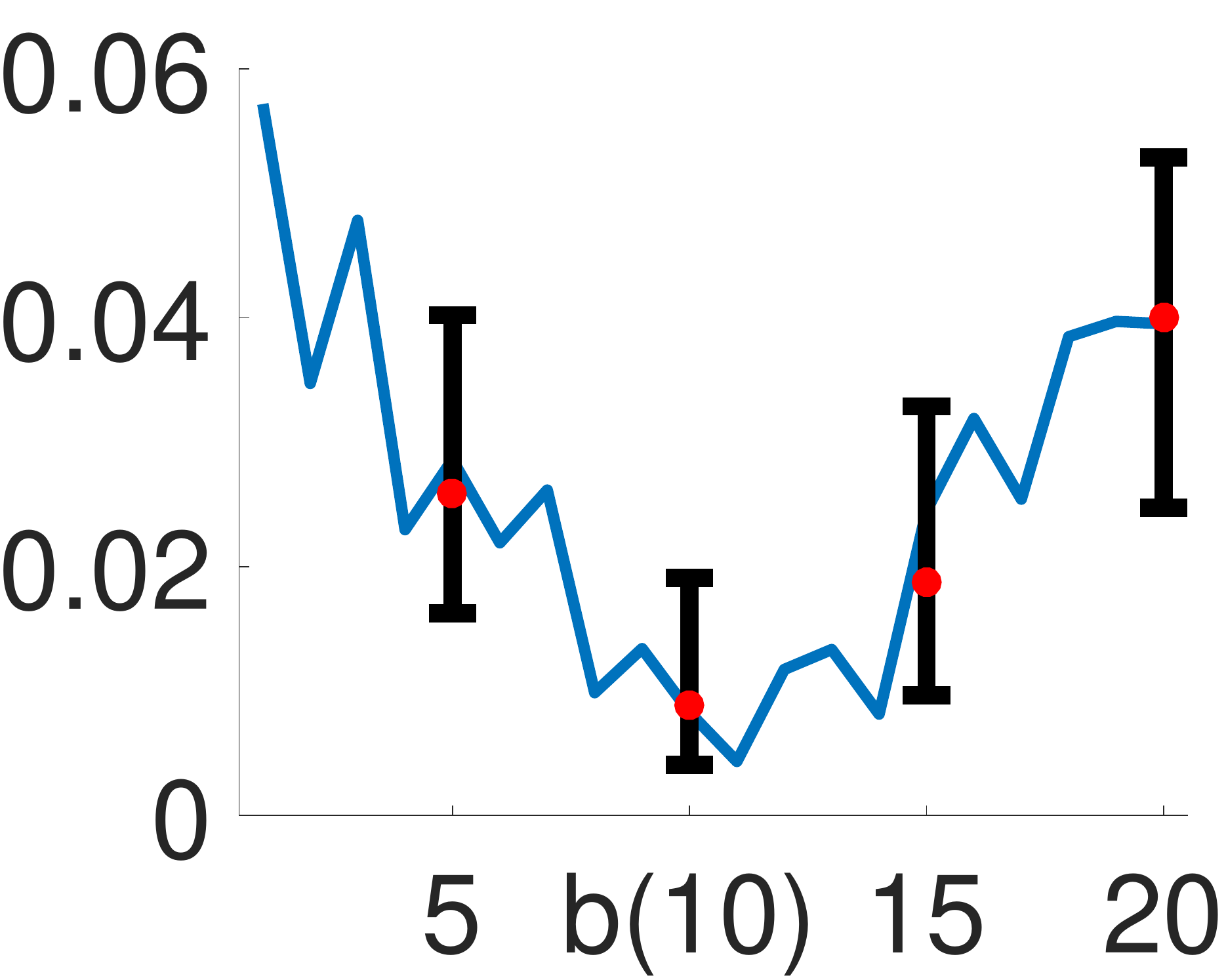}           \\ \hline
				\multicolumn{1}{|c|}{$\eta$}   &   \includegraphics[scale = 0.18]{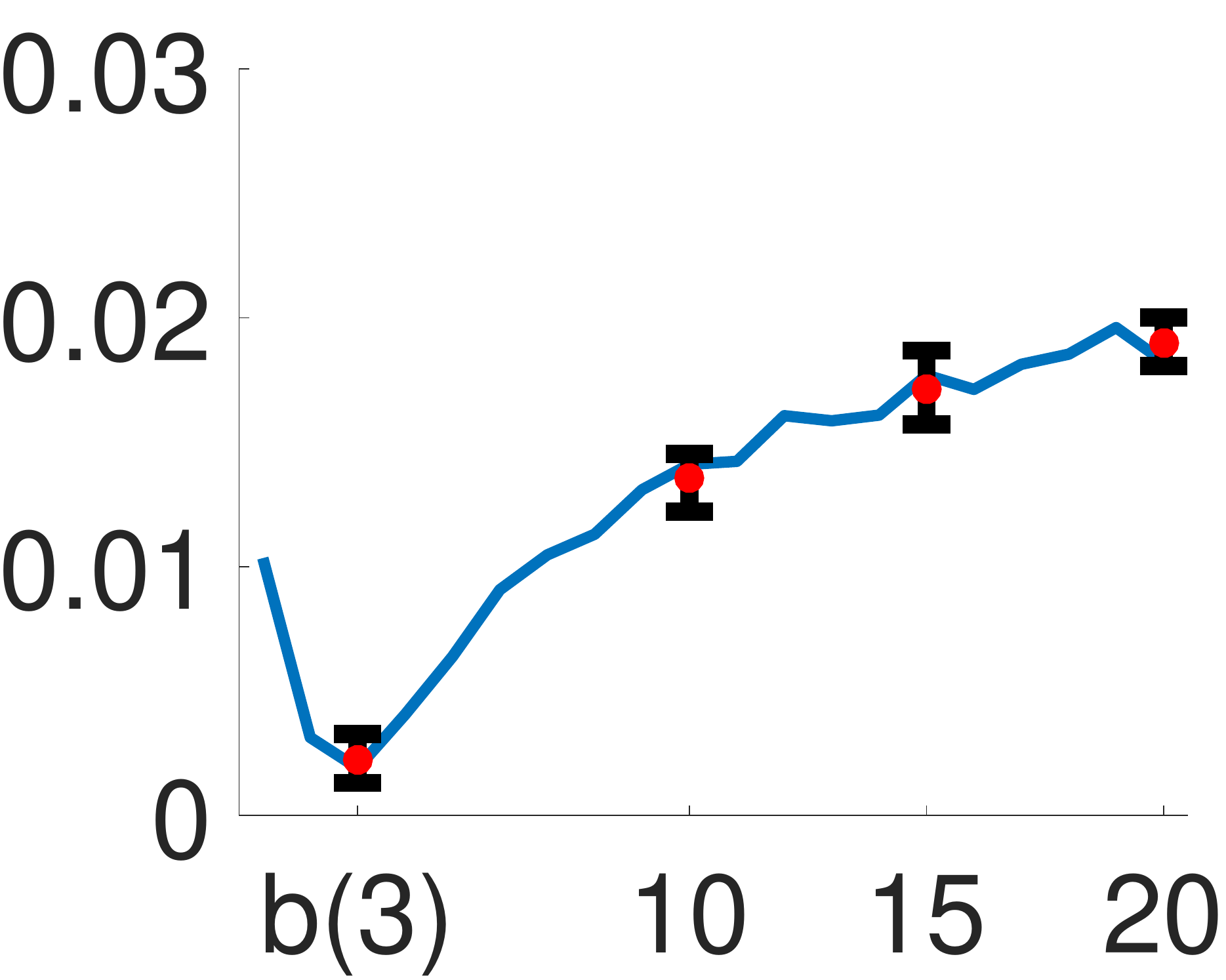}   & \includegraphics[scale = 0.18]{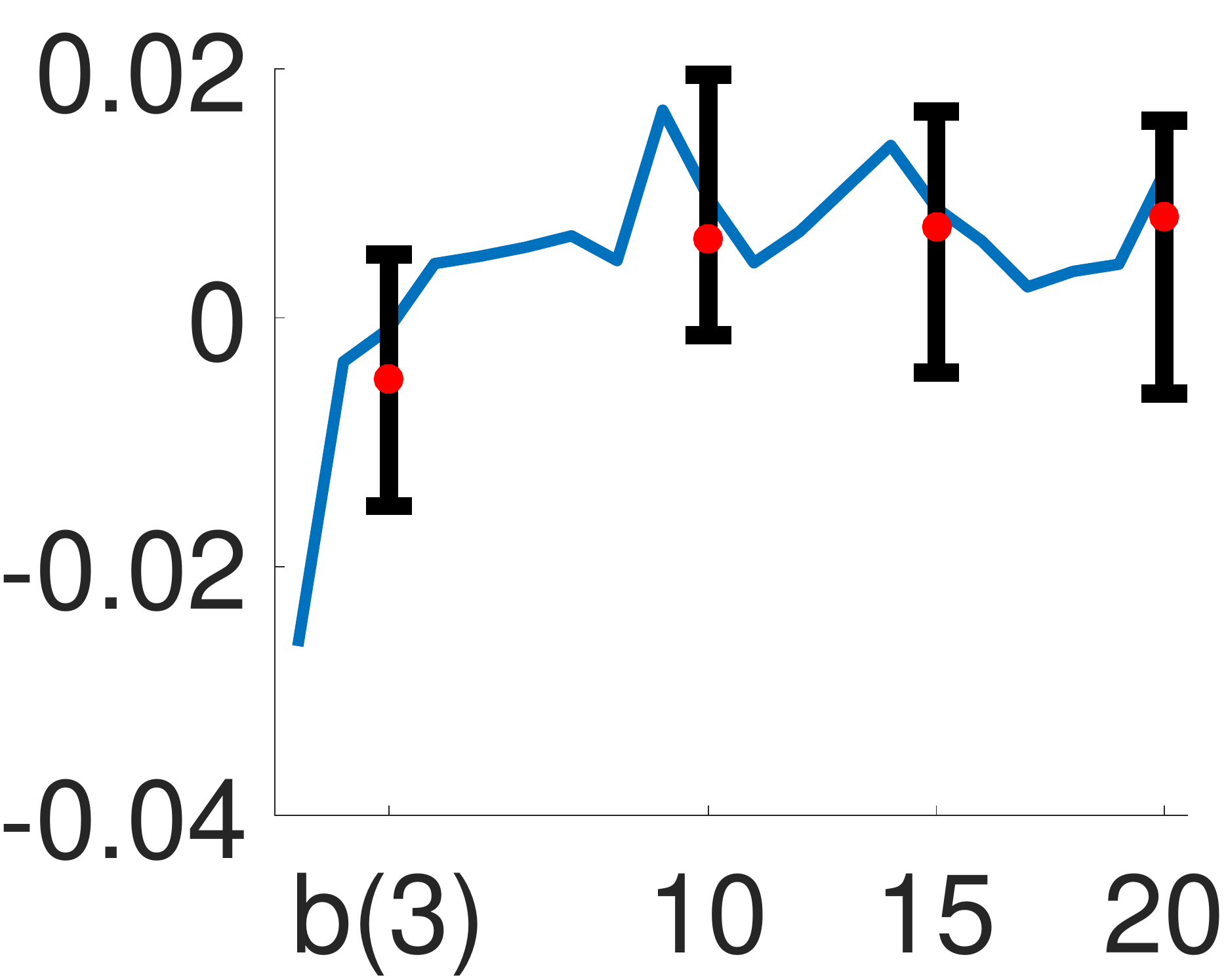} &	\includegraphics[scale = 0.18]{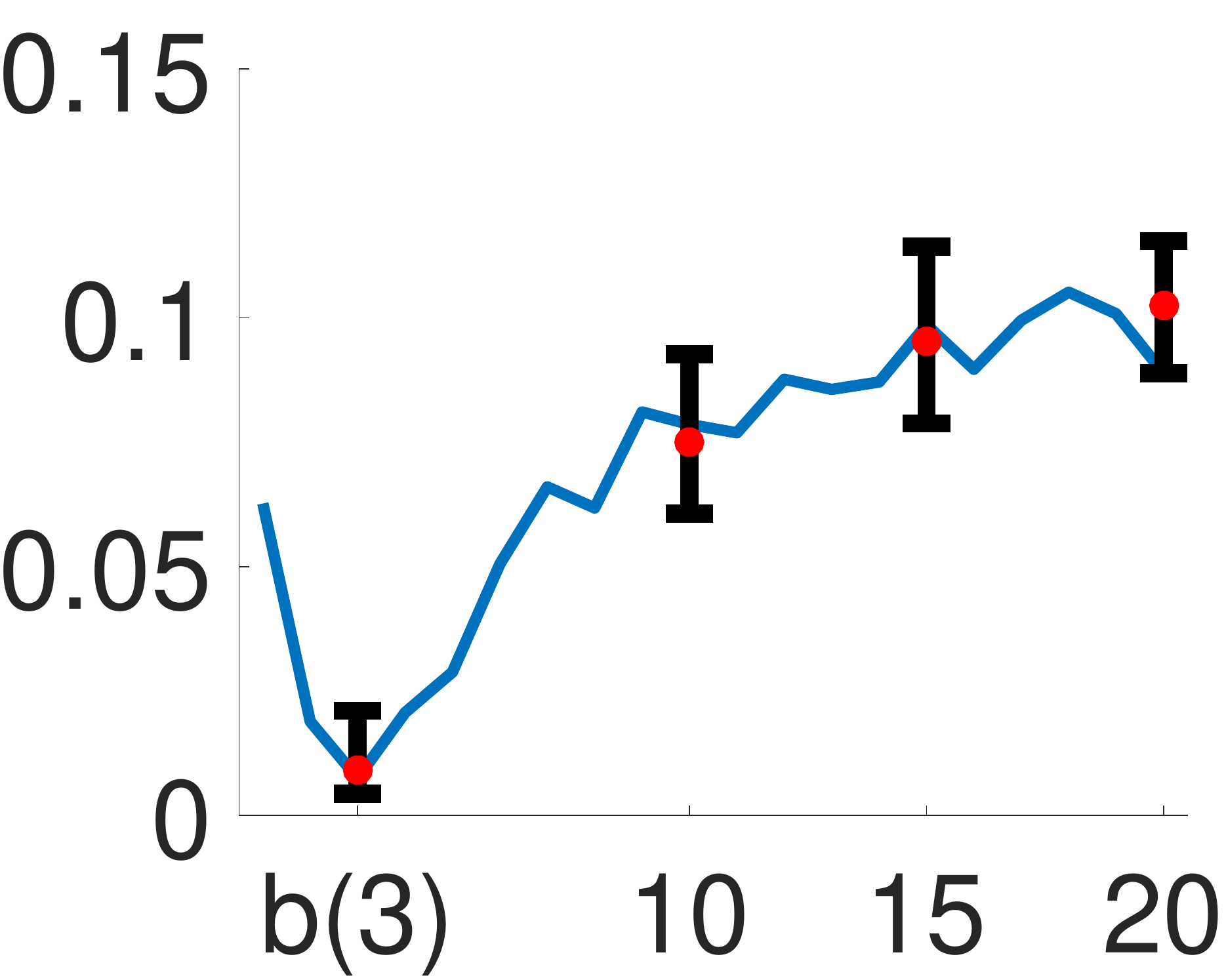}           \\ \hline
				\multicolumn{1}{|c|}{$\gamma$} &  \includegraphics[scale = 0.18]{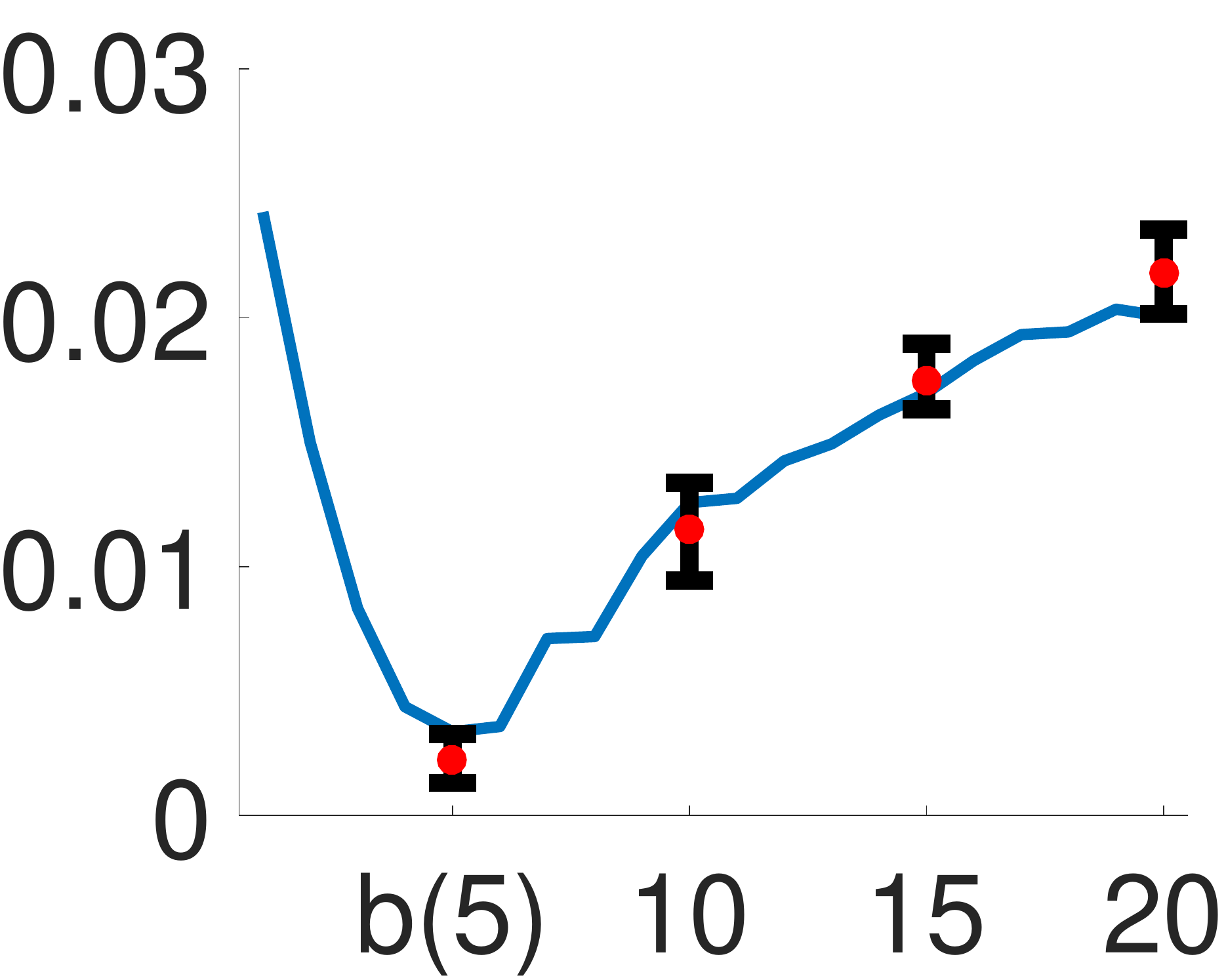}      & \includegraphics[scale = 0.18]{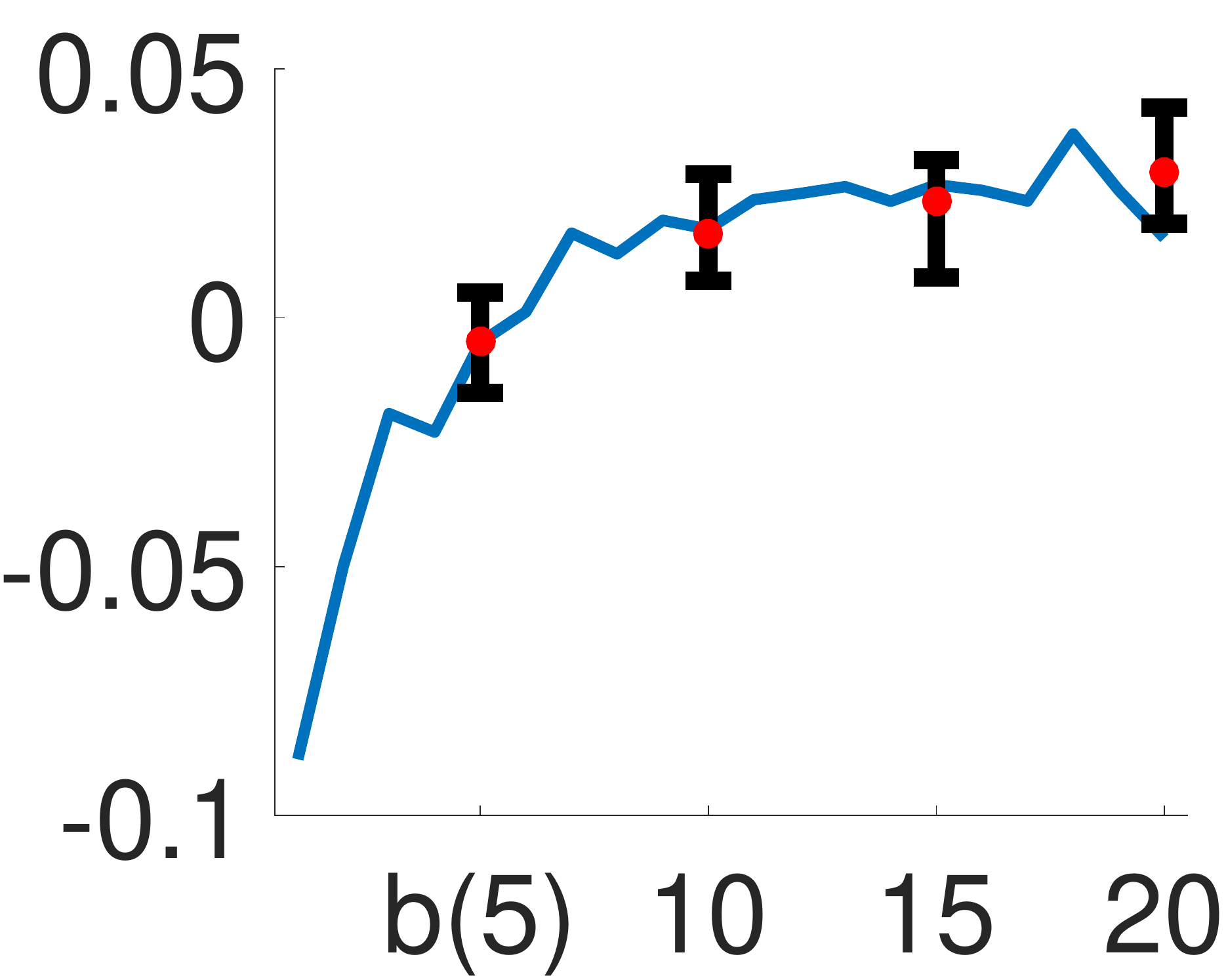}  &	\includegraphics[scale = 0.18]{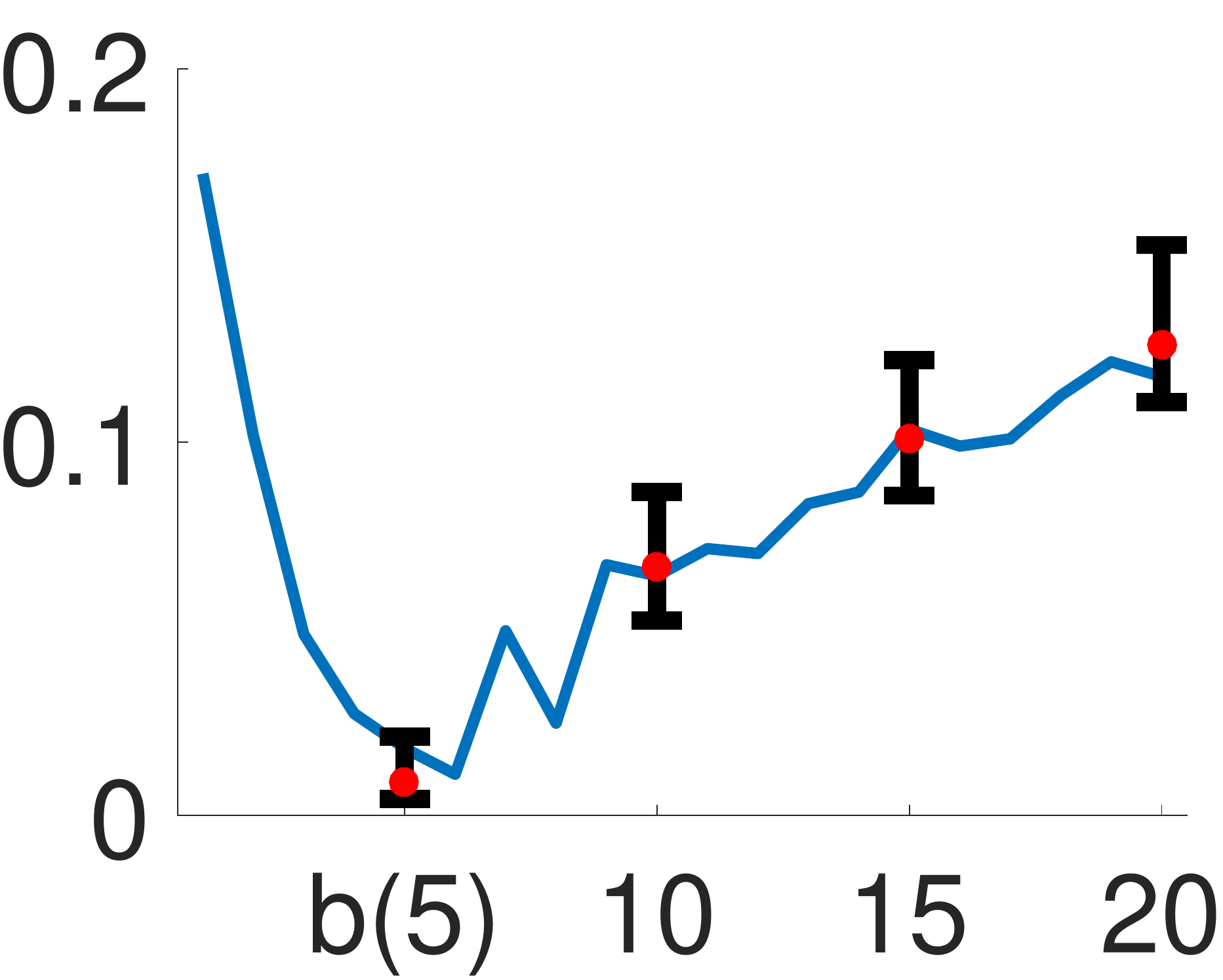}           \\ \hline
			\end{tabular}
		\end{center}
		\caption{Assessment of sensitivity using all of the proposed measures for various perturbations of prior parameters in the CCV model. The baseline setting is marked with a $b$.}
		\label{fig:CCV_sim_combined}
	\end{figure}

	\begin{figure}[!t]
		\begin{center}
			\begin{tabular}{c|c|c|c|}
				\cline{2-4}
				& $\mathbb{D}$ & $\mathbb{V}$ & $\mathbb{E}$ \\ \hline
				\multicolumn{1}{|c|}{$a_0$} & \includegraphics[scale = 0.18]{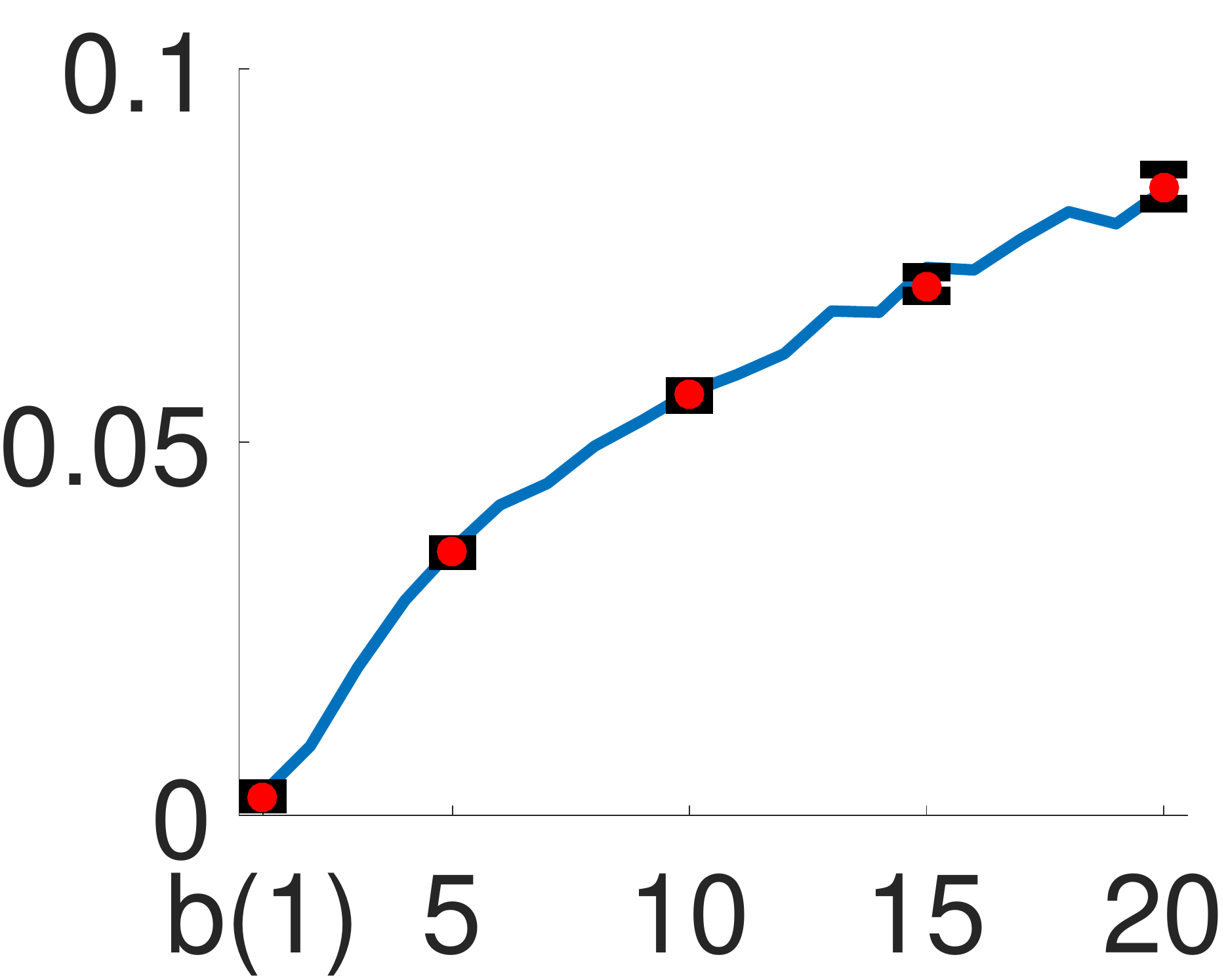}     & \includegraphics[scale = 0.18]{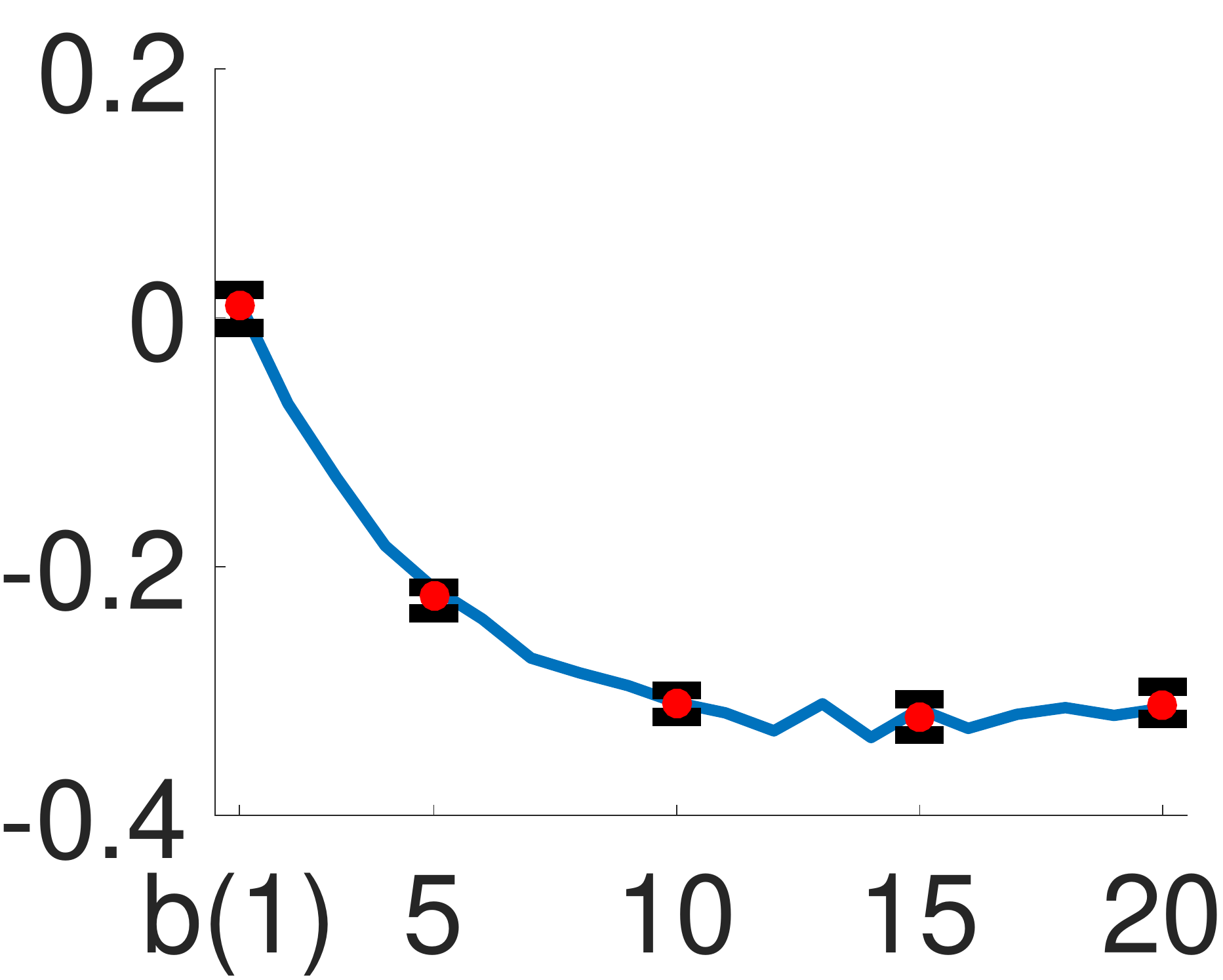}   &	\includegraphics[scale = 0.18]{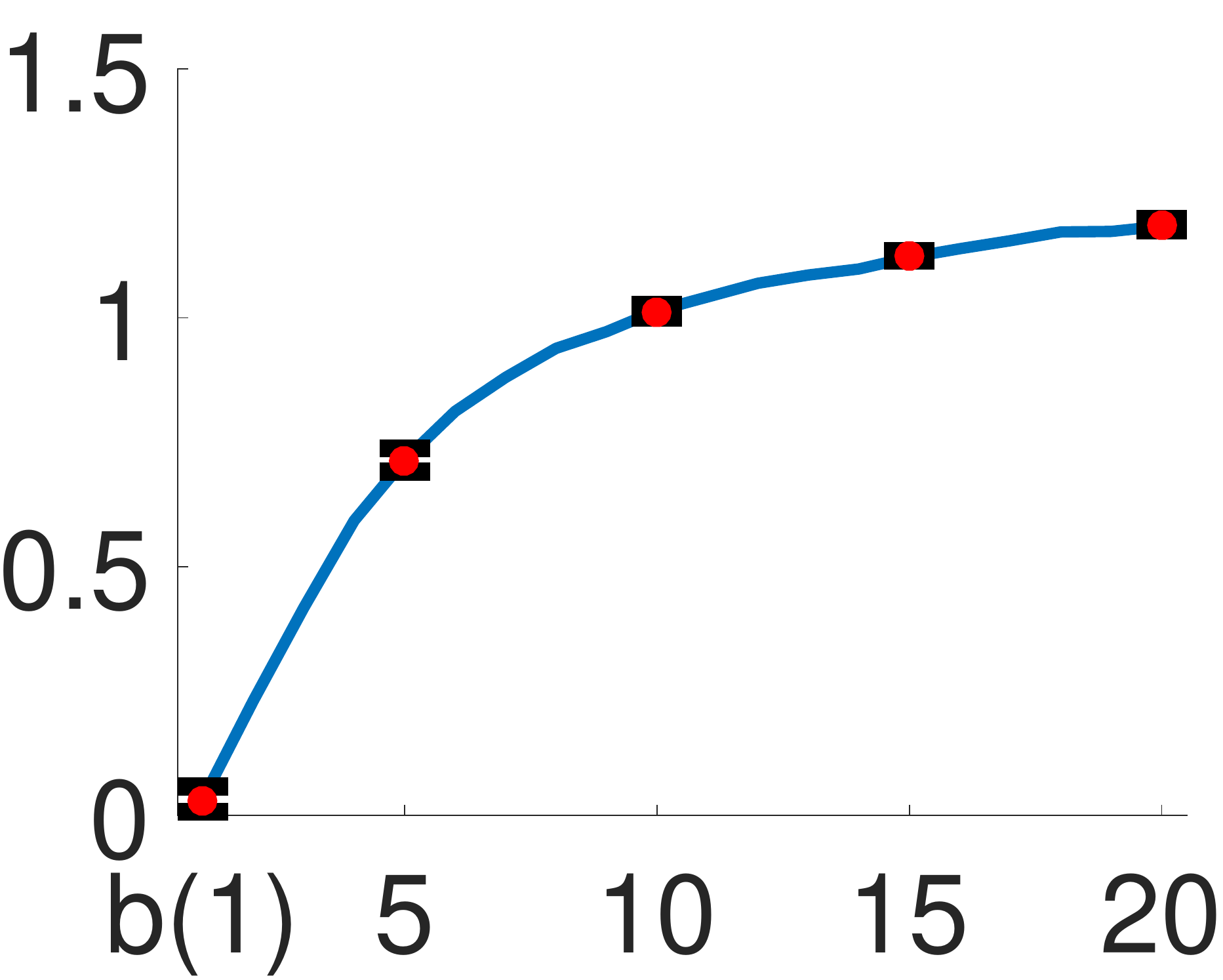}                \\ \hline
				\multicolumn{1}{|c|}{$a_1$} & \includegraphics[scale = 0.18]{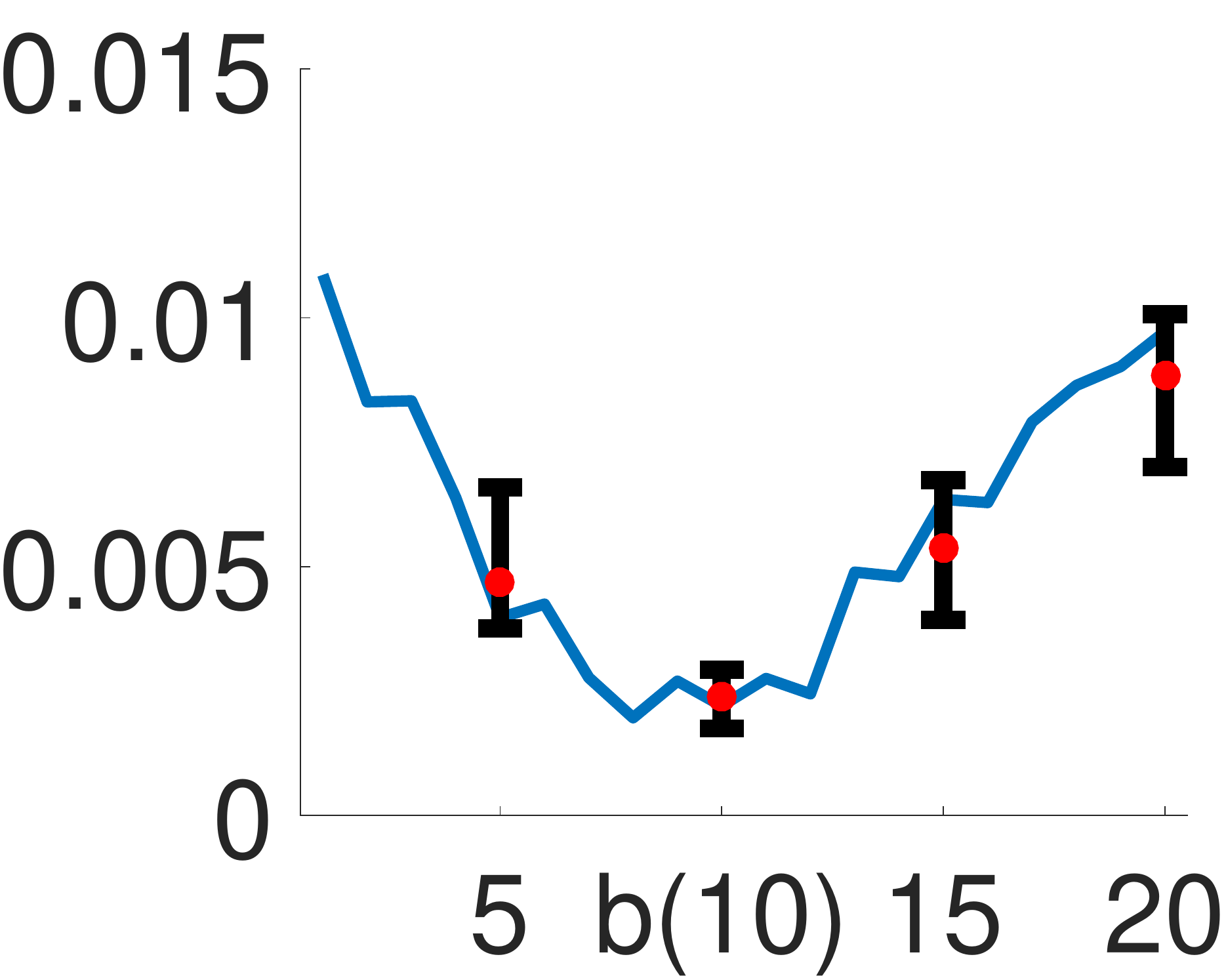}     & \includegraphics[scale = 0.18]{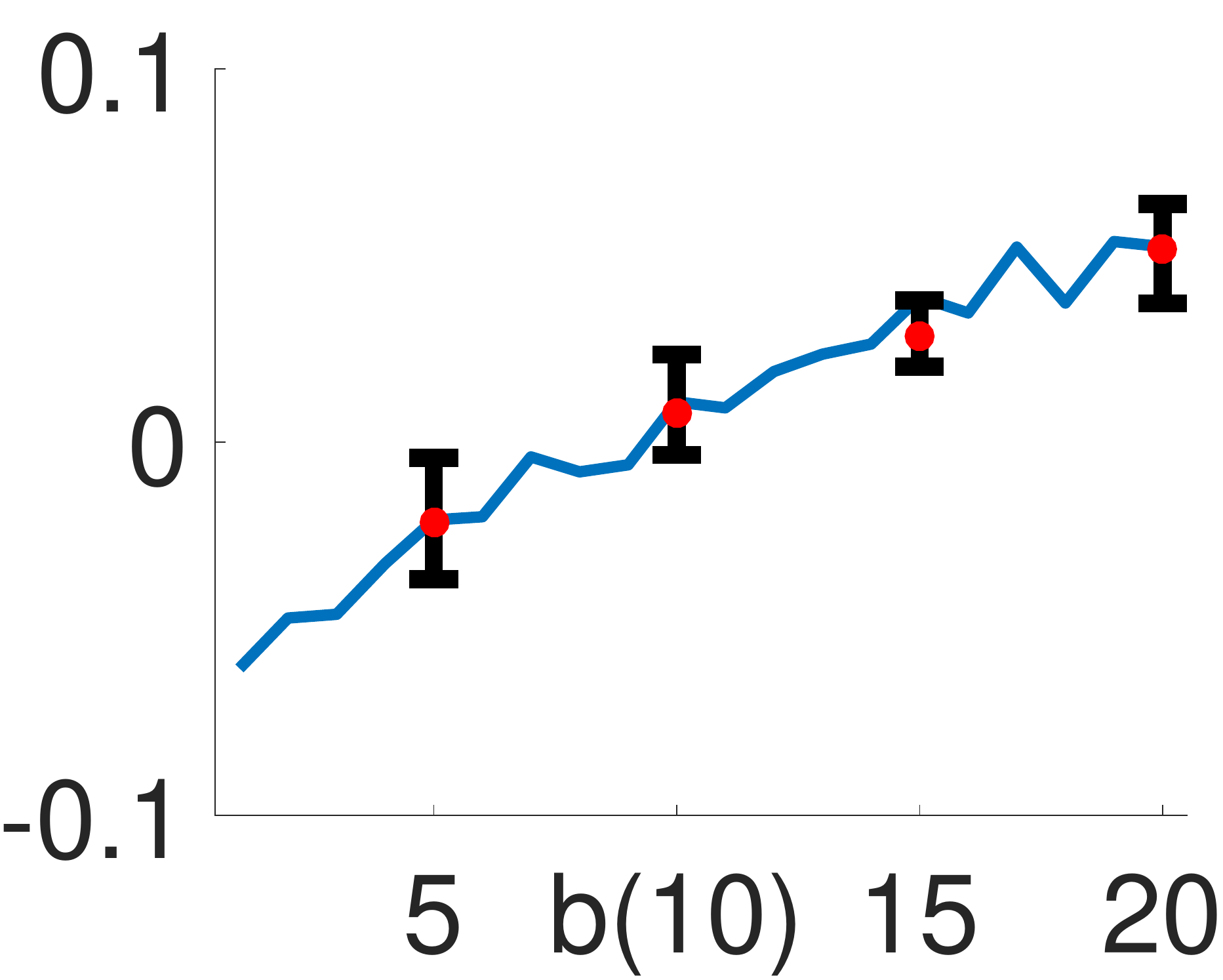}   &	\includegraphics[scale = 0.18]{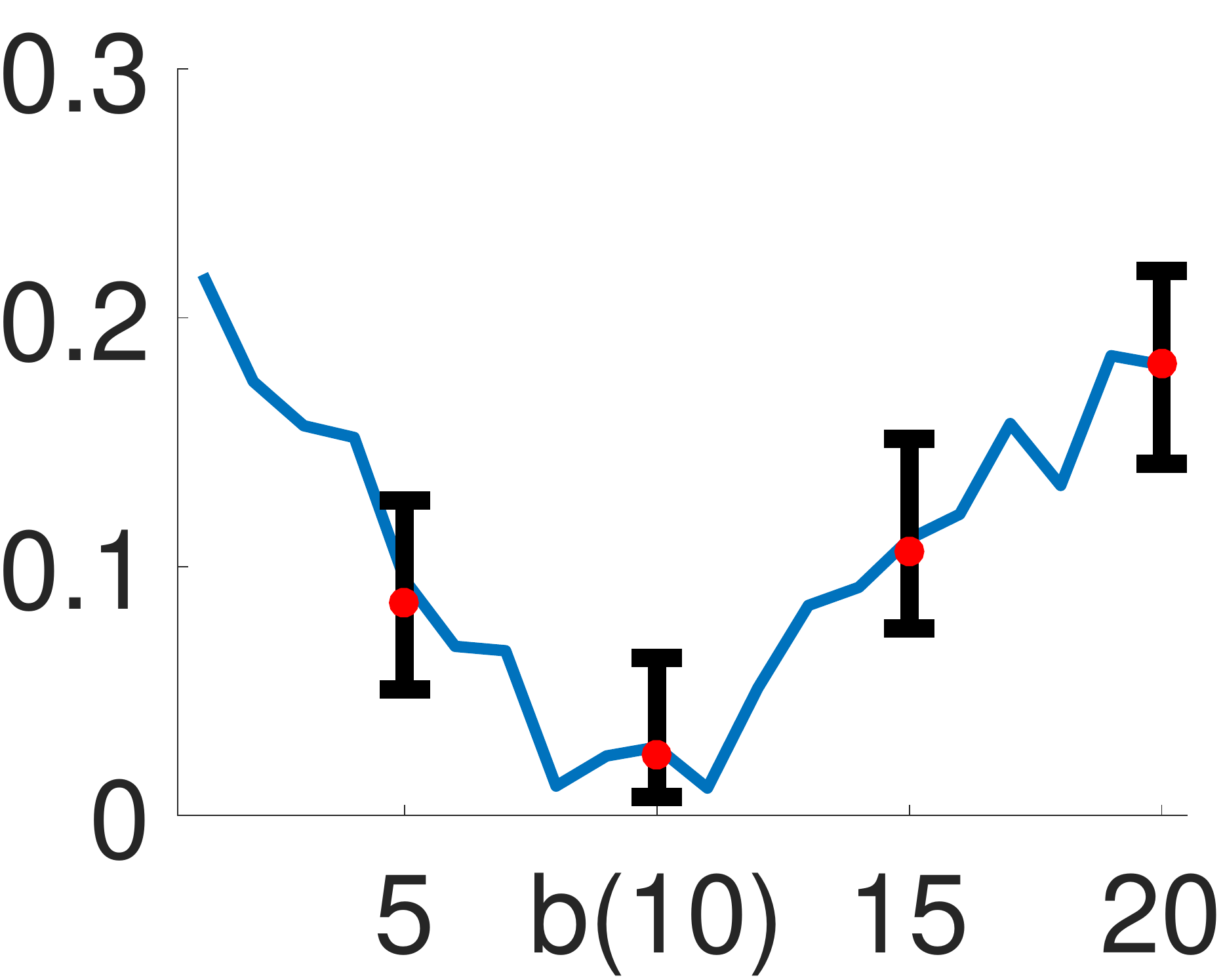}                \\ \hline
				\multicolumn{1}{|c|}{$\eta$}   &   \includegraphics[scale = 0.18]{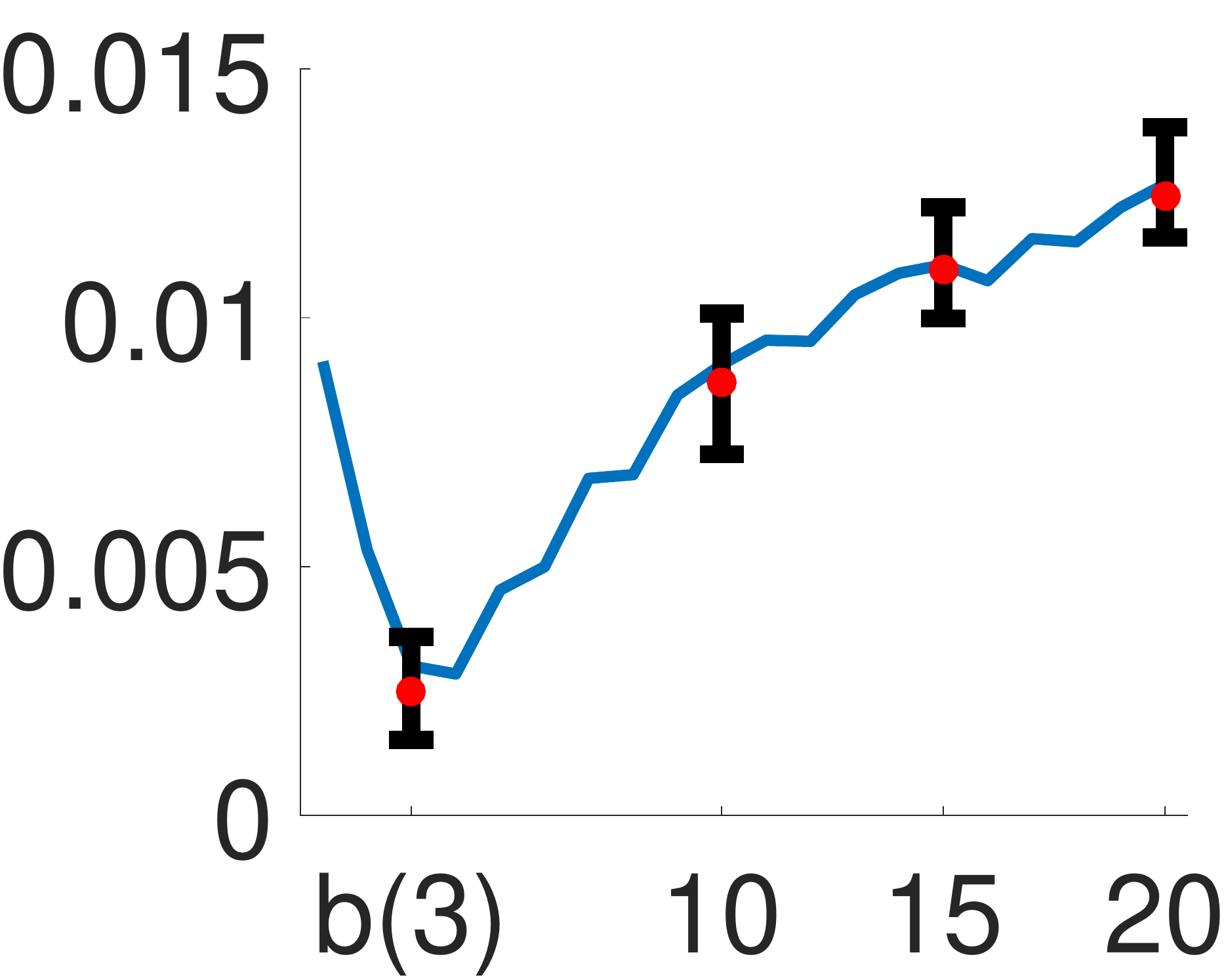}   & \includegraphics[scale = 0.18]{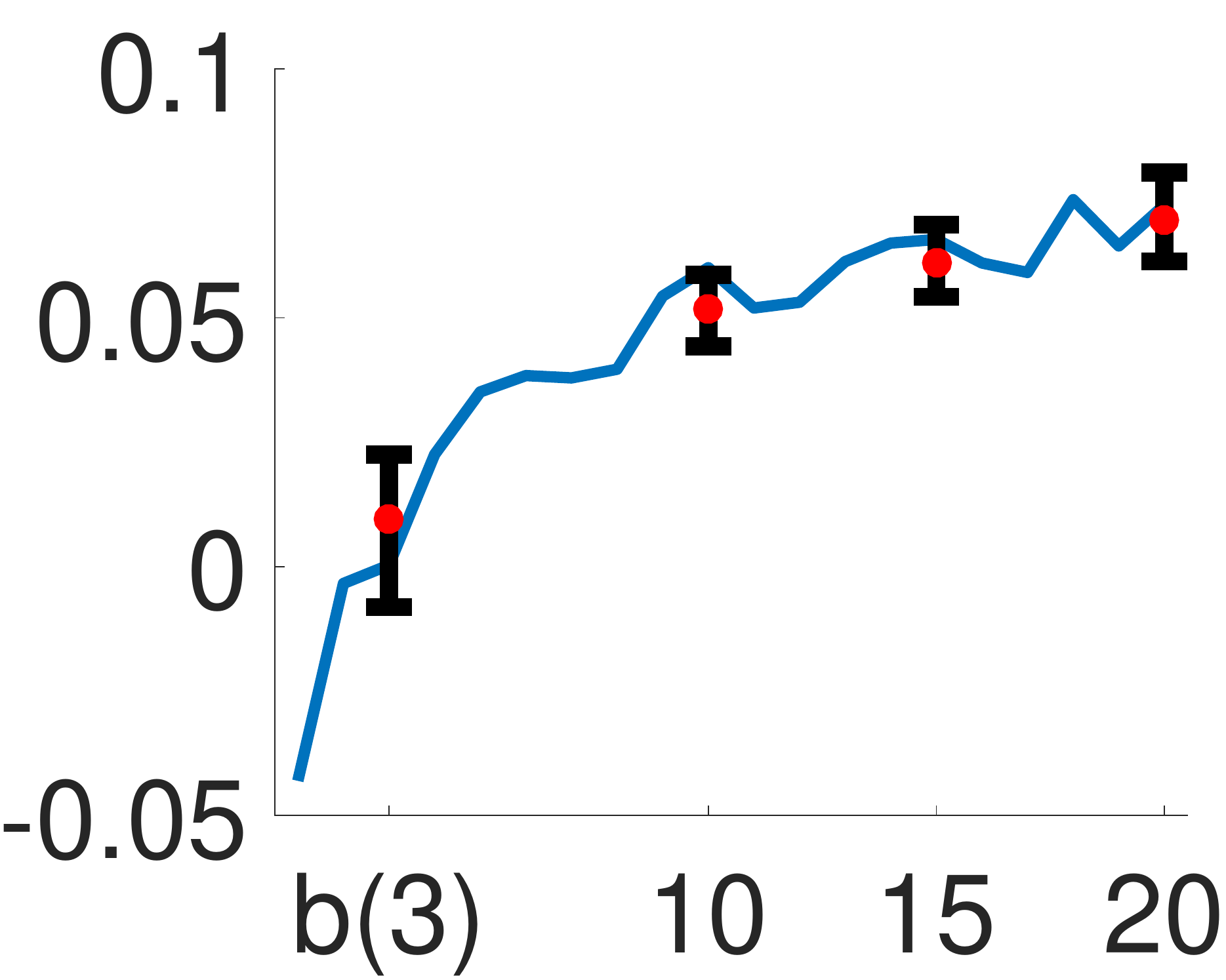} &	\includegraphics[scale = 0.18]{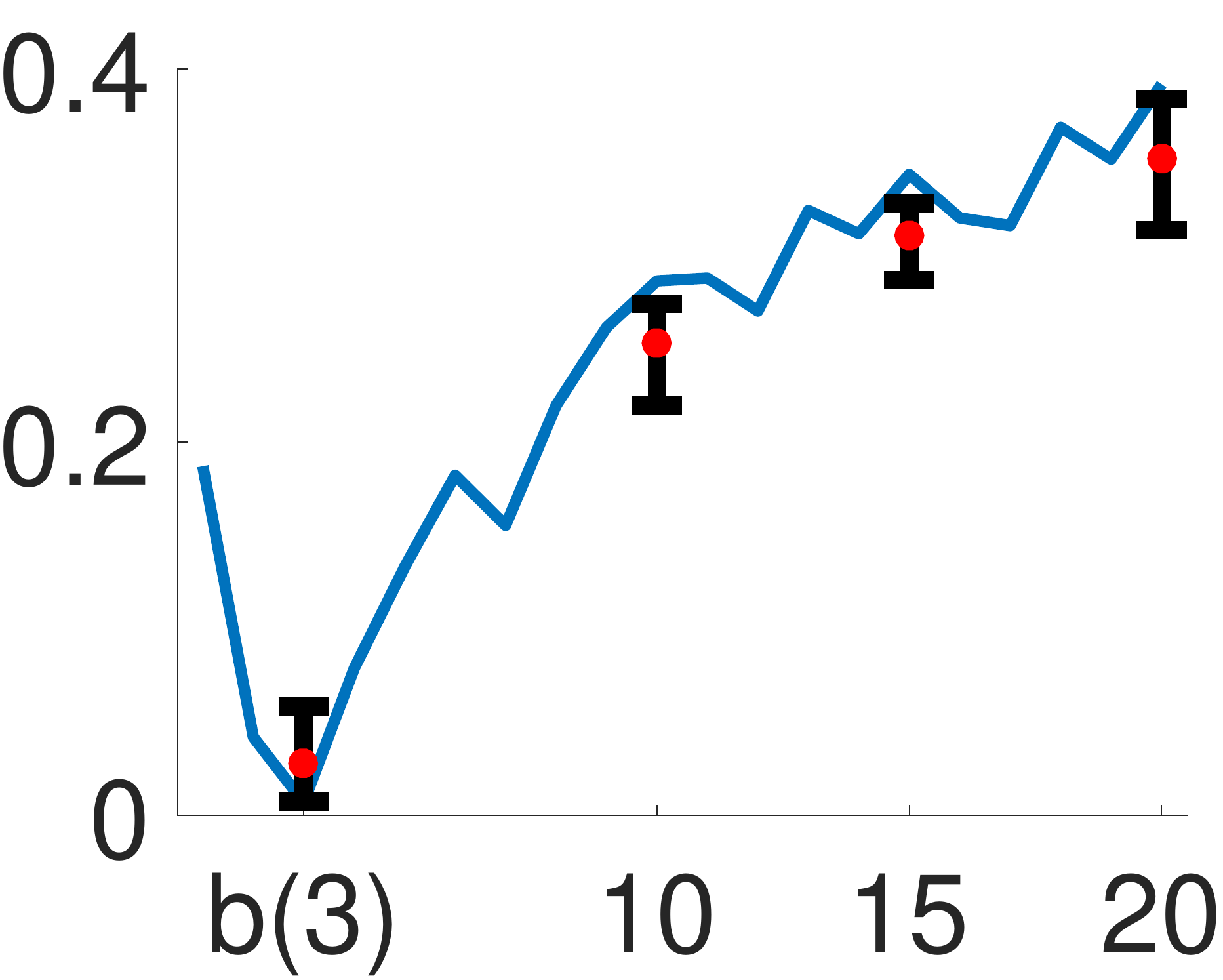}                \\ \hline
				\multicolumn{1}{|c|}{$\gamma$} &  \includegraphics[scale = 0.18]{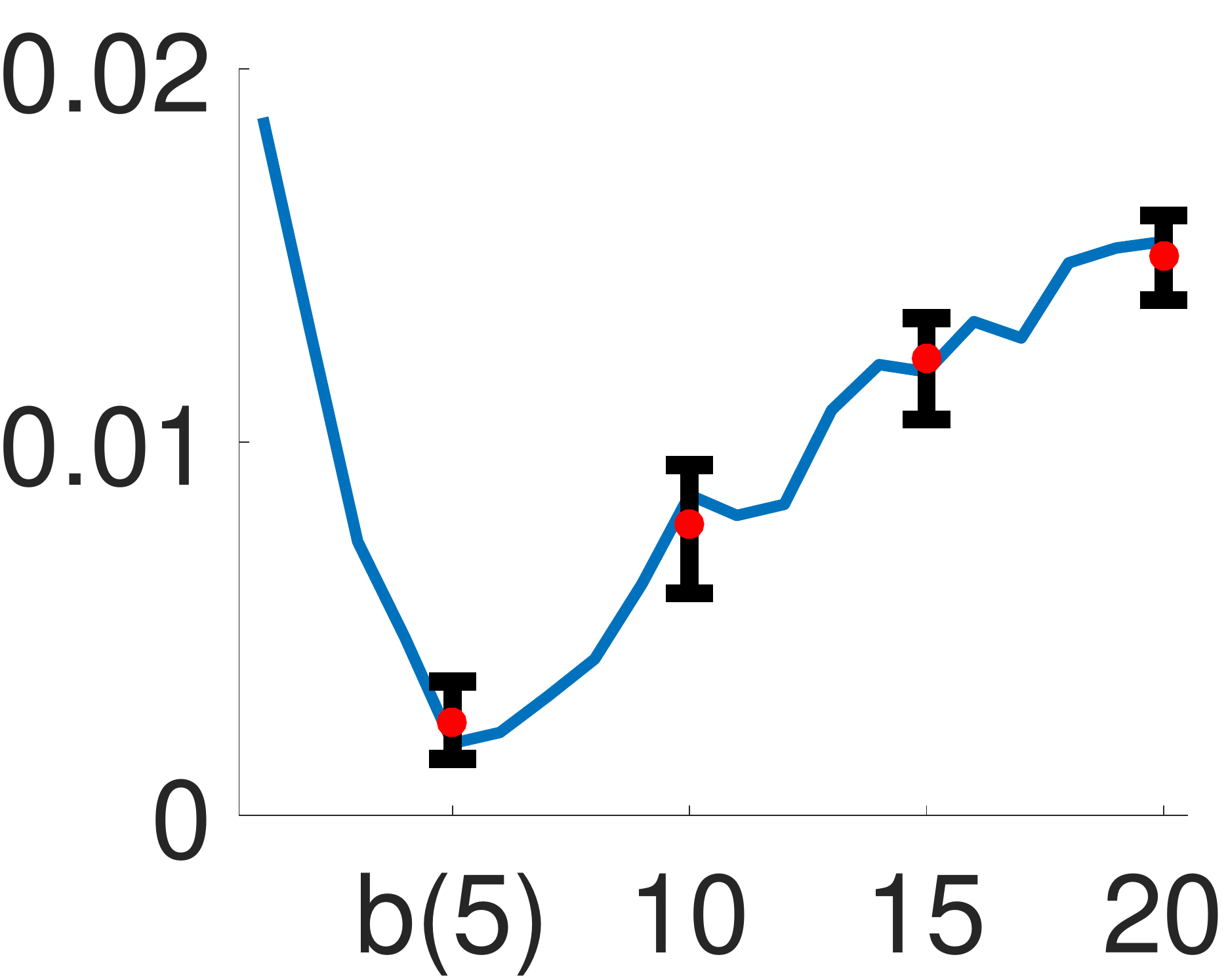}      & \includegraphics[scale = 0.18]{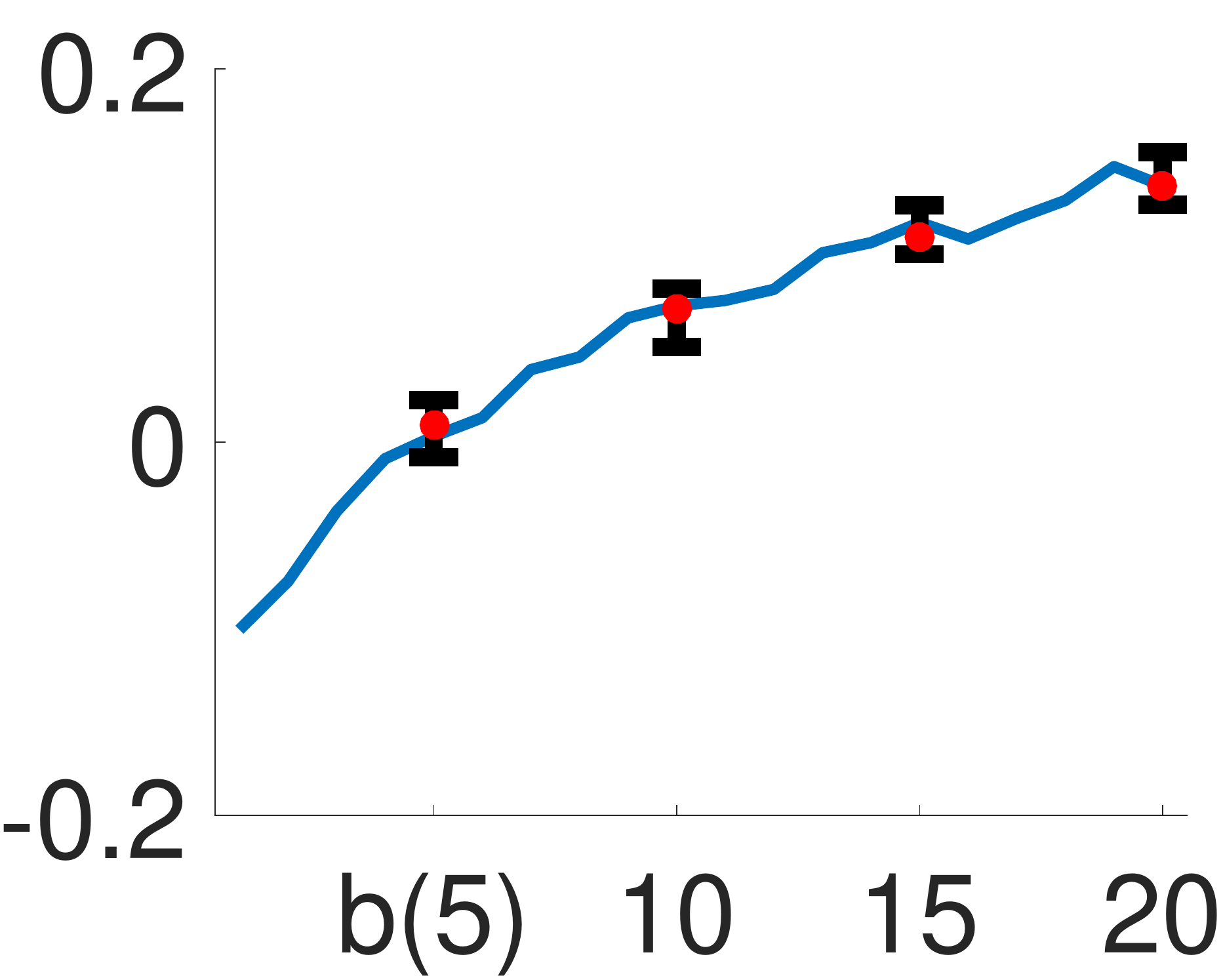}  &	\includegraphics[scale = 0.18]{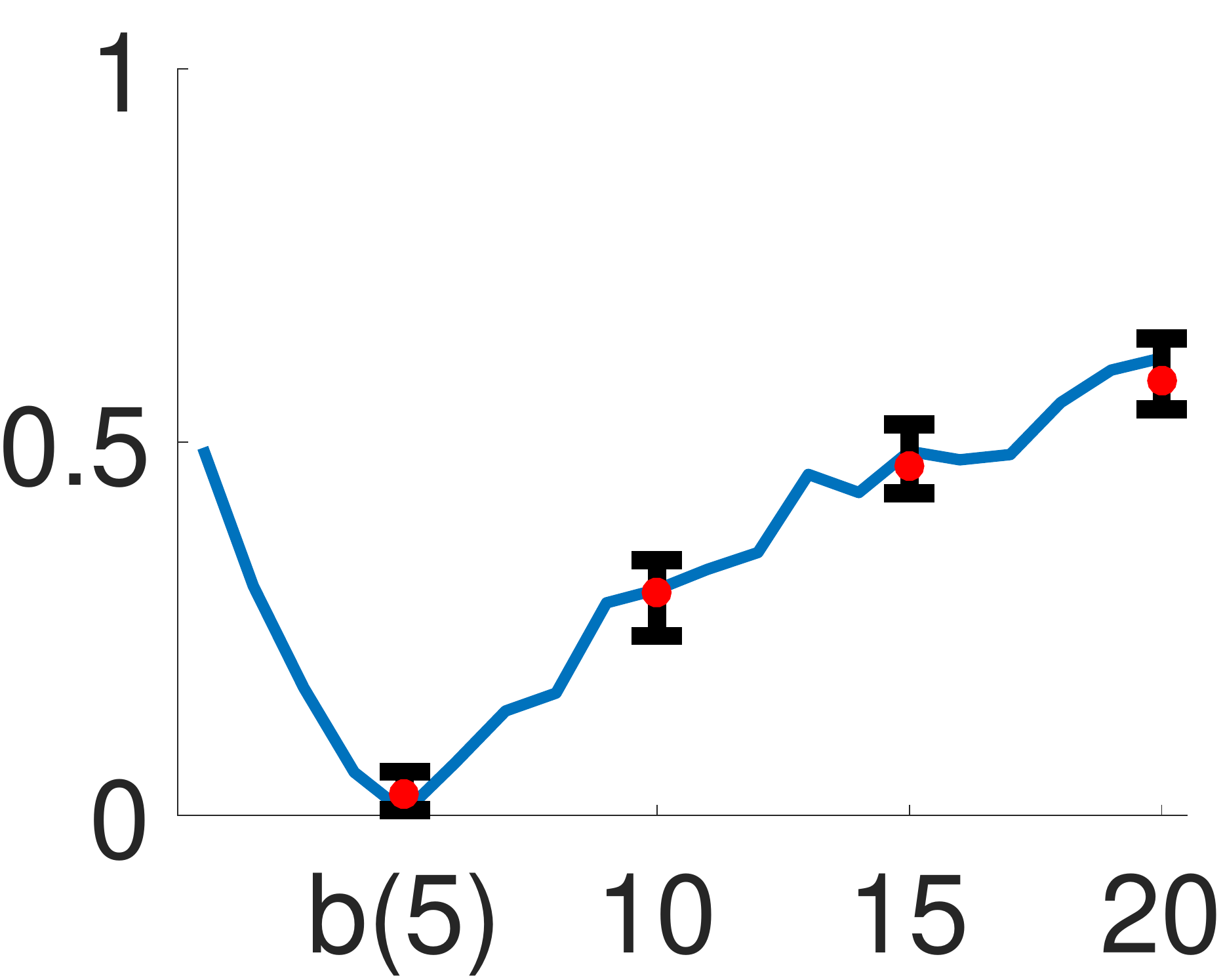}                \\ \hline
				\multicolumn{1}{|c|}{$\phi$}   &   \includegraphics[scale = 0.18]{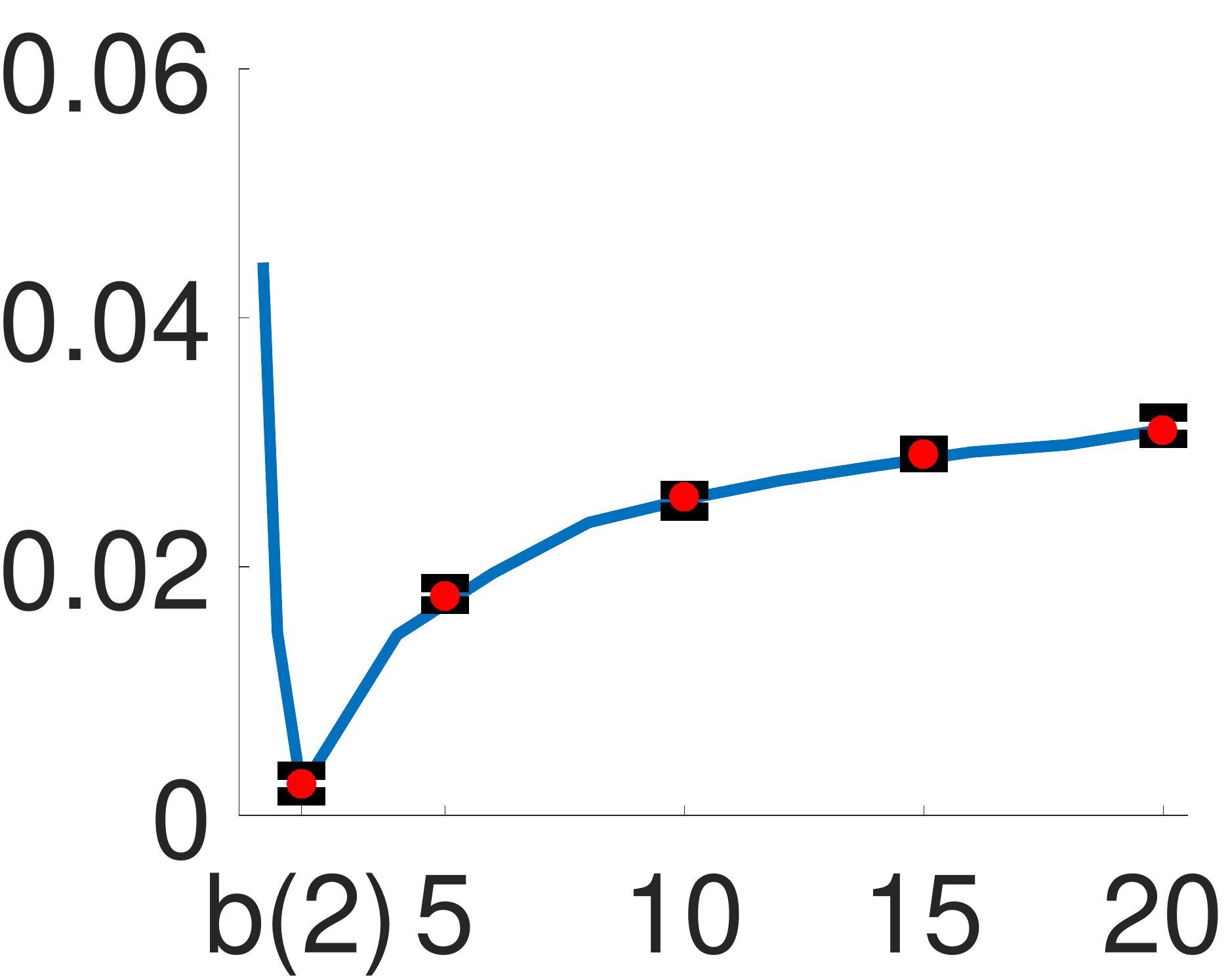}   & \includegraphics[scale = 0.18]{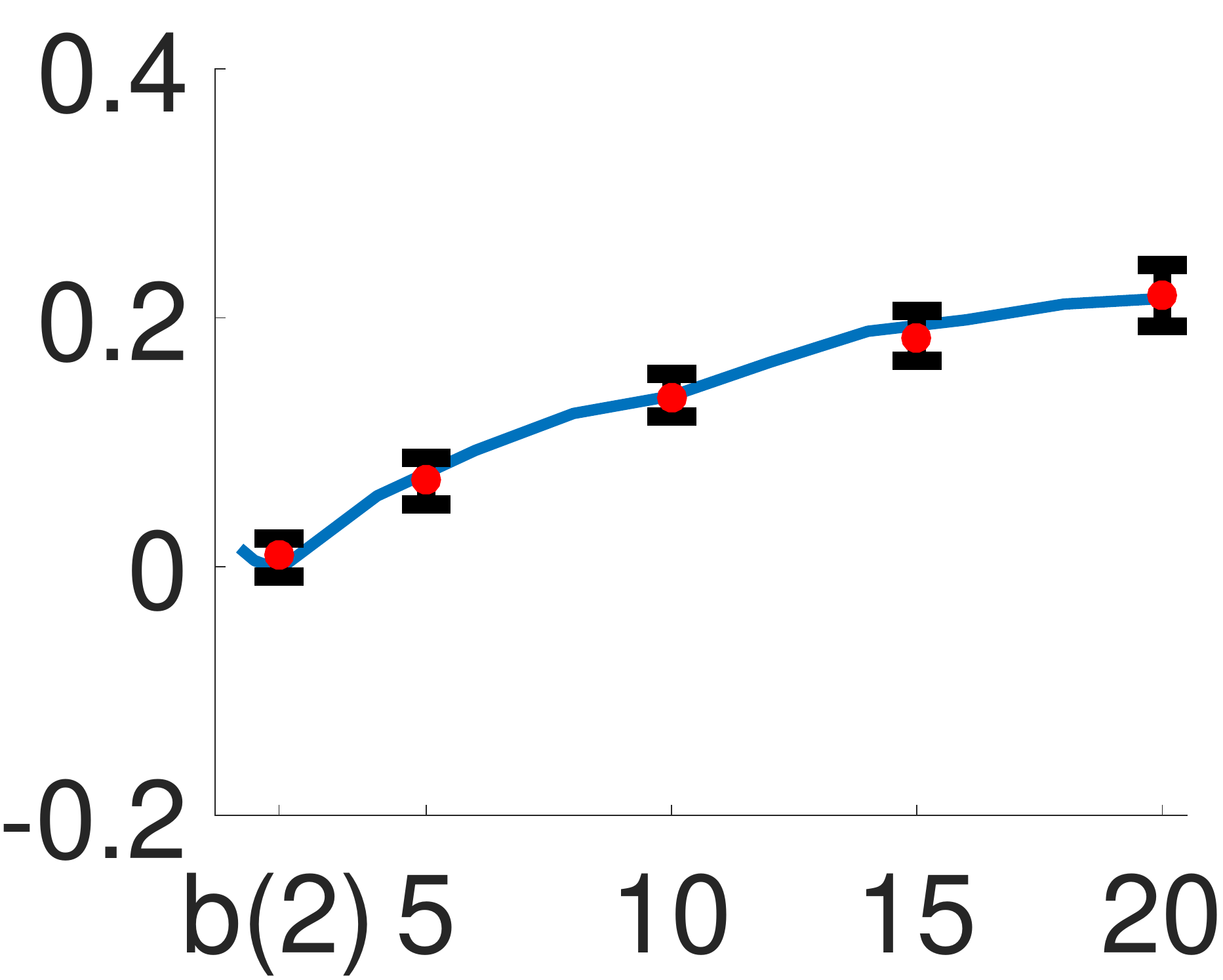}   &	\includegraphics[scale = 0.18]{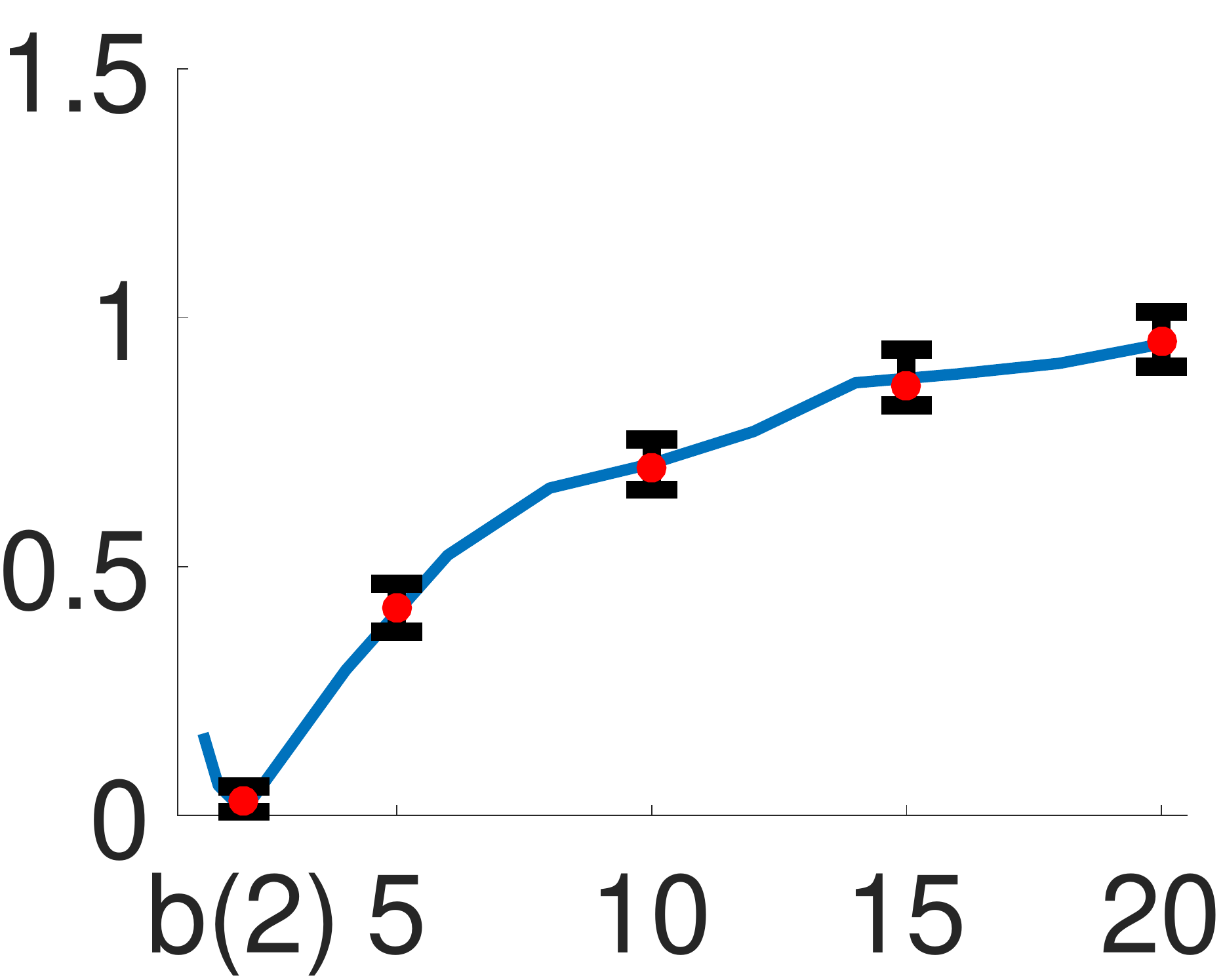}                    \\ \hline
			\end{tabular}
		\end{center}
		\caption{Assessment of sensitivity using all of the proposed measures for various perturbations of prior parameters in the DCV model. The baseline setting is marked with a $b$.}
		\label{fig:DCV_sim_combined}
	\end{figure}

	\begin{figure}[!t]
		\begin{center}
			\begin{tabular}{c|c|c|}
				\cline{2-3}
				& (a) & (b) \\ \hline
				\multicolumn{1}{|c|}{$a_0$}  & \includegraphics[scale = 0.27]{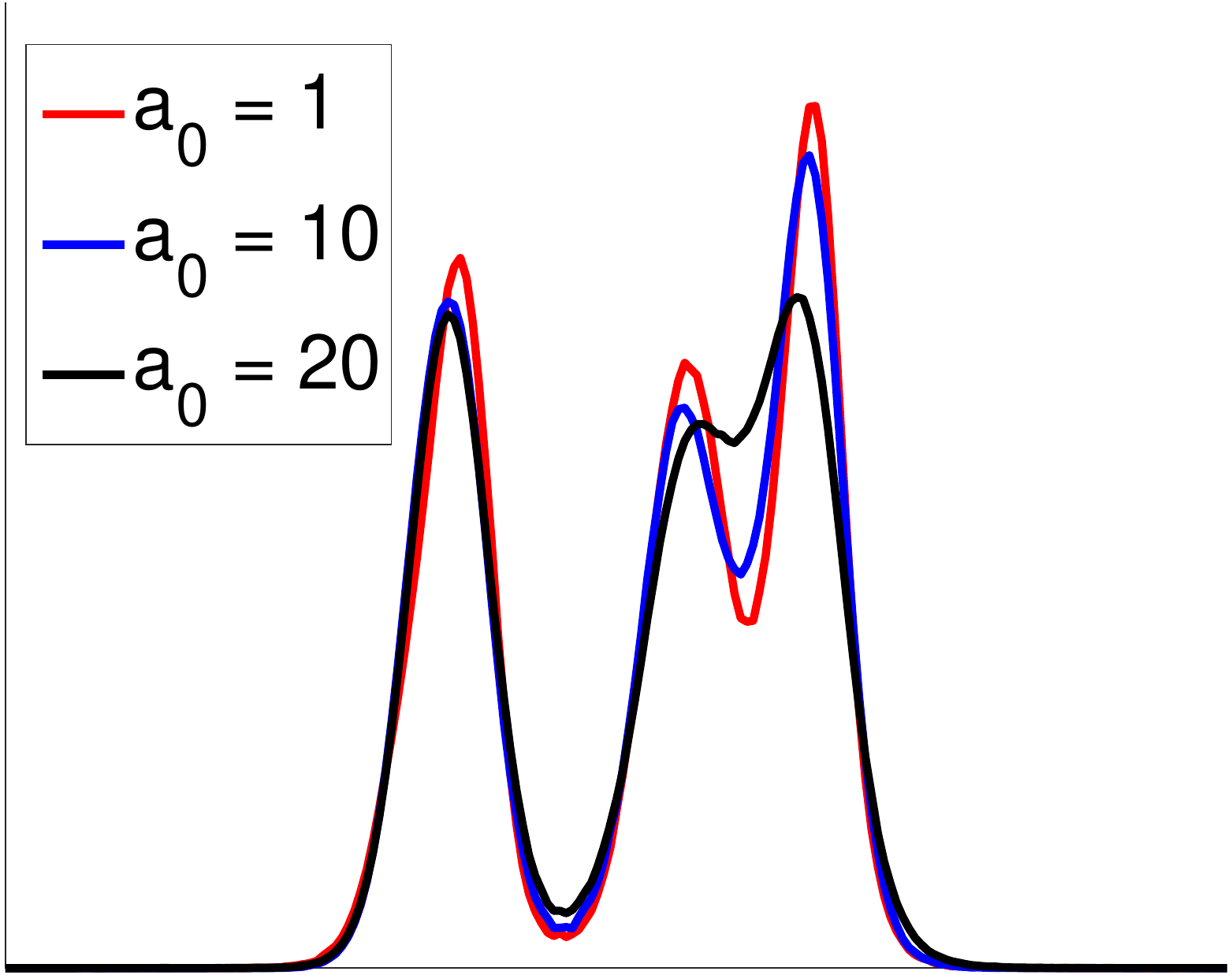}     & \includegraphics[scale = 0.24]{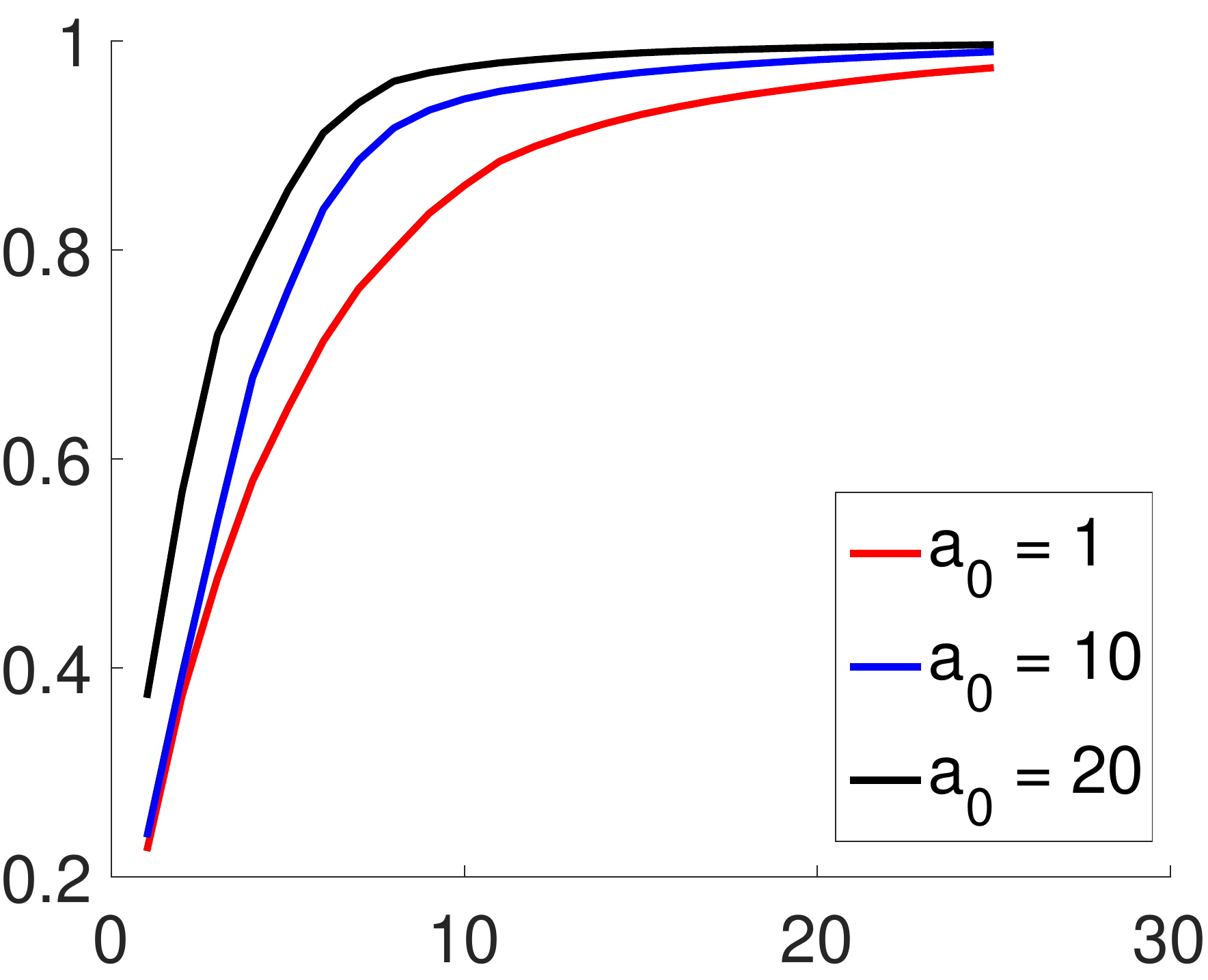}          \\ \hline
				\multicolumn{1}{|c|}{$\gamma$}  & \includegraphics[scale = 0.27]{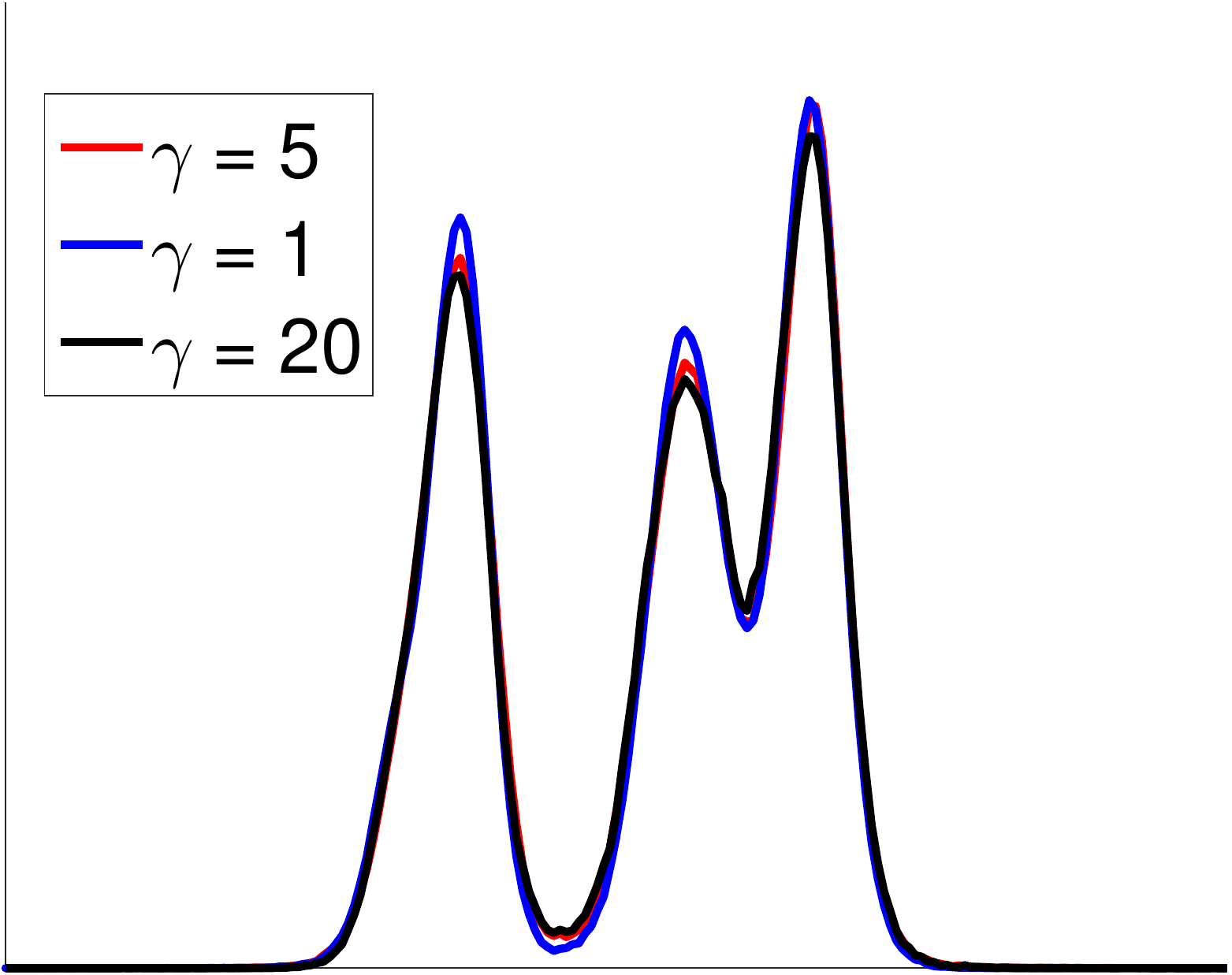}     & \includegraphics[scale = 0.24]{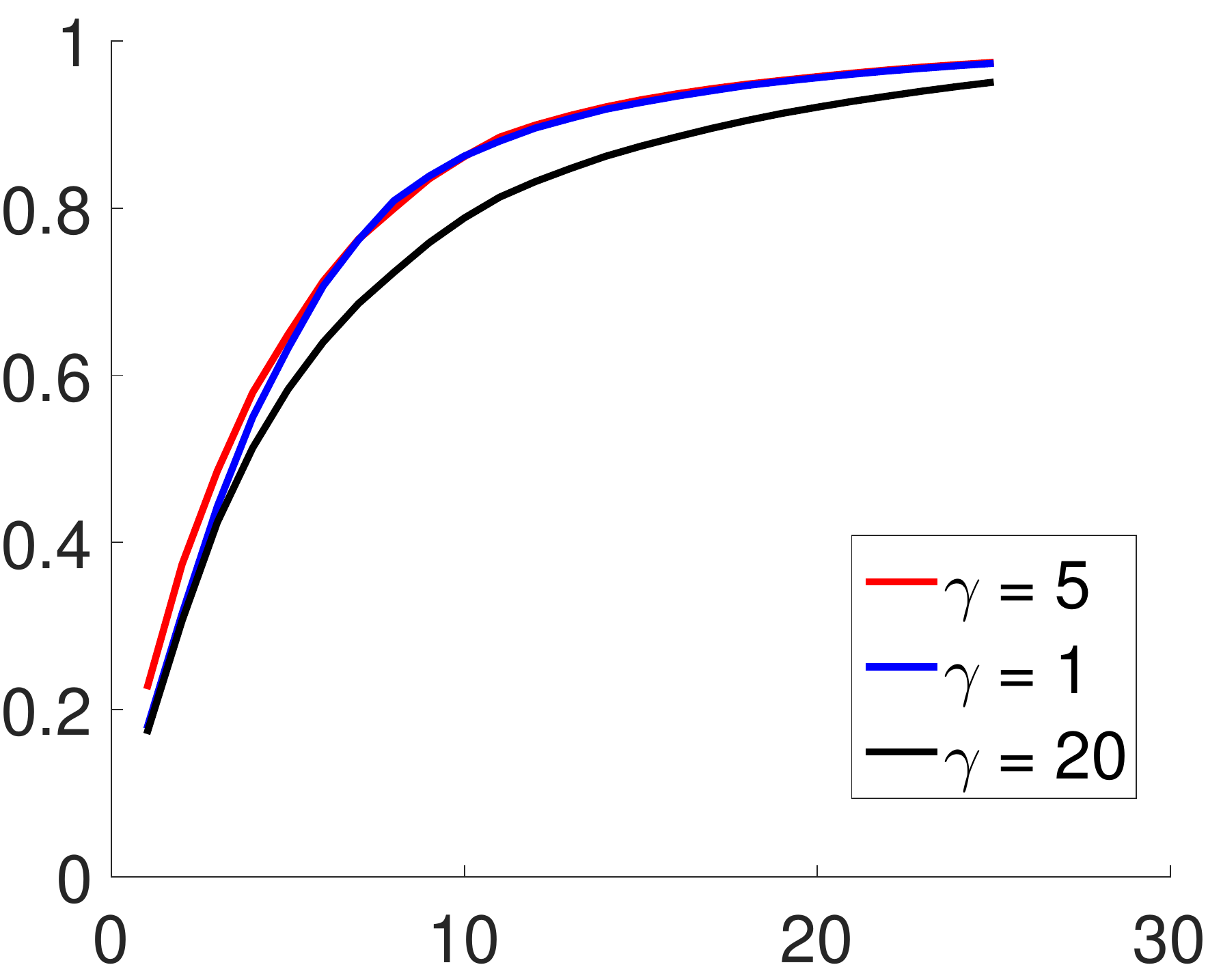}       \\ \hline
			\end{tabular}
		\end{center}
		\caption{Detailed assessment of sensitivity due to perturbations of the parameters $a_0$ and $\gamma$ in the DCV model. The baseline in all of the plots is given in red. (a) Posterior average density estimates. (b) Proportion of cumulative variance explained by the first 25 principal components.}
		\label{fig:DCV_int}
	\end{figure}

Figure \ref{fig:CCV_sim_combined} shows all of the proposed global sensitivity measures for the different types perturbations. The model parameters for the baseline model were chosen as $a_0 = 1$, $a_1 = 10$, $\eta = 3$, and $\gamma = 5$. The ranges of the various perturbations are as follows: $a_0 =$ 1 to 20, $a_1 =$ 1 to 20, $\eta =$ 1 to 20 and $\gamma = $ 1 to 20. For perturbations of the prior distribution for $a$, the CCV model is less sensitive to changes in the second shape parameter $a_1$ than the first shape parameter $a_0$, i.e., the magnitude of all of the measures is much smaller for perturbations of $a_1$ than perturbations of $a_0$. This can be easily seen in the first two rows of Figure \ref{fig:CCV_sim_combined}. This suggests that the CCV model is most robust to changes in $a_1$. Additionally, the behavior of all of the sensitivity measures for perturbations of $\eta$ and $\gamma$ are comparable, i.e., all of the plots exhibit similar patterns and very small magnitudes of sensitivity.
	
	\subsubsection{DCV Model}
	\label{subsubsec:DCVsimu}
	
	We use the exact same setting for this simulation study as the one considered for the CCV model in the previous section. We generate the same type of dataset, and fit the DCV model separately for various perturbations of a single parameter, keeping all of the other baseline settings fixed. The only change from the CCV model is the addition of the shape parameter $\phi$ in the $Inverse$ $Gamma$ prior for the variance component. The baseline value of $\phi$ is fixed at 2. Figure \ref{fig:DCV_sim_combined} displays all of the proposed sensitivity measures for different prior perturbations of the DCV model. The ranges of perturbations for all of the parameters that were previously considered in the CCV model are the same in this simulation. For $\phi$, (last row in Figure \ref{fig:DCV_sim_combined}), we consider perturbations ranging from 1.5 to 20. As expected, the sensitivity measures in this case behave similarly to the CCV model for perturbations of the parameters $a_0$, $a_1$, $\eta$ and $\gamma$. Interestingly, the DCV model seems to be more sensitive to perturbations of $a_1$ than the CCV model, especially based on the measures $\mathbb{V}$ and $\mathbb{E}$. In general, the magnitude of the sensitivity measures is higher for the DCV model than the CCV model for all of the perturbations that we consider. This indicates that the DCV model is more sensitive to the various changes in the prior. For values of $\phi$ that are close to the baseline, the DCV model is fairly sensitive, especially in terms of the spread measures. These tend to slowly level off as this parameter is increased beyond 10. It seems that the measures would not change much if we kept on increasing the value of $\phi$.
	
	To further illustrate the use of the proposed global sensitivity measures in capturing the various perturbations of the prior parameters of the DCV model, we consider a couple of interesting perturbations and study them in more detail. In particular, we consider two perturbed values of $a_0$ and $\gamma$. Figure \ref{fig:DCV_int}(a) shows the posterior averages of the samples for the baseline model and the perturbed models. In the top panel, the differences in shift of the posterior samples are clearly visible as structural changes in the posterior average density estimate, and are effectively captured by the sensitivity measure $\mathbb{D}$. On the other hand, perturbations of $\gamma$ do not appear to affect the posterior averages very much. As a result, the measure $\mathbb{D}$ is an order of magnitude smaller in this case. The cumulative variance plots are displayed in panel (b). This plot indicates that the variance/covariance structure of the posterior samples coming from perturbations of $a_0$ are more different than those coming from perturbations of $\gamma$. The proposed sensitivity measures are able to capture such differences effectively, as can be seen in the first and fourth rows of Figure \ref{fig:DCV_sim_combined}.
		
	\subsection{Real Datasets}
	\label{subsec:RealData}
	
	\begin{figure}[!t]
		\begin{center}
			\begin{tabular}{|c|c|c|}
				\hline
				Acidity & Galaxy & Enzyme \\ \hline
				\includegraphics[scale = 0.26]{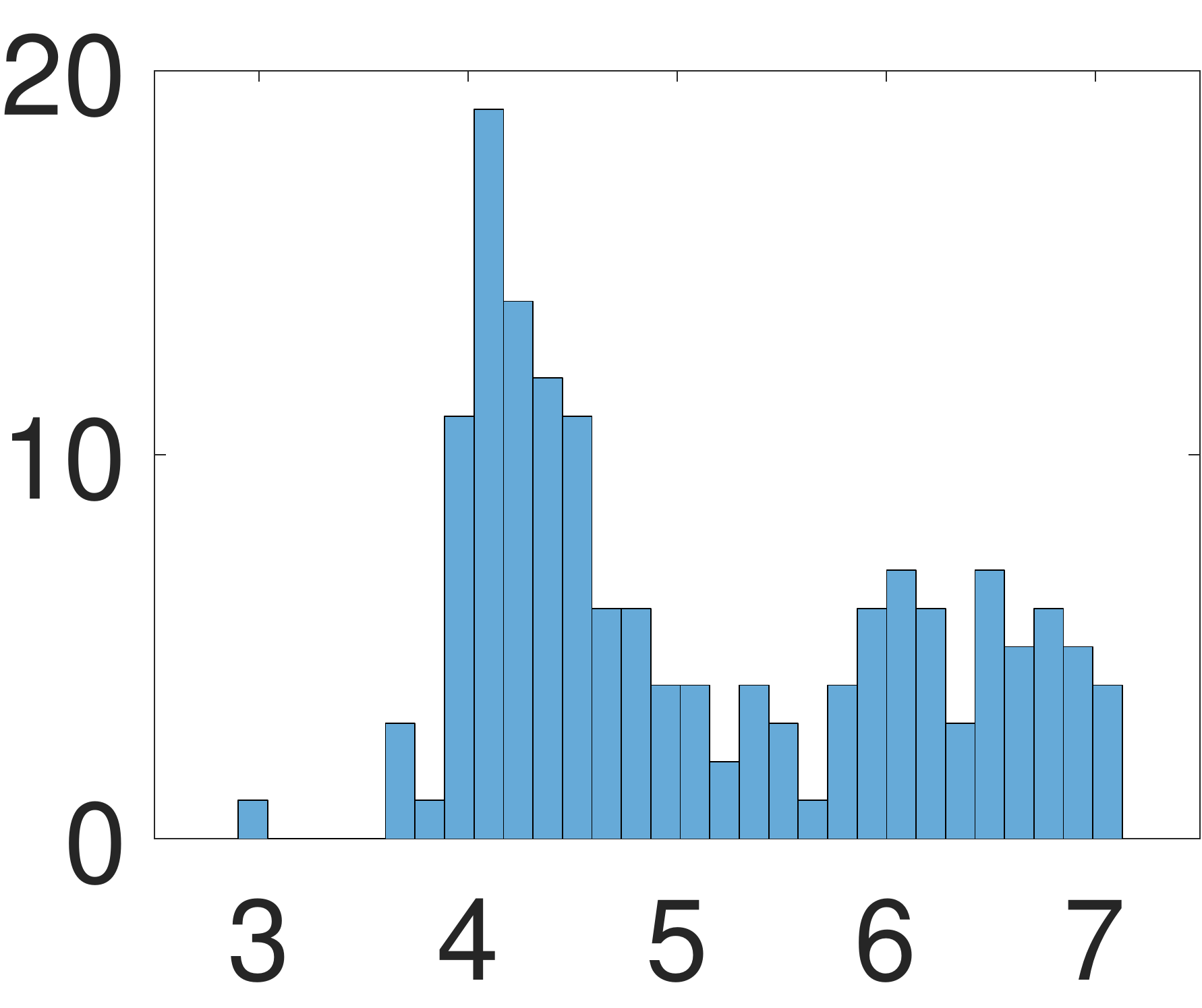}  & \includegraphics[scale = 0.26]{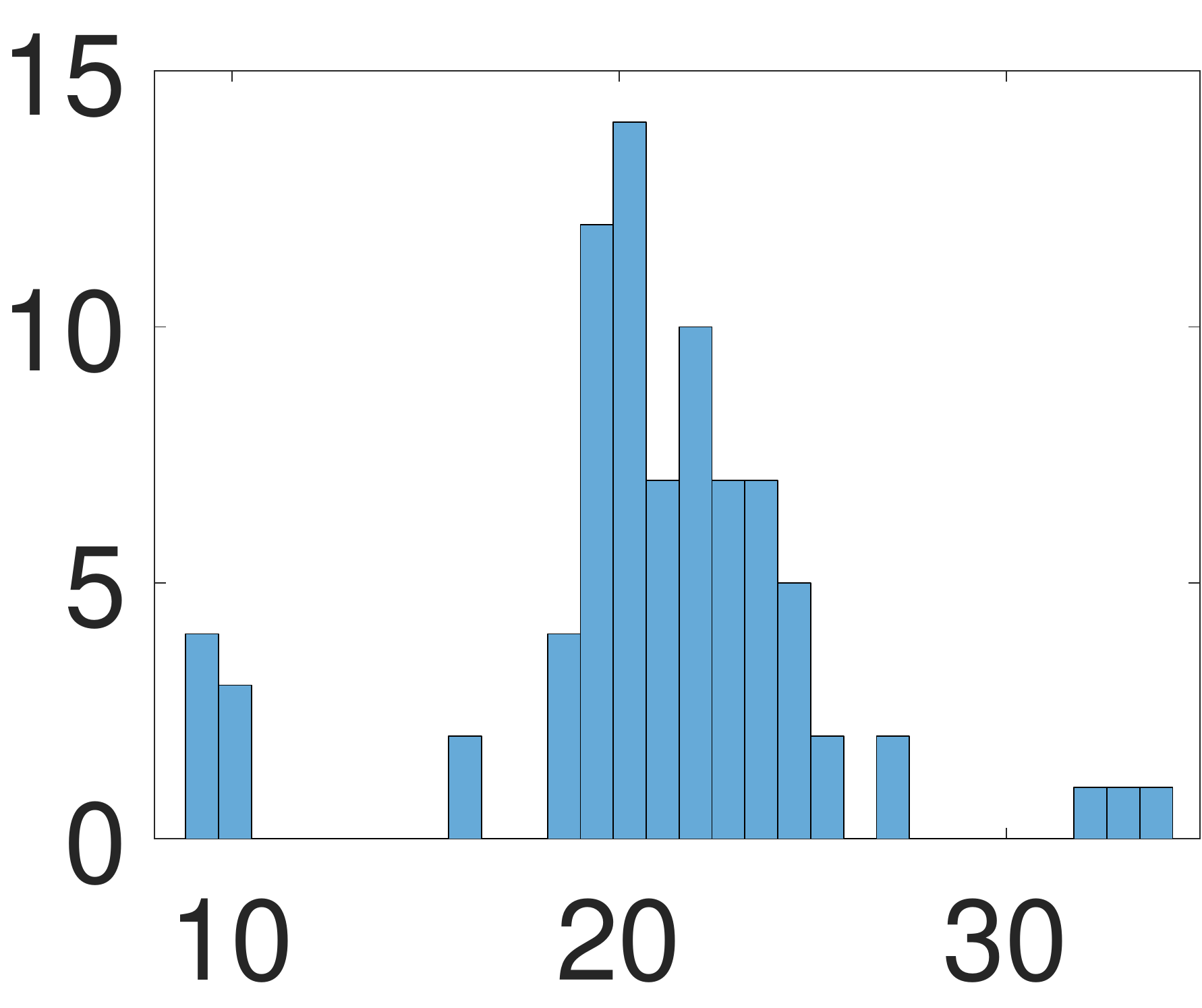}  & 	\includegraphics[scale = 0.26]{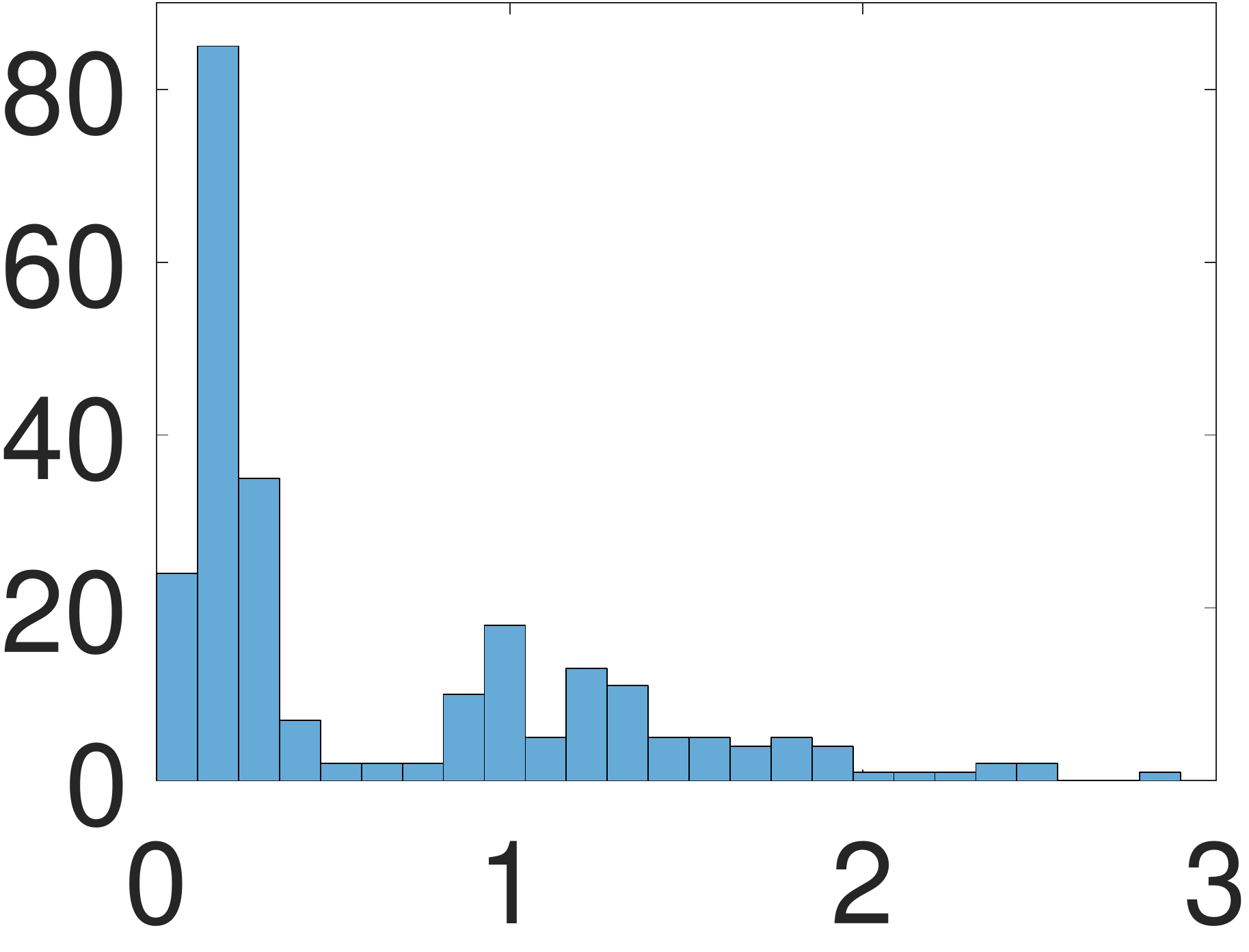}     \\ \hline
			\end{tabular}
		\end{center}
		\caption{Histograms of three real datasets: Acidity, Galaxy and Enzyme.}
		\label{fig:Hist}
	\end{figure}
	
	In this section, we analyze three popular datasets that have been previously examined in multiple studies including Richardson and Green \cite{richardson1997} and Griffin \cite{griffin2010}: Acidity, Galaxy and Enzyme. The Acidity dataset contains 155 measurements of the acidity index for lakes in north-central Wisconsin. This data has been previously analyzed on the log scale, and we follow the same transformation for fitting our models as well. The Galaxy dataset was first introduced as a density estimation problem using mixtures of normals by Roeder and Wasserman \cite{roeder1997}. This dataset records the estimated velocity of 82 distant galaxies, which are moving away from our own galaxy. Finally, the Enzyme dataset contains measurements regarding enzymatic activity in the blood for a particular enzyme for a group of 245 unrelated individuals. We display the histograms for all of the datasets in Figure \ref{fig:Hist}.
	
Based on the proposed global sensitivity measures, we show the effects of perturbing the concentration parameter $\alpha$ in the DPGMM for all three datasets. We do not consider the DP model due to its limitations for continuous density estimation. We also present results for all three datasets corresponding to the different prior perturbations of the DCV model. We do not present detailed results of the CCV model since it's a simpler version of the more flexible DCV model. For the DPGMM model, we specify the same baseline settings for the three examples as in our simulations: $\alpha = 1$, $\boldsymbol{m} = 0$, $r = \frac{1}{9}$, $\nu = 5$ and $S = 1$. For the DCV model, we follow the guidance in Griffin \cite{griffin2010} to set our baseline settings for each dataset.
	
	\subsubsection{Sensitivity Assessment for DPGMM}
	\label{subsubsec:Acidity}
	
	\begin{figure}[!t]
		\begin{center}
			\begin{tabular}{|c|c|c|}
				\hline
				$\mathbb{D}$ & $\mathbb{V}$ & $\mathbb{E}$ \\ \hline
				\includegraphics[scale = 0.18]{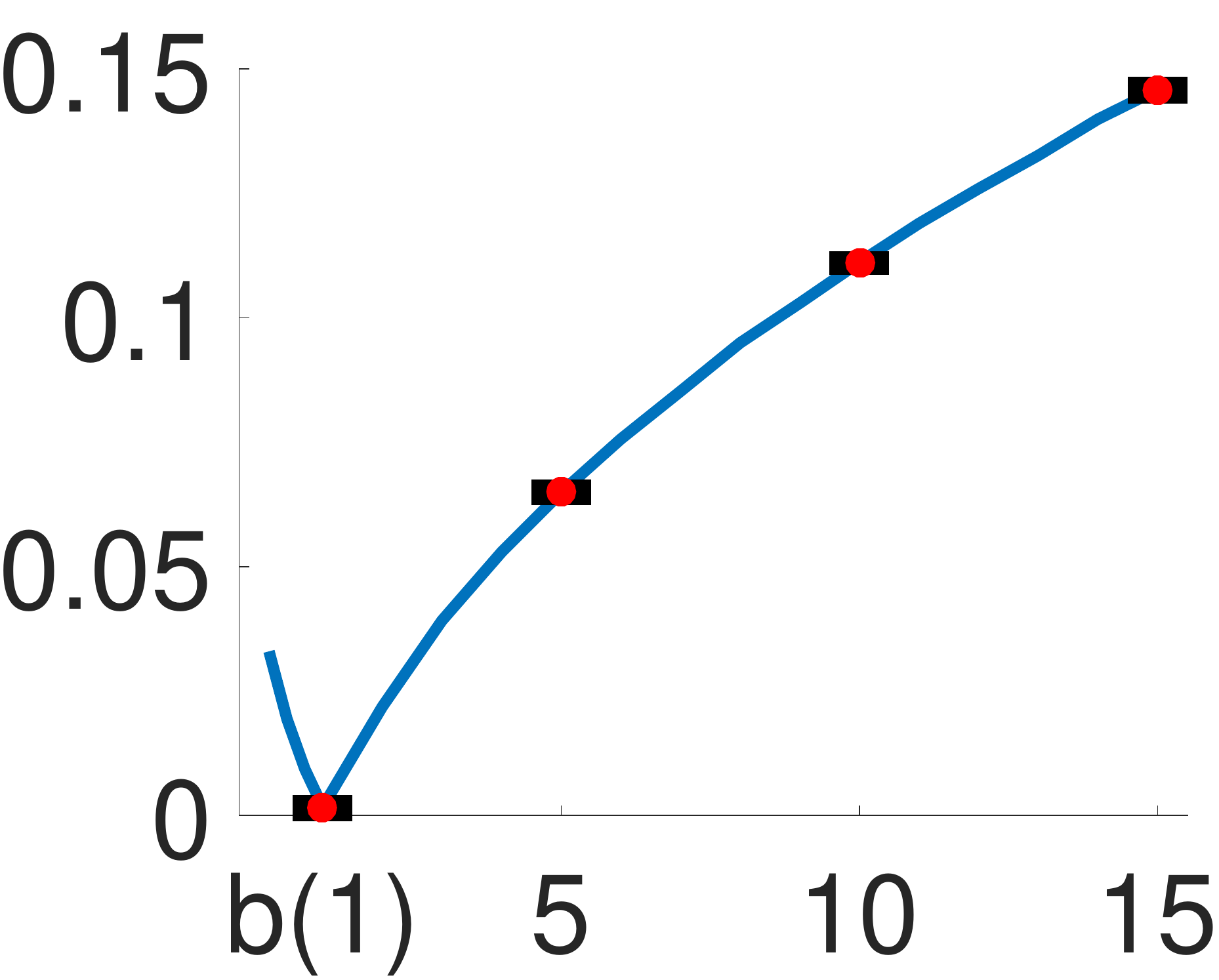}          & \includegraphics[scale = 0.18]{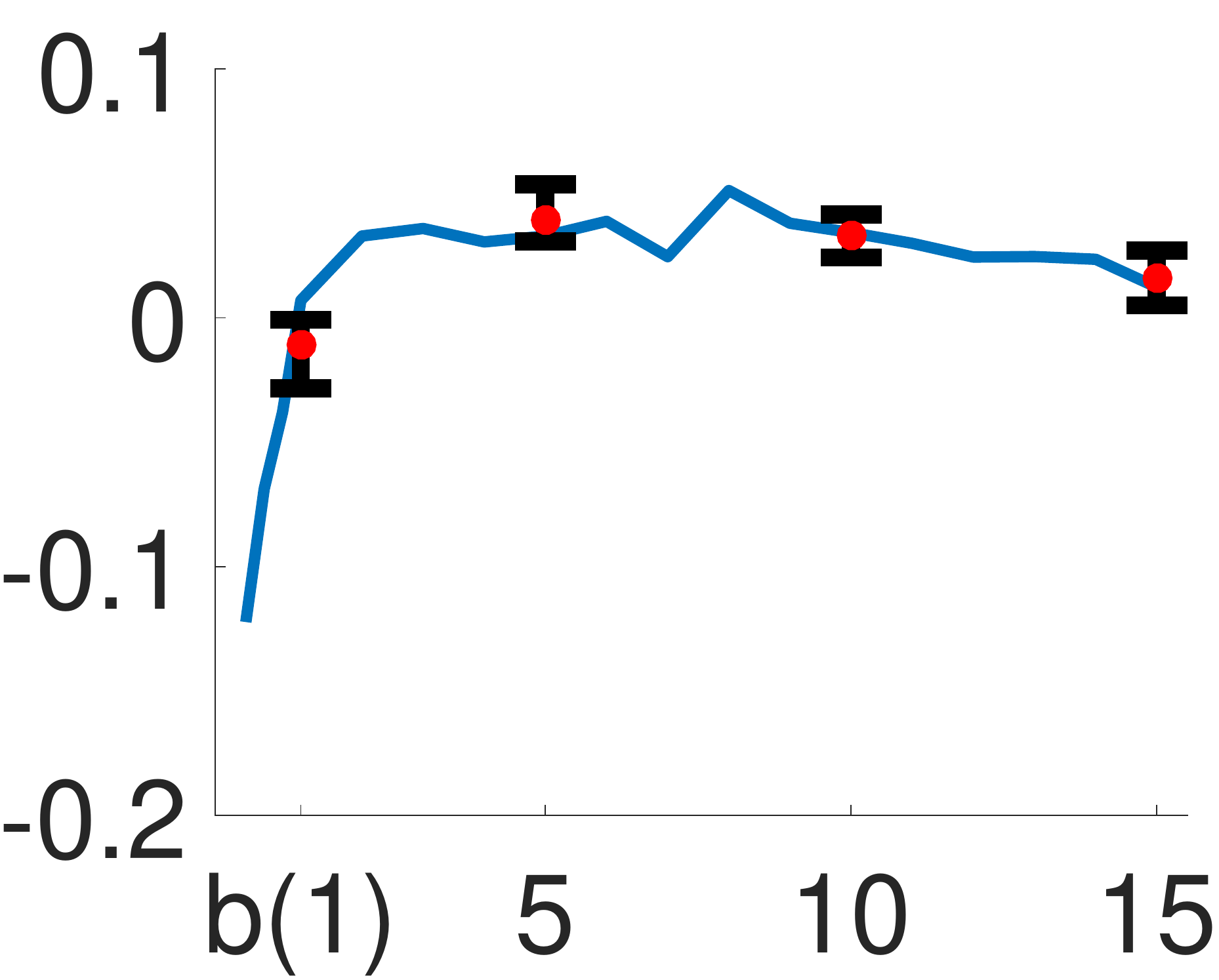}   	&	\includegraphics[scale = 0.18]{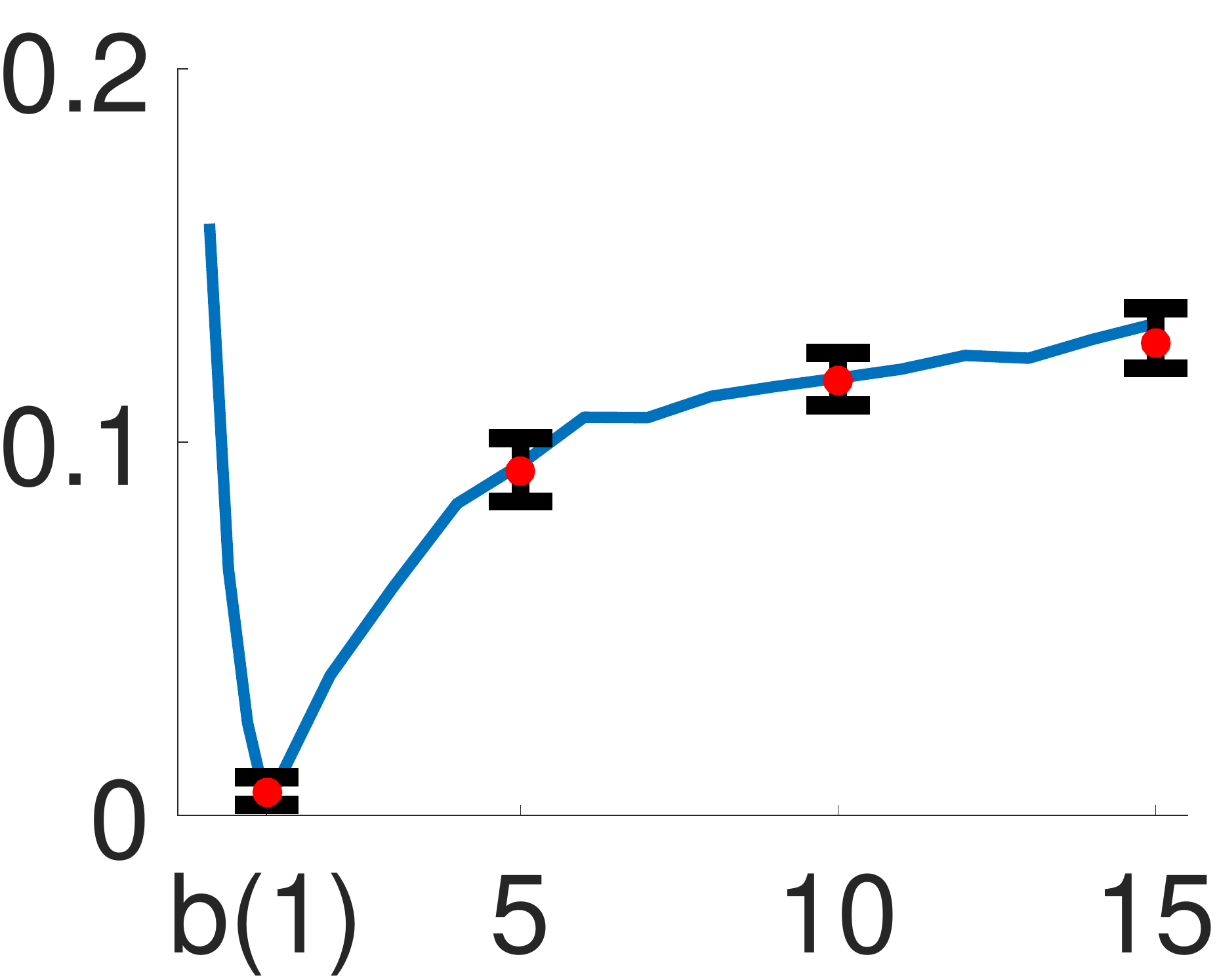}                    \\ \hline
    				\includegraphics[scale = 0.18]{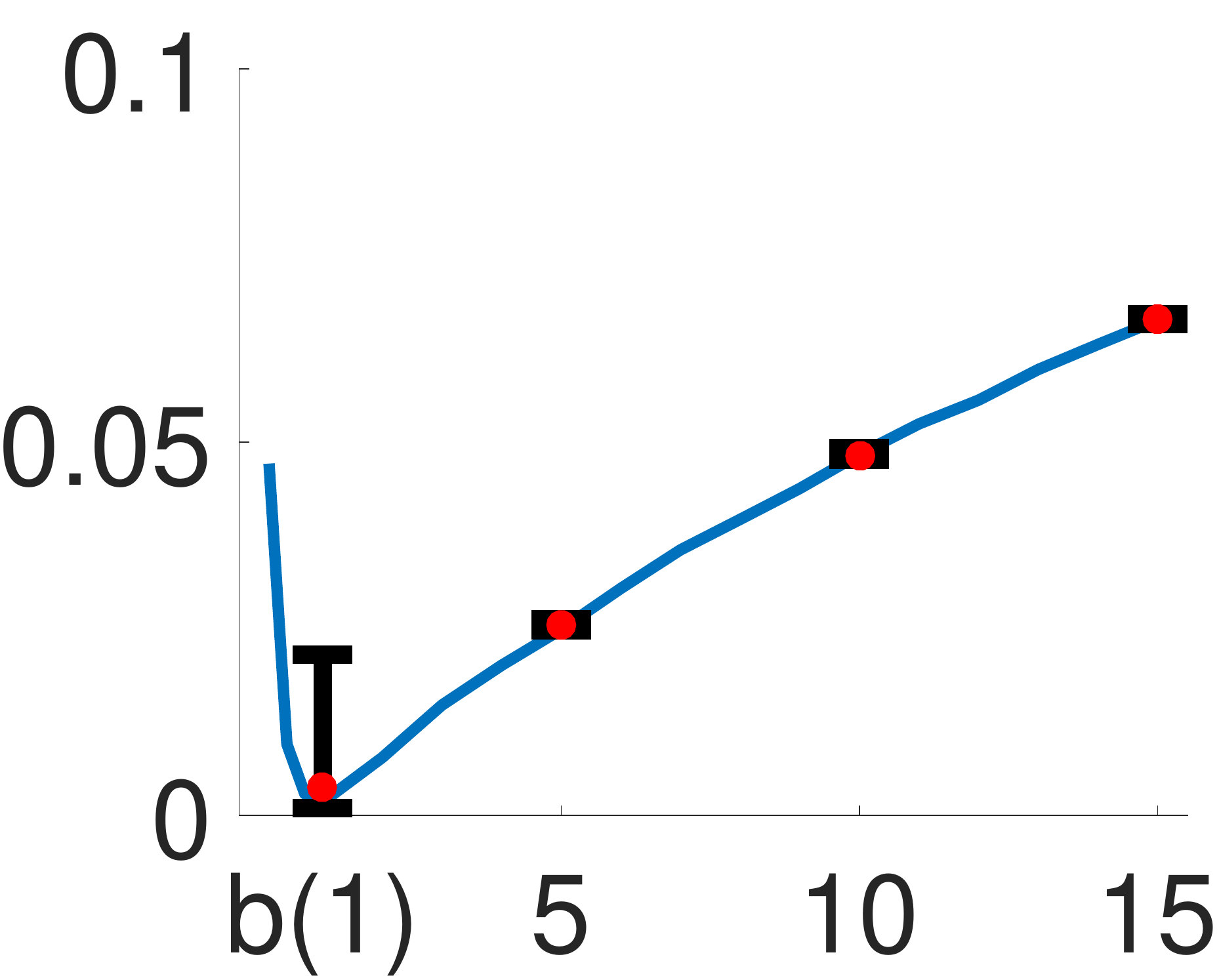}          & \includegraphics[scale = 0.18]{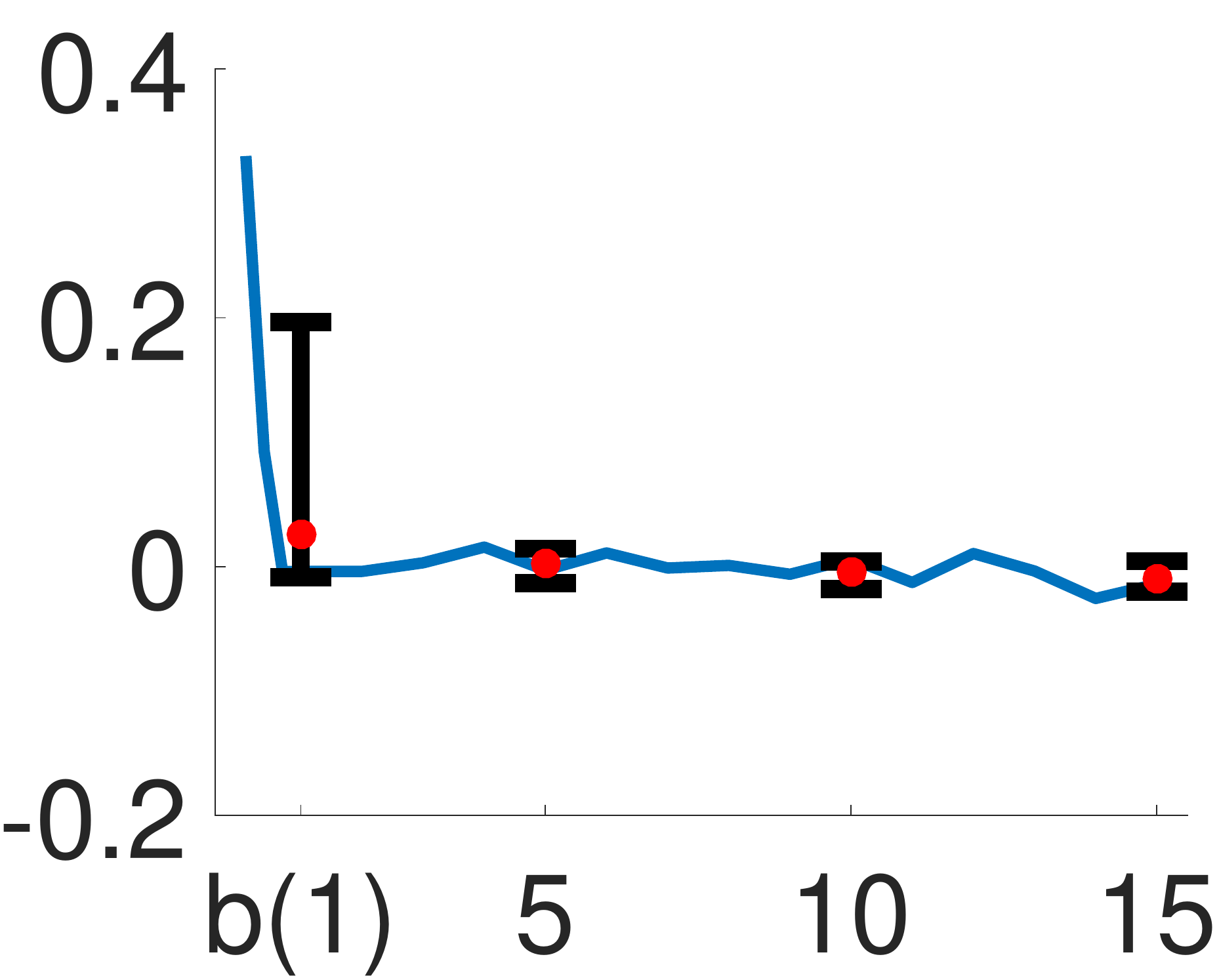}   & 		\includegraphics[scale = 0.18]{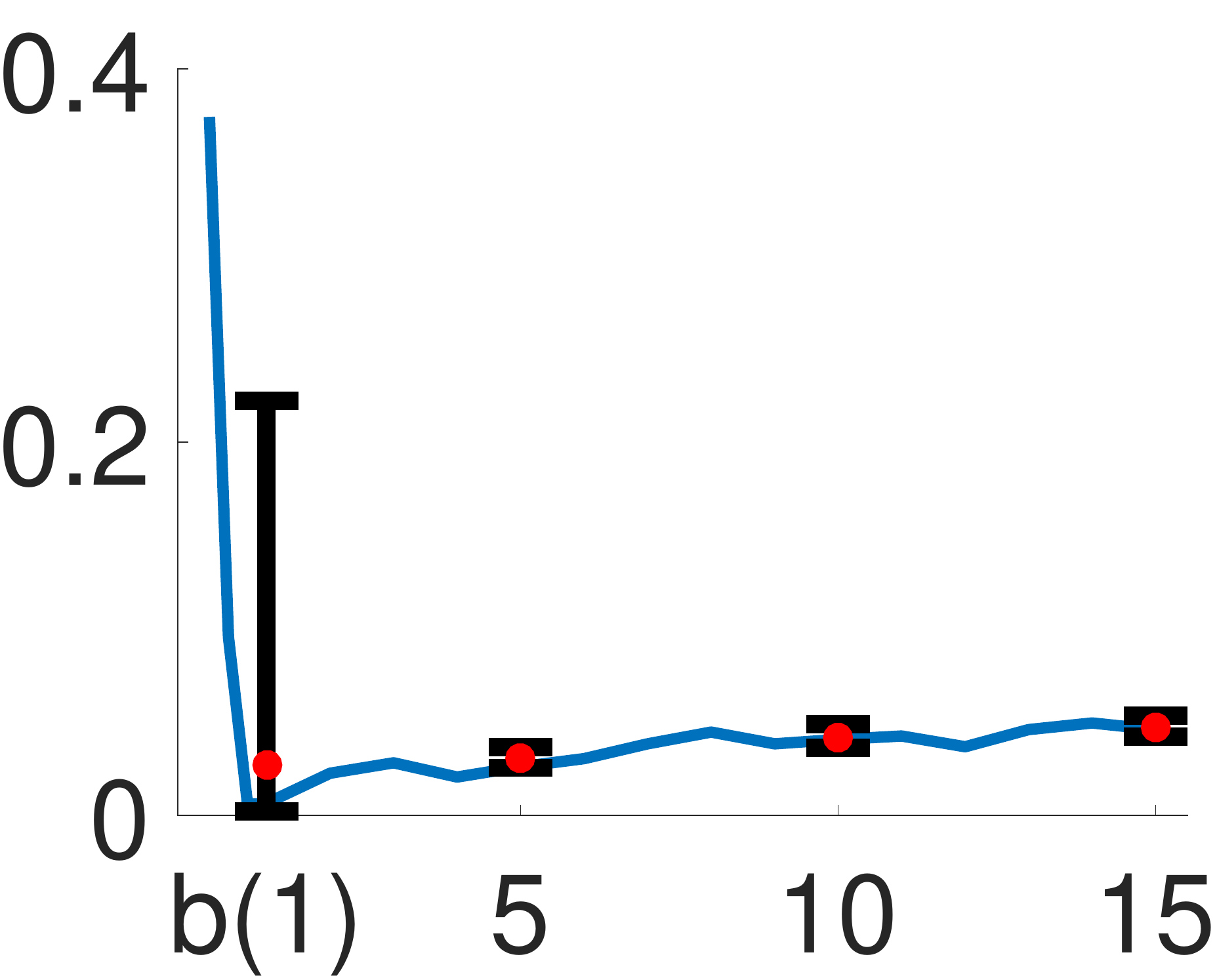}                 \\ \hline
    				\includegraphics[scale = 0.18]{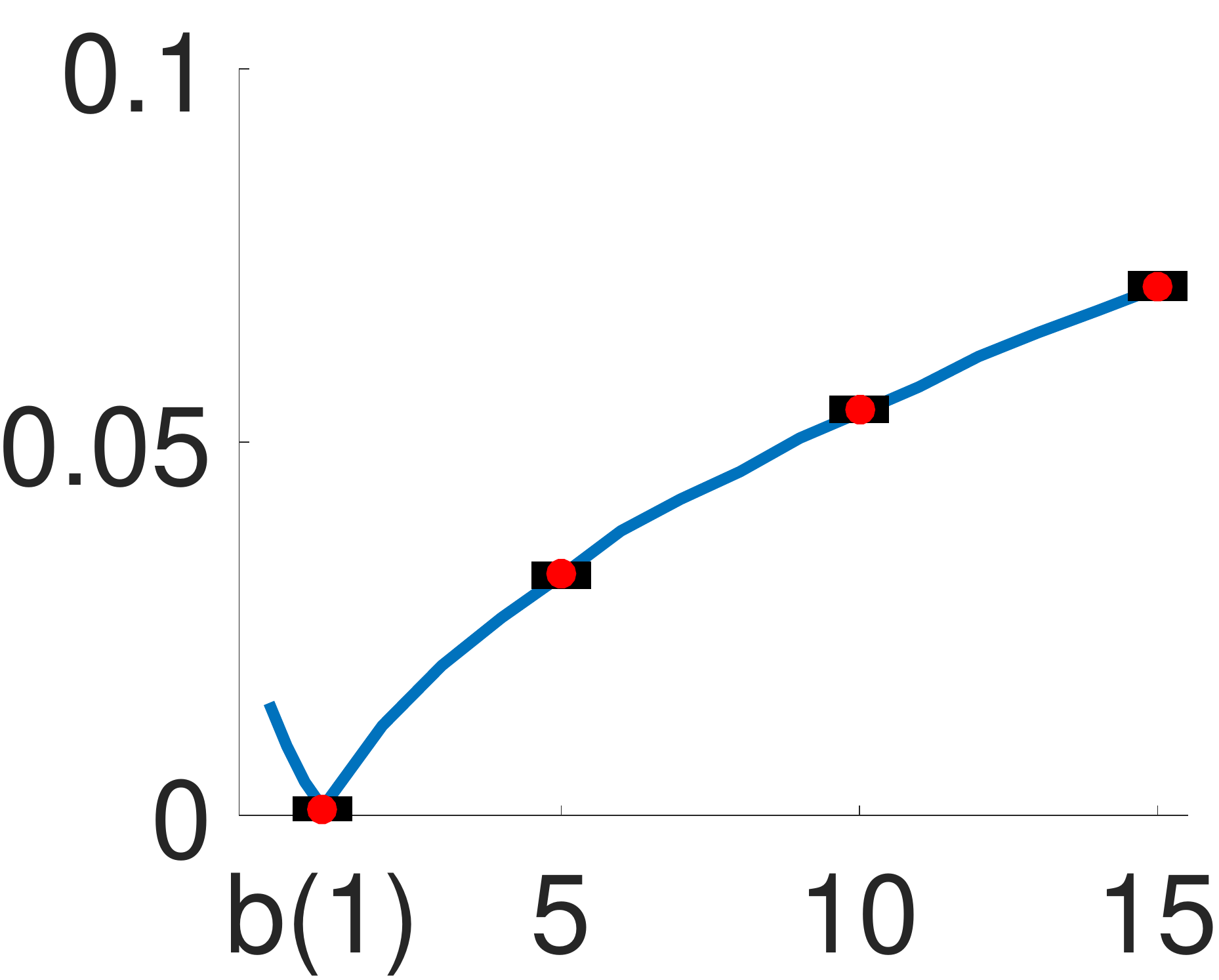}          & \includegraphics[scale = 0.18]{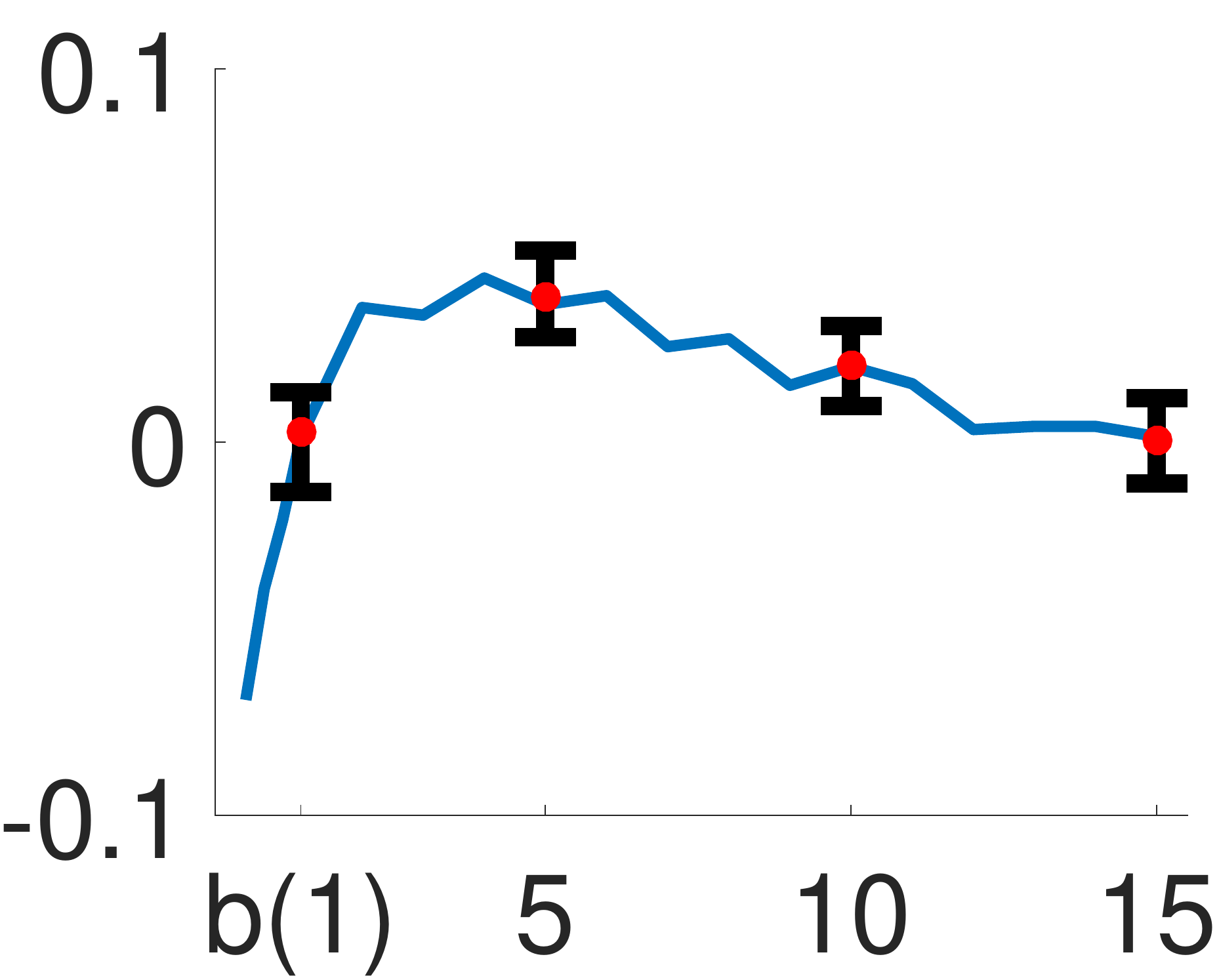}   	&	\includegraphics[scale = 0.18]{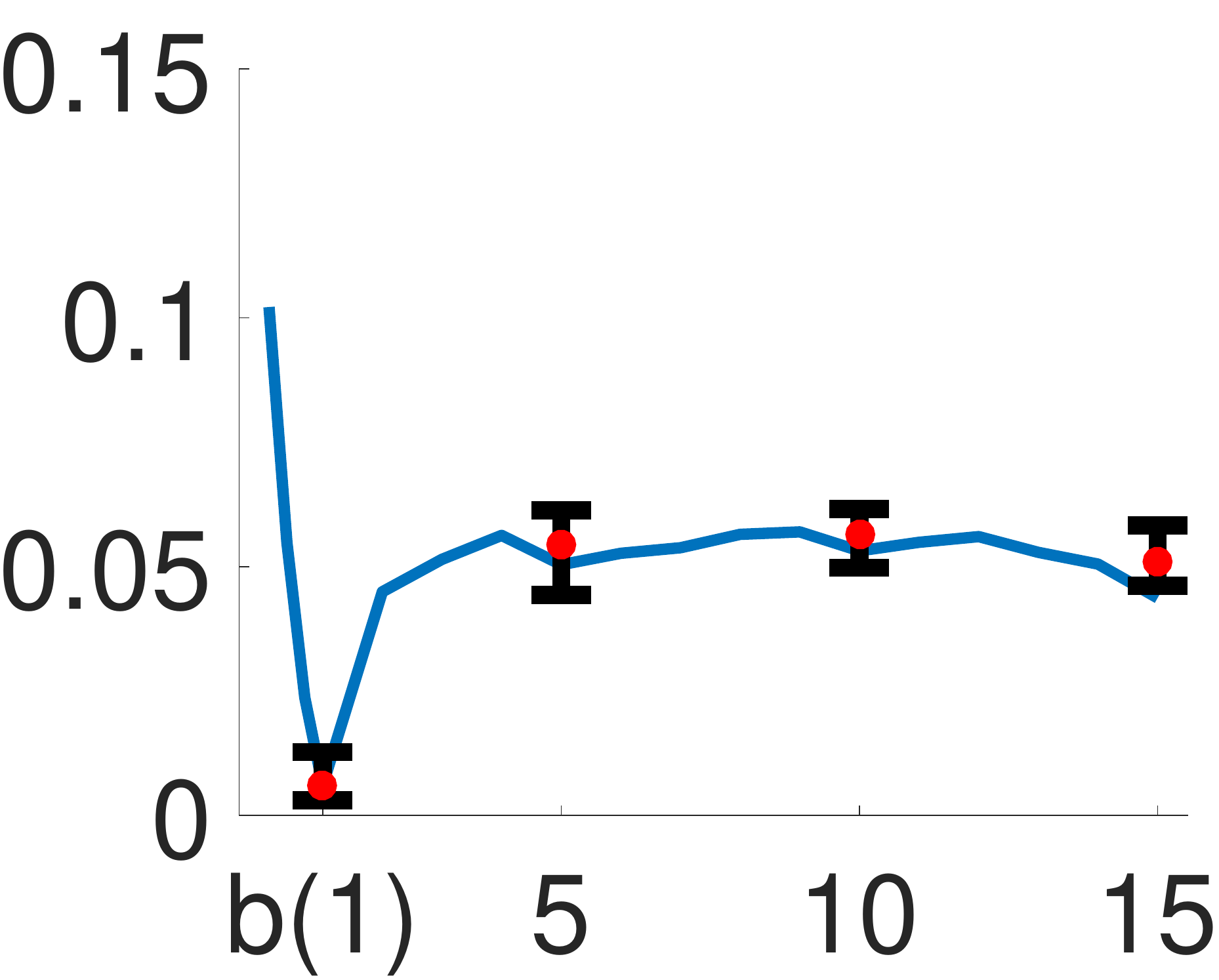}                 \\ \hline
			\end{tabular}
		\end{center}
		\caption{Sensitivity to perturbations of $\alpha$ in the DPGMM for the Acidity (top row), Galaxy (middle row) and Enzyme (bottom row) datasets.}
		\label{fig:DPM_acidity}
	\end{figure}
	
In this study, we perturb $\alpha$ for values between 0.1 and 15 while keeping all of the other parameters in the model fixed. We then compute the proposed geometric sensitivity measures to assess global effects of this perturbation on density estimation for all three datasets. The results are displayed in Figure \ref{fig:DPM_acidity} where each row considers a different dataset (top = Acidity, middle = Galaxy and bottom = Enzyme). For the Acidity dataset, the measure $\mathbb{D}$ suggests that the DPGMM model is sensitive, in terms of the shift of the posterior samples, to increases in $\alpha$, but not very sensitive to decreases. The opposite is true for the spread measures $\mathbb{V}$ and $\mathbb{E}$, although $\mathbb{E}$ suggests some sensitivity to increases in $\alpha$ as well.

For the Enzyme data, the sensitivity measures behave very similarly to the Acidity data. In fact, in both of these cases, the structure of the histograms appears similar and suggests a bimodal density estimate. The histogram for the Galaxy data is quite different, and also results in very different sensitivity patterns based on the $\mathbb{V}$ and $\mathbb{E}$ measures. The shift measure $\mathbb{D}$ behaves similarly to the previous cases.
	
	\subsubsection{Sensitivity Assessment for DCV Model}

\begin{figure}[!t]
		\begin{center}
			\begin{tabular}{|c|c|c|}
				\hline
				$\mathbb{D}$ & $\mathbb{V}$ & $\mathbb{E}$ \\ \hline
				\includegraphics[scale = 0.18]{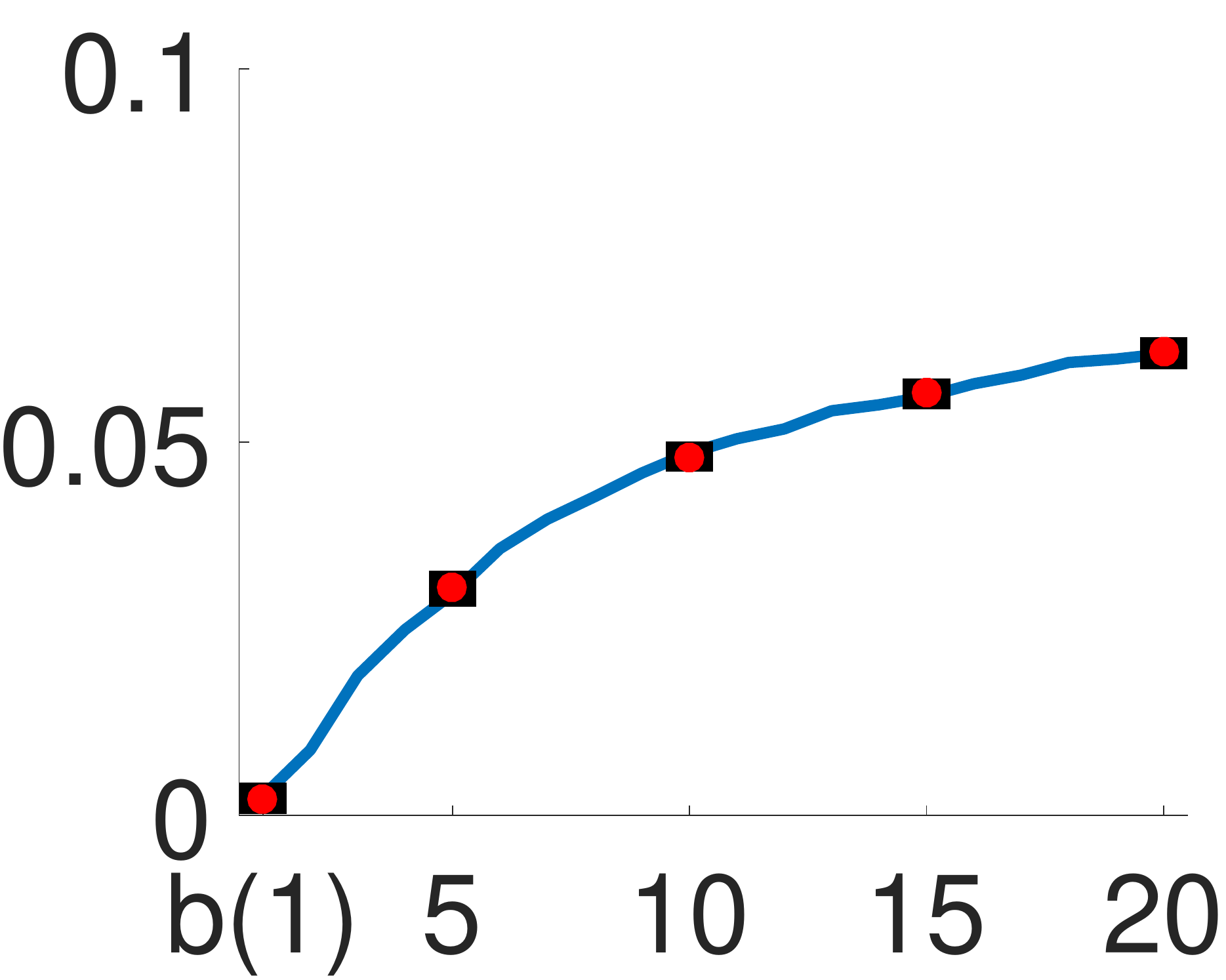}     & \includegraphics[scale = 0.18]{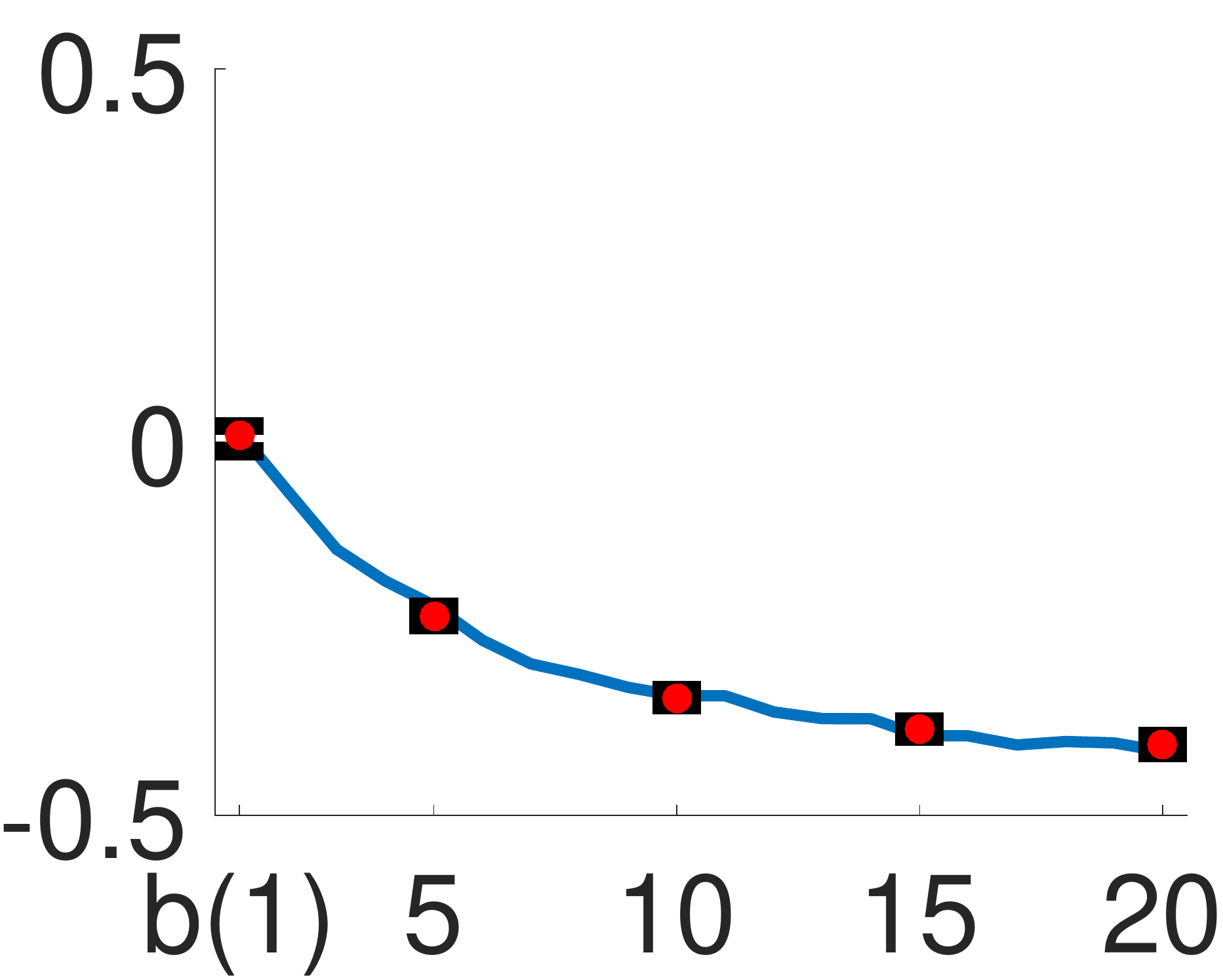}   &	\includegraphics[scale = 0.18]{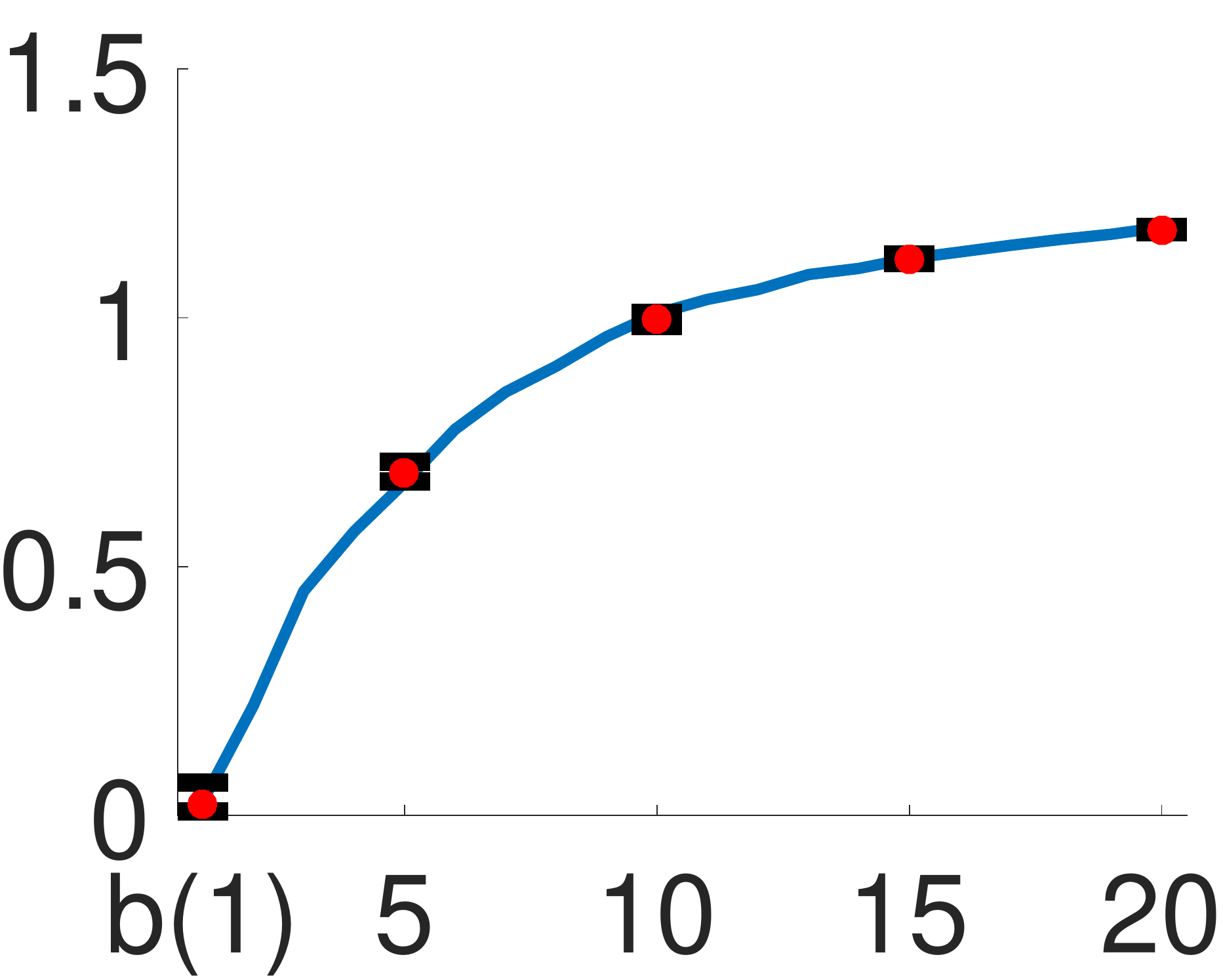}                \\ \hline
				\includegraphics[scale = 0.18]{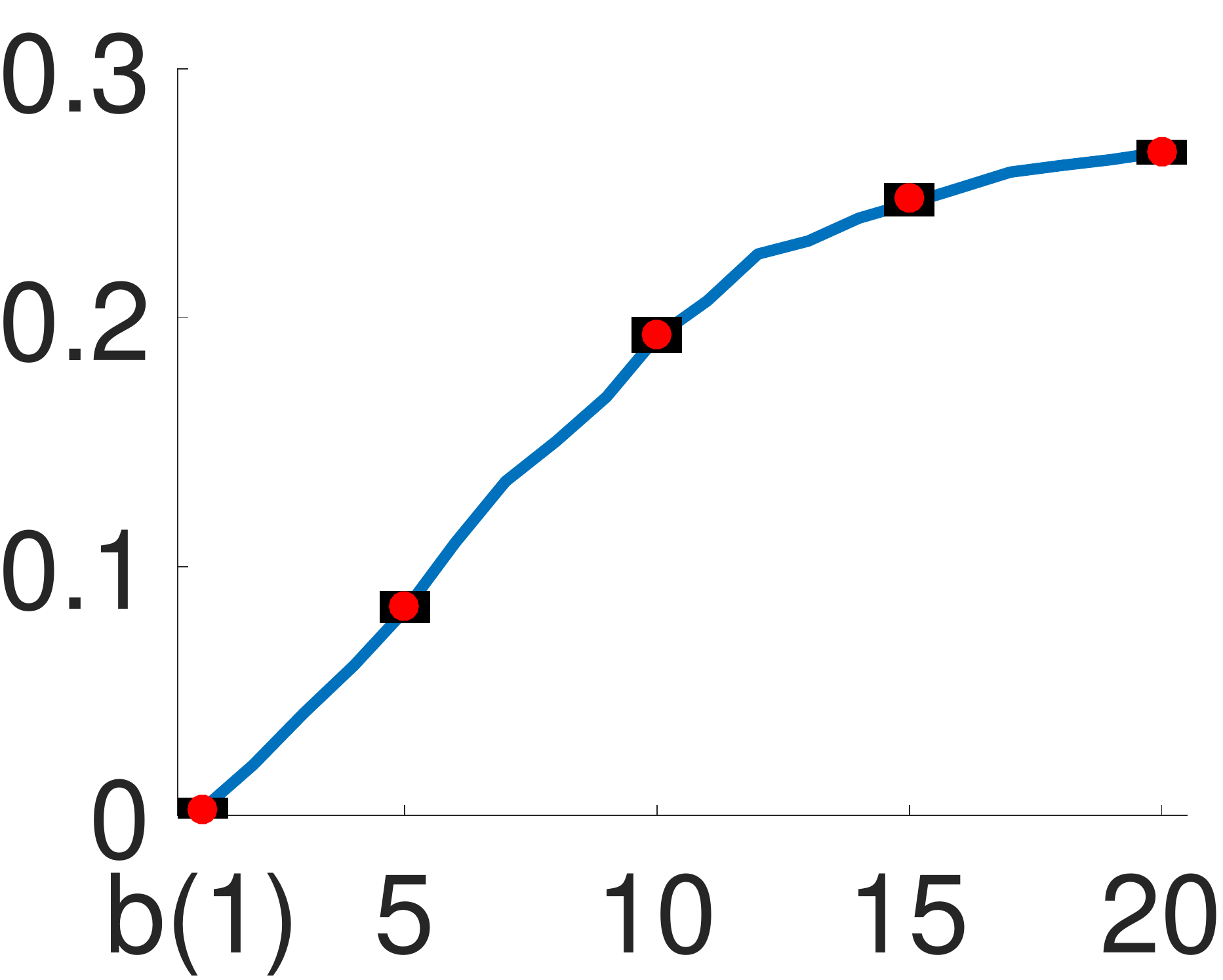}     & \includegraphics[scale = 0.18]{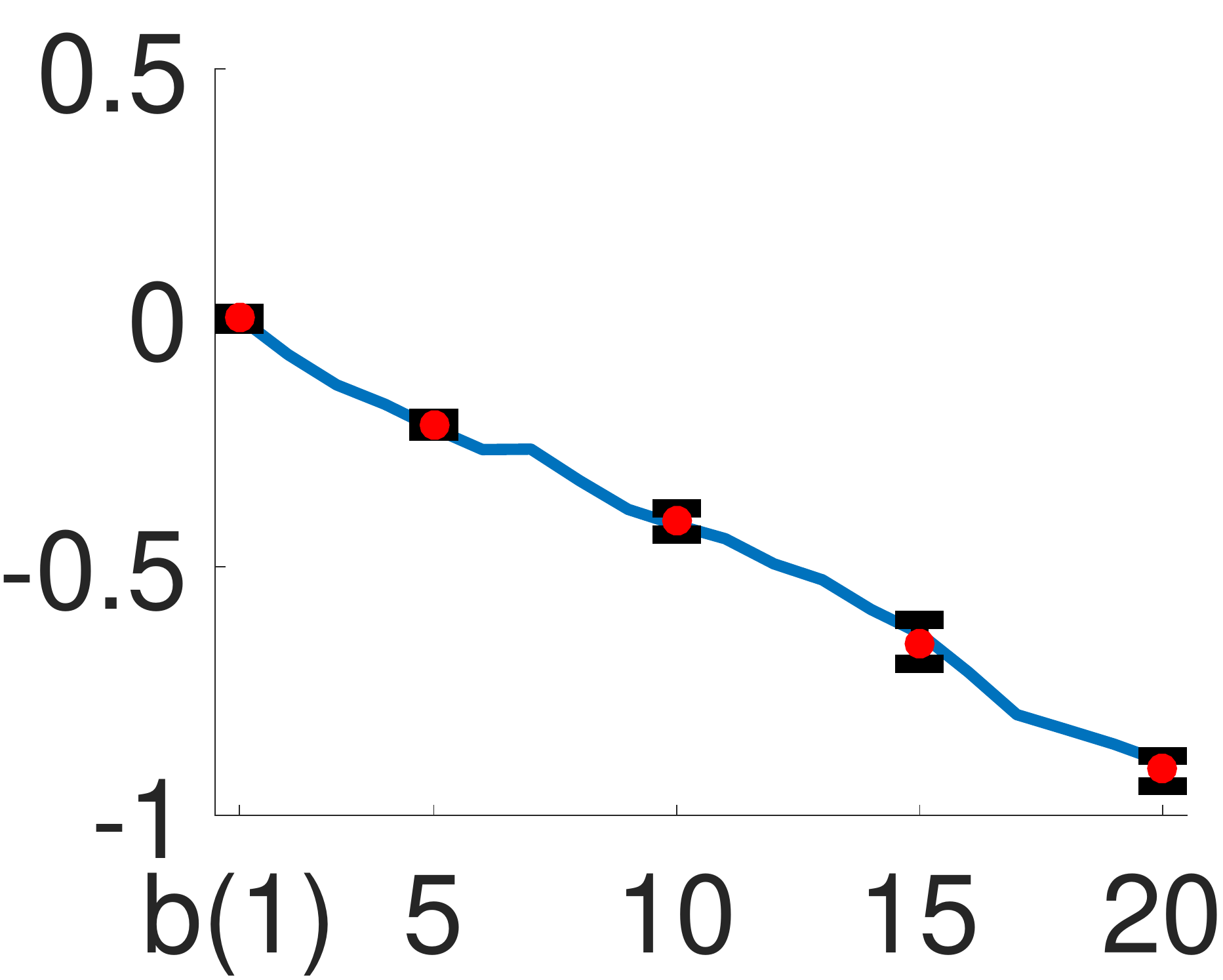}   &	\includegraphics[scale = 0.18]{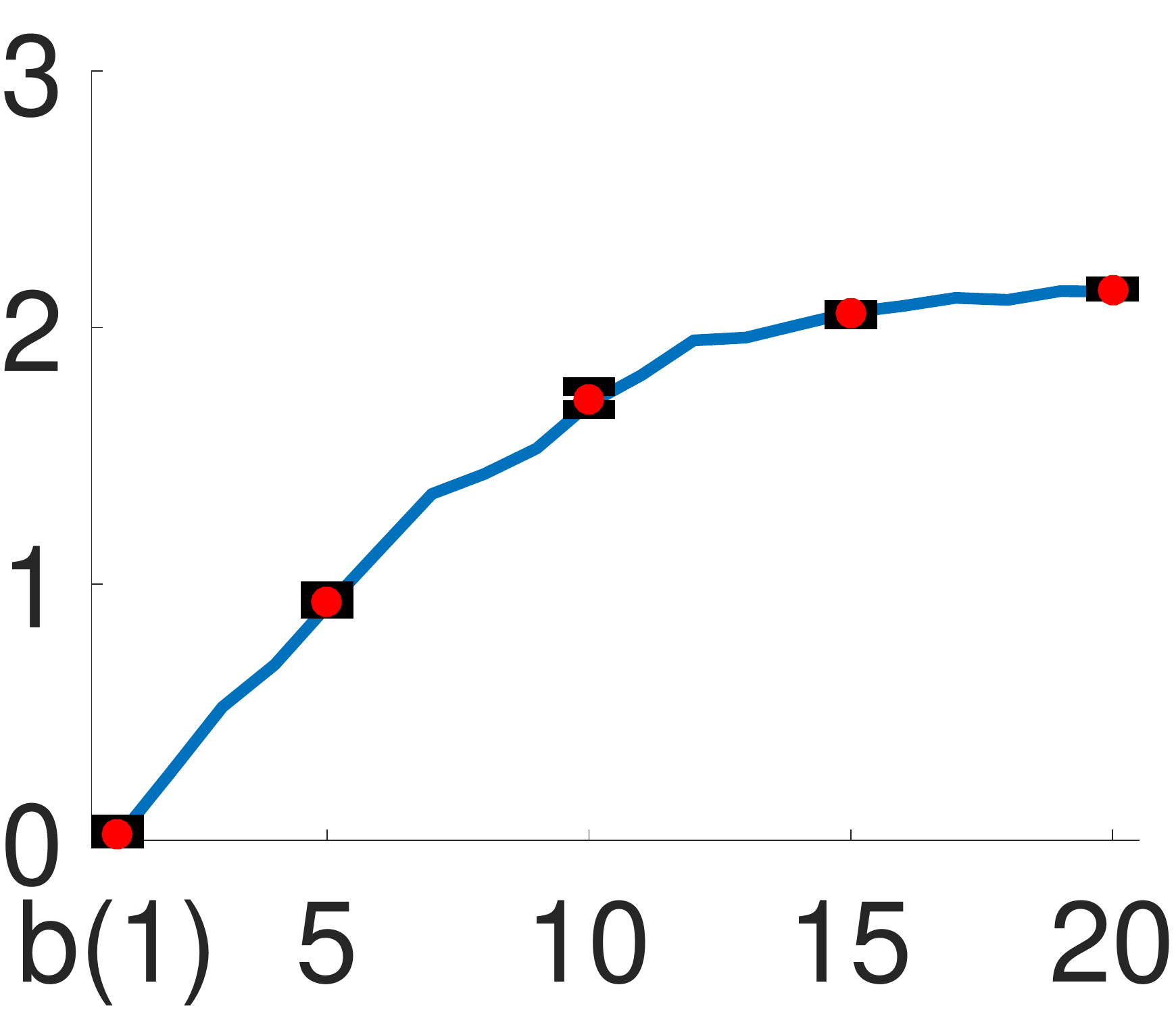}                \\ \hline
				\includegraphics[scale = 0.18]{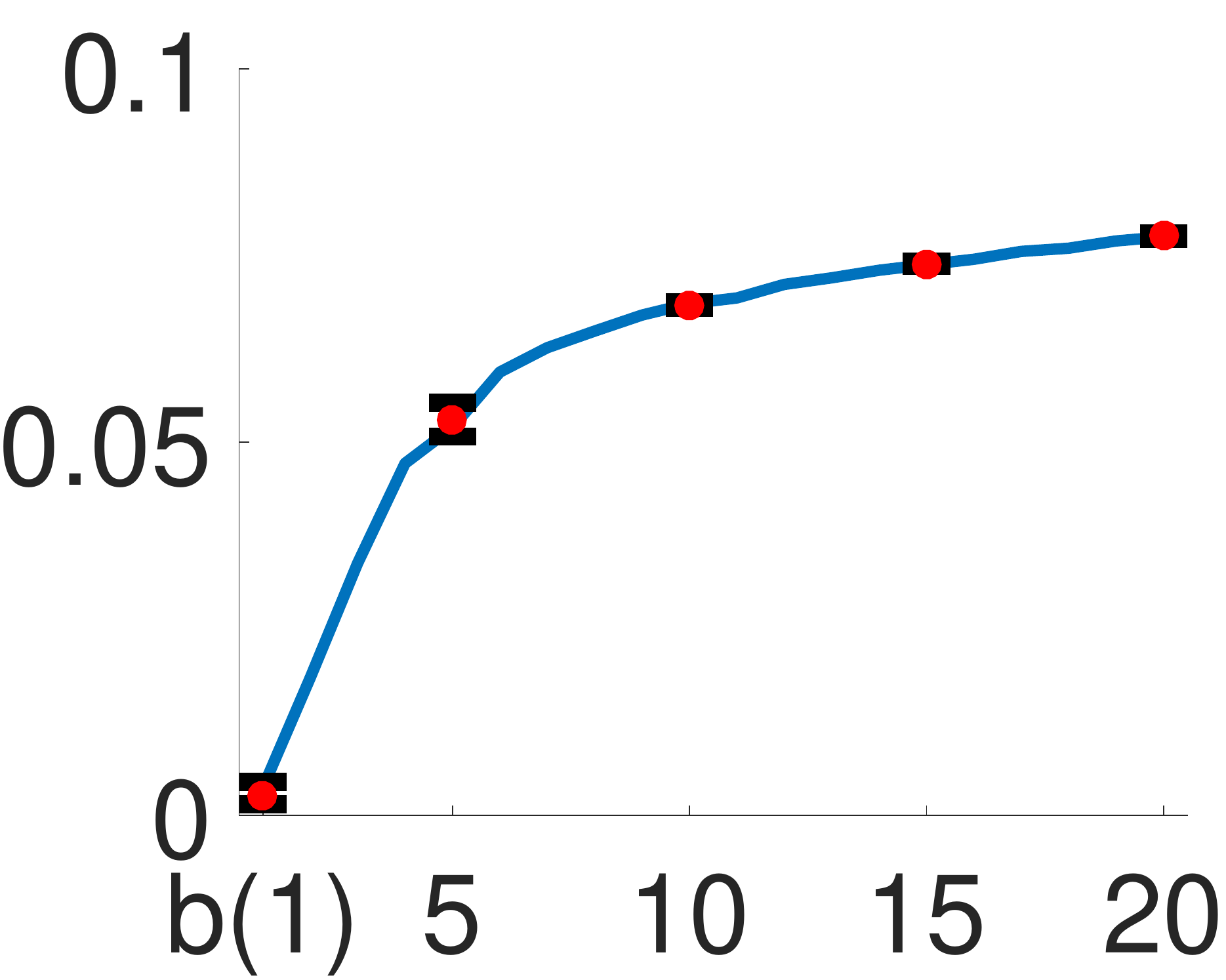}     & \includegraphics[scale = 0.18]{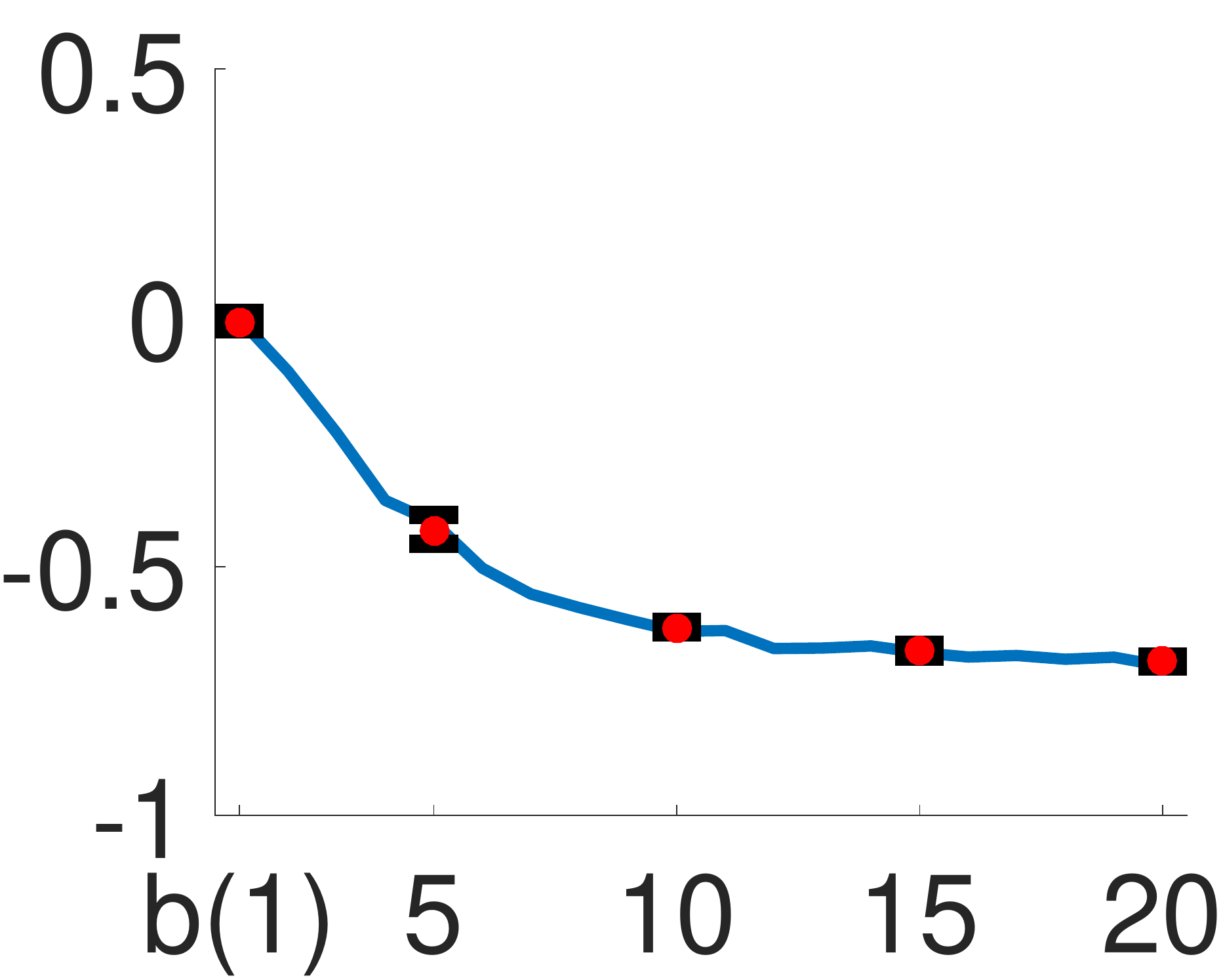}   &	\includegraphics[scale = 0.18]{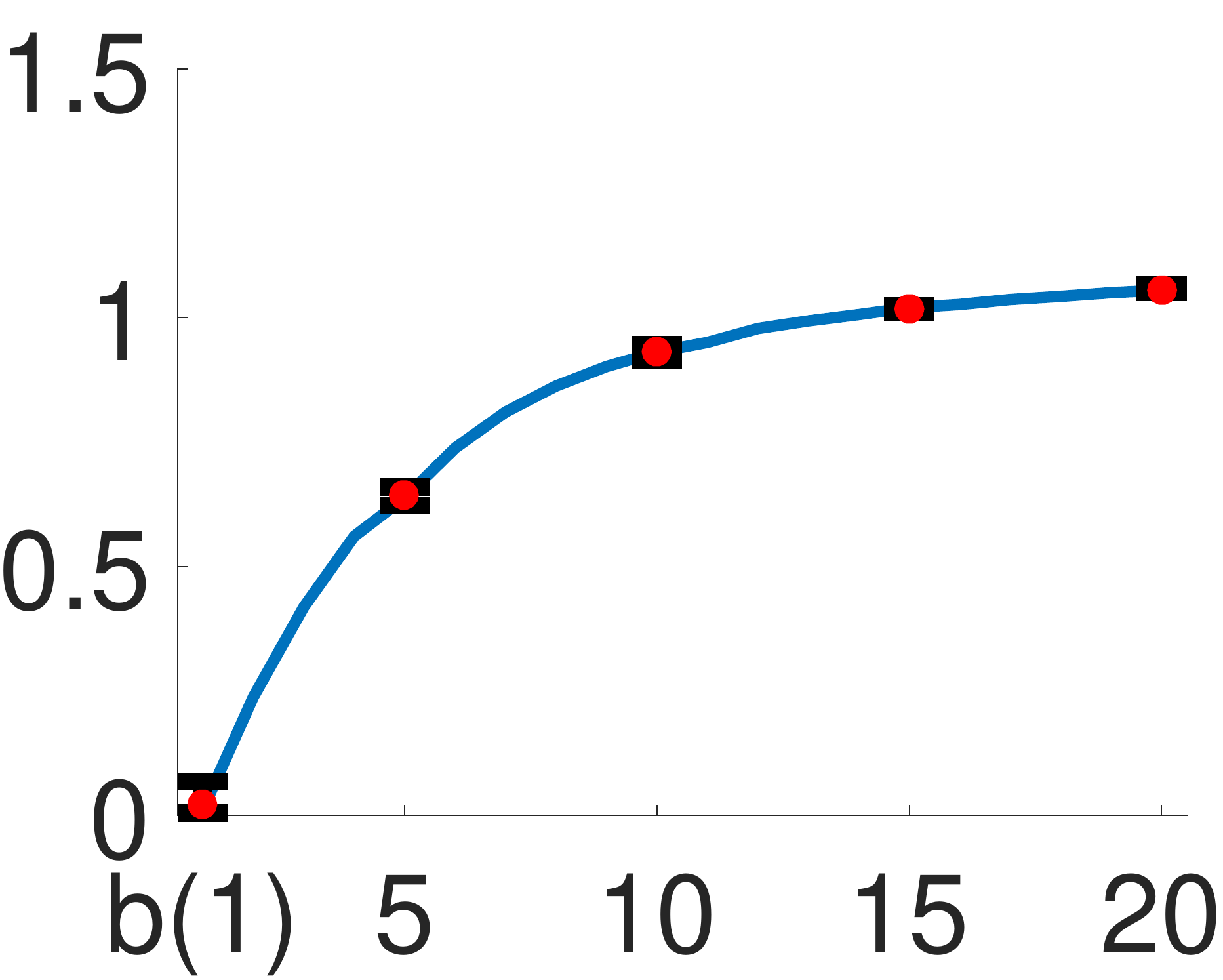}                \\ \hline
			\end{tabular}
		\end{center}
		\caption{Sensitivity to perturbations of $a_0$ in the DCV model for the Acidity (top row), Galaxy (middle row) and Enzyme (bottom row) datasets.}
		\label{fig:DCV_a0}
	\end{figure}

\begin{figure}[!t]
		\begin{center}
			\begin{tabular}{|c|c|c|}
				\hline
				$\mathbb{D}$ & $\mathbb{V}$ & $\mathbb{E}$ \\ \hline
				\includegraphics[scale = 0.18]{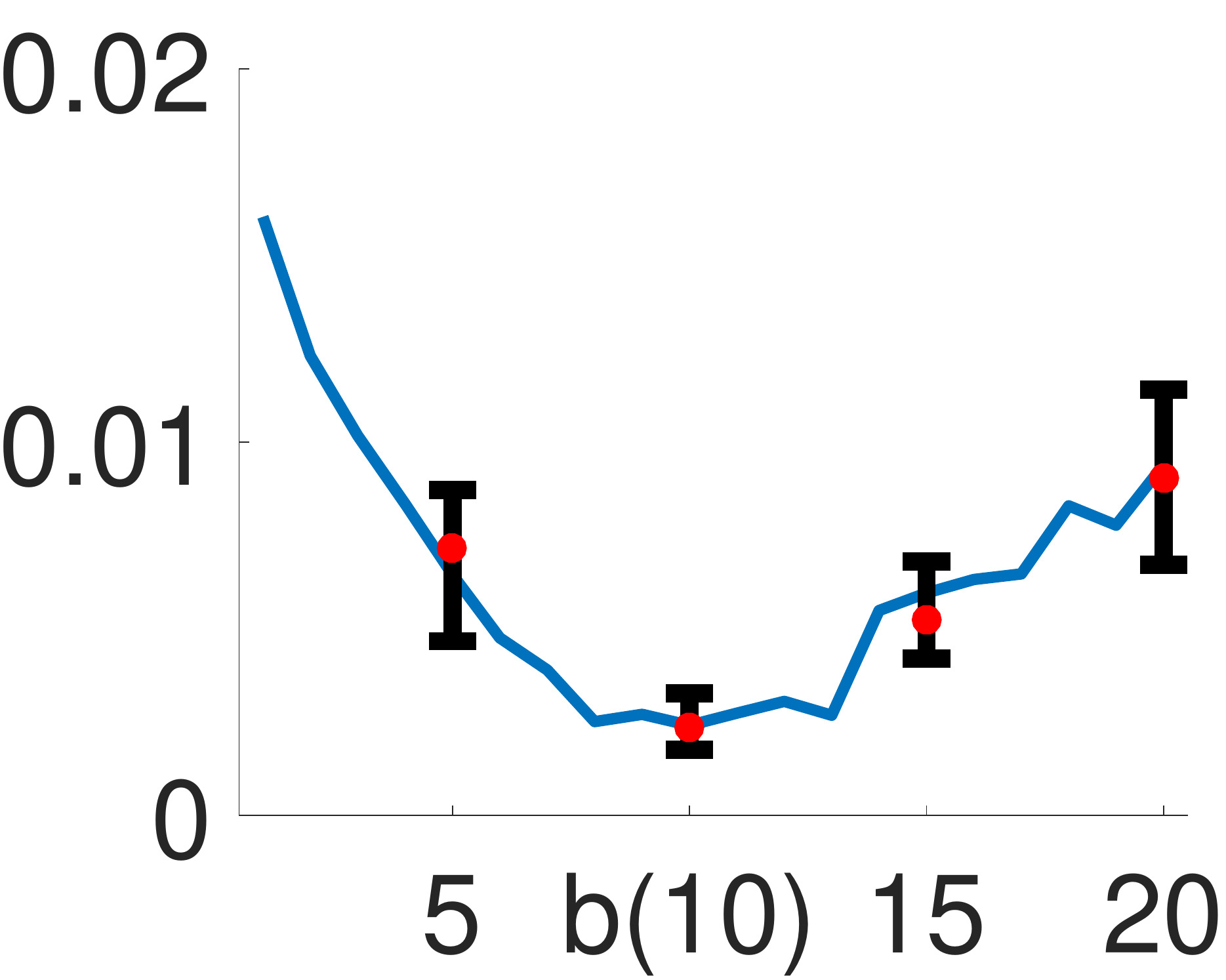}     & \includegraphics[scale = 0.18]{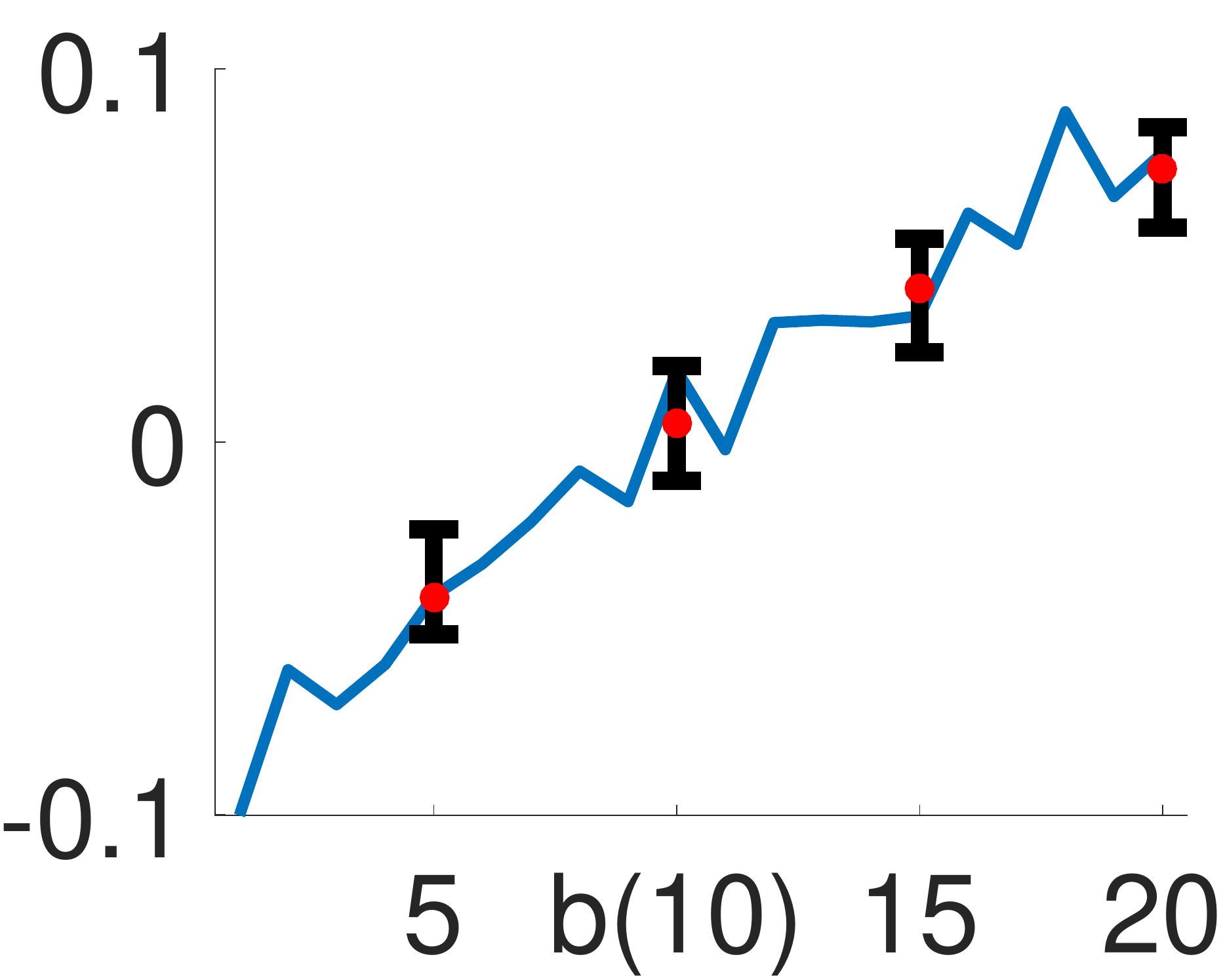}   &	\includegraphics[scale = 0.18]{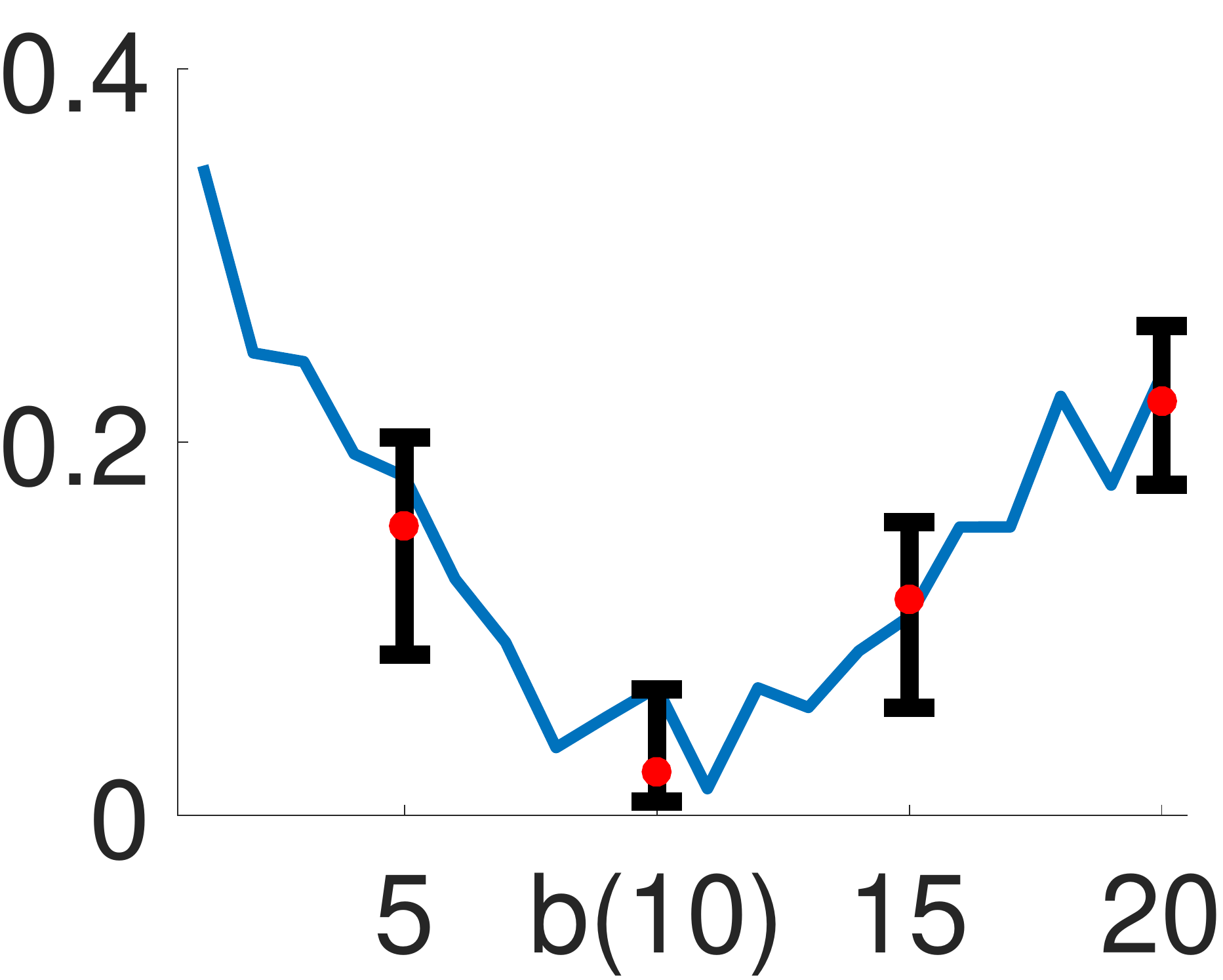}                \\ \hline
				\includegraphics[scale = 0.18]{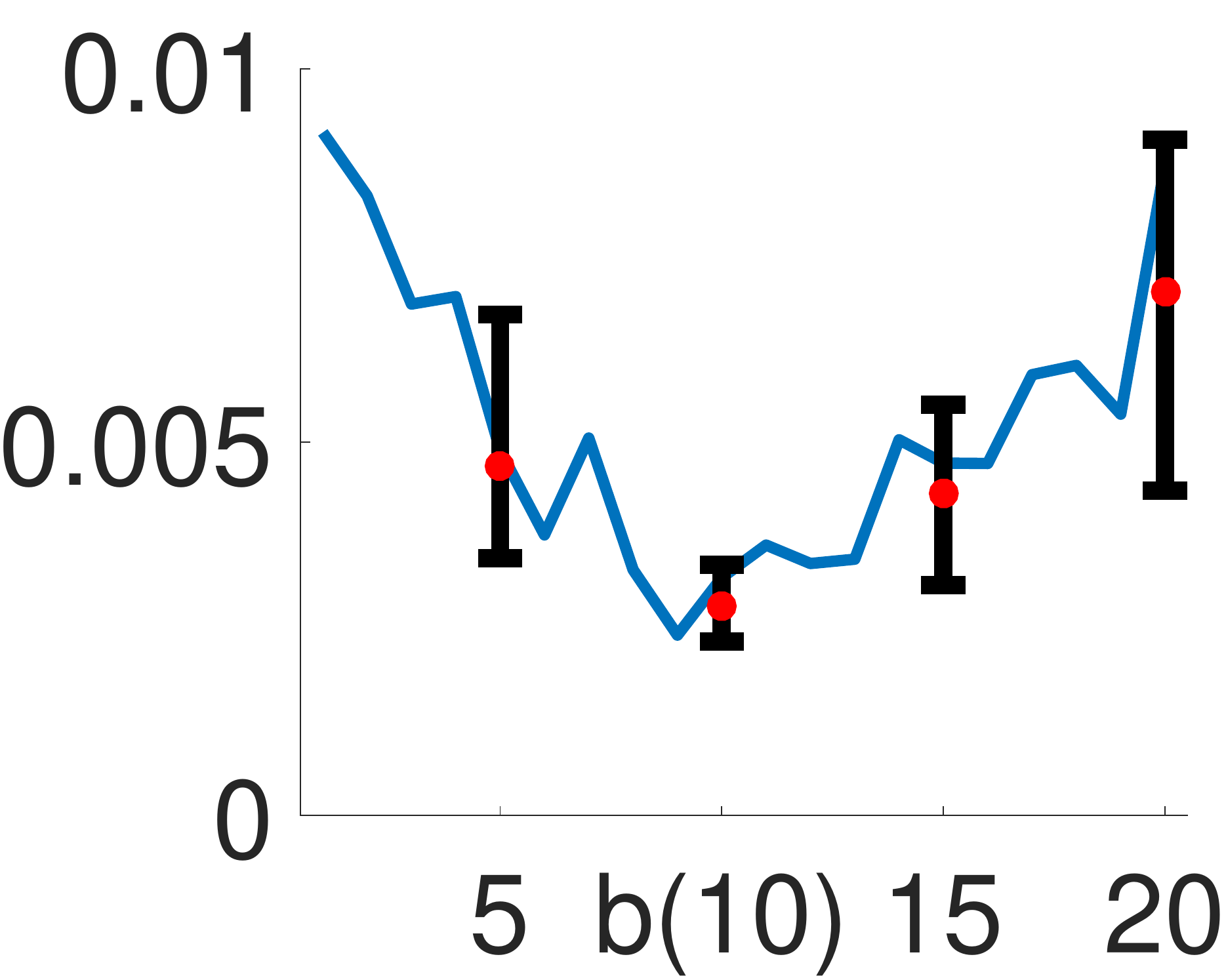}     & \includegraphics[scale = 0.18]{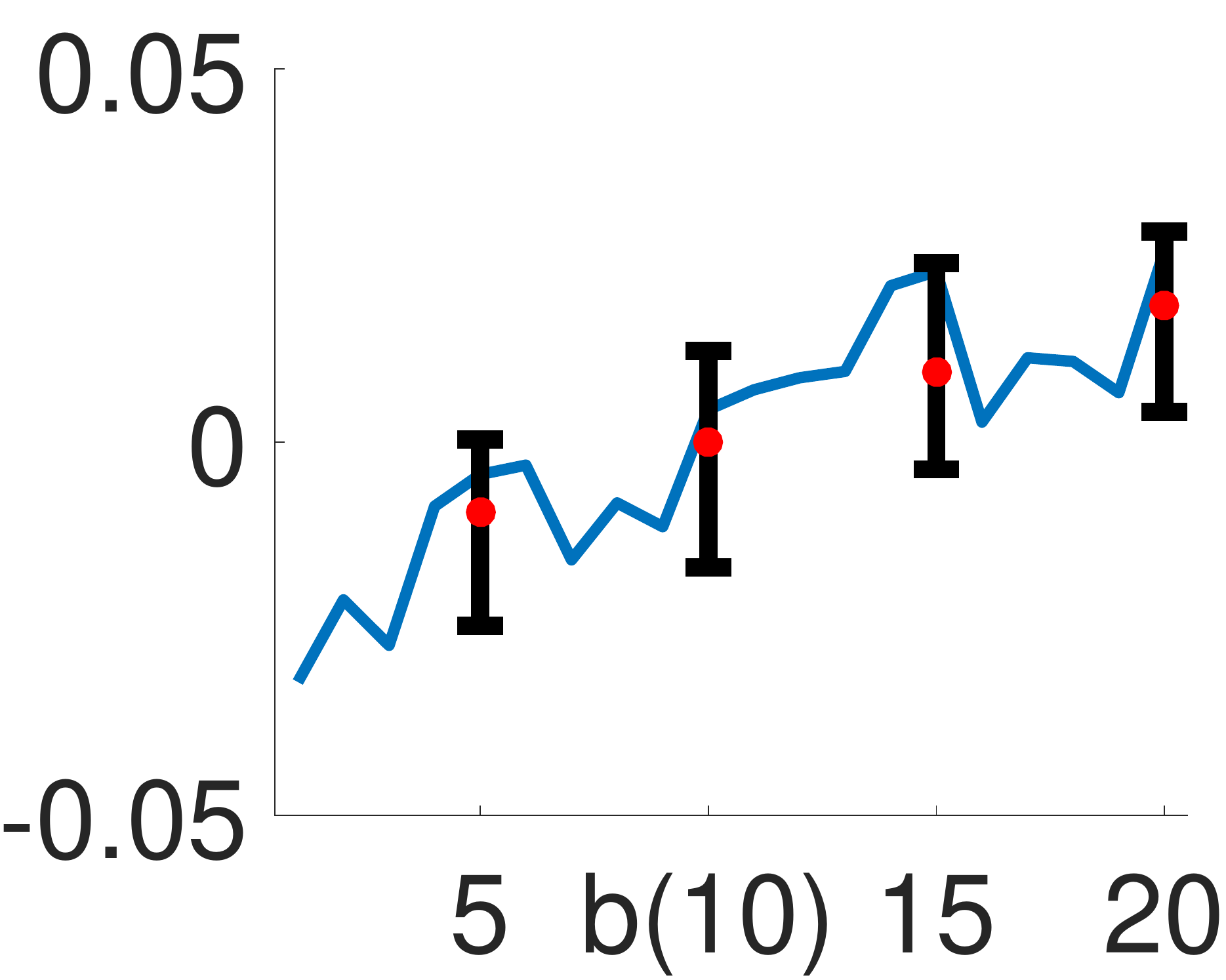}   &	\includegraphics[scale = 0.18]{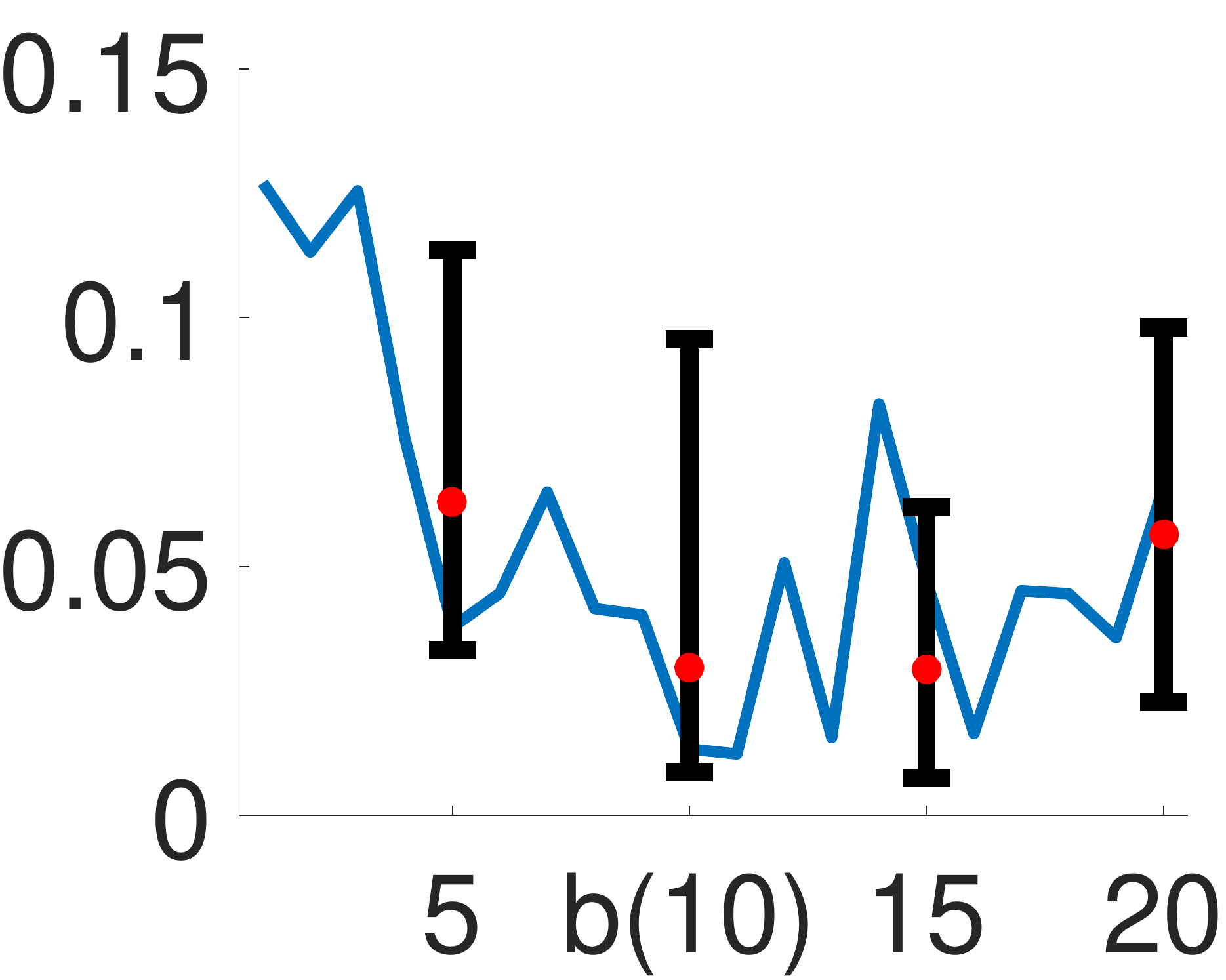}                \\ \hline
				\includegraphics[scale = 0.18]{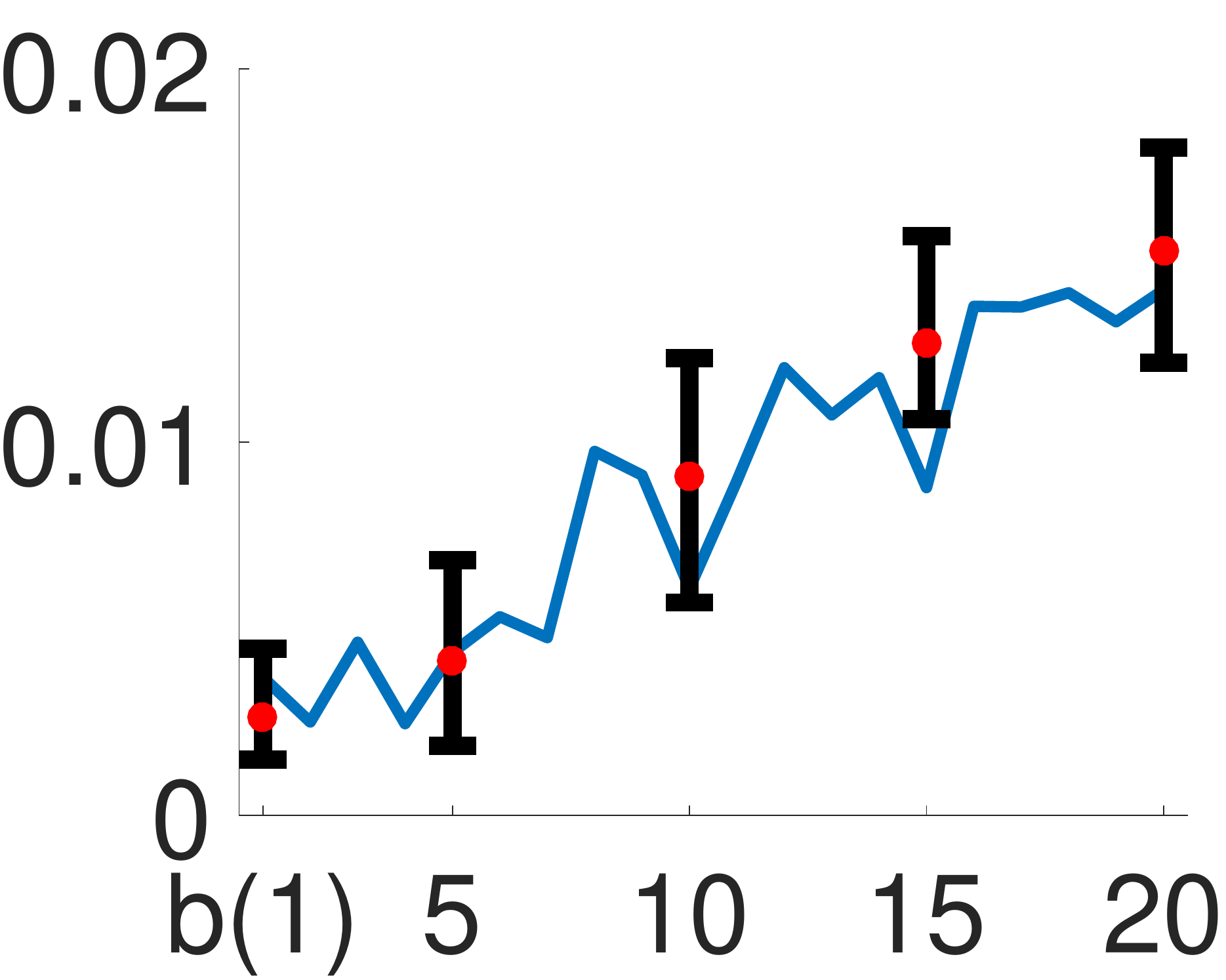}     & \includegraphics[scale = 0.18]{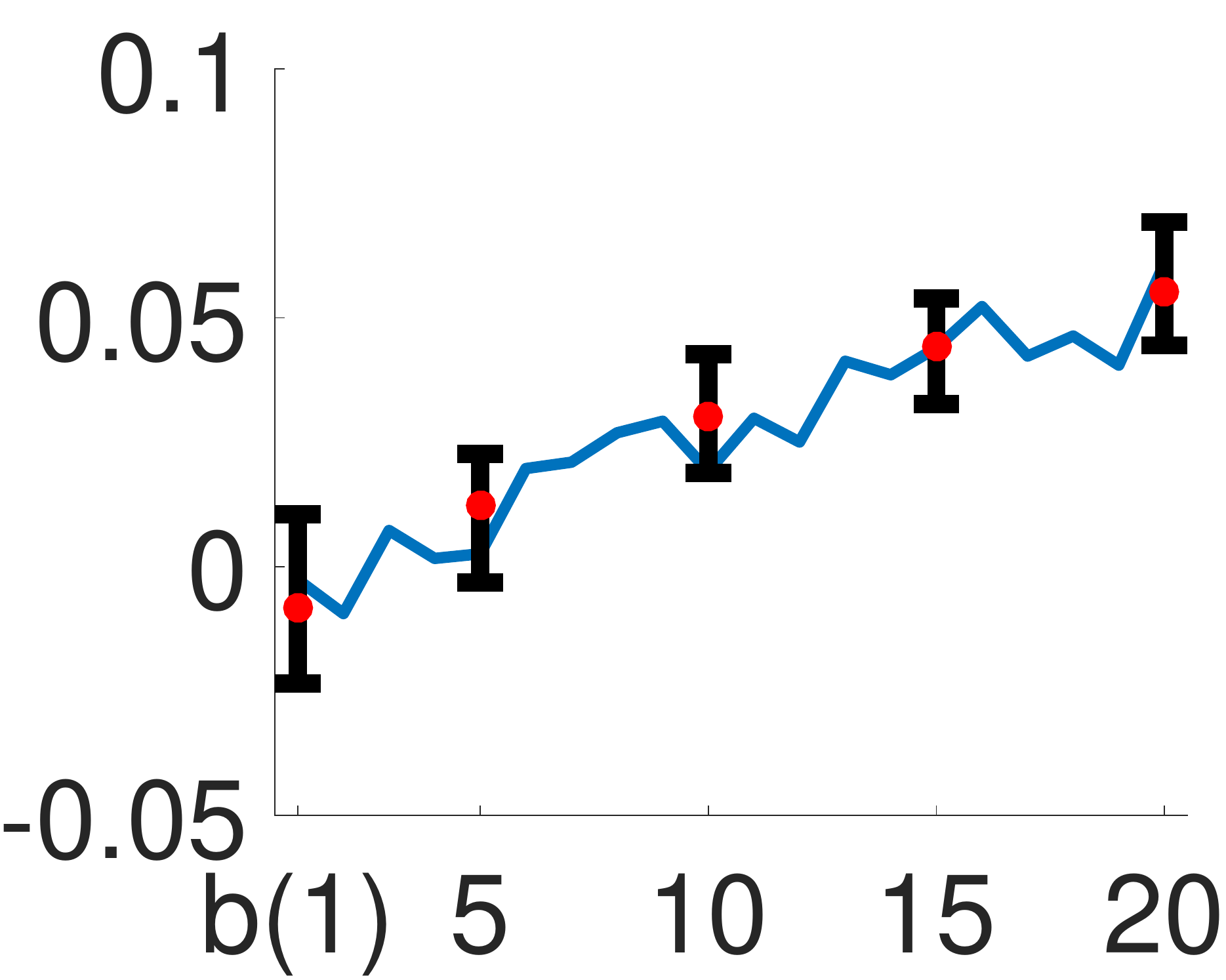}   &	\includegraphics[scale = 0.18]{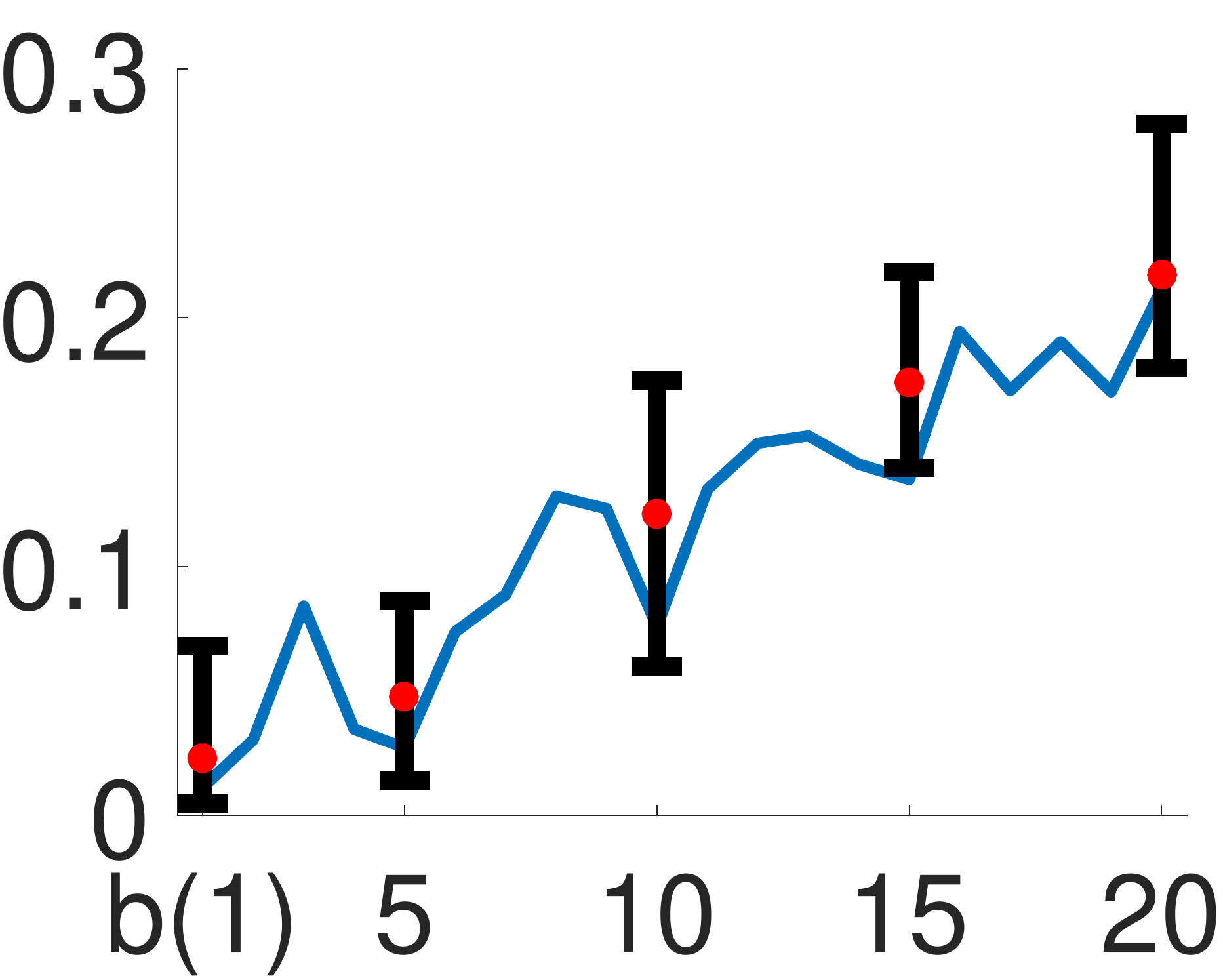}                \\ \hline
			\end{tabular}
		\end{center}
		\caption{Sensitivity to perturbations of $a_1$ in the DCV model for the Acidity (top row), Galaxy (middle row) and Enzyme (bottom row) datasets.}
		\label{fig:DCV_a1}
	\end{figure}

\begin{figure}[!t]
		\begin{center}
			\begin{tabular}{|c|c|c|}
				\hline
				$\mathbb{D}$ & $\mathbb{V}$ & $\mathbb{E}$ \\ \hline
				\includegraphics[scale = 0.18]{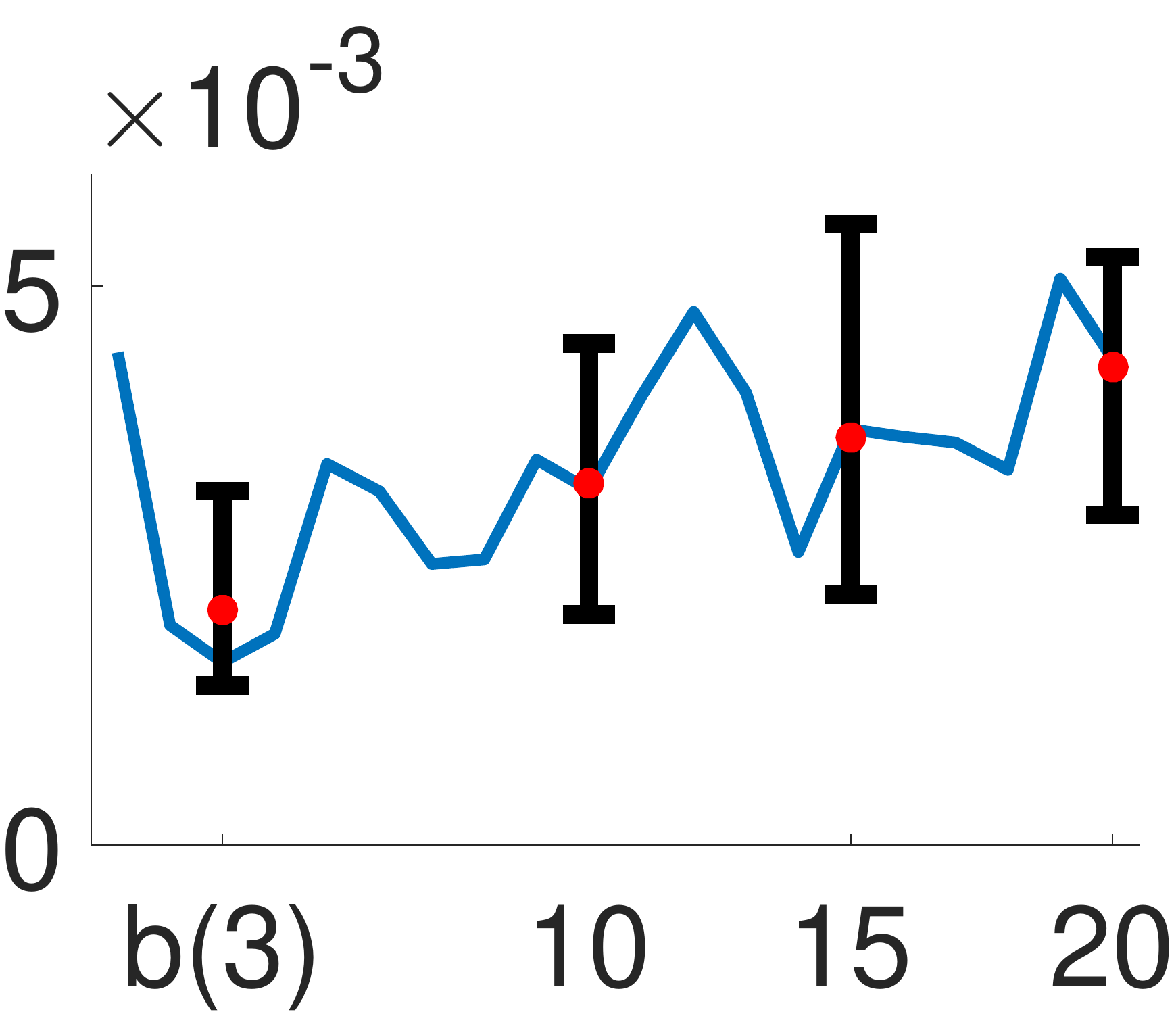}   & \includegraphics[scale = 0.18]{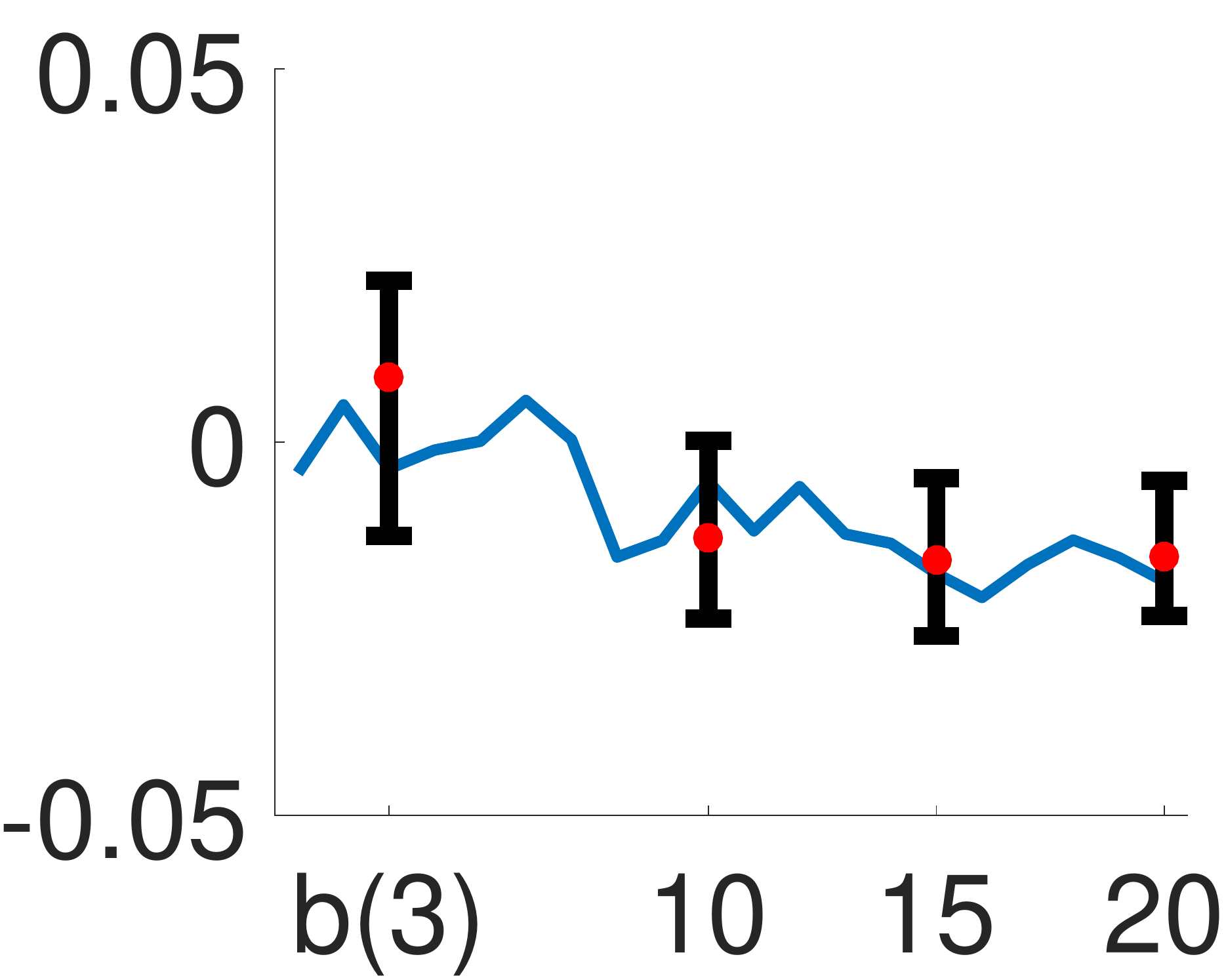} &	\includegraphics[scale = 0.18]{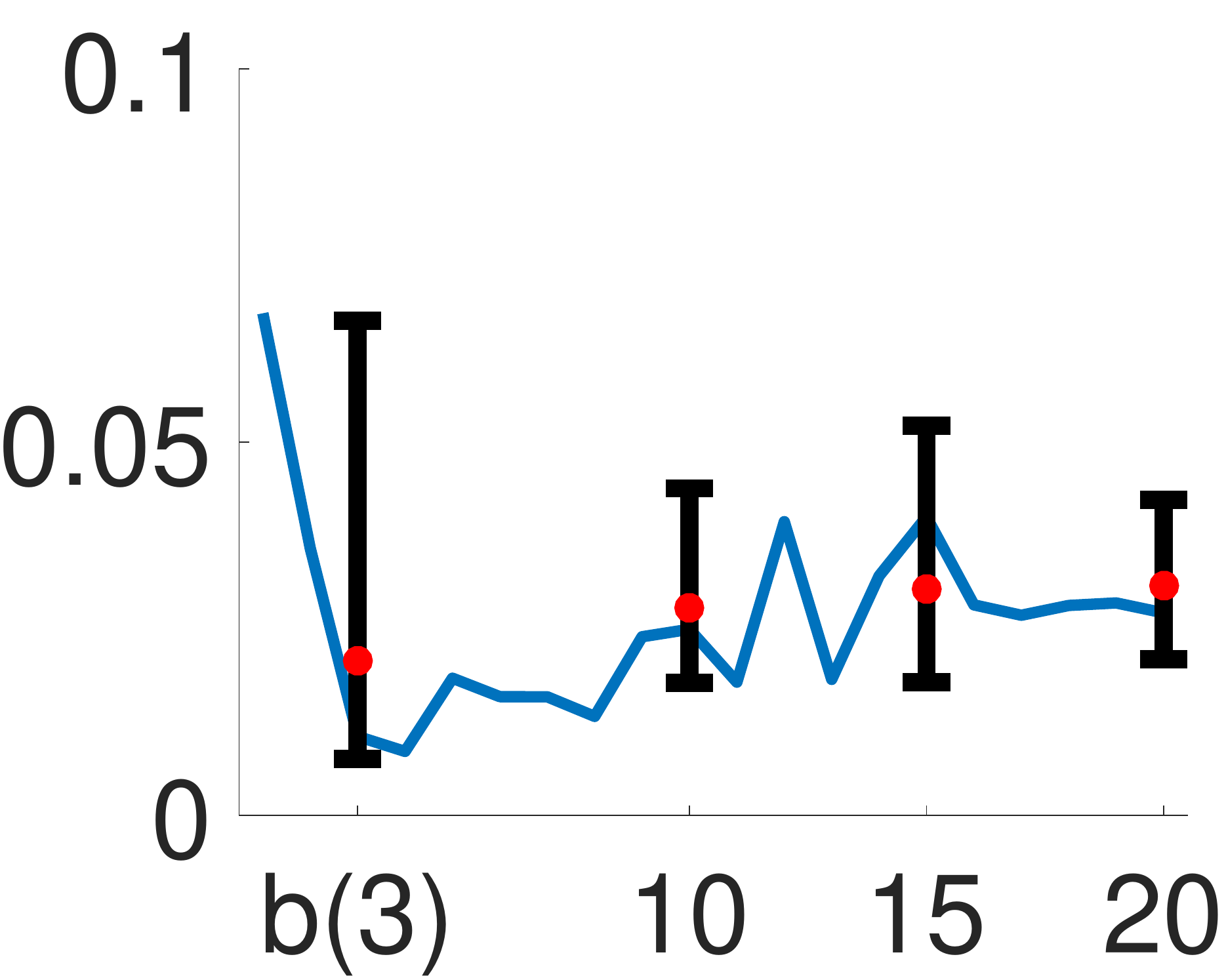}                \\ \hline
				\includegraphics[scale = 0.18]{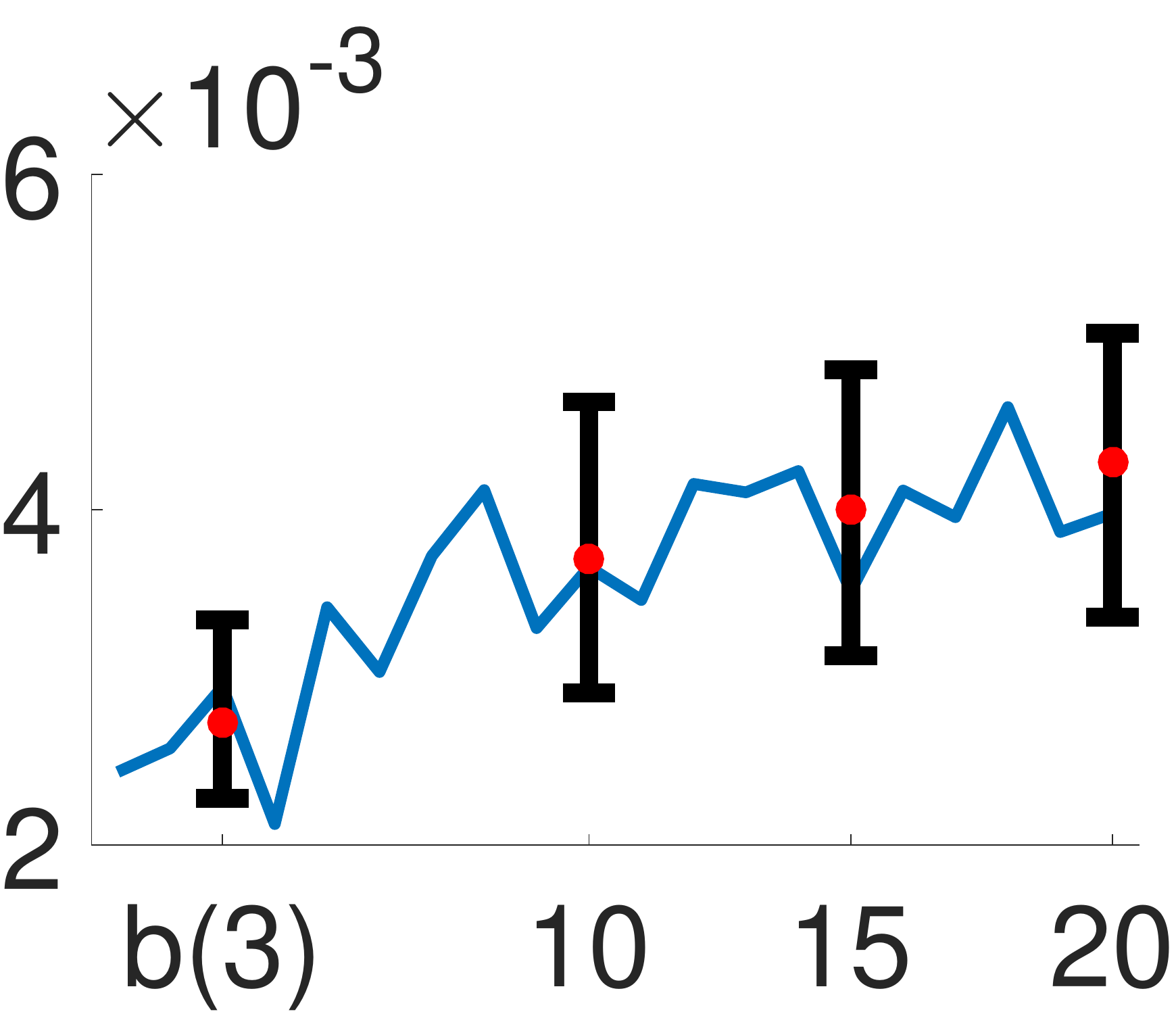}   & \includegraphics[scale = 0.18]{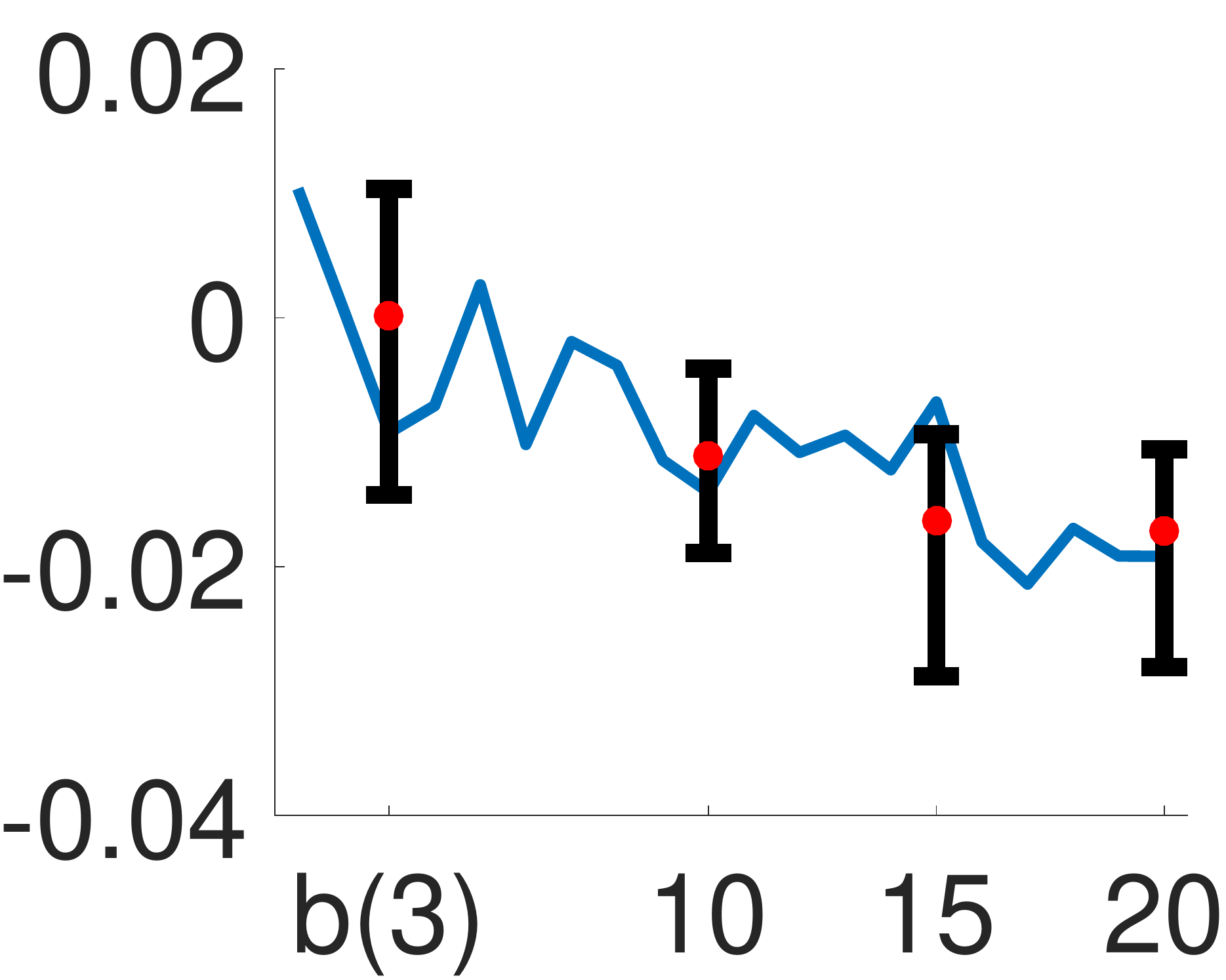} &	\includegraphics[scale = 0.18]{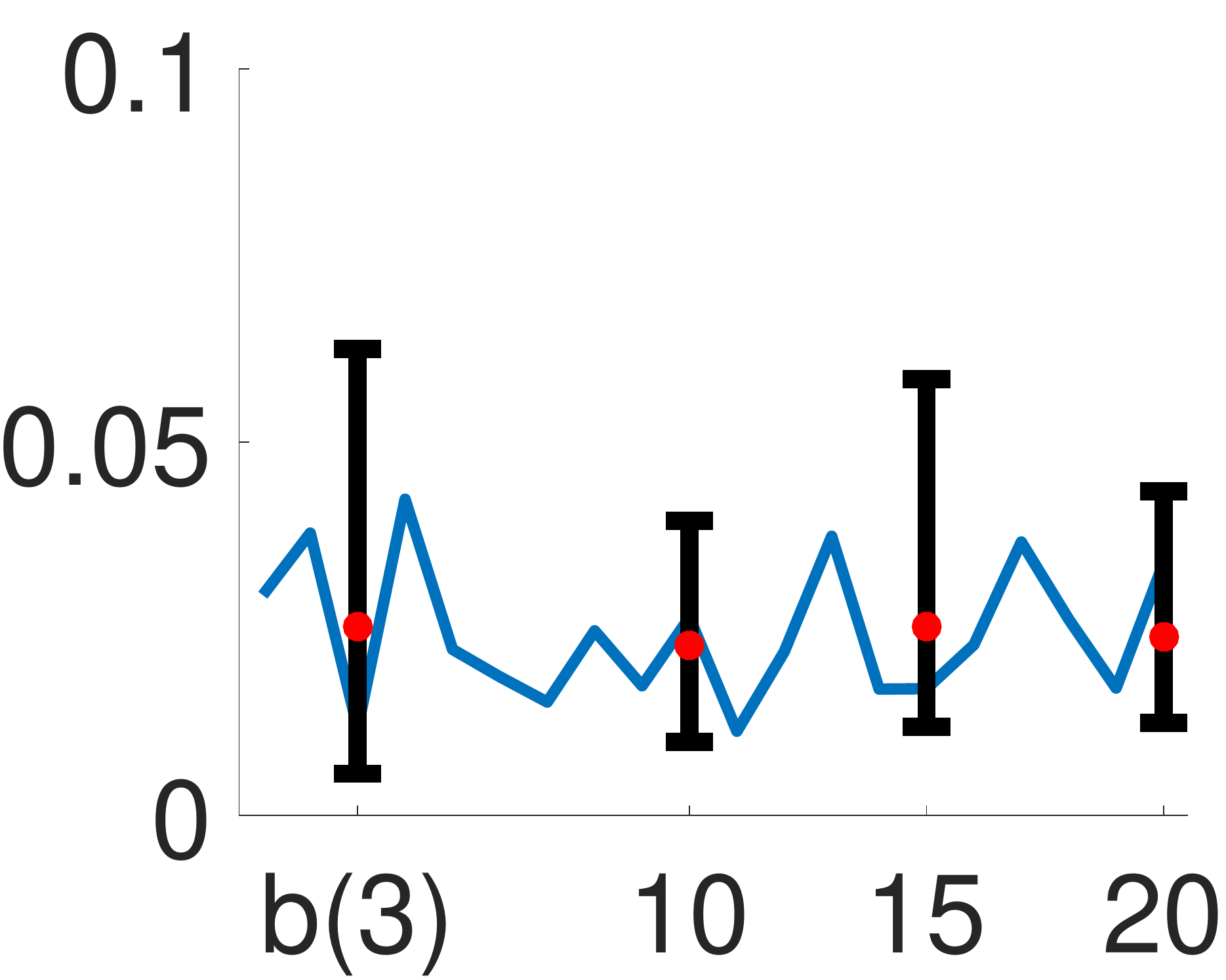}                \\ \hline
				\includegraphics[scale = 0.18]{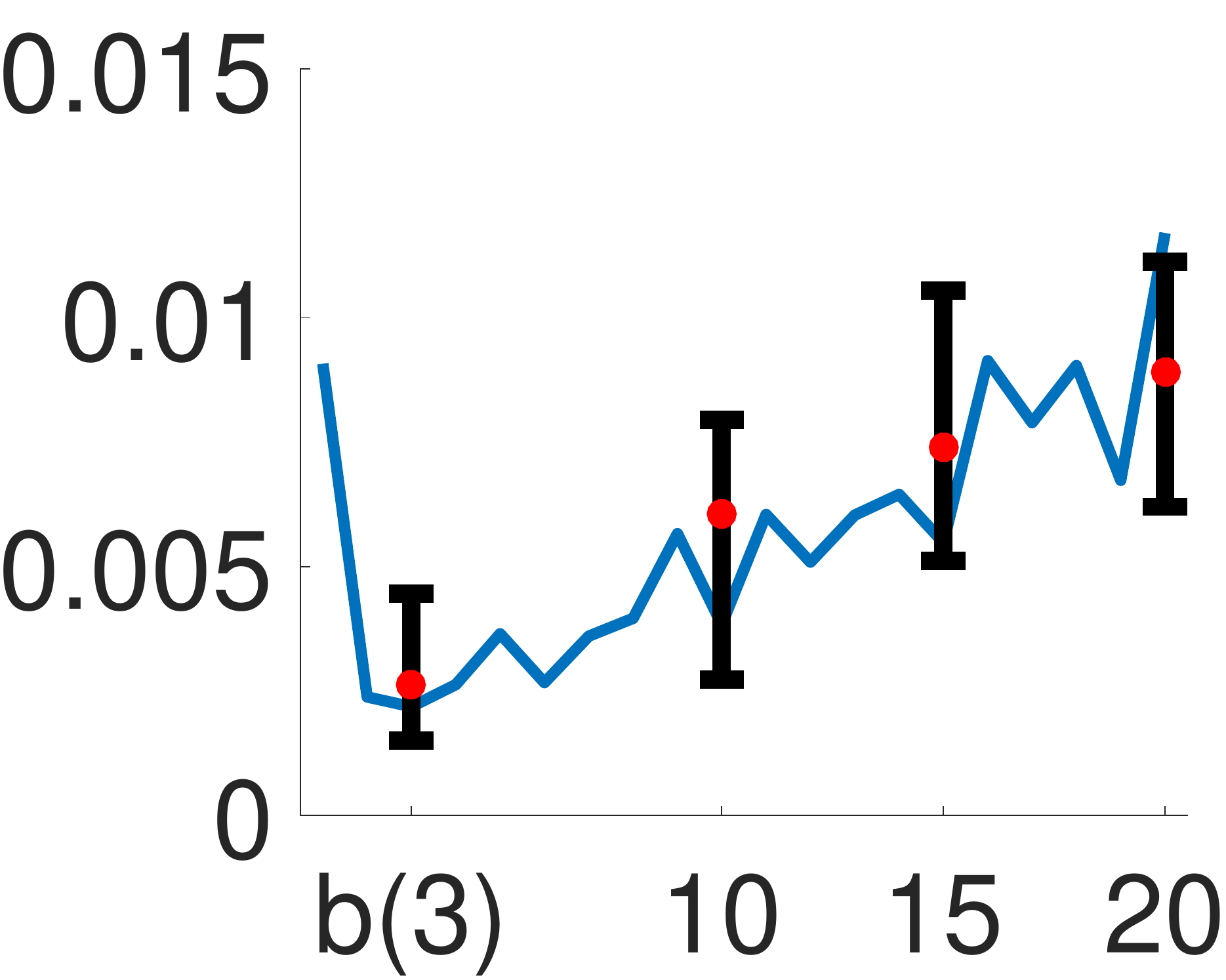}   & \includegraphics[scale = 0.18]{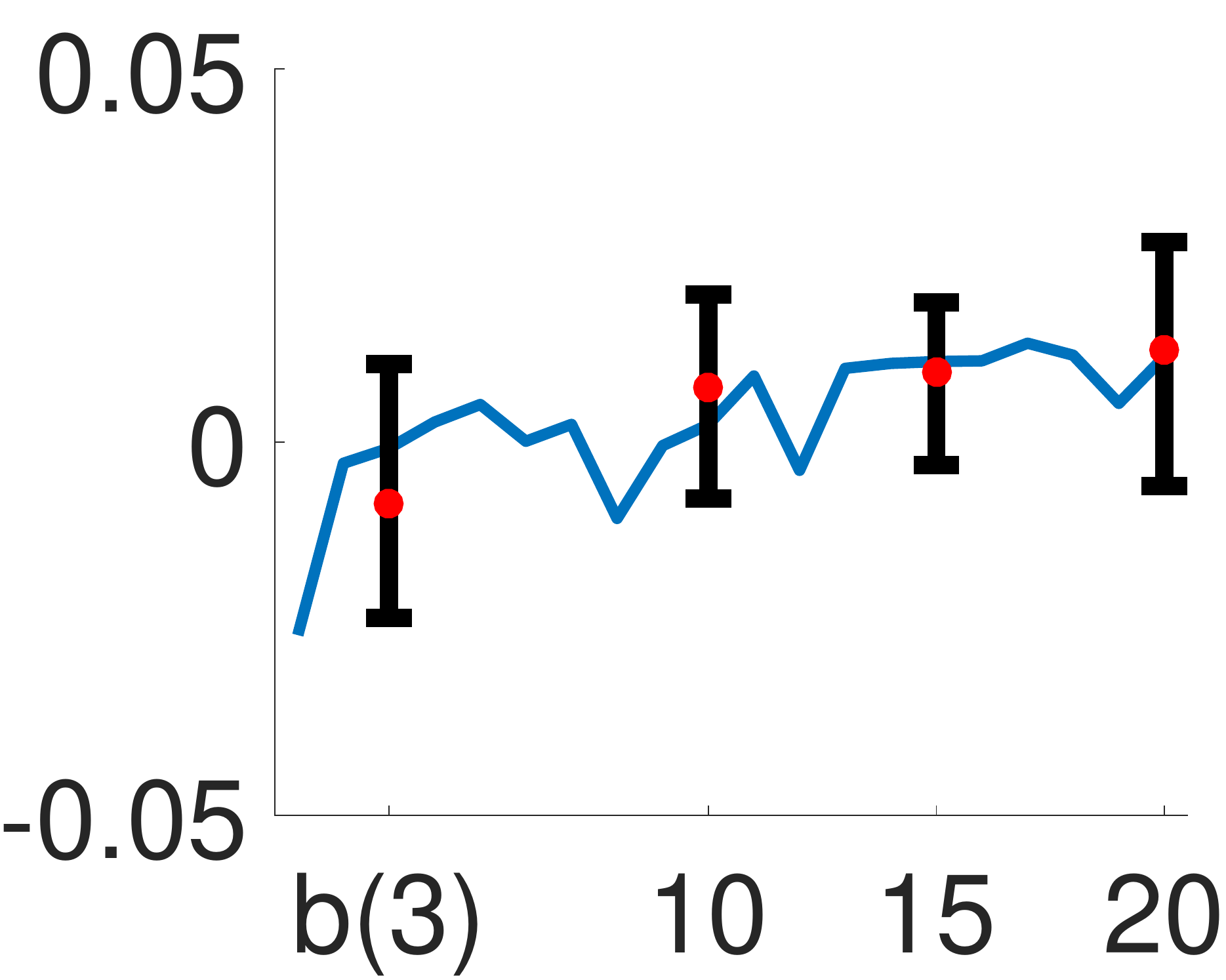} &	\includegraphics[scale = 0.18]{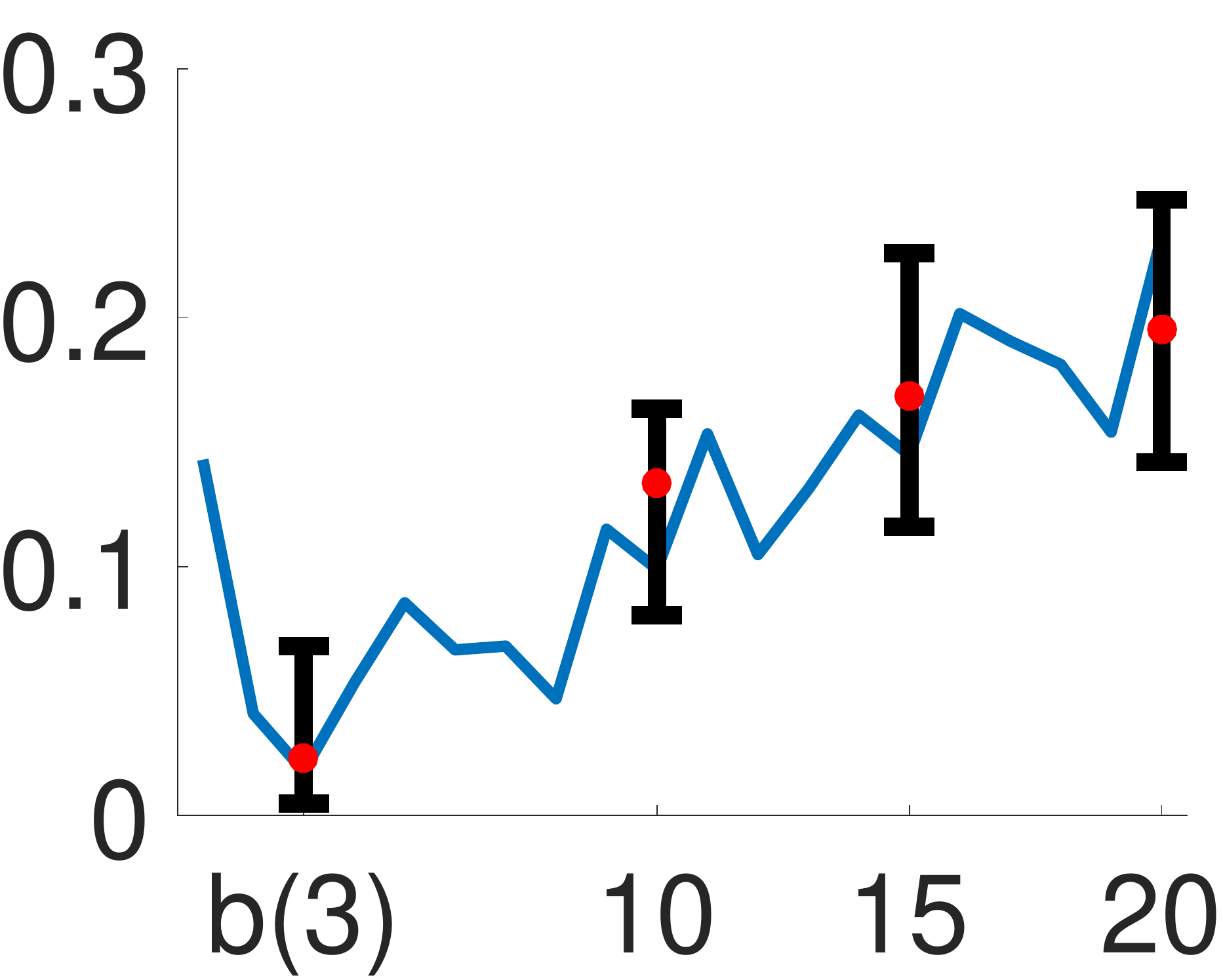}                \\ \hline
			\end{tabular}
		\end{center}
		\caption{Sensitivity to perturbations of $\eta$ in the DCV model for the Acidity (top row), Galaxy (middle row) and Enzyme (bottom row) datasets.}
		\label{fig:DCV_eta}
	\end{figure}

\begin{figure}[!t]
		\begin{center}
			\begin{tabular}{|c|c|c|}
				\hline
				$\mathbb{D}$ & $\mathbb{V}$ & $\mathbb{E}$  \\ \hline
				\includegraphics[scale = 0.18]{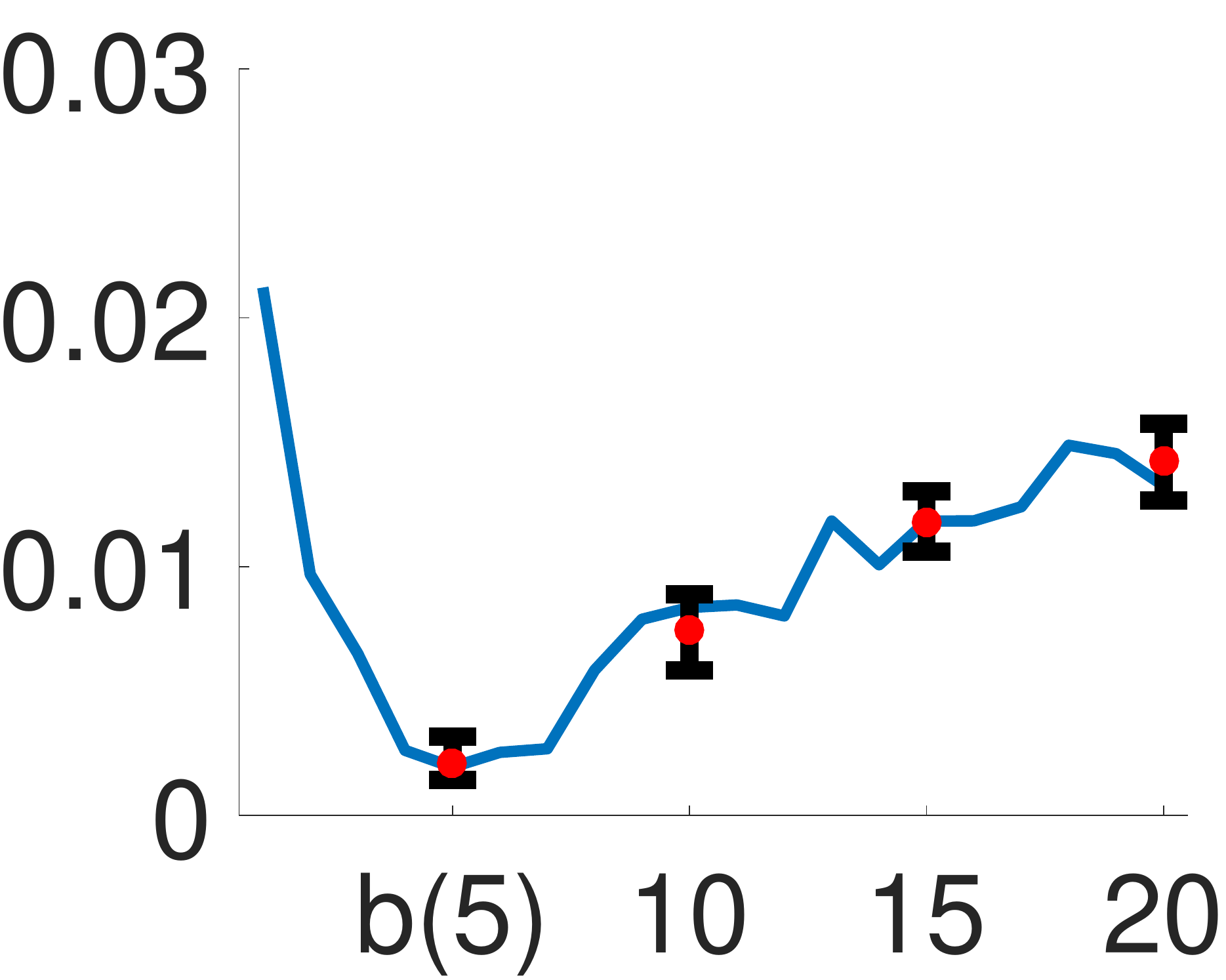}      & \includegraphics[scale = 0.18]{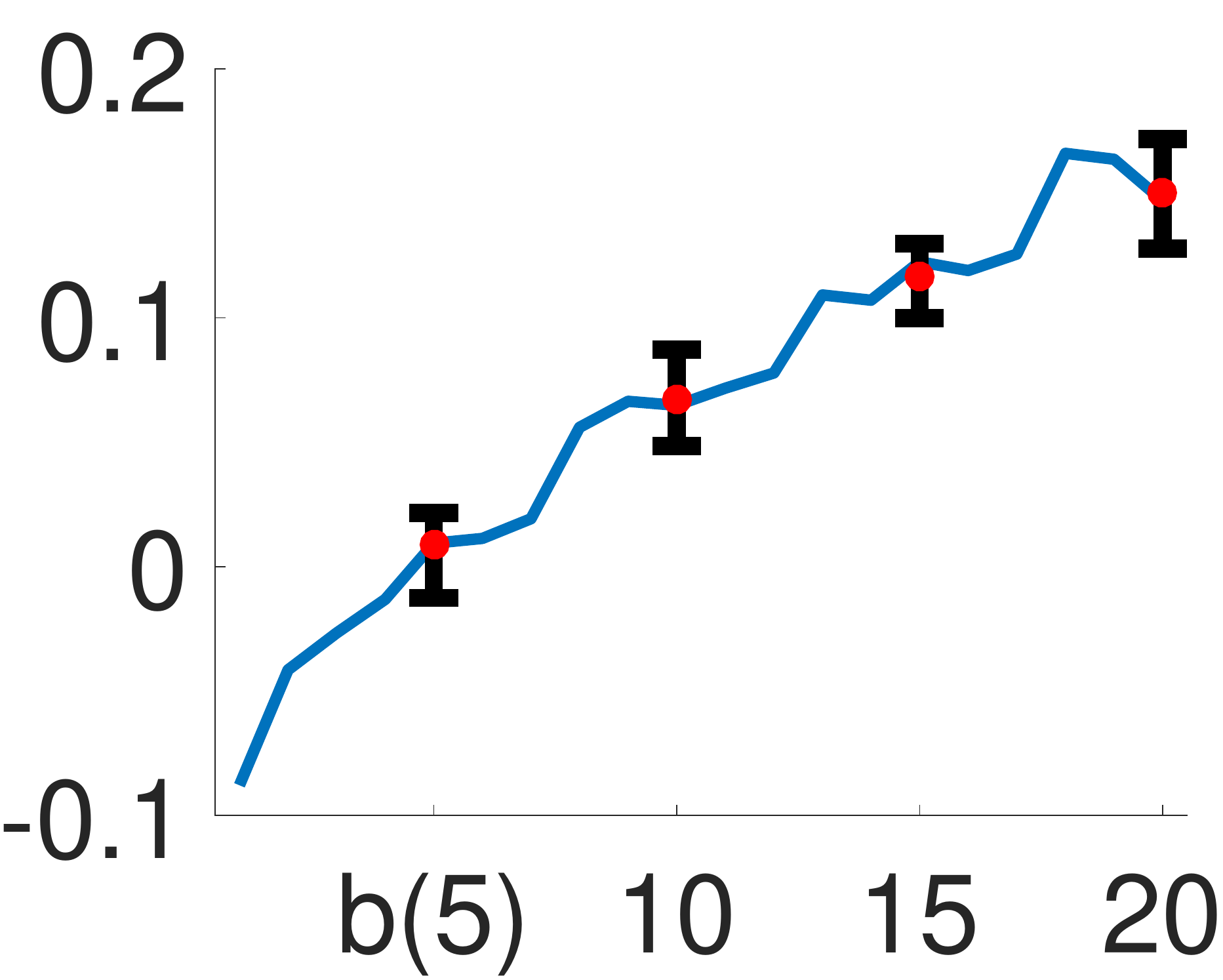}  &	\includegraphics[scale = 0.18]{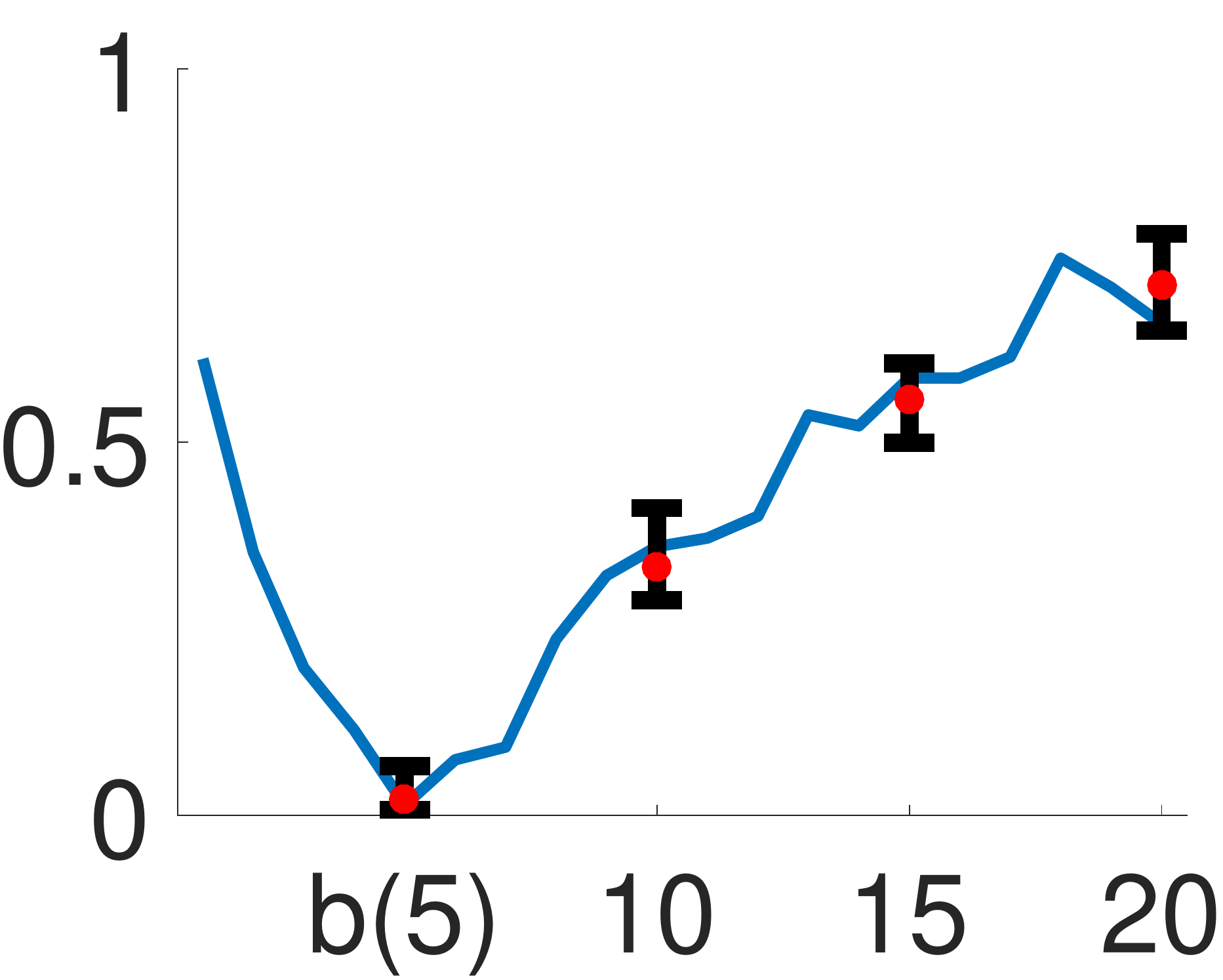}              \\ \hline
				\includegraphics[scale = 0.18]{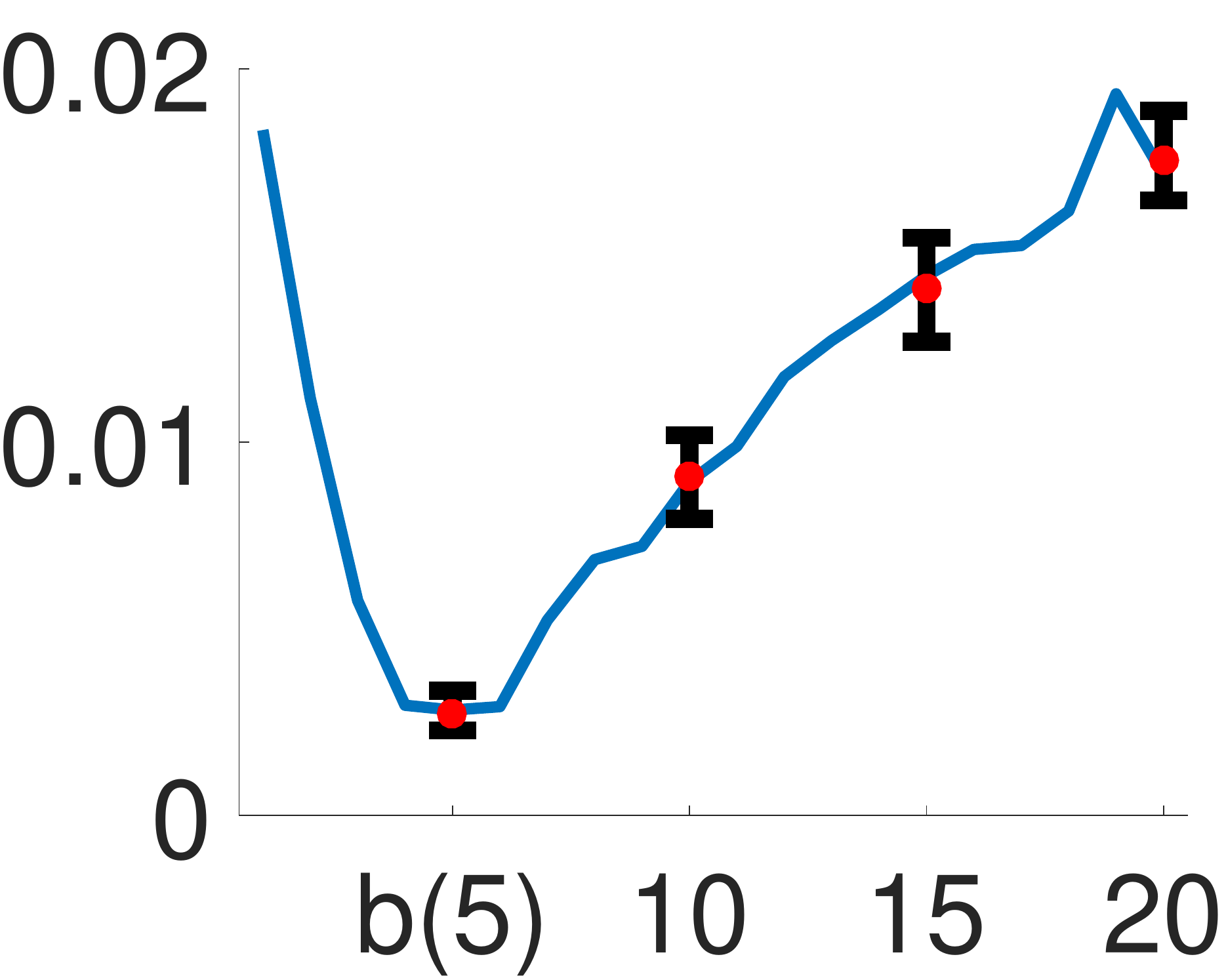}      & \includegraphics[scale = 0.18]{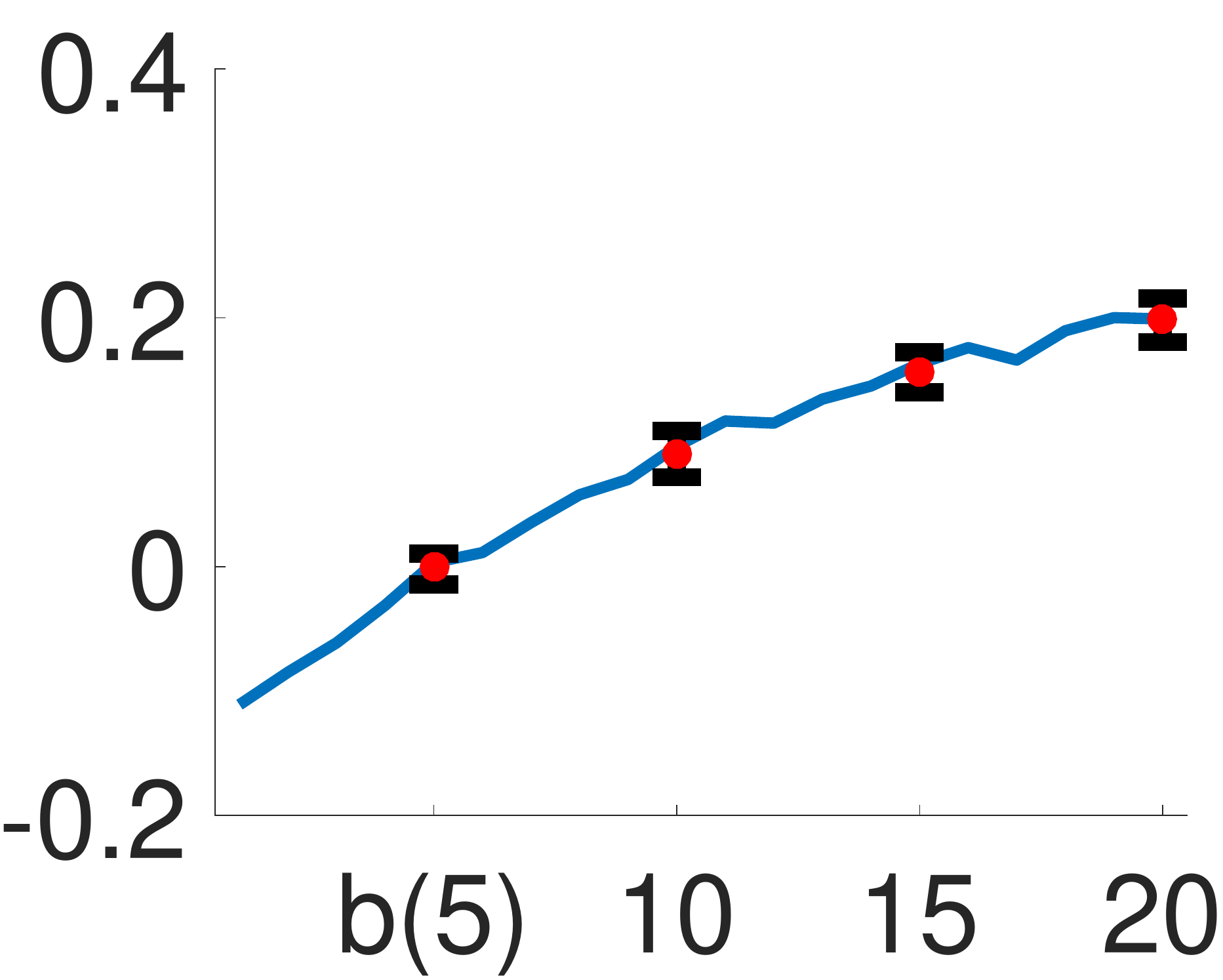}  &	\includegraphics[scale = 0.18]{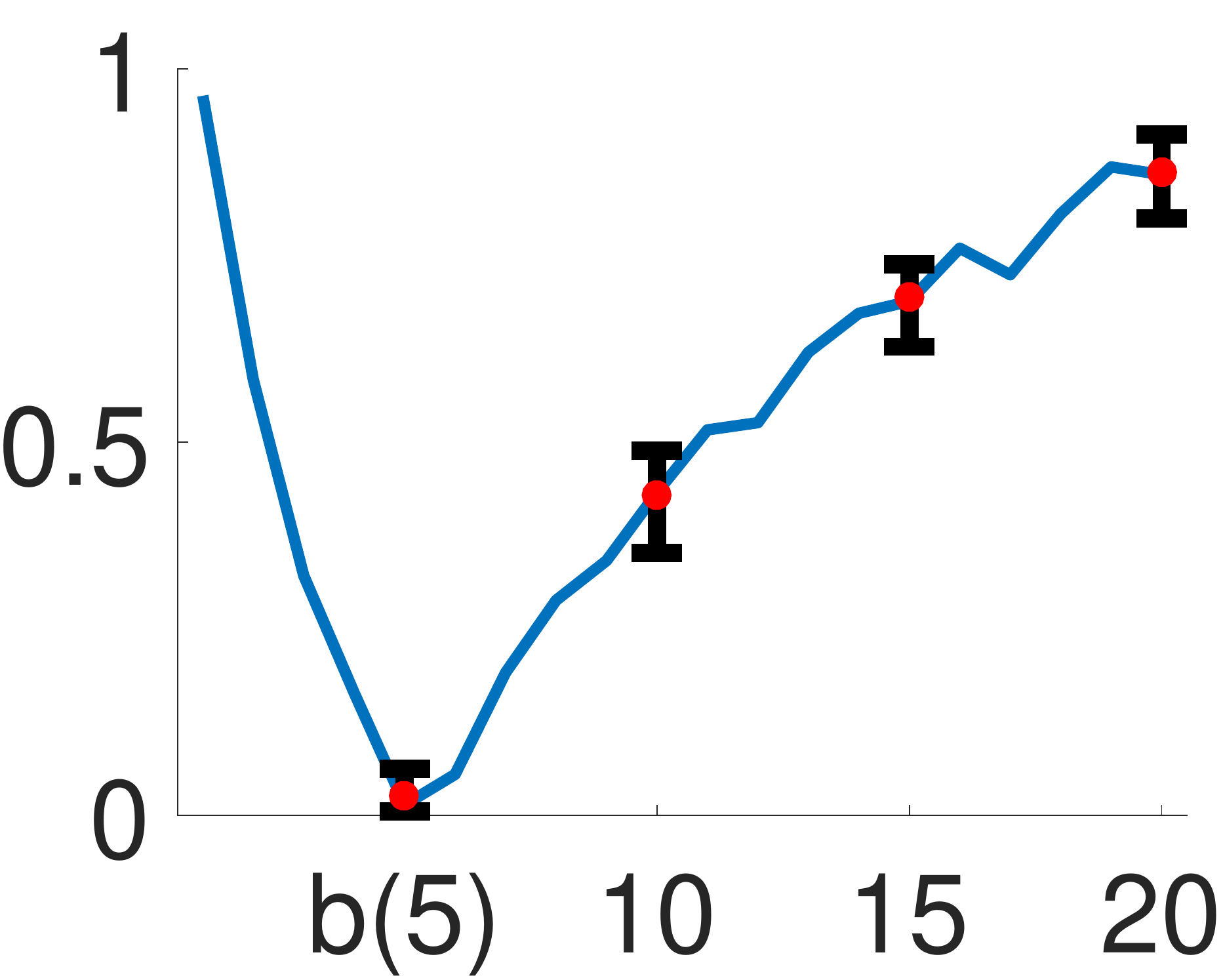}              \\ \hline
				\includegraphics[scale = 0.18]{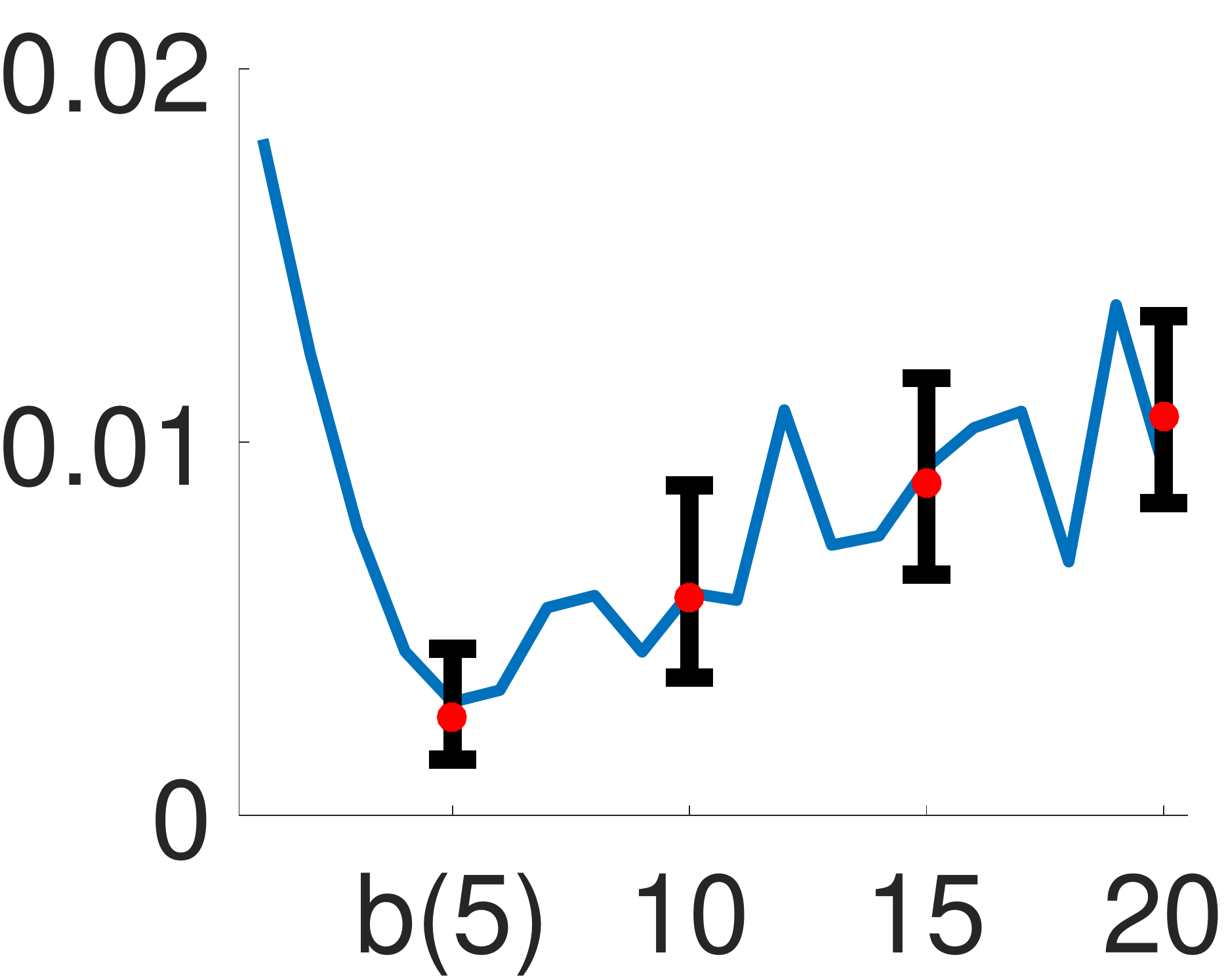}      & \includegraphics[scale = 0.18]{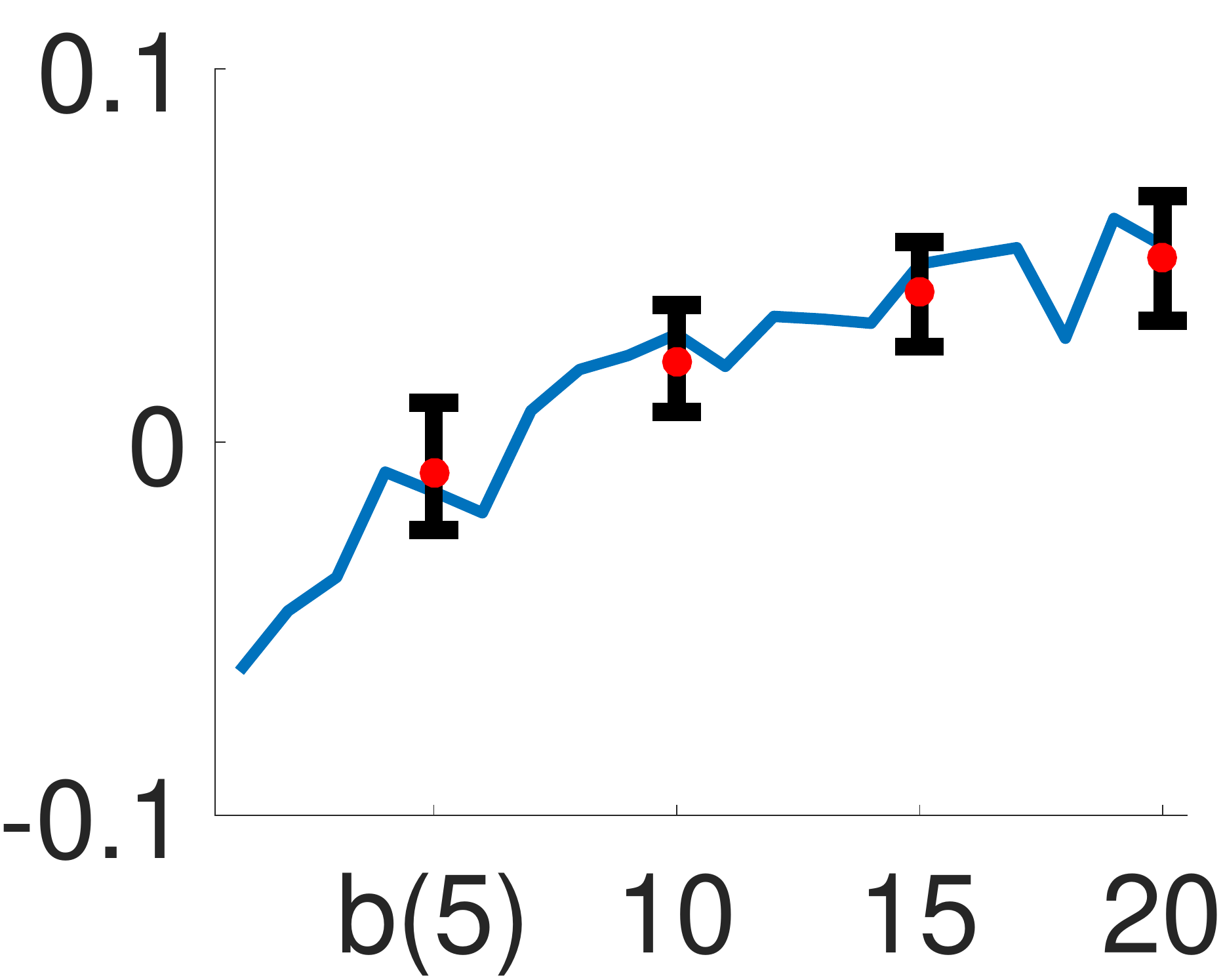}  &	\includegraphics[scale = 0.18]{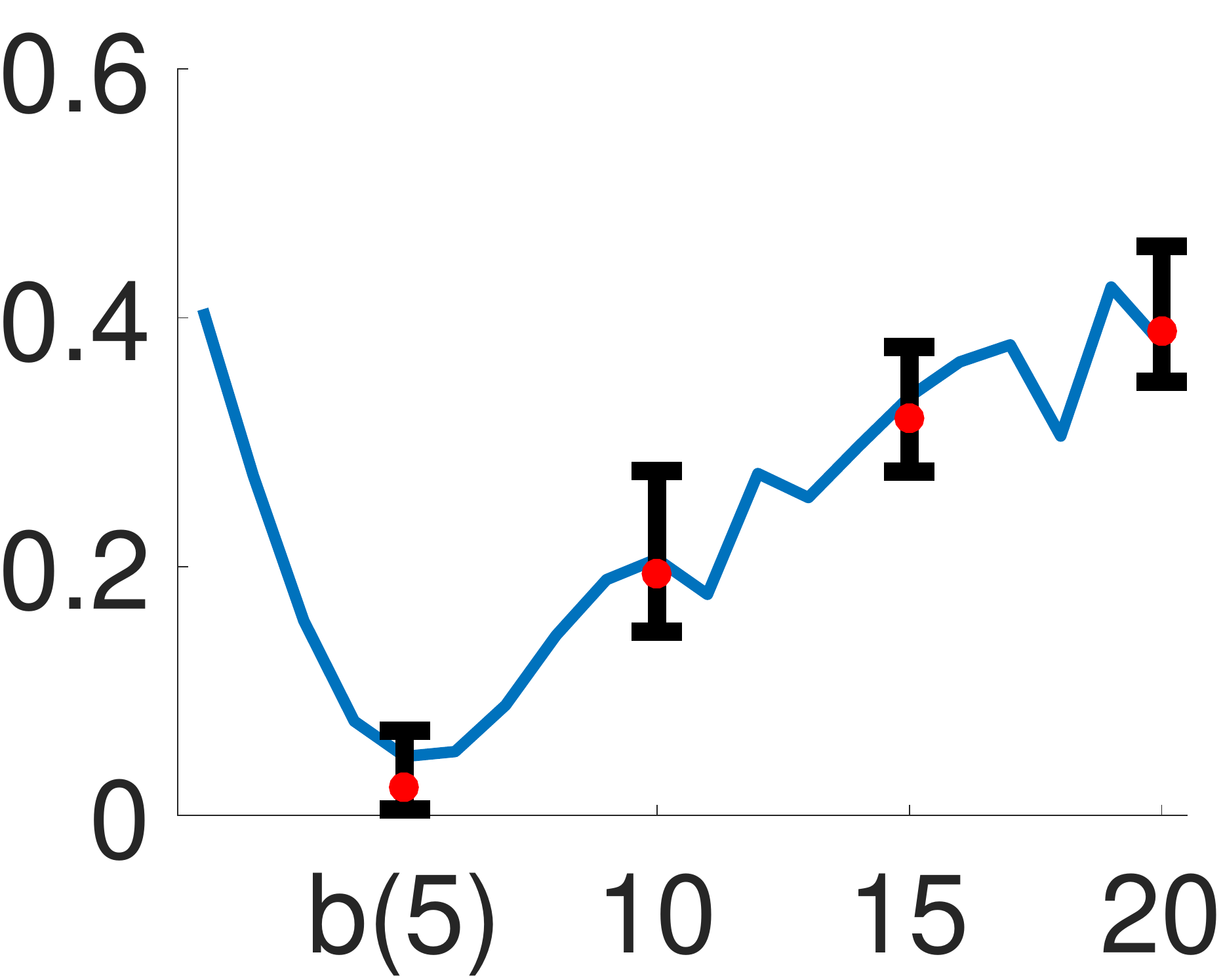}              \\ \hline
			\end{tabular}
		\end{center}
		\caption{Sensitivity to perturbations of $\gamma$ in the DCV model for the Acidity (top row), Galaxy (middle row) and Enzyme (bottom row) datasets.}
		\label{fig:DCV_gamma}
	\end{figure}

\begin{figure}[!t]
		\begin{center}
			\begin{tabular}{|c|c|c|}
				\hline
				$\mathbb{D}$ & $\mathbb{V}$ & $\mathbb{E}$  \\ \hline
				\includegraphics[scale = 0.18]{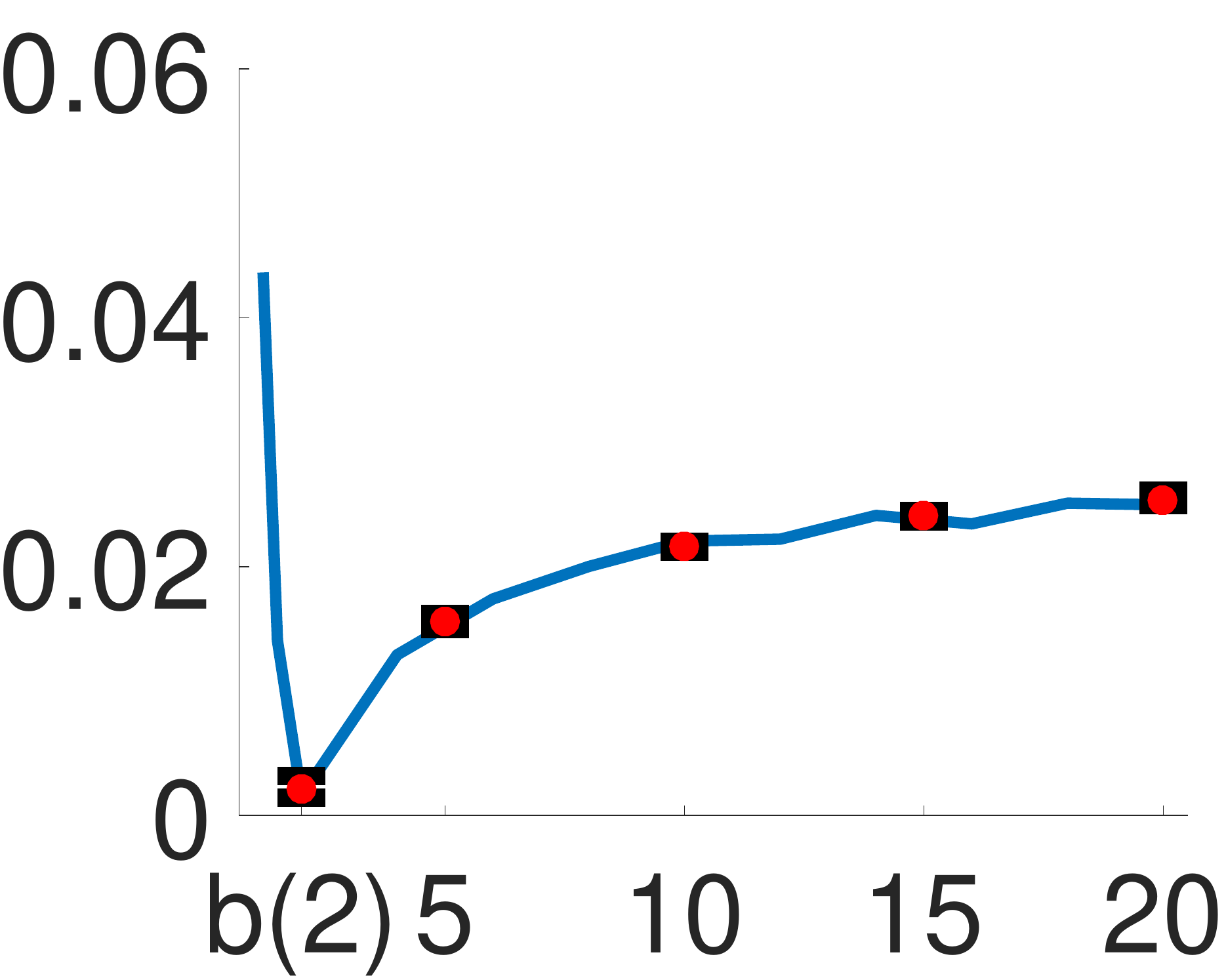}   & \includegraphics[scale = 0.18]{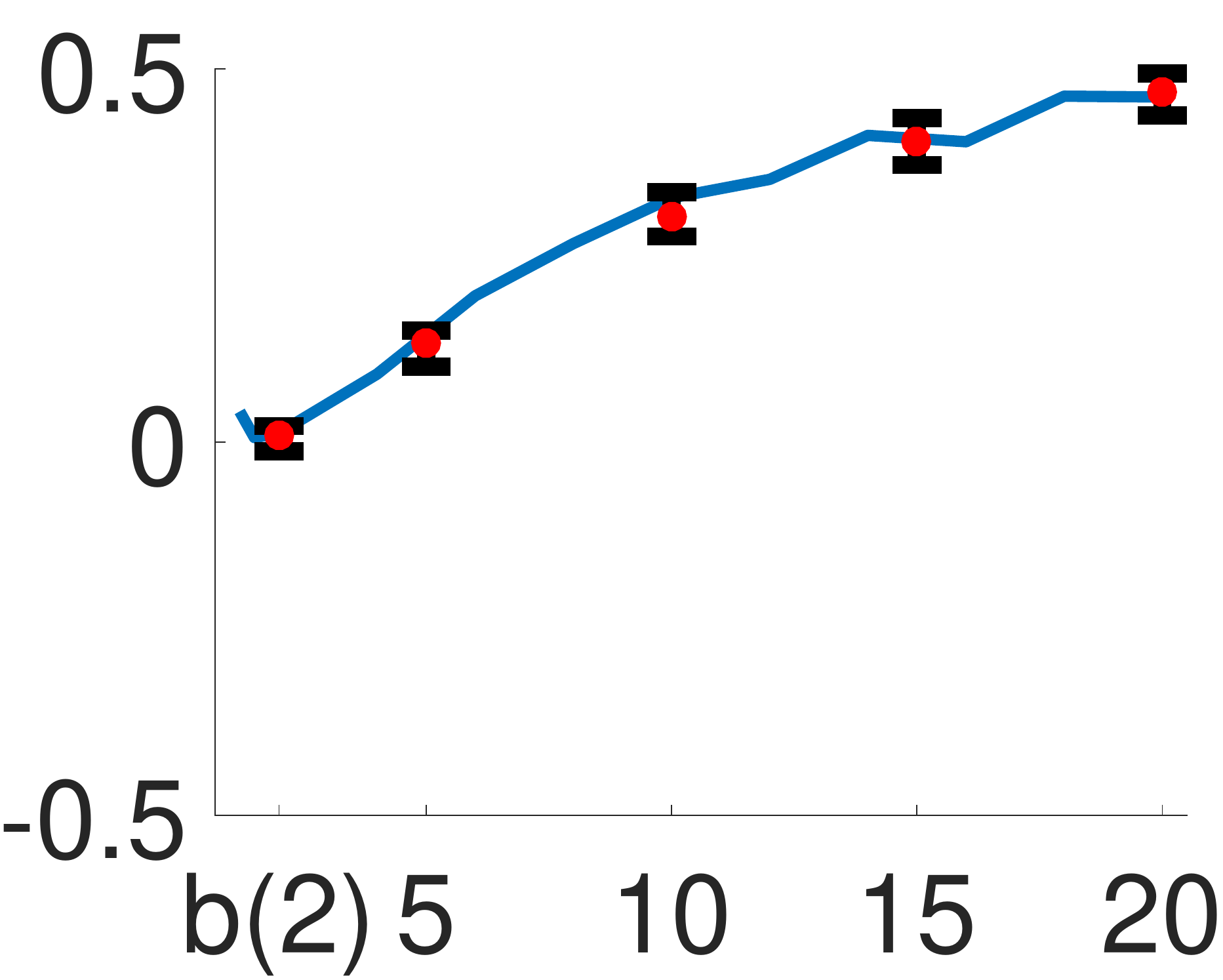}   &	\includegraphics[scale = 0.18]{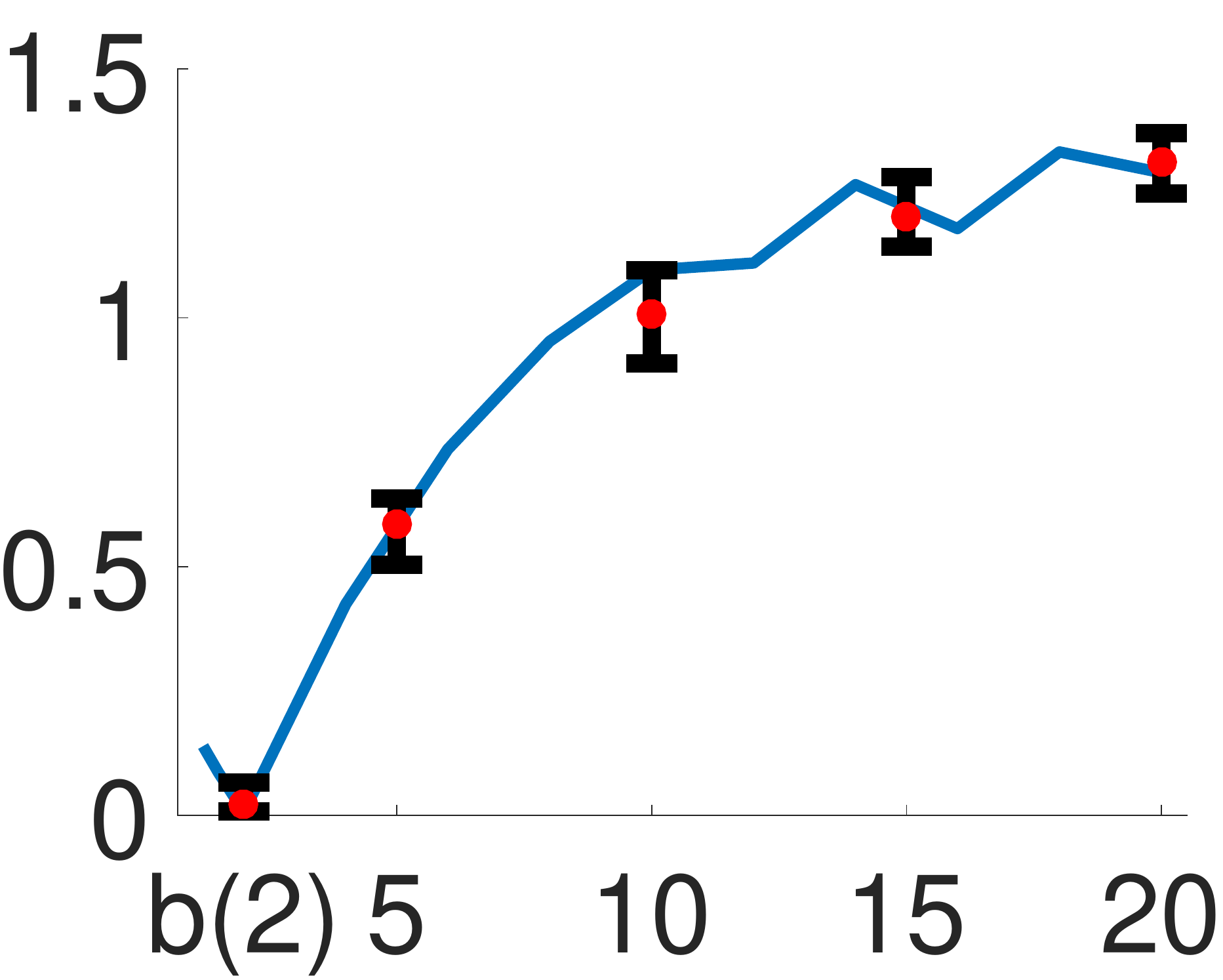}                \\ \hline
				\includegraphics[scale = 0.18]{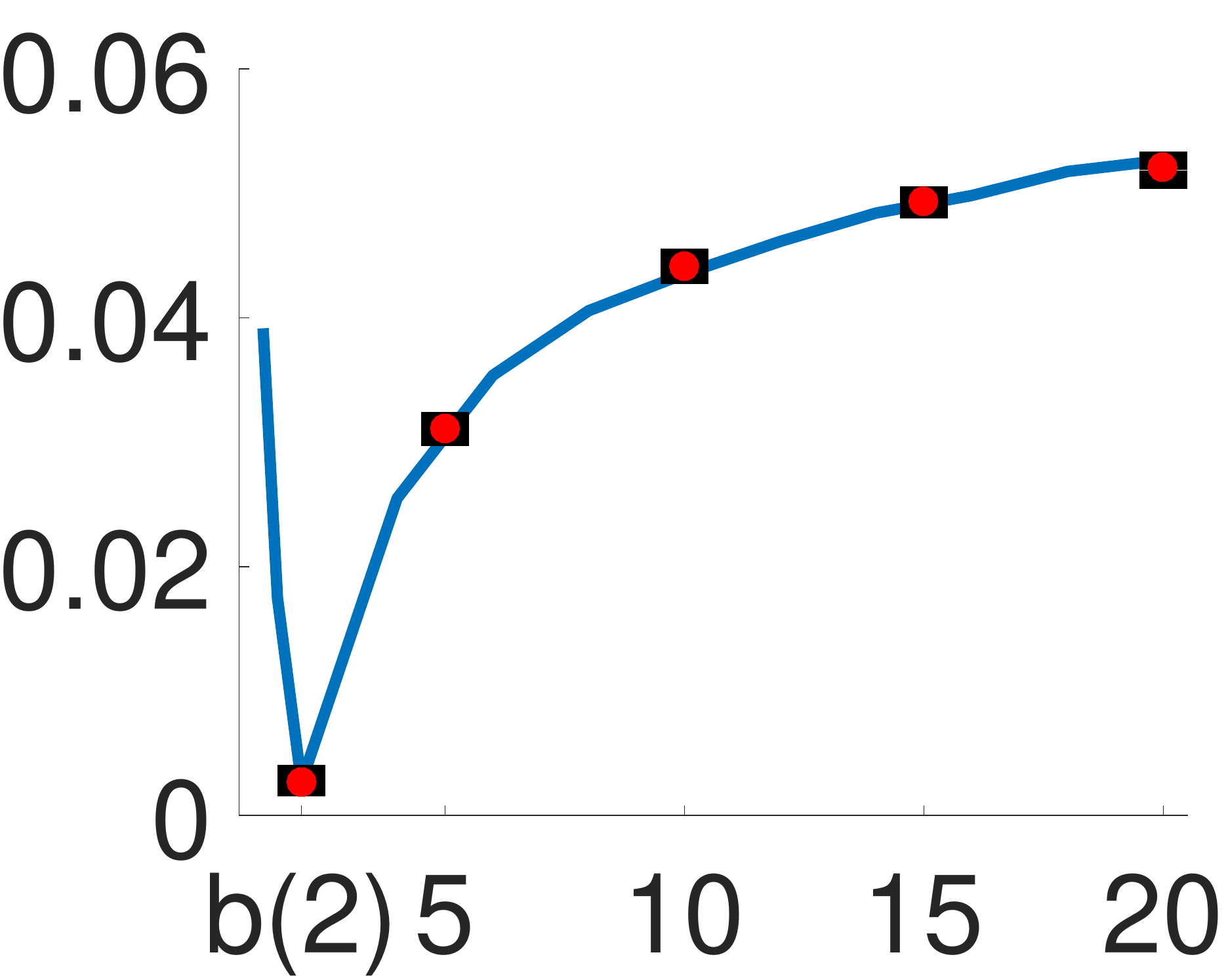}   & \includegraphics[scale = 0.18]{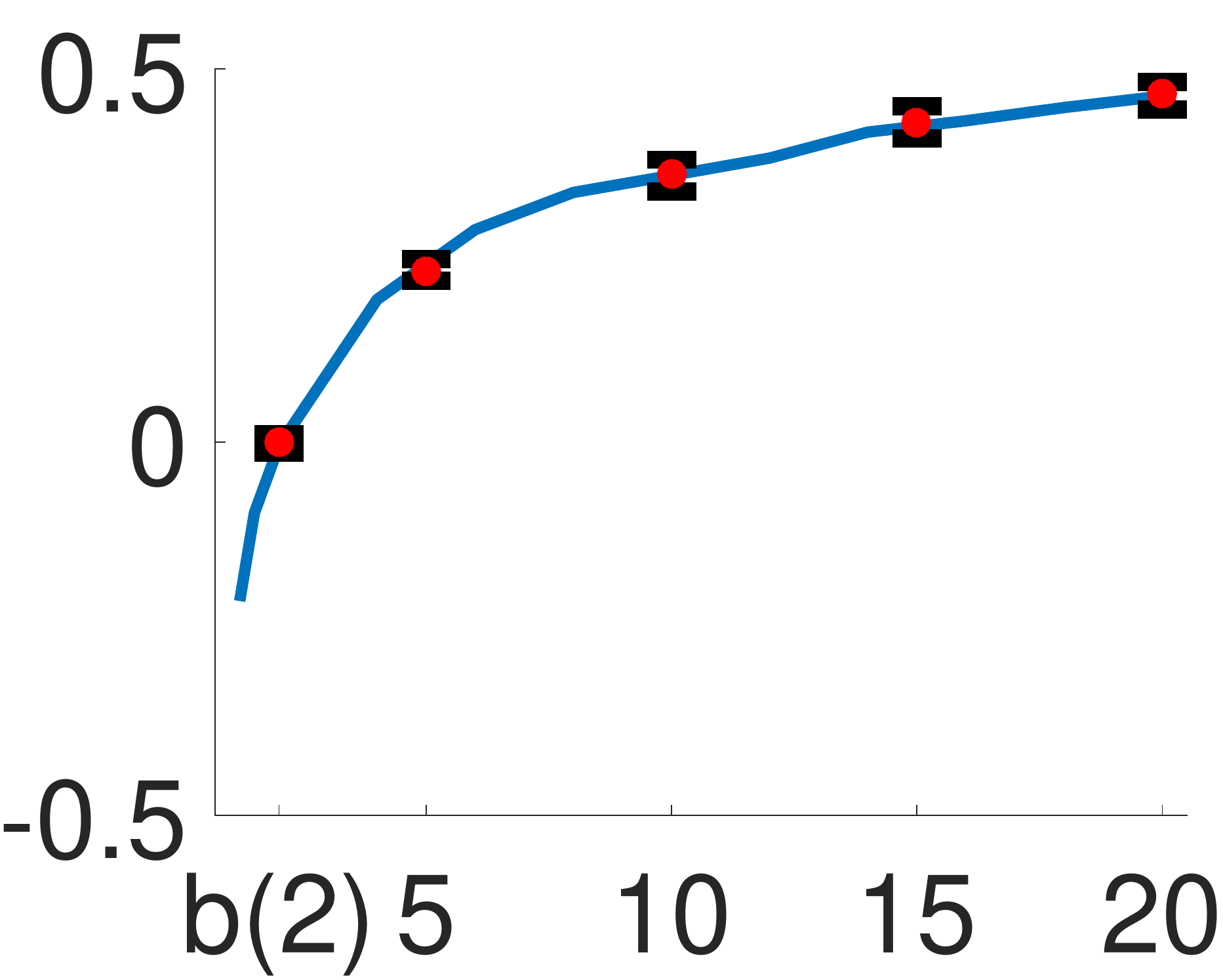}   &	\includegraphics[scale = 0.18]{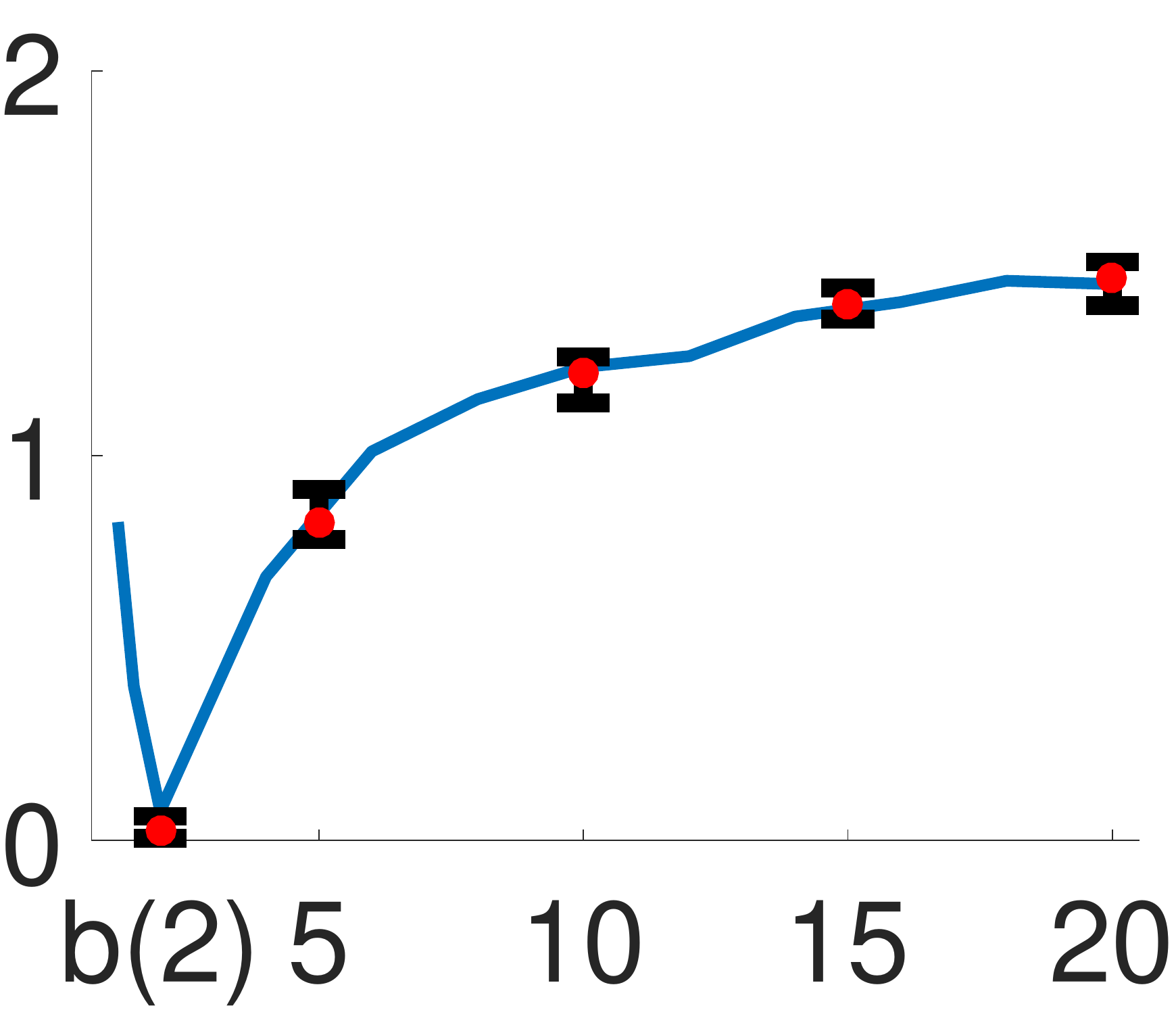}                \\ \hline
				\includegraphics[scale = 0.18]{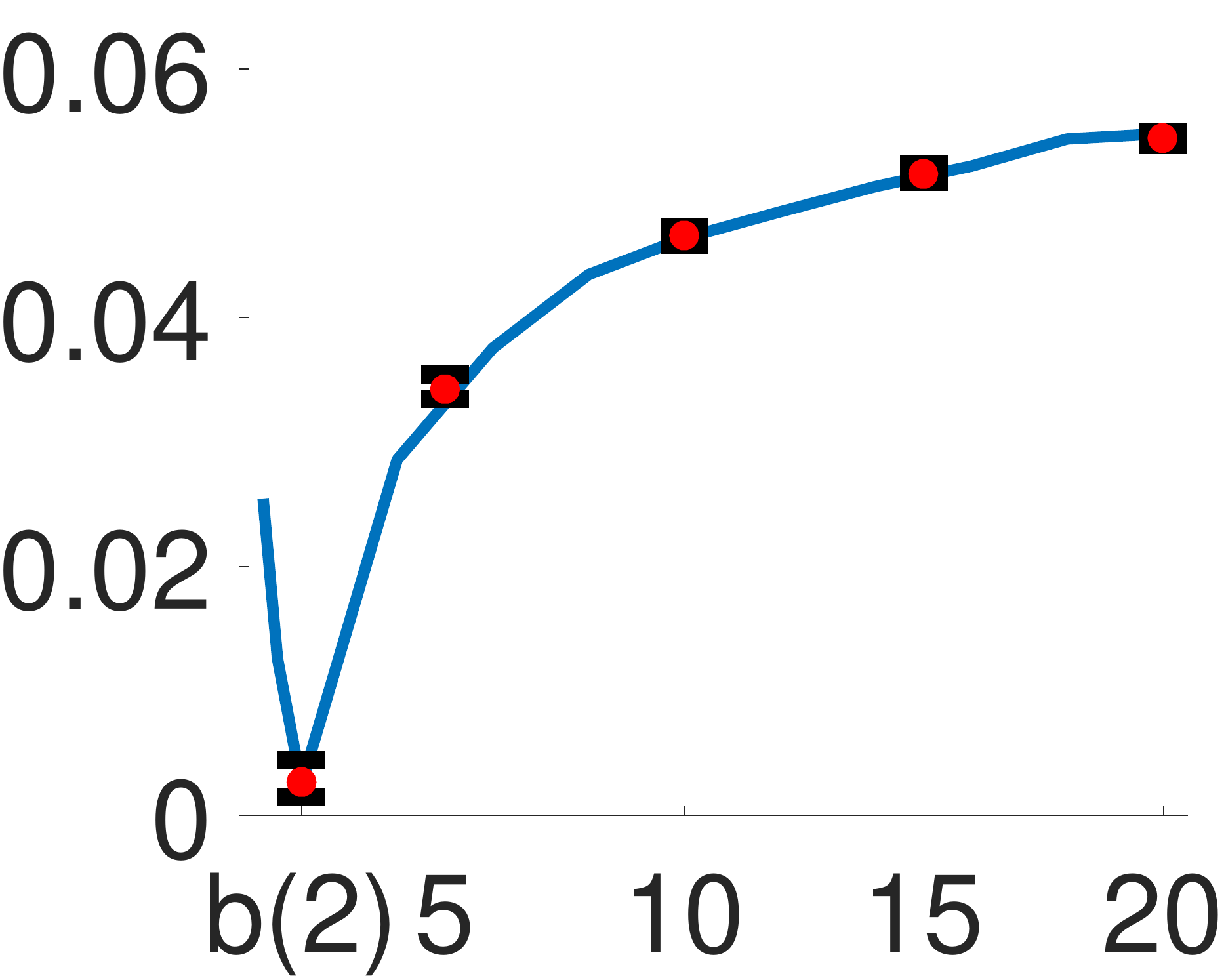}   & \includegraphics[scale = 0.18]{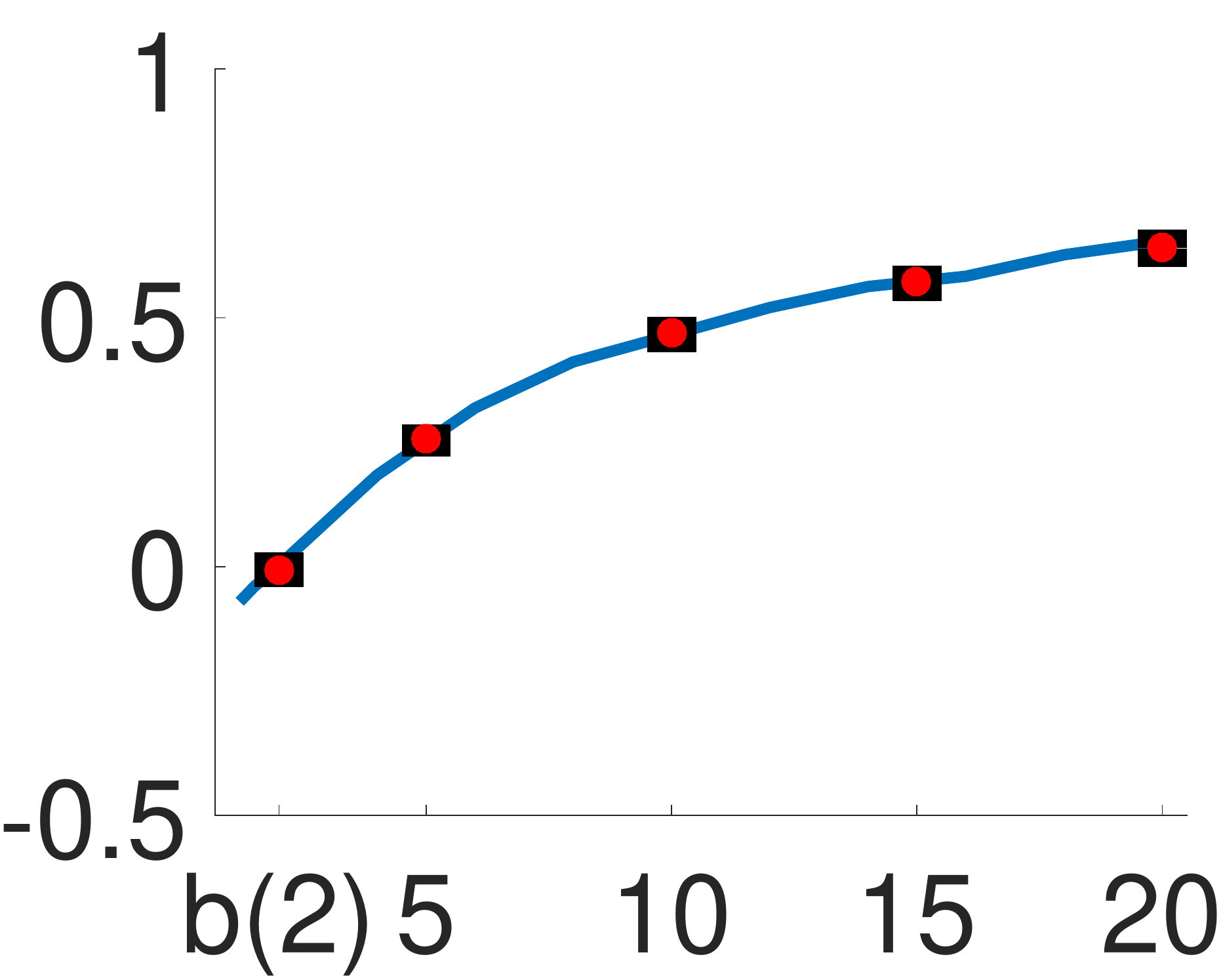}   &	\includegraphics[scale = 0.18]{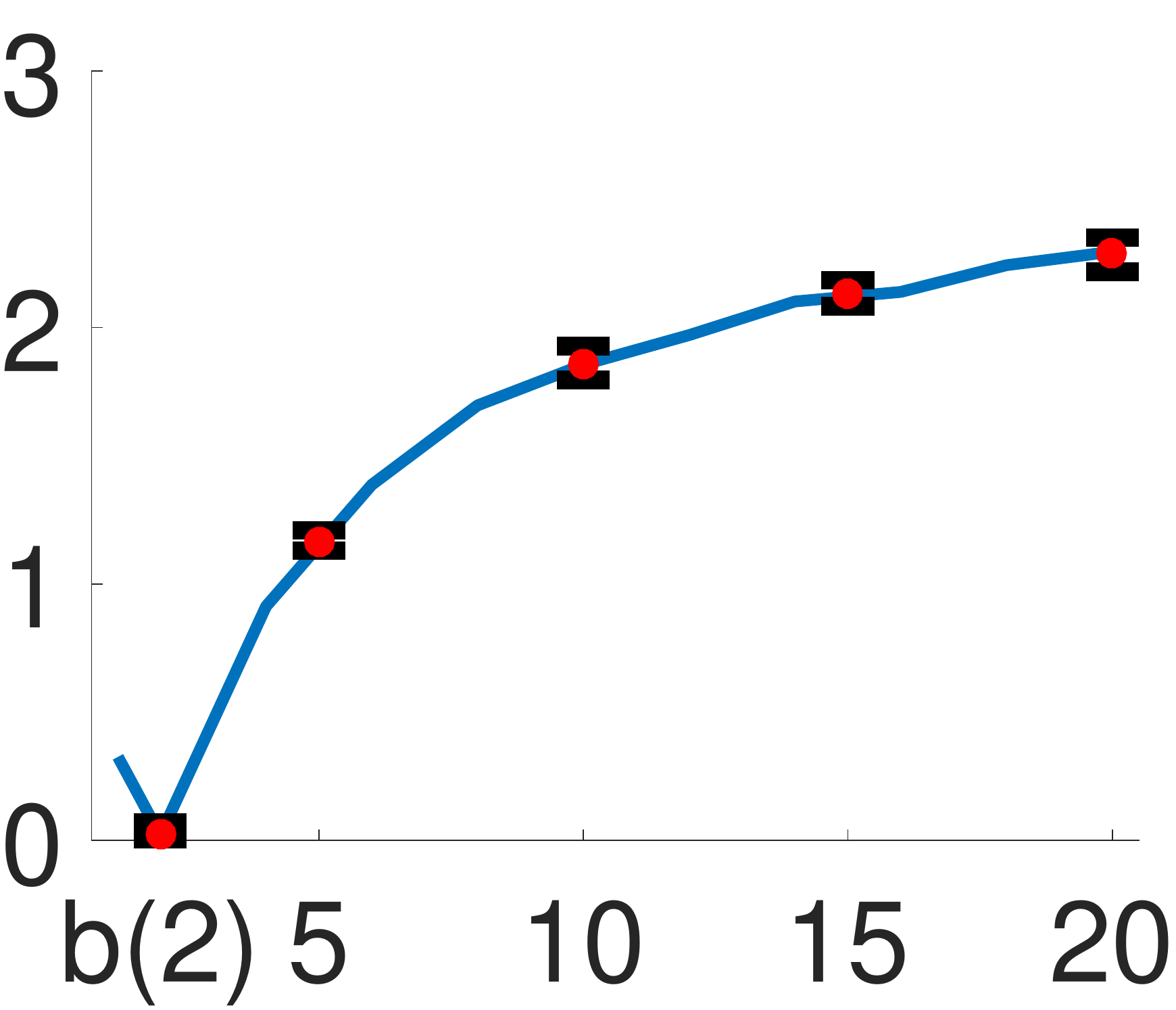}                \\ \hline
			\end{tabular}
		\end{center}
		\caption{Sensitivity to perturbations of $\phi$ in the DCV model for the Acidity (top row), Galaxy (middle row) and Enzyme (bottom row) datasets.}
		\label{fig:DCV_phi}
	\end{figure}

Figures \ref{fig:DCV_a0}--\ref{fig:DCV_phi} display the three geometric sensitivity measures for various perturbations of the parameters of the DCV model for the Acidity, Galaxy and Enzyme datasets. The baseline parameter settings for the Acidity and Galaxy datasets were chosen to be $a_0 = 1$, $a_1 = 10$, $\eta = 3$, $\gamma = 5$ and $\phi = 2$. The baseline settings were the same for the Enzyme dataset except now $a_1 = 1$. Each figure assesses sensitivity to perturbations of one parameter, holding all of the other parameters in the model fixed, and each row in these figures shows the results for the different datasets (top = Acidity, middle = Galaxy, bottom = Enzyme).

We begin with the effects of perturbing the parameter $a_0$, the results of which are displayed in Figure \ref{fig:DCV_a0}. The overall sensitivity trends based on all three measures are very similar for the three datasets. The main result of note here is that the model is much more sensitive for the Galaxy dataset based on all three measures. The results for perturbations of the parameters $a_1$ and $\eta$ are given in Figures \ref{fig:DCV_a1} and \ref{fig:DCV_eta}, respectively; the DCV model is not sensitive to perturbations of these parameters. This is evident based on all three measures. In Figure \ref{fig:DCV_gamma}, we display the effects of perturbing the parameter $\gamma$. The shift sensitivity measure $\mathbb{D}$ indicates that the posterior sample averages do not change very much due to perturbations of this parameter. The magnitude of the overall spread measure $\mathbb{V}$ is also fairly small. On the other hand, perturbations of $\gamma$ seem to affect the covariance structure of the posterior samples. This can be seen through the measure $\mathbb{E}$ in the last column of Figure \ref{fig:DCV_gamma}. Furthermore, the sensitivity appears greater when the DCV model is fitted to the Galaxy dataset. Finally, Figure \ref{fig:DCV_phi}, displays the results when the parameter $\phi$ in the DCV model is perturbed. Again, it appears that the posterior average densities are not very sensitive to such perturbations (see the first column of the figure). On the other hand, it appears that the spread measures $\mathbb{V}$ and $\mathbb{E}$ are affected. In all three cases, the overall variance increases as $\phi$ is increased. Furthermore, it appears that the proportion of cumulative variance from tPCA also changes with $\phi$. We note that the magnitude of sensitivity is now bigger when the DCV model is fitted to the Enzyme data.

	\section{Summary and Future Work}
	\label{sec:Discuss}

In this work, we have defined three geometrically-motivated, easy to compute global sensitivity measures for assessing robustness of Bayesian nonparametric density estimation models. These measures build on the FR Riemannian geometric framework for analyzing PDFs. Using this geometry, we are able to compute various geometric quantities of interest analytically, making the computation of our measures very efficient. Since posterior samples in the types of models we consider are exactly PDFs, we are able to define sensitivity in terms of various differences between samples generated using a baseline nonparametric Bayesian model and various perturbed versions; all of the models we consider use Dirichlet-type priors, and thus, perturbations are applied to either the precision parameter or the base probability measure. In particular, we defined a shift sensitivity measure based on the FR distance between posterior sample average densities. Additionally, we defined two spread-based sensitivity measures that assess differences in overall variance of the samples and covariance structure. We performed simulation studies in the context of four different Bayesian nonparametric density estimation models and showed the effectiveness and intuition behind our measures. Then, we studied sensitivity of two models in the context of three real datasets that have been analyzed previously in the literature.

There exist other dissimilarity measures, e.g., the Kullback-Leibler divergence or Hellinger distance, that could be used to define similar sensitivity measures. However, the KL divergence (and many others) is not a proper metric on the space of PDFs, and thus cannot be used to compute sample statistics, making its use limited in this context. The Hellinger distance is an extrinsic distance on the space of PDFs and can be used to compute the extrinsic sample average and variance, which can then be used to define sensitivity measures similar to $\mathbb{D}$ and $\mathbb{V}$. However, we are additionally interested in computing sensitivity measures based on the structure of the sample covariance matrix of posterior density samples. In this case, the interpretability of the covariance matrix and subsequent PCA is not clear in the context of sensitivity analysis. For example, principal directions of variation traverse positive functions, but ones that don't necessarily integrate to one. Thus, we feel that the proposed approach based on the intrinsic Fisher-Rao Riemannian geometry of the space of PDFs provides a convenient setting for sensitivity analysis. In particular, it allows one to efficiently compute different sample statistics such as the mean and covariance, and to explore variability in posterior samples via tPCA.

The FR geometric framework opens the door for defining other sensitivity measures based on posterior samples. The three measures defined in this paper are effective at capturing specific characteristics of the samples, and may not always capture all of the effects of the model perturbations. To capture all effects, one could try to define a measure based directly on some cumulative distance between the posterior samples without summarizing them via the mean or covariance. Such an approach may however be sensitive to sampling variability. While in this paper we focused on \emph{global sensitivity to prior perturbations}, in the future we will generalize this framework to also consider (1) local sensitivity, and (2) identification of influential observations. Finally, we plan to generalize the proposed approach to other nonparametric Bayesian models, not necessarily ones for PDF estimation.

\noindent\textbf{Acknowledgments:} The authors would like to thank Karthik Bharath for valuable discussions and suggestions. SK was partially supported by NSF DMS 1613054, NSF CCF 1740761 and NIH R01-CA214955.

\singlespacing
\bibliographystyle{elsarticle-num}

\bibliography{References}

\end{document}